%% file: thesis_arxiv.tex
\documentclass[twoside=semi,paper=a4,fontsize=10pt]{scrbook}  

\input{preamble.tex}
\input{add_preamble_networks.tex}

\title{Symmetries and self-similarity of many-body wavefunctions}
\author{\textbf{Piotr Migdał}\\advisor: Maciej Lewenstein\\co-advisor: Javier Rodr\'{\i}guez-Laguna}

\date{submitted: Sept 2014, defended: Dec 2014}

\titlehead{A Thesis submitted for the degree of Doctor of Philosophy}

\publishers{\includegraphics[width=8cm]{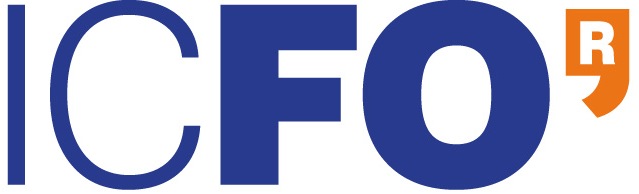}\\
ICFO--Institut de Ci\`{e}ncies Fot\`{o}niques,\\08860
Castelldefels, Spain}

\begin{document} 

    \maketitle

    \frontmatter
    \tableofcontents  

    \include{ackn_abstract}

    \mainmatter

    \include{introduction}

    \include{invariants}

    \include{quantum_sequences}

    \include{networks}

    \include{conclusion}

    \backmatter

\input{thesis.bbl}
\end{document}

%% file: preamble.tex
\usepackage{amsthm}
\usepackage[utf8]{inputenc}
\usepackage[T1]{fontenc}
\usepackage[polish,english]{babel}
\usepackage{indentfirst}
\usepackage{graphicx}
\usepackage{amsmath,bbm}
\usepackage{amsfonts}
\usepackage{url,hyperref}
\hypersetup{breaklinks=true, colorlinks=true, urlcolor=blue}
\usepackage{color}
\usepackage{todonotes}

\usepackage[sort&compress,numbers]{natbib}
\usepackage{doi}

\usepackage{letltxmacro}

\LetLtxMacro{\ORIGselectlanguage}{\selectlanguage}
\makeatletter
\DeclareRobustCommand{\selectlanguage}[1]{%
  \@ifundefined{alias@\string#1}
    {\ORIGselectlanguage{#1}}
    {\begingroup\edef\x{\endgroup
       \noexpand\ORIGselectlanguage{\@nameuse{alias@#1}}}\x}%
}
\newcommand{\definelanguagealias}[2]{%
  \@namedef{alias@#1}{#2}%
}
\makeatother

\definelanguagealias{en}{english}
\definelanguagealias{EN}{english}

\newcommand{\bra}[1]{\langle #1 |}
\newcommand{\ket}[1]{| #1 \rangle}
\newcommand{\braket}[2]{\langle #1 | #2 \rangle}
\newcommand{\Tr}{\hbox{Tr}}

\newcommand{\Renyi}[0]{R\'enyi}

\def\C{{\mathbbm C}}

\def\N{{\mathbbm N}}
\def\P{{\mathbbm P}}
\newcommand{\openone}{{\mathbbm I}} 
\newcommand{\vi}{\vec{n}}

\newtheorem{theorem}{Theorem}
\newtheorem{lemma}{Lemma}
\newtheorem{corollary}{Corollary}

\renewcommand{\(}{\left(}
\renewcommand{\)}{\right)}

%% file: add_preamble_networks.tex
\usepackage{nicefrac}
\usepackage[caption=false]{subfig}

\newcommand{\half}{\mbox{$\textstyle \frac{1}{2}$}}

\newcommand{\brakets}[2]{\langle\, #1\,|\,#2\,\rangle}
\newcommand{\bracket}[3]{\left\langle #1 \left| #2 \right| #3 \right\rangle}
\newcommand{\brackets}[3]{\langle #1 | #2 | #3 \rangle}
\newcommand{\proj}[1]{\ket{#1}\bra{#1}}

\newcommand{\tr}{\textrm{tr}}

\DeclareMathOperator{\Span}{span}
\DeclareMathOperator{\real}{Re}

\newcommand{\ee}{\mathrm{e}}
\newcommand{\ii}{\mathrm{i}}
\newcommand{\dd}{\mathrm{d}}
\newcommand{\identity}{\mathbf{1}}

\renewcommand{\Re}{\mathfrak{Re}}
\renewcommand{\Im}{\mathfrak{Im}}

\newcommand{\LL}{L}

\newcommand{\VV}{\mathcal{V}}

\newcommand{\tave}[1]{\widehat{#1}}

\newcommand{\sym}[1]{\widetilde{#1}}

\newcommand{\E}{E}
\newcommand{\HPr}{\Lambda}

\newcommand{\SPr}{\Pi}

\newcommand{\TM}{R}

\newcommand{\T}{T}
\newcommand{\F}{F}
\renewcommand{\P}{P} 

\newcommand{\eqr}[1]{Eq.~(\ref{#1})}
\newcommand{\fir}[1]{Fig.~\ref{#1}}
\newcommand{\secr}[1]{Sec.~\ref{#1}}

\newcommand{\AAA}{\mathcal{A}}
\newcommand{\BBB}{\mathcal{B}}
\newcommand{\CCC}{\mathcal{C}}
\newcommand{\HHH}{\mathcal{H}}

\newcommand{\NNN}{\mathcal{N}}

\newcommand{\CC}{X}

\usepackage[dua]{acronym}
\newacro{ctqw}[CTQW]{continuous time quantum walk}
\newacro{cm}[CPM]{coherent paths matrix}
\newacro{mm}[MM]{mixing matrix}
\newacro{ba}[BA]{Barab\'asi-Albert}
\newacro{er}[ER]{Erd\H{o}s-R\'{e}nyi}
\newacro{ws}[WS]{Watt-Strogatz}
\newacro{rg}[RG]{random geometric}
\newacro{rr}[RR]{random regular}
\newacro{st}[ST]{star}
\newacro{ca}[CA]{coauthorship}
\newacro{em}[EM]{e-mail}
\newacro{kc}[KC]{karate club}
\newacro{ce}[CE]{\emph{C. elegans} metabolic}
\newacro{nmi}[NMI]{\emph{normalized mutual information}}


%% file: ackn_abstract.tex
\chapter*{Acknowledgments}

Science, even in its naked form, is a collective, social process \cite{Fleck1979}.
For science any theory, observation or discovery is related to people and their relations.
For a scientist it is all above plus much more.

I am indebted to Maciek\footnote{Q: Is it misspelled? A: No. Keys [j] and [k] are neighbors, but this time it is misleading: \emph{Maciek} is a casual form of \emph{Maciej}.} Lewenstein, especially for providing me a lot of freedom for pursuing my diverse scientific and educational interests, and his belief in me.
I am convinced that there is no better gift for independence and creativity that never saying ``no''.

I am grateful to Javi Rodr\'{\i}guez-Laguna for his countless insights into anything, from scientific remarks on current projects (and unrelated ones), through pieces of advice on academic writing and workflow, to comments on education, society and, well, anything.
It was encouraging, inspiring and fruitful.

I would like to thank Jake Biamonte for inviting me to Turin for an intensive and fascinating research collaboration.
Even tough it was a short stay, it was wonderful on so many axes. 

I am happy for hospitality and help of the administration: primarily my home institute, ICFO in Castelldefels,
and also guest institutes, such as ISI Foundation in Turin and IFT in Madrid.
It is thank to your work and attitude that I felt welcomed, and free from paperwork burden.

I am grateful to my family and close friends for the support and encouragement.
Out of many lessons learnt the most important one is that, in a long run, happiness is as important as intellectual prowess.

The list only starts here.
All coauthors, discussion partners, lecturers, fellow PhD students and friends --- thank you!

\newpage

This PhD was supported by Spanish MINCIN/MINECO project TOQATA (FIS2008-00784), EU Integrated Projects AQUTE and SIQS, CHISTERA project DIQUIP, ERC grants QUAGATUA and OSYRIS.

\chapter*{Abstract}



The study of the structure of quantum states can provide insight into the possibilities of quantum mechanics applied to quantum communication, cryptography and computations, as well as the study of condensed matter systems.
For example, it shows the physical restrictions on the ways how a quantum state can be used and allows us to tell which quantum states are equivalent up to local operations.
Therefore, it is crucial for any analysis of the properties and applications of quantum states.

This PhD thesis is dedicated to the study of the interplay between symmetries of quantum states and their self-similar properties.
It consists of three connected threads of research: polynomial invariants for multiphoton states, visualization schemes for quantum many-body systems and a complex networks approach to quantum walks on a graph.

First, we study the problem of which many-photon states are equivalent up to the action of passive linear optics.
We prove that it can be converted into the problem of equivalence of two permutation-symmetric states, not necessarily restricted to the same operation on all parties. 
We show that the problem can be formulated in terms of symmetries of complex polynomials of many variables, and provide two families of invariants, which are straightforward to compute and provide analytical results.
Furthermore, we prove that some highly symmetric states (singlet states implemented with photons) offer two degrees of robustness --- both against collective decoherence and against a photon loss.
Additionally, we provide two proposals for experiments, feasible with an optical setup and current technology: one related to the direct measurement of a family of invariants using photon-counting, and the other concerting the protection of transmitted quantum information employing the symmetries of the state. 

Second, we study a family of recursive visualization schemes for many-particle systems, for which we have coined the name ``qubism''.
While all many-qudit states can be plotted with qubism, it is especially useful for spin chains and one-dimensional translationally invariant states.
This symmetry results in self-similarity of the plot, making it more comprehensible and allowing
to discover certain structures. 
This visualization scheme allows to compare states of different particle numbers (which may be useful in numerical simulations when particle number is an open parameter) and puts emphasis on correlations between neighboring particles.
The visualization scheme can be used to plot probability distribution of sequences, e.g. related to series of nucleotides in RNA and DNA or --- aminoacids in proteins. 
However, unlike classical probabilistic ensembles of sequences, visualizing quantum states offers more --- showing entanglement and allowing to observe quantum phase transitions.

Third, we study quantum walks of a single particle on graphs, which are classical analogues of random walks.
Our focus in on the long-time limit of the probability distribution.
We define ``quantumness'' to be the difference between the probability distributions of the quantum and related classical random walks.
Moreover, we study how (especially in the long-time limit) off-diagonal elements of the density matrix behave.
That is, we measure coherence between different nodes,
and we use them to perform quantum community detection --- splitting of a graph into subgraphs in such a way that the coherence between them is small.
We perform a bottom-up hierarchical aggregation, with a scheme similar to modularity maximization, which is a standard tool for the, so called, community detection for (classical) complex networks. 
However, our method captures properties that classical methods cannot --- the impact of constructive and destructive interference, as well as the dependence of the results on the tunneling phase.

%% file: introduction.tex







\chapter{Introduction}
\label{ch:introduction}

\section{Background}

\begin{figure}[!ht]
	\centering
	\includegraphics[width=\textwidth]{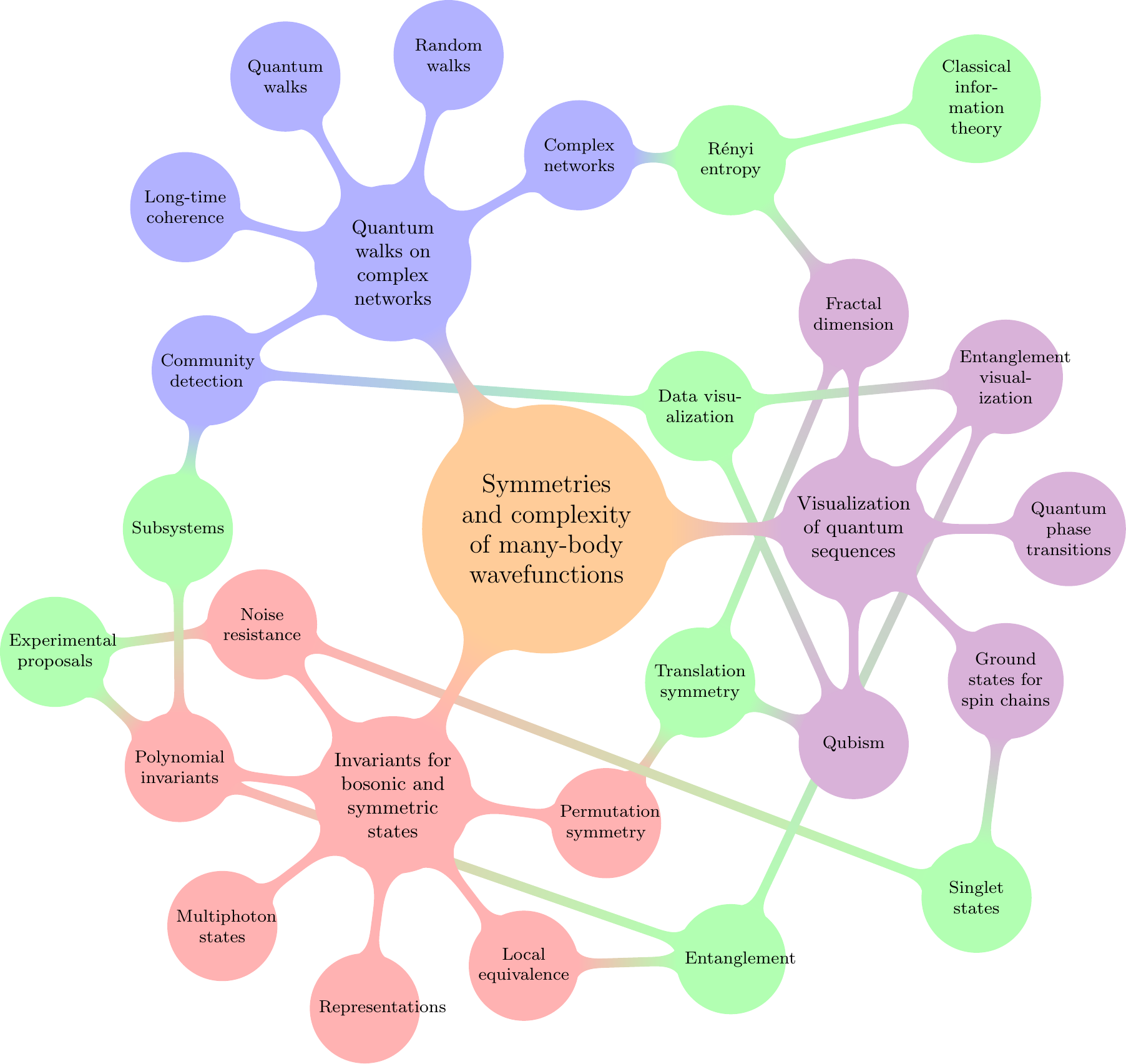}

	\caption{A graph of the main concepts of this thesis and their relations.
	\label{fig:thesis-diagram}
	}
\end{figure}

This PhD thesis is divided into three chapters, each one describing a distinct thread of research:
\begin{itemize}
\item \textbf{\nameref{ch:invariants}},
\item \textbf{\nameref{ch:qubsim}},
\item \textbf{\nameref{ch:networks}}.
\end{itemize}
Yet, these threads are connected through common concepts and methods related to study of entanglement, symmetry with respect to interchange of particles and self-similarity of quantum systems.
An illustrative graph of these concepts and their relations is depicted in
Fig.~\ref{fig:thesis-diagram}.

\subsection{Historical background}

While quantum mechanics dates back to the beginning of the 20th century \cite{DiracBook},
only in the last decades it was started being considered as a tool for information processing, that is, communication, cryptography and computation \cite{NielsenChuang2000}.
One of the first research in that line was a study of the information capacity for transmission of information with quantum \cite{Holevo1973} rather than classical states \cite{Shannon1948original}.
The result, now known as Holevo's bound, is that regardless if we are operating with classical or quantum $d$-level states, we can transmit up to $\log_2(d)$ bits of information per state.
That is, for this particular task there is no advantage of using quantum states over classical states.
However, many other works show significant differences between classical and quantum states.
Perhaps the most striking is Bell's theorem \cite{Bell1964}, putting bounds on certain correlations, which cannot be broken by any probability distribution stemming from classical mechanics.
Much to the surprise of the author, it turned out that some quantum states break the bound.
Thus, in particular, quantum mechanics cannot be thought as a classical theory with yet unknown parameters, settling a dispute, which looked as purely philosophical \cite{Einstein1935,WheelerZurek1983book}.
Stronger-than-classical correlations inspired the FLASH paper \cite{Herbert1982}, a protocol for faster-than-light communication.
While, as expected, this result could not hold, the flaw in it was so subtle, that it inspired further progress \cite{Peres2003,Kaiser2011book}, starting from the no-cloning theorem \cite{Wootters1982,Dieks1982}, a proof that there can be no machine making copies of an arbitrary quantum state.

Impossibility to clone an unknown state offers certain advantages --- we may use quantum states to transmit information that we do not want to be copied.
If one party (let us call her Alice) sends a quantum message to a friend (let us call him Bob), then any trial of our enemy
to make a copy of the message will disturb the message Bob receives.
This property is utilized in protocols for generating shared secret keys, allowing for perfectly secure cryptography \cite{Bennett1984,Ekert1991}.

Another application of quantum information is quantum computation.
Feynman is attributed with the first idea to use the principles of quantum mechanics for computation \cite{Feynman1982}.
Moreover, his intuition that quantum systems can simulate other quantum system turned out to be correct \cite{Lloyd1996}.
The first quantum algorithms offered significant speedup for
testing whether a function is constant \cite{Deutsch1992},
the computation of the discrete logarithm \cite{Shor1994},
and database access \cite{Grover1996}.
Unfortunately, the power of quantum information and computation comes at the price of another intrinsic quantum mechanic problem: its extreme sensitivity to noise.
As the number of parameters grows exponentially with the number of particles, even a small noise, attenuation or uncertainty of the setting may result in drastic changes of subtle parameters of the state. 
The typical classical approach to overcome this problem is to amplify the signal, so that it becomes much stronger than noise and presents redundancy against losses.
However, in quantum information this strategy is disallowed, because of the no-cloning property, which allows secure communication.
Moreover, on the one hand the pervasiveness of interaction between particles makes creation of entangled states feasible, but on the other hand it makes it easy to have an uncontrollable evolution and to entangle our system with the environment.
These uncontrolled interactions are operationally the same as the loss of quantum properties in the form of decoherence \cite{Zurek2003}.


\subsection{Properties of quantum mechanics}


Mathematically, quantum states are described as vectors in a complex Hilbert space.
Pure quantum states of $d$-levels, or \emph{qudits}, are represented by vectors of $d$ dimensions and complex entries,
\begin{equation}
	\ket{\psi} =
	\begin{bmatrix}
		\psi_1\\
		\psi_2\\
		\vdots\\
		\psi_d
	\end{bmatrix}.
\end{equation}
When we perform a measurement in the computational basis, the probability of obtaining a given outcome is the absolute value squared of the respective vector entry
\begin{equation}
	P(i) = \psi_i^* \psi_i = |\psi_i|^2,
\end{equation}
also know as the Born rule.
The only linear operations that preserve probability are unitary operations. 
Consequently, when describing a purely quantum evolution, we restrict ourselves to using only these operations.

When considering a composite system of many distinguishable subsystems, the recipe to construct its wavefunction is given by the tensor product of wavefunctions representing each subsystem
\begin{equation}
	\ket{\psi} = \ket{\phi_1} \otimes \ket{\phi_2} \otimes \cdots \otimes \ket{\phi_N}.
	\label{eq:product-states}
\end{equation}
Thus, the probability of getting a particular outcome is independent from other measurements
\begin{equation}
	P(i_1, i_2, \ldots, i_N) = P_1(i_1) P_2(i_2) \cdots P_N(i_N).
\end{equation}
Tensor product acts as a Cartesian product on the Hilbert space basis.
So for subsystems of dimensions $d_1$, $d_2$, $\ldots$ and $d_N$ the dimension of the global Hilbert space is $d_1 d_2 \cdots d_N$.
Not all quantum states of many particles can be written as a \emph{product state} \eqref{eq:product-states}.
These states are called \emph{entangled states} \cite{Horodecki2009}.
If each subsystem has dimension $d$, then product states constitute a manifold of $N (d-1)$ complex dimensions.
However, if we consider all possible states, we get a manifold of $d^N - 1$ complex parameters.
Consequently, from a measure-theoretic perspective, almost all pure states are entangled.

In many scenarios we need to deal with statistical mixtures of pure states.
This can be done by using density matrices.
For a pure state $\ket{\psi}$ it is defined as 
\begin{equation}
	\rho_{\ket{\psi}} = \ket{\psi}\bra{\psi}.
\end{equation}
That is, its entries are $\rho_{ij}=\psi_i \psi_j^*$.
The diagonal of density matrix consists of the probabilities of the different outcomes, given the measurement is performed in the computational basis.
The statistical mixture of two states, with probabilities $\mu$ and $(1-\mu)$ can be written as
\begin{equation}
	\rho = \mu \rho_1 + (1-\mu) \rho_2.
\end{equation}
This description encapsulates the fact that different mixtures of pure states can yield the same quantum correlations.
Moreover, it allows straightforward calculation of expectations values of operators $A$, that is
\begin{equation}
	\langle A \rangle = \Tr[ A \rho ].
\end{equation}

For mixed states the notion of entanglement is more complicated.
The most standard approach is to define separable states as states being in the convex hull of $\rho_{\ket{\psi}}$, where $\ket{\psi}$ are product states, i.e.
\begin{equation}
	\rho = \sum_i p(i) \rho_{\ket{\psi_i}}.
\end{equation}
Yet, unlike for pure states, the problem to tell whether a given state is separable or not is NP-hard \cite{Gurvits2003}.

While mixed states are harder to analyze than pure states, they are essential to study quantum mechanics itself.
That is, even if we study a pure state of two particles, its subsystems are generally in a mixed state.
When we study any state described by $\rho$, the state of its subsystem $A$ after ignoring subsystem $B$ reads
\begin{equation}
\rho_A = \Tr_A\left[ \rho \right],
\end{equation}
where $\Tr_A$ is the partial trace.
This operation traces out everything but the system $A$ (in this case, it traces out the subsystem $B$), that is
\begin{equation}
[\rho_A]_{i_A; j_A} = \sum_{i_B, j_B} [\rho]_{i_A, i_B; j_A, j_B}.
\end{equation}
Use of a mixed state to analyze a subsystem is not only done because of our ignorance, i.e. lack of knowledge, of subsystem $B$.
The other party can be light years away, and whatever we do cannot be affected by operations performed on the remote subsystem.
Or even, the other party may have crossed the event horizon of a black hole, so even in principle its information may not be accessible to us.

The notion of entanglement does depend on the choice of subsystems with respect to which we want to assess entanglement.
Consider a single photon that passes through a $50\%:50\%$ beam splitter. 
If we choose to represent our quantum states with particles, the state is given by
\begin{equation}
	\frac{\ket{A} + \ket{B}}{\sqrt{2}},
\end{equation}
with the following meaning: a photon is in a superposition of mode $A$ and mode $B$, with equal amplitudes.
As any state of a single particle it is always in the form a product state \eqref{eq:product-states}, thus is not entangled.
However, if we move to the second quantization picture, describe our state with the occupation of modes, $A$ and $B$, then the state is
\begin{equation}
	\frac{\ket{0,1} + \ket{1,0}}{\sqrt{2}},
\end{equation}
that is, a superposition of
\begin{itemize}
\item having no photons in mode $A$ and a photon in mode $B$,
\item having a photon in mode $A$ and no photons in mode $B$,
\end{itemize}
which is entangled.
Depending on the problem we study, we may want to use one representation or the other.

\subsection{Entanglement}

The non-local character of quantum effects provides a motivation for defining entanglement as quantum correlations that cannot be generated by local operations, even if assisted by classical communication.

As a simple example, a product state $\ket{00}$ can be converted by local operations to $\ket{10}$, another product state.
However, there are no local operations that would allow transforming $\ket{00}$ into an entangled state $(\ket{00}+\ket{11})/\sqrt{2}$.

Since product states can be simulated by classical devices, practically all intrinsically quantum protocols need to rely on entanglement.
However, even arbitrary small entanglement is sufficient for universal quantum computation \cite{Nest2012}.
In quantum information, we are typically interested in properties up to the choice of local basis.
Consequently, typical entanglement measures are defined up to local unitary operations\footnote{
A generic quantum information scientist will not tell a difference between $(\ket{01}-\ket{10})/\sqrt{2}$ and $(\ket{00}+\ket{11})/\sqrt{2}$.}.
This notion is formalized by \emph{entanglement monotones} \cite{Vidal2000} --- non-increasing quantities under local operations.

The easiest case to study entanglement is a bipartite system in a pure state.
Let us call the parts $A$ and $B$, each of size $N$. 
In this case all entanglement properties can be studied by choosing a convenient pair of local bases.
This procedure, called the \emph{Schmidt decomposition}, reads
\begin{equation}
	\ket{\psi} = \sum_k \lambda_k \ket{\phi_k} \otimes \ket{\varphi_k},
\end{equation}
where $\lambda_k$ are non-negative real numbers and $\{\ket{\phi_k}\}_k$ is a set of orthonormal vectors for subsystem $A$ (and analogously for $\{\ket{\varphi_k}\}_k$ and $B$).
Technically, the Schmidt decomposition is the \emph{singular value decomposition} of matrix $\ket{\psi}_{i_A, i_B}$, that is
\begin{gather}
	\begin{bmatrix}
		\psi_{11}  & \hdots & \psi_{1 N} \\
		\vdots     & \ddots & \vdots     \\
		\psi_{N 1} & \hdots & \psi_{N N}
	\end{bmatrix}
	=\\
	\begin{bmatrix}
		&  &  \\
		\ket{\phi_1} & \cdots \vphantom{\ddots} & \ket{\phi_N}\\
		&  & 
	\end{bmatrix}
	\begin{bmatrix}
		\lambda_1 &        & 0  \\
		          & \ddots &    \\
		0         &        & \lambda_N
	\end{bmatrix}
	\begin{bmatrix}
		&  &  \\
		\ket{\varphi_1} & \cdots \vphantom{\ddots} & \ket{\varphi_N}\\
		&  & 
	\end{bmatrix}^T.
\end{gather}
Since it is equivalent to changing local bases, the only quantities related to entanglement is the set of \emph{Schmidt values}, i.e. $\{\lambda_k\}_k$.
For a product state there is only one non-zero Schmidt value.
Two states present the same entanglement if and only if they set of Schmidt values is the same.
The Schmidt values contain the same information as
\begin{equation}
	\Tr[\rho_A^q] = \Tr[\rho_B^q] = \sum_k \lambda_k^q.  
\end{equation}
Consequently, the study of reduced density matrices is related to invariants for a quantum state.
That is, the study of a subsystem is an important tool for studying properties of the global system.

Multipartite entanglement is significantly more difficult to analyze.
In similarity with the bipartite case, one approach is to consider polynomials in the wavefunction coordinates such that they are invariant with respect to local unitary operations.
While this method is general, there is no easy procedure to find a complete set of independent invariants.
Mathematically, these invariants can be expressed as expectation values of many copies of the initial state \cite{Grassl1998}.
Thus, also taking a supersystem plays a role in the study of the properties of quantum states.


In this thesis we restrict to the study of entanglement for pure states.
While a lot of research in quantum information is focused on qubits, we work mainly on systems of finite, but arbitrary, dimension. 
In a number of cases the qudit case is significantly harder than the qubit case and either requires qualitatively different proofs to show the same properties or have properties that cannot be reduced to the qubit case.

\subsection{Quantum information with photons}

One practical and promising tool for quantum information are excitations of electromagnetic field, that is, photons.
They are massless particles, with spin $1$ and Bose statistics.
Their main advantage is the ease of creation, transmission and measurement.
Photons travel in transparent media without interaction among themselves or entangling to the environment. Consequently, a quantum state created with photons in one place can be processed and measured in another place.
In fact, many hallmark properties of quantum information were first demonstrated using photons, for example BB84 protocol for cryptography \cite{Bennett1992}, quantum teleportation \cite{Boschi1998} and Bell test \cite{Aspect1982}.

Photon pure states, as any bosons, can be described as a polynomial of creation operators acting on the electromagnetic vacuum, for example
\begin{equation}
\left(\tfrac{1}{3\sqrt{2}}a_1^{\dagger 3} + \tfrac{1}{\sqrt{3}} a_1^\dagger a_2^\dagger + \tfrac{1}{\sqrt{3}} a_3^\dagger \right)\ket{\Omega},
\end{equation}
where by $\ket{\Omega}$ we denote the vacuum state, and the state being described reads is superposition of
\begin{itemize}
\item three photons in mode $1$,
\item one photon in mode $1$ and one photon in mode $2$,
\item one photon in mode $3$. 
\end{itemize}
Since creation operators commute, the permutation symmetry of bosonic states is ingrained in the polynomial representation.

Not every multiphoton state can be easily created.
The easiest ones are coherent states, which are naturally created by lasers, and squeezed states --- states of light which can be created by the propagation of a strong light beam through a nonlinear crystal.
More difficult methods of state creation is via cavity quantum electrodynamics \cite{Brattke2001,Haroche2013}.
Further processing can be done using linear optics, i.e. beam splitters and phase retarders.
Other operations are significantly harder to perform, e.g. quantum non-demolition measurements \cite{Braginsky1980,Brune1992}, or give only probabilistic results, e.g. conditional measurement \cite{Lvovsky2001}.

\subsection{Entanglement invariants for symmetric states}

Passive linear optics can be understood as the set of operations on multi-photon states restricted to many-particle interference \cite{Tichy2013}, but without interaction among the particles.
Even this small subset of all conceivable operations is useful for quantum communication and
cannot be efficiently simulated by classical computers \cite{Aaronson2010}.
Along with conditional measurement, linear optics is as powerful as a universal quantum computer \cite{Knill2001}.

Since it is easy to apply linear optics operations in a laboratory, the difficulty of quantum state generation, processing and measurement is related to transformations that cannot be performed in this framework.
In this thesis, we have undertaken the task of finding out which photon states can be reached from a given state, using only linear optics. 

Our key contribution is the introduction of two families of invariants \cite{Migdal2014ffdag}, which are straightforward to compute and provide both numerical and analytical insight into the geometry of permutation-symmetric states \cite{BengtssonZyczkowski2006book}.
They are expressed as the expectation values of polynomials in annihilation and creation operators, and are related to particular symmetries of the state.
Moreover, we show an experimental scheme, using an optical setting, to directly measure the values for one of the families of invariants.
We show that our problem is equivalent to the problem of assessing the equivalence between permutation-symmetric states of distinguishable particles \cite{Migdal2013aaa}.
This in particular builds a bridge between invariants for linear operations acting on bosonic states and entanglement properties of distinguishable states.

Even if photons are relatively uncoupled from transparent media, at distances suitable for practical applications particle loss and interaction with the environment is inevitable. 
We show a way to overcome these problems by transmitting quantum information encoded in singlet states built with photons \cite{Migdal2011dfs}.
These states are invariant under collective decoherence.
At the same time, the quantum information they carry is immune against all one-particle losses.
We propose an experimental protocol as a proof-of-principle demonstration of these properties.

\subsection{Quantum sequences and qubism}

\begin{figure}[!ht]
	\centering
	\begin{tabular}{cc}
		\includegraphics[width=0.4\textwidth]{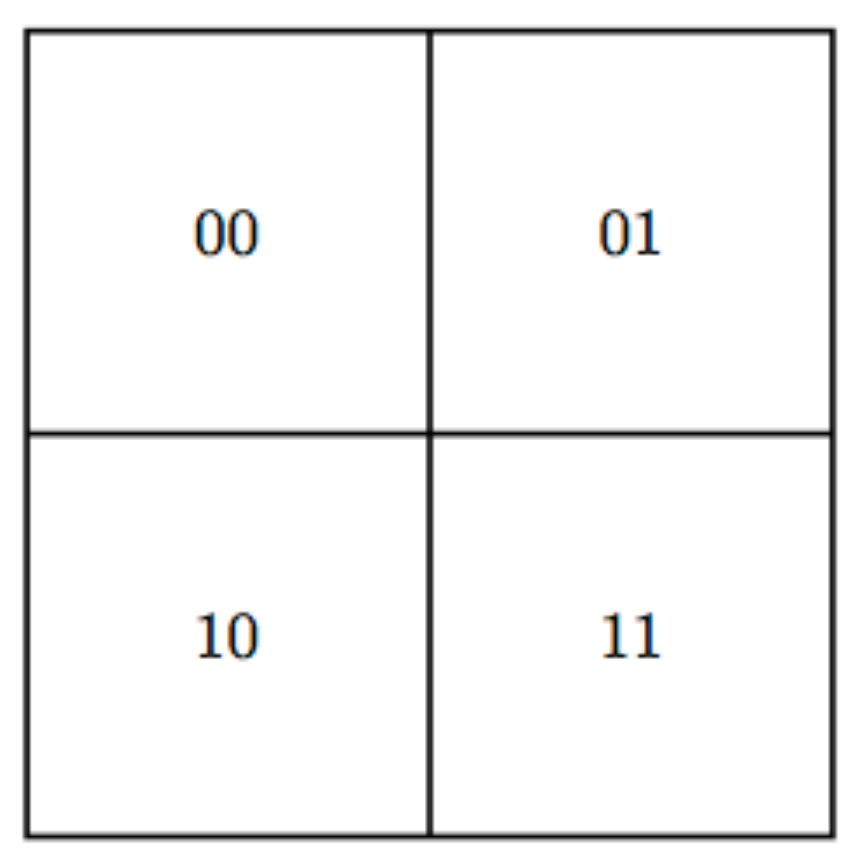} &
		\includegraphics[width=0.4\textwidth]{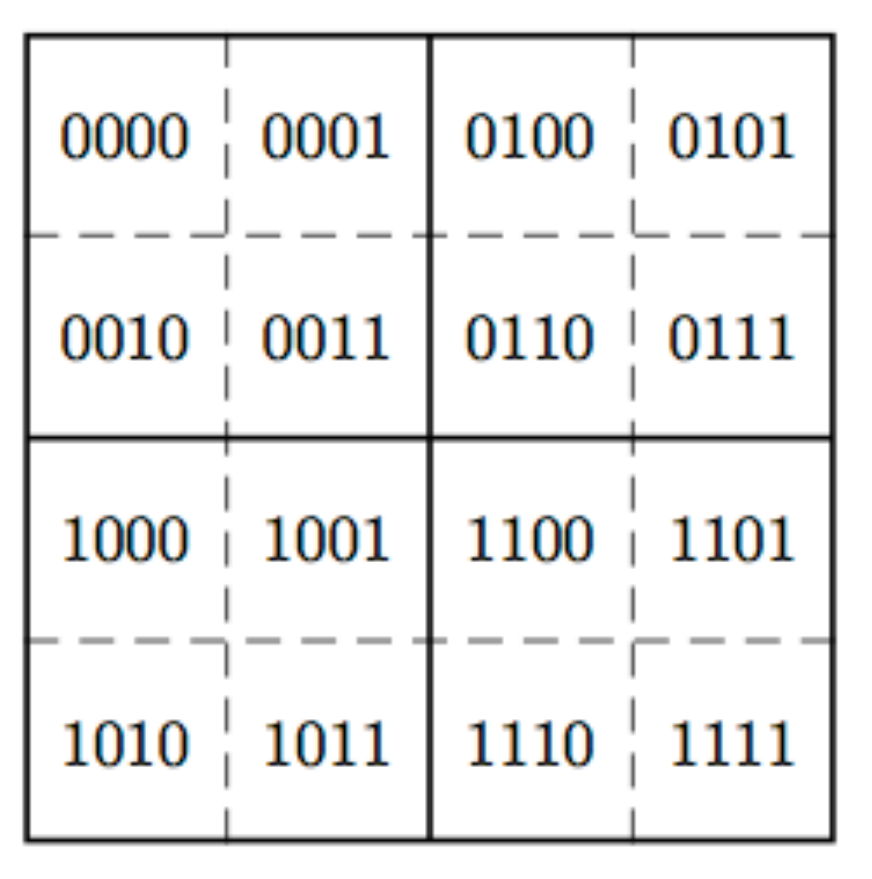} \\
		\includegraphics[width=0.4\textwidth]{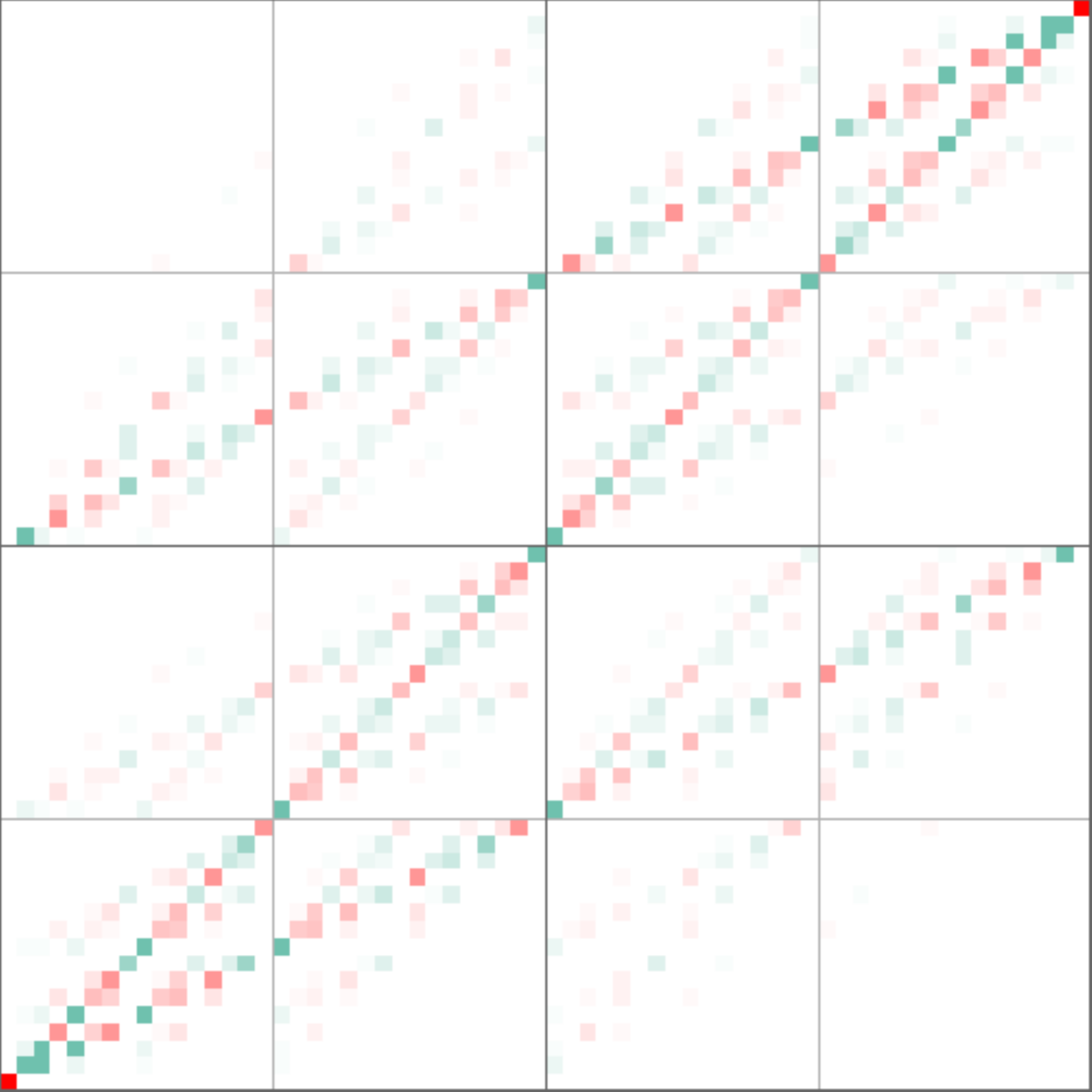} &
		\includegraphics[width=0.4\textwidth]{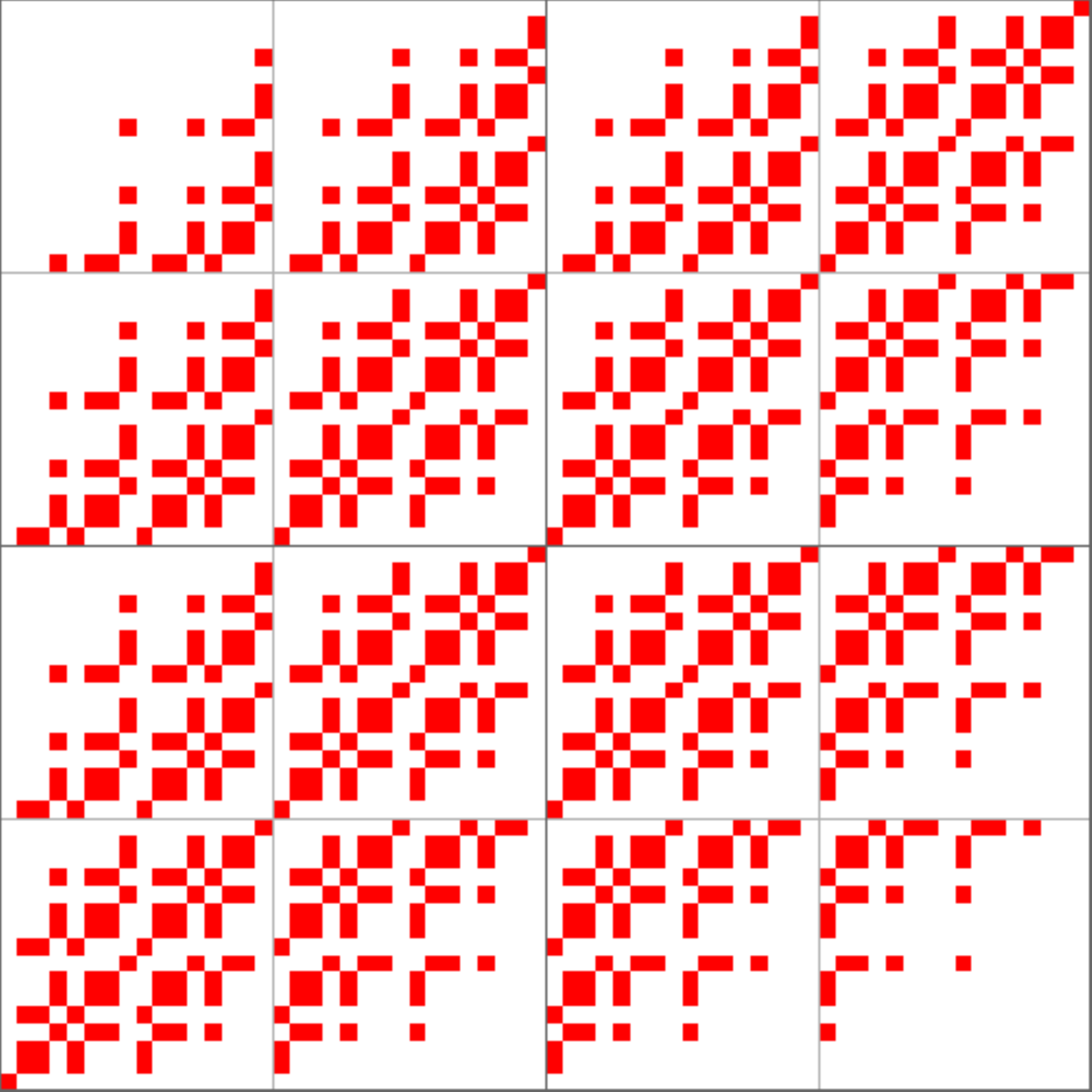}
	\end{tabular}

	\caption{Qubism, a 2D plotting scheme of many-body wavefunctions.
	Top:
	A recipe for the recursive visualization scheme.
	For the first iteration level the basis for the first two particles, 
	$|00\rangle$, $|01\rangle$, $|10\rangle$ and $|11\rangle$,
	is mapped into one of the four smaller squares (left).
	Next, we  repeat the procedure for the next two particles, taking the smaller squares as the starting point (right).
	Bottom: Examples of qubistic plots for qubit states of $N=12$ particles. 
	We plot the ground state for the Heisenberg Hamiltonian with the periodic boundary conditions (left) and the half-filled Dicke state (right).
	Saturation represents the absolute value of the amplitude and color represent the sign.
	\label{fig:intro-qubism}
	}
\end{figure}

Quantum entanglement of many particles is difficult to describe and quantify even for pure states.
If we want to get insight into the structure of a given state, one approach is to calculate its various entanglement measures.
In order to analyze the full state, we introduce a visualization scheme, called \emph{qubism}, for pure states of many $d$-level particles, which generates two-dimensional images \cite{Rodriguez-Laguna2011}.
We plot all amplitudes of a given state in the computational basis, and arrange them in a specific way, which makes certain quantum properties visible, see Fig.~\ref{fig:intro-qubism}.
In particular, due to the recursive nature of the plot, translational symmetry shows up as self-similarity of the plot, while entanglement shows up as a type of this self-similarity. 

One-dimensional spin chains constitute an interesting class of Hamiltonians, which play a role as toy models and have proved to be a fruitful ground for developing techniques for analyzing many-body states. 
Properties such as ferromagnetism, block entanglement, transport properties \cite{Lewenstein2012book} and correlation length were studied in such systems, especially in the context of quantum phase transitions.
One of the most relevant task of quantum many-body physics is to investigate how does the ground state change with the parameters in the Hamiltonian \cite{Sachdev1999book}.
%
We employ qubism to exhibit particular properties of spin chains and show how the plot can be used to make conjectures about the structure of the state.
We show that phase transitions are usually apparent, and can be seen without previous knowledge of the order parameter.
For numerous physical systems the ground state is a singlet state \cite{Auerbach1994}, that is, belongs to subspace of zero total angular momentum --- also this property is visible in the plot.

\subsection{Complexity and the R\'enyi entropy}

Self-similarity of qubistic plots can be not only seen, but also quantified, using their fractal dimensions.
As we deal with a probability distribution derived from quantum mechanics, rather than a set,
its fractal dimension can be characterized as a function, rather than a single number. 
We use the R\'enyi entropy \cite{Renyi1961} of the order $q$, which is defined as:
\begin{equation}
	H_q = \frac{1}{1-q} \log_2 \left( \sum_{i=1}^N p_i^q \right),
	\label{eq:renyi}
\end{equation}
where $(p_1,\ldots,p_N)$ is a probability distribution.
Its scaling properties with coarse-graining of the plot can be quantified as the multifractal dimensions \cite{Halsey1986,Theiler1990}.
That is, for fractals the entropy \eqref{eq:renyi} is expected to grow linearly as we are doubling the resolution.
The fractal dimension is defined as the linear coefficient.
The parameter $q$ is related to the sensitivity to low and high probability densities.
In particular, for $q \to 0$ we obtain the fractal dimension of the support, i.e. all non-zero probabilities, whereas for $q \to \infty$ we obtain the fractal dimension of the set of the most probable outcomes.

The R\'enyi entropy has applications to other quantum problems, in particular for the entropic uncertainty principle \cite{Bialynicki-Birula1975,Bialynicki-Birula2006}, a generalization of Heisenberg's uncertainly principle.
Moreover, a quantum variant of the R\'enyi entropy, where instead of summing probabilities we perform $\Tr[\rho^q]$, has applications in assessing the purity of a mixed state.
Consequently, when applied to reduced density matrices, these entropies are entanglement invariants.

Additionally, the R\'enyi entropy has applications in the study of other complex systems, e.g. probability distribution and degree distribution on complex networks.
Its low value implies high heterogeneity of a network.

\subsection{Quantum complex networks}

Many complex systems can be represented in an abstract way as a graph, that is, a set of nodes connected by edges.
When we analyze real systems these graphs are called complex networks \cite{Albert2002,Newman03thestructure}.
The structure of a complex networks can be characterized with a number of parameters.
The simplest one beyond the node and edge count is the degree distribution, that is, the distribution of nodes with respect to the number of outgoing edges.
This parameter allows us to tell how homogeneous or heterogeneous are nodes with respect to their connection to other nodes.
Sometimes the degree distribution function can be identified as a power law.

In quantum mechanics, a Hamiltonian can be viewed as a graph, with edges between nodes being weighted by the respective transition amplitudes.
The unitary evolution of a quantum state can be interpreted as a quantum walk, in which the walker tunnels to its neighboring nodes.
Unlike for a classical random walk, in which the probability distribution converges to a steady state, in a quantum walk the long-time behavior does depend on the initial state and oscillates rather than converges to a steady state. 
Nonetheless, a natural question would be to compare classical and quantum behavior.
We have found that after averaging out oscillations, the probability distributions are close to each other, and their difference depends on the degree distribution of the network \cite{Faccin2013}.

\begin{figure}
	\centering
	\includegraphics[width=\textwidth]{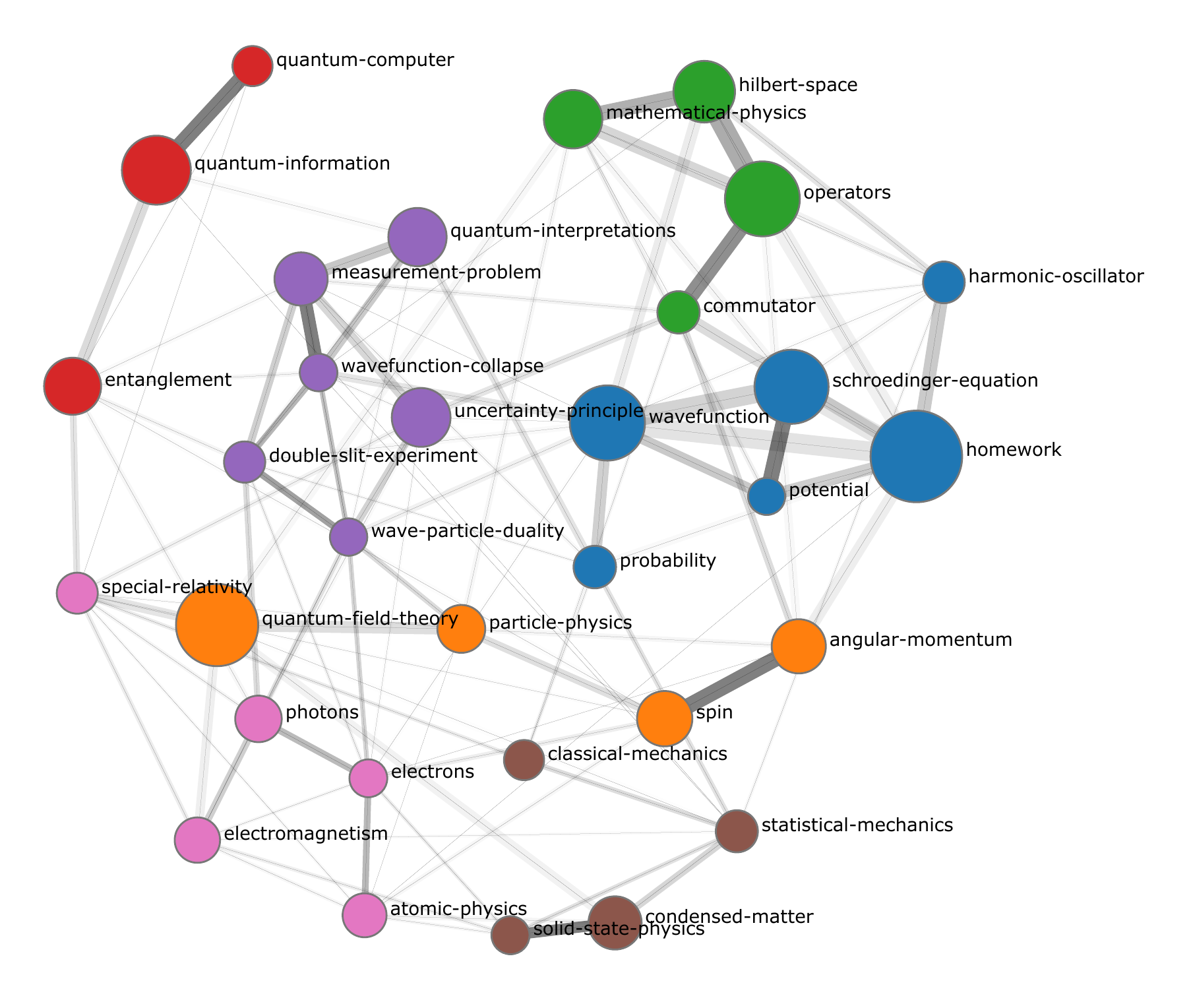}
	\caption{\label{fig:tagoverflow}
	An example of community detection applied for graph visualization: network of tags associated with \texttt{quantum-mechanics} on Physics - Stack Exchange, a questions and answers site.
	The area of each node is proportional to the number of questions labeled by a given tag.
	Edge widths are related to the number of questions with both tags.
	Edge shades are related to correlation between two tags: $P(\text{tag1} \cap \text{tag2})/[P(\text{tag1})P(\text{tag2})]$.
	Each color represents a distinct community, based on a greedy modularity maximization.
	This open-source project by the author is accessible via the following link: \url{http://stared.github.io/tagoverflow/}.}
\end{figure}

In order to get insight into the structure of a complex network, we can split a graph into communities \cite{Fortunato2010}, that is subgraphs, each of them with nodes more densely connected inside that with the rest of the graph.
It allows both to study in depth each subgraph and to analyze a coarse-grained graph, with communities being the new nodes.
An archetypal community would be a clique --- a subgraph with all nodes connected within itself and with no outgoing edges. 
However, in real-world data communities are typically less pronounced, see Fig.~\ref{fig:tagoverflow}.

There is no universal recipe for \emph{community detection}, i.e. splitting a graph into communities.
A typical approach is to define some target function, which is maximal for the best splitting.
Perhaps the most common one is modularity \cite{Newman2004,Clauset2004} --- a quantity measuring how much more nodes are connected inside a community that with the outside.
But even for modularity, an exact maximization is an NP-complete problem \cite{Brandes2008}, so heuristic methods are important, for example greedy or hierarchical models, inspired by renormalization \cite{Blondel2008}. 

Another important question is related to the strength of quantum effects.
At small scales quantum effect are apparent --- energies of molecules are directly based on quantum mechanics.
However, biological systems do not seem to exhibit long-range quantum effects.
So, even at the level of a single excitation, how can we assess a typical scale for quantum effects?
We focus on a physical system, \emph{light harvesting complex II}, utilized by plants to absorb light.

To investigate regions with strong quantum effects, we introduce a community detection method suited for quantum walks \cite{Faccin2013community}. 
Our approach is to look at the behavior of a single excitation, performing a quantum walk from a given site. 
We distinguish between nodes for which interference effects are relevant and those for which they are not. 
The distinction is established via the density matrix.
We define the splitting of a graph into communities as removing all coherence between communities. Our target functions, instead of the modularity, are typical quantum informational measures of the purity of a state or the fidelity of a quantum channel.

\section{Structure}

This PhD thesis is divided into three chapters, each one describing a distinct thread of research,
connected through common concepts and methods related to study of entanglement, symmetry and self-similarity of quantum systems.
All chapters are aimed to be self-contained, and they can be read in any order.

In Chapter~\ref{ch:invariants}:~\textbf{\nameref{ch:invariants}}
we study the problem of which many-photon states are equivalent up to the action of passive linear optics.
We prove that it can be converted into the problem of equivalence of two permutation-symmetric states, not necessarily restricted to the same operation on all parties. 
We show that the problem can be formulated in terms of symmetries of complex polynomials of many variables, and provide two families of invariants, which are straightforward to compute and provide analytical results.
Furthermore, we prove that some highly symmetric states (singlet states implemented with photons) offer two degrees of robustness --- both against collective decoherence and a photon loss.
Additionally, we provide two proposals for experiments, feasible with an optical setup and current technology: one related to the direct measurement of a family of invariants using photon-counting, and the other on protecting transmitted quantum information employing the symmetries of the state. 

In Chapter~\ref{ch:qubsim}:~\textbf{\nameref{ch:qubsim}}
we study a family of recursive visualization schemes for many-particle systems, for which we have coined the name \emph{qubism}.
While all many-qudit states can be plotted with qubism, it is especially useful for spin chains and one-dimensional translationally invariant states.
This symmetry results in self-similarity of the plot, making it more comprehensible and allowing us to discover certain structures from it. 
This visualization scheme allows to compare states of different particle numbers (which may be useful in numerical simulations when particle number is an open parameter) and puts emphasis on correlations between neighboring particles.
The visualization scheme can be used to plot probability distribution of sequences, e.g. related to series of nucleotides in RNA and DNA or --- amino acids in proteins. 
However, unlike classical probabilistic ensembles of sequences, visualizing quantum states offers more --- showing entanglement and allowing us to observe quantum phase transitions.

In Chapter~\ref{ch:networks}:~\textbf{\nameref{ch:networks}}
we study quantum walks of a single particle on graphs, which are quantum analogues of classical random walks. 
Our focus in on the long-time limit of the probability distribution.
We define ``quantumness'' to be the difference between the probability distributions of the quantum and related random walks.
Moreover, we study how (especially in the long-time limit) off-diagonal elements of the density matrix behave.
That is, we measure coherence between different nodes,
and we use this coherence to perform quantum community detection --- splitting of a graph into subgraphs in such a way that the coherence between them is small.
We perform a bottom-up hierarchical aggregation, with a scheme similar to modularity maximization, which is a standard tool for the, so called, community detection for (classical) complex networks. 
However, our method captures properties that classical methods cannot --- the impact of constructive and destructive interference, as well as the dependence of the results on the tunneling phase.

\section{Contribution}

This PhD thesis is based on the following peer-reviewed papers and preprints, in the chronological order:
\begin{itemize}
\item \cite{Migdal2011dfs},
	Chapter~\ref{ch:invariants}, especially Sec.~\ref{s:singlet-space}:\\
	P. Migdał, K. Banaszek,\\
	Immunity of information encoded in decoherence-free subspaces to particle loss,\\
	Phys. Rev. A 84, 052318 (2011), arXiv:1107.3786,
\item \cite{Rodriguez-Laguna2011},
	Chapter~\ref{ch:qubsim}:\\
	J. Rodriguez-Laguna, P. Migdał, M. Ibanez Berganza, M. Lewenstein, G. Sierra,\\
	Qubism: self-similar visualization of many-body wavefunctions,\\
	New J. Phys. 14 053028 (2012), arXiv:1112.3560,
	appeared in the New Journal of Physics Highlights of 2012,
\item \cite{Migdal2013aaa},
    Chapter~\ref{ch:invariants}, especially Sec.~\ref{s:slocc-equivalence}:\\
	P. Migdał, J. Rodriguez-Laguna, M. Lewenstein,\\
	Entanglement classes of permutation-symmetric qudit states: symmetric operations suffice,\\
	Phys. Rev. A 88, 012335 (2013), arXiv:1305.1506,
\item \cite{Migdal2014ffdag},
    Chapter~\ref{ch:invariants}, especially Sec.~\ref{s:polynomial-invariants}:\\
	P. Migdał, J. Rodríguez-Laguna, M. Oszmaniec, M. Lewenstein,\\
	Multiphoton states related via linear optics,\\
	Phys. Rev. A 89, 062329, arXiv:1403.3069,
	selected by PRA as Editors' Suggestion,
\item \cite{Faccin2013},
	Chapter~\ref{ch:networks}, especially Sec.~\ref{sec:walks}:\\
	M. Faccin, T. Johnson, J. Biamonte, S. Kais, P. Migdał,\\
	Degree Distribution in Quantum Walks on Complex Networks,\\
	Phys. Rev. X 3, 041007 (2013), arXiv:1305.6078,
\item \cite{Faccin2013community},
	Chapter~\ref{ch:networks}, especially Sec.~\ref{sec:community}:\\
	M. Faccin, P. Migdał, T. Johnson, J. Biamonte, V. Bergholm,\\
	Community Detection in Quantum Complex Networks,\\
	Phys. Rev. X 4, 041012 (2014), arXiv:1310.6638.
\end{itemize}
Moreover, the author wrote a paper in mathematical psychology \cite{Migdal2011twochoice}.
Additionally, the author contributed to open source projects related to the thesis, in particular:
\begin{itemize}
\item QuTiP (\href{http://qutip.org/}{\texttt{qutip.org}}, a Python package for quantum physics): implemented qubism and other visualizations of many body quantum states, improved the Bloch sphere visualization \cite{Johansson2012a},
\item Wikipedia article on matrix product states \cite{wiki:matrix_product_state}.
\end{itemize}



%% file: invariants.tex







\chapter{Invariants for bosonic and symmetric states}
\label{ch:invariants}

\section{Introduction}

In quantum physics, one of the fundamental symmetries is the symmetry with respect to exchange of particles.
At least in three dimensional space, particles need either to be symmetric with respect to interchange of particles (bosons) or antisymmetric (fermions).
Even for distinguishable particles (i.e. fermions and bosons, but when each particle occupy exclusive sets of modes, e.g. different spatial positions) symmetry with respect to exchange of particles still plays a role.
A permutation-symmetric state is a state of the maximal total spin \cite{Dicke1954}; 
on the opposite end there is singlet subspace (i.e. the subspace with total spin zero), having some antisymmetric properties.

In quantum information, we are almost always interested in properties up to the choice of local basis.
Consequently, typical entanglement measures are defined up to local unitary operations.
In particular, \emph{entanglement monotones} \cite{Vidal2000} --- defined as quantities non-increasing by local operations.

In this chapter we study the relation between symmetries of the state (especially: permutation symmetry and singlet state) and its capabilities to be used in information theory.
We attempt to answer the question which pairs of states can be transformed into each other within a fixed set of operations.

The content of this chapter is the following.
In Sec.~\ref{sec:symmetry-overview} we present an overview and background knowledge related to entanglement and its relation to symmetries, in particular --- local unitary equivalence.
Sec.~\ref{sec:preliminaries} introduces some notation and basic mathematical facts that we use through this chapter. In particular we introduce basics of the geometry of Hilbert space, equivalences with respect to local operations and provide simple examples.
In Sec.~\ref{s:slocc-equivalence} we show rigorously the relation between geometry of bosonic states subjected to linear operations and the local equivalence of permutation-symmetric states.
Moreover, as a byproduct of the methods we apply, we introduce a discrete family of states, which contains W and GHZ states as special cases.
In Sec.~\ref{s:polynomial-invariants} we introduce two families of polynomial invariants to test whether two many-photon states can be related via linear optics. 
We show their relation to the geometry of bosonic states and propose an experimental setup for their direct measurement.
In Sec.~\ref{s:singlet-space} it is shown that creating singlet states from bosons does not only protect the information against collective decoherence, but also makes it immune to one-particle loss.
Unless explicitly stated, we work on pure states with fixed number of particles $n$, each with the same number of levels $d$.

\section{Overview}
\label{sec:symmetry-overview}

Entanglement is perhaps the  most important resource for quantum information (for a review see \cite{Horodecki2009}), and its characterization is one of the key tasks of quantum theory.
Particularly difficult is the problem of characterizing entangled mixed states (for a recent review of various necessary criteria see \cite{Guehne2009}).
The problem for pure states is much simpler.
But even in this case, only a few settings are completely understood --- in particular, bipartite entanglement, where the Schmidt decomposition provides a method of classification of pure entangled states of two parties \cite{Horodecki2009}.
In a multipartite scenario very little is known about the different classes of entanglement. Typical questions that one would like to answer concern entanglement classes of pure states which are invariant with respect to local operations, typically assumed to constitute a group (unitary, general linear, etc.).
The corresponding classes of states are called then LU-, SLOCC-equivalent, etc., where LU denotes local unitary, and SLOCC --- stochastic local operation and classical communications. Only a few rigorous results are known concerning these questions, which we list below 
\begin{itemize}
\item For three qubits a generalization of the Schmidt decomposition has been formulated (see \cite{acin2000generalized} and references therein) --- this result provides a classification of invariant states with respect to local unitaries. There is a considerable amount of work regarding this and the related problem of geometrical invariants by the Sudbery group \cite{Carteret2000, Williamson2011}. 
\item Classification of entanglement of three qubit states according to LU and SLOCC has been presented in Ref. \cite{Sudbery2000,DeVicente2012} and \cite{Dur2000}, respectively. 
\item Classification of entanglement of 4 qubits according to SLOCC has been presented in Ref. \cite{Verstraete2002} (see also a series of papers by Miyake \cite{Miyake2002, Miyake2003}).
\item For many qudits a multiparticle generalization of the Schmidt decomposition \cite{Carteret2000, Verstraete2003} provides a general way to answer whether two states are LU-equivalent.
\end{itemize}

There is also a considerable amount of work on many-qubit states cf. \cite{Miyake2004, Miyake2004a,Kraus2010a, Kraus2010b}, but very little is known about general many qudit states.
The difficulty of classifying entanglement for multipartite pure states is one of the motivations for considering restricted families of states.
Such restrictions are typically introduced by considering symmetries \cite{Sawicki2012}, which might be physically motivated. In this spirit many authors considered totally permutation-symmetric pure states of $n$ qubits (cf. \cite{Aulbach2010,Devi2010,Mathonet2010,Aulbach2011,Cenci2010a, Ganczarek2011}), since such states naturally describe systems of many bosons, and appear frequently in the context of quantum optics. Similarly, quantum correlations in totally antisymmetric states (as representative states of fermions) have been intensively investigated (for a review see \cite{Eckert2002} and references therein). In the next introductory subsection we focus on symmetric 
states and their particular role in physical applications.  

A many-qudit wavefunction can be permutation-symmetric for two reasons. One is when it describes a system of  bosons, so
that the particles are indistinguishable on a fundamental level. Second is when the particles are distinguishable but, because of a particular setting (e.g. a Hamiltonian for which the particles form an eigenstate), they happen to be in permutation-symmetric state. The latter situation occurs for instance for the Lipkin-Meshkov-Glick model \cite{Lipkin1965} of nuclear shell structure, and related models of quantum chaos \cite{Gnutzmann2000}.
It is worth stressing that the two situations are not the same.
In the later case we are able to manipulate each particle separately in a different way, while in the first we are restricted to operations modifying each boson in the same way. The question is whether those two settings give rise to same entanglement classes, i.e. if for symmetric states the classification can be reduced to studying operations that act in the same way on all particles.
Moreover, the entanglement geometry of permutation-symmetric states is interesting and relevant, e.g. for quantum computation using linear optics \cite{Aaronson2010}. 
As mentioned above, this question was widely studied in the qubit case \cite{Aulbach2010a, Devi2010, Ganczarek2011}, but most of the obtained results are not-applicable for qudit systems of dimension $d>2$, the general case which we are going to address.

In the field of {\em quantum optics}, this question can be recast in this way: can a given multi-photon state, $\ket{\psi_1}$, be transformed into another one, $\ket{\psi_2}$, using only {\em linear optics}? By (passive) linear optics we mean the use of beam-splitters and wave plates, which is known to be equivalent to the action of arbitrary unitary operations on each mode \cite{Lee2002}.
This question bears special relevance both in theory and practical applications. On the theoretical side, linear optics with postselection has been proved to be able to efficiently realize a universal quantum computer \cite{Knill2001, Kok2007}. But even without postselection, linear optics transformations of multi-photon states constitute an intermediate stage between classical and full-fledged quantum computation \cite{Aaronson2010}. On the practical side, our ability to generate decoherence-free states \cite{Bourennane2004, Wu2011} relies on our ability to transform multi-photon states.
Operation by linear optics can be viewed as multi-particle interference, as opposed to multi-particle interaction. But beyond a generic interference phenomenon, it bears effects which are specific to bosons \cite{Tichy2012,Tichy2013}.

In this chapter we consider equivalence under linear optics transformations of pure states of $n$ photons in $d$ modes, disregarding the possibility of postselection. We show that the problem is equivalent to the LU-equivalence of bosonic states.

As an illustration, let us consider a state of two photons occupying two different modes or channels, $\ket{\psi_1}=\ket{1,1}$ -- i.e. one photon in each mode.
It is possible to transform this state into $\ket{\psi_2}=(\ket{2,0}-\ket{0,2})/\sqrt{2}$ using Hong-Ou-Mandel interference \cite{Hong1987} (i.e.: two-photon interference in a $50\%:50\%$ beam splitter), but it is not possible to place both photons in the same channel with $100\%$ efficiency, i.e.: $\ket{\psi_3}=\ket{2,0}$ is not achievable.
Of course, it is always possible to perform postselection, measuring the number of photons in the second channel and retaining only the states that contain none, but the efficiency will drop to $50\%$.
Translating the problem to LU-equivalence, we can say that a state $(\ket{01}_P + \ket{10}_P)/\sqrt{2}$ (one particle in one mode and one in the other, symmetrized) is not equivalent to state $\ket{00}_P$ (both particles in the same state).

Moreover, quantum systems are powerful yet fragile carriers of information. The ability to create and manipulate superposition states offers verifiably secure cryptography \cite{Gisin2002,Scarani2009}, reduces the complexity of certain computational problems \cite{Ekert1996}, and enables novel communication protocols \cite{Buhrman2010}. However, in practical settings one needs to protect the quantum states carrying information against decoherence, i.e. uncontrolled interactions with the environment. This is accomplished by building redundancy into the physical implementation.
Compared to the classical case, this task is much more challenging \cite{Gottesman2010} due to limitations in handling quantum information, boldly exemplified by the no-cloning theorem \cite{Wootters1982}.
When an ensemble of elementary quantum systems decoheres through symmetric coupling with the environment, one can identify collective states that remain invariant in the course of evolution. These states span a so-called \emph{decoherence-free subspace} (DFS) that is effectively decoupled from the interaction with the environment  \cite{Duan1997,Zanardi1997,Lidar1998}. More generally, it is possible to identify subspaces that can be formally decomposed into a tensor product of two subsystems, one of which absorbs decoherence, while the second one, named  a \emph{noiseless subsystem} or \emph{a decoherence-free subsystem}, remains intact \cite{Kempe2001}.

\section{Preliminaries}
\label{sec:preliminaries}

\subsection{Representations and notation for symmetric states}

Let us consider the system of $n$-photons in $d$ modes. There are, at least, two possible descriptions of the Hilbert space ${\cal S}_n^d$ describing the system. In a {\em mode description}, i.e.: the second quantization picture (see for example \cite{FetterBook1971}), ${\cal S}_n^d$ is treated as a subspace of the full Fock space $\mathrm{Fock}\left(\C ^d \right)$. Let $\vec n\equiv(n_1,\cdots,n_d)$ be a multi-index denoting the photon count for each mode and let $|\vec n| = \sum_{k=1}^d n_k$. The basis states spanning ${\cal S}_n^d$ are specified by the photon count on each mode,
\begin{equation}
\ket{\vec n} =
\frac{(a^{n_1}_1)^\dagger\cdots(a^{n_d}_d)^\dagger}{\sqrt{(n_1!)\cdots(n_d!)}}
\ket{\Omega} 
\equiv 
{\tilde a}^\dagger_{\vec n} \ket{\Omega},\ |\vec n| =n \ .
\label{eq:multia}
\end{equation}
In the above expression $\ket{\Omega}$ is the Fock vacuum, $a_1,\ldots,a_d$ are annihilation operators, and ${\tilde a}^\dagger_{\vec n}$ is a normalized monomial defined as above, creating $\ket{\vec n}$ from vacuum.

In a {\em particle description}, Hilbert space ${\cal S}_n^d$ is treated as the permutation-symmetric subspace of $(\C^d)^{\otimes n}$,  $\mathrm{Sym}^n \left(\C ^d\right)$. Let us fix the basis vectors of $\C ^d$: $\ket{1},\ket{2},\ldots,\ket{d}$. Basis states of $\left(\C ^d\right)^{\otimes n}$ with a simple tensorial form,
\begin{equation}
\ket{\phi}=\ket{i_1}\otimes \ket{i_2}\otimes \cdots \otimes \ket{i_n} \ ,\ i_k \in \left\{1,\ldots,d\right\}\ ,
\end{equation}
are not {\em permutation symmetric}. A basis for $\mathrm{Sym}^n\left(\C ^d\right)$ is obtained from product vectors in $(\C^d)^{\otimes n}$ by symmetrization over all factors in the tensor product. Let us define an asymmetric state from given mode counts $\vi=\{n_1,\cdots,n_d\}$:
\begin{equation}
\ket{\vi}_A \equiv \ket{1}_P^{\otimes n_1} \otimes \ket{2}_P^{\otimes n_2}
\otimes \cdots \otimes \ket{d}_P^{\otimes n_d}\ .
\label{eq:ordered-state}
\end{equation}
In the above expression we explicitly put the subscript $P$ to emphasize that we deal with tensor product of states in particle representation.
The state $\ket{\vi}_A$ can be thought of as a naive state in particle representation with the corresponding photon counts for each mode.
The corresponding normalized symmetric state is given by:
\begin{equation}
\ket{\vi} = 
\frac{\sqrt{n_1! \ldots n_d!}}{\sqrt{n!}}
\sum_{\mathrm{perm}} \ket{1}_P^{\otimes n_1} \otimes \ket{2}_P^{\otimes n_2}
\otimes \cdots \otimes \ket{d}_P^{\otimes n_d}\ ,
\label{eq:fock_in_particle_representation}
\end{equation}
where the sum is over the different permutations of the factors appearing in the tensor product. Notice the required normalization factor. There exists another way of expressing the state $\ket{\vi}$ in particle basis
\begin{equation}
\ket{\vi} = N(\vi)\; \P^{(n)}_{sym} \ket{\vi}_A \ ,
\end{equation}
where $\P^{(n)}_{sym}$ is the projector onto the completely symmetric subspace of $(\C^d)^{\otimes n}$ and the normalization factor $N(n)$ is given by
\begin{equation}
N(n) = \sqrt{\frac{n!}{n_1!\cdots n_d!}}\ .
\end{equation}

During most of this chapter, we will work within the mode description, as it is more natural for dealing with boson states. However, in some parts of this paper we will use also the particle representation and we will proceed between them both, when it is convenient.
States written in the particle representation will have a subscript $P$.
States written in mode representation will have commas between modes.

An arbitrary pure state of the system can be written as: 
\begin{equation}
\ket{\psi} = \sum_{|\vec n|=n} \alpha_{\vec n}\; \ket{\vec n} \ ,
\label{def.f}
\end{equation}
where $\alpha_{\vec n}$ are complex amplitudes and $\ket{\vec n}$ are normalized states with fixed number of photons in each mode. To each state $\ket{\psi}$ we associate a unique homogeneous polynomial in the creation operators according to the recipe:
\begin{equation}
\ket{\psi}=\sum_{|\vec n|=n} \alpha_{\vec n}\; \ket{\vec n}= f^\dagger\ket{\Omega}  
\quad\to\quad
f^\dagger \equiv 
\sum_{|\vec n|=n} \alpha_{\vec n} \; {\tilde a}^\dagger_{\vec n} \ .
\label{eq:state-polynomial}
\end{equation}

\subsection{Local unitary equivalence}

In quantum information processing one practical questions is which states can be achieved from state $\ket{\psi}$, with quantum operations applied only to individual particles.

There are two important operation protocols (see \cite{Aulbach2011} for a reference):
\begin{itemize}
\item Local Operations and Classical Communication (LOCC),
\item Stochastic Local Operations and Classical Communication (SLOCC).
\end{itemize}

Both of them allow local operations (such as unitary operations or measurements) and exchange of classical information --- perhaps with measurements or operations being conditional on already obtained outcomes.
The difference between LOCC and SLOCC is that in the first case we require deterministic success, while in the later --- success with some non-zero (however small) probability.

A typical example of a LOCC protocol would deterministically distinguishing two orthogonal states with only local projection measurements (with basis of subsequent measurement being dependent on the previous measurement outcomes) \cite{Walgate2000}.

One key characteristics of states, derived from LOCC and SLOCC, are classes states that are equivalent with respect to them --- i.e. for each pair of states from a class there is a S(LOCC) protocol that does $\ket{\psi_1}\mapsto\ket{\psi_2}$ and another one performing the reverse operation $\ket{\psi_2}\mapsto\ket{\psi_1}$.
Not every (S)LOCC transformation can be reversed.
For example, we can map an entangled state into a product state (by measurements and applying appropriate unitary operations), but the inverse is not possible.

For LOCC, states are equivalent if and only if there exist local unitary operations $U_1,\ldots,U_n$ such that:
\begin{equation}
U_1 \otimes U_2 \otimes \cdots \otimes U_n \ket{\psi_1} = \ket{\psi_2}.
\label{eq:lu_equivalence_u1u2u3}
\end{equation}
This should not be surprising: we cannot make non-trivial measurements (not to disturb the state) and consequently, we cannot get any information for communication.
The only operations that we are free to perform are local unitary operations.
That is, LOCC-equivalence is the same as local unitary equivalence, or LU-equivalence, and we will use the later term.
For qubits, the problem was solved in \cite{Kraus2010a, Kraus2010b} using normal forms..

For SLOCC, pure states are equivalent if and only if there exist local invertible operations $A_1,\ldots,A_n$ (not necessarily unitary, normal or diagonalizable) such that:
\begin{equation}
A_1 \otimes A_2 \otimes \cdots \otimes A_n \ket{\psi_1} = \ket{\psi_2}.
\label{eq:slocc_equivalence_a1a2a3}
\end{equation}
The original proof is in \cite{Dur2000}. 
It can be expressed as follows:
on each particle we perform a positive operator-valued measurement (POVM).
That is, for the $i$-th particle we use operators $\{\tilde{A}_i^{k}\}_k$, such that
\begin{equation}
\sum_k \tilde{A}_i^{k\dagger} \tilde{A}_i^{k} = \mathbb{I}
\end{equation} 
and for each $k$ we get an outcome
\begin{equation}
\frac{\tilde{A}_i^{k} \ket{\psi}}{\sqrt{\bra{\psi} \sum_k \tilde{A}_i^{k\dagger} \tilde{A}_i^{k} \ket{\psi}}}
\end{equation}
with probability
\begin{equation}
p_k = \bra{\psi} \tilde{A}_i^{k\dagger} \tilde{A}_i^{k} \ket{\psi}.
\end{equation}.
We set $\tilde{A}_i^{1}$ equal to $A_i$, up to normalization.
Then, by conditioning our result on getting outcome "1" for every particle, we obtain $\ket{\psi_2}$ as in \eqref{eq:slocc_equivalence_a1a2a3}.
Moreover, we need to impose that $A_i$ is invertible, so as to guarantee that we are able to go back from $\ket{\psi_2}$ to $\ket{\psi_1}$.
As we see, also in this case we were able to avoid communicating:
we are able to set local operations in advance and condition the result on a particular measurement outcome.

To give some taste of equivalence classes both with respect to LOCC and SLOCC,
we perform the Schmidt decomposition on a pure state of two particles
\begin{equation}
\ket{\psi_1} = \sum_k \lambda_k \ket{\phi_k} \otimes \ket{\varphi_k}, 
\end{equation}
where $\{\ket{\phi_k}\}$ and $\{\ket{\varphi_k}\}$ are sets of orthonormal one-particle states and the Schmidt values $\lambda_k$ are non-negative real numbers, in decreasing order.
Then state $\ket{\psi_1}$:
\begin{itemize}
\item is LU-equivalent to $\ket{\psi_2}$ iff they have the same Schmidt values,
\item is SLOCC-equivalent to $\ket{\psi_2}$ iff they have the same number of non-zero Schmidt values.
\end{itemize} 
In particular, for two qubits all equivalence classes are represented by
\begin{itemize}
\item LU-equivalence: one-parameter family $\sqrt{1-\lambda^2} \ket{00} + \lambda \ket{11}$
for $0 \leq \lambda \leq 1/\sqrt{2}$,
\item SLOCC-equivalence: a discrete family of two states --- a product state $\ket{00}$ and an entangled state $(\ket{00} + \ket{11})/\sqrt{2}$.
\end{itemize}
In general, entanglement classes for more than two particles are much more involved, even for the symmetric qubit ($d=2$) states with three \cite{Dur2000} or four \cite{Verstraete2002} particles.

Let us see how this problem can be stated for bosonic systems. 
First, let us describe the action of (passive) linear optics on pure states described in different representations. Within the mode representation, the action of linear optics is expressed mathematically as the application of {\em unitary operations} on the creation operators, i.e.:
\begin{align}
a_i'^\dagger = \sum_{j=1}^{d} U_{ij} a_j^\dagger \ ,
\label{eq:u-transf}
\end{align}
where $U \in SU(d)$. Conversely, all $SU(d)$ operations among the modes can be achieved with a sequence of two-mode operations, such as beam-splitters and wave plates, in a way which resembles the action of Euler angles \cite{Reck1994}.
We use the word \emph{passive}, since we want to exclude operations that are linear in both creations and annihilation operators, but do mix them (i.e. squeezing).
Alternatively, in particle representation, transformation \eqref{eq:u-transf} is equivalent to the action of the same $U$ on each particle:
\begin{equation}
\ket{\psi'}_P = U^{\otimes n} \ket{\psi}_P \ .
\label{eq:u-transf2}
\end{equation}
The equivalence between both representations corresponds to the equivalence between the first and second quantization pictures for bosonic states \cite{FetterBook1971}.
Instead of a unitary operation $U$, we can use any $d\times d$ complex matrix $A$.
We also use passive linear optics, and add ancillary modes (empty on input, conditioned to be empty on output) so that
\begin{equation}
\mathcal{U} =
\begin{bmatrix}
t A & B\\
C & D
\end{bmatrix}
\end{equation}
is unitary for some number $t$ and matrices $B$, $C$ and $D$.
For the equivalence relation we need to assume that $A$ is invertible.

We are ready to state the problem of equivalence between two bosonic pure states under the action of linear optics. The problem is formulated as follows. Given two pure states, $\ket{\psi_1}=f_1^\dag \ket{\Omega}$ and $\ket{\psi_2}=f_2^\dag \ket{\Omega}\in {\cal S}_n^d$,
we ask whether there exists a unitary transformation on the modes $U\in SU(d)$ such that $f_1$ and $f_2$ are related by a rotation among the variables
\begin{equation}
f_2(\vec{a})^\dagger = f_1(U^* \vec{a})^\dagger \ . \label{lu.equivalence}
\end{equation}
Alternatively, in the particle description, \eqref{lu.equivalence} is equivalent to
\begin{equation}
\ket{\psi_2}_P = U\otimes U \otimes \cdots \otimes U \ket{\psi_1}_P \ .
\label{eq:lu_equivalence_uuu}
\end{equation}
Both problems can be directly translated to their stochastic variants, with $A\in GL(d)$.
Formula \eqref{eq:lu_equivalence_uuu} is a special case of \eqref{eq:lu_equivalence_u1u2u3}, with all unitary operations being the same, i.e.
\begin{equation}
U \equiv U_1 = U_2 = \ldots = U_n,
\end{equation}
and analogously with \eqref{eq:slocc_equivalence_a1a2a3}.
Certainly, if states are related by \eqref{eq:lu_equivalence_uuu}, they are also LU-equivalent. But, if two pure permutation-symmetric states are LU-equivalent, does it mean that they are related by linear optics?
We show it in Sec.~\ref{s:slocc-equivalence} (see also our paper \cite{Migdal2013aaa}), both for LU-equivalence and SLOCC-equivalence.


\subsection{Invariants}

As it was stated in the introduction, our approach to the equivalence problem \eqref{lu.equivalence} is based on the construction of particular classes of invariants of the local unitary group representing linear optics. Let us consider the action of a group $G$ on some set $X$. For $x\in X$ and $g\in G$, let us denote the action of $g$ on $x$ by $g\cdot x$, which again belongs to $X$. A function $h: X\mapsto X$ is invariant under the action of $G$ if and only if
\begin{equation}
h(g\cdot x)=h(x)\ \text{for all $x\in X$ and all $g\in G$\ .}
\end{equation}

In our case we have $X=\mathcal{S}^{d}_n$, $G=SU(d)$ and the action of $G$ is given by \eqref{eq:u-transf} or, equivalently, by \eqref{eq:u-transf2}. A theorem by Hilbert states that, for a compact group $G$ acting in a unitary fashion on a finite dimensional vector space, there exists a finite number of independent invariants (which are polynomial in the coordinates of $\ket{\psi}$) that are able to distinguish whether two vectors belong to the same orbit of $G$ \cite{Weyl1997, Kraft1996, Grassl1998}.
A convenient way to write down the invariants involves using tensor diagrams \cite{Biamonte2013invariants} --- they make it explicit why certain polynomials are invariant and allow us to avoid multiple index contractions.
Thus, the LU-equivalence problem can be solved completely once the minimal set of independent polynomial invariants is known. This problem is in general unsolved.
For recent developments in the theory of invariants in the context of entanglement theory see \cite{Vrana2011}.
In our work we do not attempt to study all invariants of the action of $SU\left(d\right)$ on $\mathcal{S}_{n}^d$. Instead, we focus on two families of invariants, analyzing their usefulness and physical relevance.

There are relevant differences between structures of LU and and SLOCC invariants.
Since LU-equivalence is characterized by a compact group acting on vector space,
\begin{itemize}
\item there is a finite number of polynomial invariants that are necessary to distinguish orbits,
\item all invariant polynomials can be written as a sum of polynomials of a particular form \cite{Sudbery2000}.
\end{itemize}
In SLOCC, or equivalently --- equivalence up to local invertible operations, the previous statements do not hold.
For example, for one SLOCC invariant, the Schmidt number (i.e. number of non-zero Schmidt values), there is no continuous function (let alone polynomial), as $\ket{00}+\epsilon \ket{11}$ has Schmidt rank $2$ for arbitrary small $\epsilon>0$.




\subsection{Simple examples}

Before considering the general problem, let us focus on simple cases for LU-equivalence of symmetric states:
\begin{itemize}
\item only two particles ($n=2$) in an arbitrary number of modes,
  or, alternatively,
\item an arbitrary number of particles in just two modes ($d=2$),
  i.e.: permutation-symmetric states for qubits,
\item multi-particle squeezed coherent states (i.e. Gaussian states). 
\end{itemize}
In particular, for the first two cases, we want to show explicitly how the bosonic mode description is related to the particle representation.

\subsubsection{Two particles}
\label{s:two_particles}

For two particles it suffices to perform a variant of the Schmidt decomposition, for symmetric states \cite{Li2001}, i.e.:
\begin{align}
\ket{\psi} = \sum_{i=1}^{d} \lambda_i \ket{\phi_i}_P \otimes \ket{\phi_i}_P \ ,
\end{align}
where $\lambda_i\geq 0$ and $\ket{\phi_i}$ are pairwise orthogonal states, the same for both particles. Thus, two pure states of two photons are related by linear optics if and only if they have the same sets of Schmidt values $\{\lambda_i\}$.
In this case, the polynomial $f(\vec{a})$ (as in \eqref{eq:state-polynomial}) is formally a quadratic polynomial in the number of modes, $d$. The Schmidt decomposition allows us to rewrite it as:
\begin{equation}
f = \sum_{i=1}^d \frac{\lambda_i}{\sqrt{2}} b_i^2
\label{factorize.n2}
\end{equation}
for a certain set of new variables $\vec{b}$, such that $b_i^\dagger \ket{\Omega} = \ket{\phi_i}$, which are related to $\vec{a}$ by a unitary rotation, i.e. $\vec{b} = U \vec{a}$.

Bear in mind that the operation we had performed is not diagonalization, since:
\begin{itemize}
\item $M_{\mu\nu} \equiv \ket{\psi}_{\mu \nu}$ is a symmetric matrix, not necessarily Hermitian,
\item $U \otimes U \ket{\psi}$ is equivalent to $U M U^T$, not $U M U^\dagger$.
\end{itemize}

\subsubsection{Two modes and Majorana representation}
\label{sec:majorana}

When there are just two modes ($d=2$), it is possible to use the {\em Majorana stellar representation} (see e.g. \cite{BengtssonZyczkowski2006book, Aulbach2011a, Markham2010} for a short introduction) and write the state as:
\begin{equation}
\ket{\psi} = A \prod_{i=1}^{n} \left(\cos\(\frac{\theta_i}{2}\) a_1^\dagger
+ e^{i \varphi_i} \sin\(\frac{\theta_i}{2}\) a_2^\dagger \right) \ket{\Omega},
\label{majorana}
\end{equation}
where pairs $(\theta_i, \varphi_i)$ can be interpreted as coordinates of points on the Bloch sphere, and $A$ is a normalizing factor. Equation (\ref{majorana}) is equivalent to a factorization of the homogeneous polynomial defined in eq. (\ref{eq:state-polynomial}) in the following form:
\begin{equation}
f(a_1,a_2)= \tilde{A} a_2^n
\prod_{i=1}^{n}
\left( \tfrac{a_1}{a_2} - x_i \right)
\label{factorize.d2},
\end{equation}
where $\tilde{A}$ is the coefficient of $a_1^n$, and we have introduced variables $x_i = - e^{i \varphi_i} \tan(\theta_i/2)$.
Linear optics acts on this representation as a rotation of the Bloch sphere as a whole. Consequently, two states are related by linear optics if and only if their Majorana representations are related by rotation \cite{Mathonet2010}.

Let us show how to decide whether two symmetric $n$-qubit states are equivalent under linear operations.
First, we apply the Majorana stellar representation to both states, resulting in two sets of vectors, $\{\vec{v}_i\}_{i\in \{1,\ldots,n\}}$ and $\{\vec{u}_i\}_{i\in \{1,\ldots,n\}}$. They may differ by a rotation (i.e. an element of SO(3)) and a permutation.
Let us choose an ordered pair of two non-parallel vectors $(\vec{u}_1, \vec{u}_2)$.
Then, for every ordered pair from the first set $(\vec{v}_i, \vec{v}_j)$ for $i \neq j$, if their scalar products match ($\vec{v}_i\cdot\vec{v}_j=\vec{u}_1\cdot\vec{u}_2$) we can construct a unique rotation that rotates the first pair into the second.
Then we check whether such rotation rotates every $\vec{v}_i$ into a distinct vector $\vec{u}_{\sigma{i}}$. If it does, states are equivalent.
If if it does not for all pairs, two states are not equivalent.
As the number of ordered pairs of two different vectors is $n^2-n$, the algorithm complexity is the maximum of $n^2$ and the complexity of an algorithm for factorization of an $n$-degree polynomial of one variable (to get the Majorana stellar representation).


\subsubsection{Gaussian states}
\label{sec:gaussian-states}

Gaussian states of light are states for which the Wigner function is Gaussian (as well as other quasi-probability distributions \cite{Cahill1969}).
Pure Gaussian states can be written as \cite{Braunstein2005}
\begin{equation}
\ket{\psi} =
\exp\left(\sum_i \alpha_i a_i^\dagger - \alpha_i^* a_i \right)
\exp\left(  \sum_{ij} Q_{ij} a_i^\dagger a_j^\dagger \right)
\ket{\Omega},
\label{eq:pure-gaussian-def}
\end{equation}
where a complex symmetric matrix $Q$ is related to squeezing and a complex vector $\vec{\alpha}$ is related to displacing the center.

They are of special importance, since they are both simple to generate and analyze \cite{Braunstein2005rmp,Weedbrook2012,Adesso2014,Giedke2001thesis}.
Unlike other states we analyze in this chapter, Gaussian states have an unbounded number of photons, with the sole exception of the vacuum state.
Their entanglement was studies with respect to modes \cite{Kraus2003,Adesso2006,Giedke2014}, with applications to standard quantum information operations \cite{NielsenChuang2000} such as quantum teleportation \cite{Enk1999} or quantum key distribution \cite{Gottesman2001}. 
Gaussian states are the only pure states described by non-negative Winger function. 
Since the behavior of Gaussian states is similar to some classical states, the negativity of Wigner function \cite{Kenfack2004} can be used to measure certain aspects of non-classicality of quantum states.

Continuous-variable systems are typically studied using canonical position and momentum or equivalently --- creation and annihilation operators
\begin{equation}
	\begin{bmatrix}
	a_i\\
	a_i^\dagger
	\end{bmatrix}
	=
	\frac{1}{\sqrt{2}}
	\begin{bmatrix}
	1 & i\\
	1 & -i
	\end{bmatrix}
	\begin{bmatrix}
	x_i\\
	p_i
	\end{bmatrix}
	\label{eq:aadag-xp}
\end{equation}
for each mode $i$.
Multiplying the wavefunction by a phase rotates all $x$ and $p$.

General affine operations on creation and annihilation, called linear unitary Bogoliubov transformations, are of the form
\begin{equation}
\vec{a} \mapsto A \vec{a} + B \vec{a}^\dagger + \vec{\alpha},
\label{eq:bogoliubov}
\end{equation}
where $A$ and $B$ are responsible for rotation and squeezing and $\vec{\alpha}$ for displacing.
$A$ and $B$ are related, so that the output modes fulfill the canonical commutation relations.
For passive linear optics $A\equiv U^*$ is unitary and both $B$ and $\vec{\alpha}$ are zero.
That is, it does not mix creation with annihilation operators, or displace them.
In particular, passive linear optics preserves the number of particles.

Let us stick to creation and annihilation operations, since they can easily represent the action of linear optics, that is
\begin{equation}
	\begin{bmatrix}
		\vec{a}\\
		\vec{a}^\dagger
	\end{bmatrix}
	\mapsto
	\begin{bmatrix}
		U^* & 0\\
		0 & U
	\end{bmatrix}
	\begin{bmatrix}
		\vec{a}\\
		\vec{a}^\dagger
	\end{bmatrix}
\end{equation}
whereas for the canonical position and momentum it is slightly more complicated
\begin{equation}
	\begin{bmatrix}
		\vec{x}\\
		\vec{p}
	\end{bmatrix}
	\mapsto
	\begin{bmatrix}
		\Re [U] & \Im [U]\\
		-\Im [U] & \Re [U]
	\end{bmatrix}
	\begin{bmatrix}
		\vec{a}\\
		\vec{a}^\dagger
	\end{bmatrix},
\end{equation}
what can be derived using \eqref{eq:aadag-xp}.

All pure Gaussian states are related by \eqref{eq:bogoliubov}.
But which states are related only by passive linear optics?
Let us focus on $\vec{\alpha}=\vec{0}$.
The problem has the same geometry as the two-particle case studied in Sec.~\ref{s:two_particles}, thus can be solved with the same method (mathematically speaking, Gaussian states \eqref{eq:pure-gaussian-def} are exponents of two-particle states).
However, we take another route, which sheds light on physical properties of Gaussian states.

The key observation is that each squeezing \eqref{eq:bogoliubov} can be decomposed as a series of operations: a passive optics, a one-mode squeezing and another passive optics \cite{Braunstein2005} (see also \cite[A.2]{Giedke2001thesis}), using so-called Bloch-Messiah decomposition, which can be used to analyze modes of link squeezing, e.g. as in \cite{Migdal2010}.
One-mode squeezings are operations of the form 
\begin{equation}
	x_i \mapsto \exp(-r_i) x_i, \quad p_i \mapsto \exp(r_i) p_i
\end{equation}
for real $r_i\geq0$ called squeezing parameters.
If we start from the vacuum state, then such squeezing results in
\begin{equation}
	\bra{\psi} x_i^2 \ket{\psi} = \exp(-2r)/2, \quad \bra{\psi} p_i^2 \ket{\psi} = \exp(2r)/2,
	\label{eq:squeezed-quadratures}
\end{equation}
which saturate the Heisenberg's uncertainty principle.
We may conclude that with passive optics operations we can build a normal form of a state,
i.e one with no correlations between modes and modes squeezed as in \eqref{eq:squeezed-quadratures}.

Let us calculate a correlator for a Gaussian state:
\begin{equation}
	\rho_{ij} = \bra{\psi} a_i^\dagger a_j \ket{\psi},
	\label{eq:gaussian-1particle-dm}
\end{equation}
which can be thought to a generalization of a one-particle reduced density matrix.
This matrix bears exact information to decide whether two states are related via linear optics.
By diagonalizing \eqref{eq:gaussian-1particle-dm}
\begin{equation}
	\rho \mapsto U \rho U^\dagger
\end{equation}
we get new, pairwise uncoupled modes. On the diagonal we get eigenvalues, i.e.:
\begin{equation}
	\bra{\psi} a_i^\dagger a_i \ket{\psi}
	=
	\tfrac{1}{2} \bra{\psi} \left( x_i^2 +p_i^2 - 1 \right)  \ket{\psi}
	= \tfrac{1}{2}\cosh(2r_i) - \tfrac{1}{2},
\end{equation}
where we used \eqref{eq:squeezed-quadratures}, and which is in one-to-one correspondence with the squeezing parameter.
Consequently, if two non-displaced Gaussian states have the same spectrum of \eqref{eq:gaussian-1particle-dm}, they are related by passive linear optics, and the exact transformation is $DU$,
where $U$ is a unitary matrix diagonalizing \eqref{eq:gaussian-1particle-dm}, and $D$ is a diagonal matrix with phases, rotating each position-momentum pairs.

\section{SLOCC and LOCC equivalence of permutationally symmetric pure states}
\label{s:slocc-equivalence}

In this section we present two results, main for our paper \cite{Migdal2013aaa}.
The first one is that, when testing whether two permutation-symmetric $n$ qudit states are equivalent under local transformations, the search can be restricted to operators which act in the same way on every particle.
This property was proven for qubits \cite{Bastin2009, Mathonet2010, Bastin2010} in the SLOCC variant (though the unitary version can be deduced from their proof).
For general qudit system it remained so far an open question
\cite[Sec.5.1.1.]{Aulbach2011}.
That is, in the course of this work, we prove that:
\begin{theorem}\label{thm:main}
Let us consider two permutation-symmetric states of $n$
qudits (i.e. $d$-level particles), $\ket{\psi}$ and $\ket{\varphi}$, for which
there exist invertible $d\times d$ matrices $A_1, \ldots, A_n$ such that
\begin{equation}
A_1 \otimes A_2 \otimes \ldots \otimes A_n \ket{\psi} =
\ket{\varphi}\label{eq:a1a2an}.
\end{equation}
Our result implies that then there exists an invertible $d\times d$ matrix $A$ such that
\begin{equation}
A \otimes A \otimes \ldots \otimes A \ket{\psi} = \ket{\varphi}.\label{eq:aaa}
\end{equation}

If we restrict ourselves to unitary matrices $A_1, \ldots, A_n$, then $A$ is
unitary.
\end{theorem}

For $A_i$ unitary, \eqref{eq:a1a2an} is a condition of equivalence of states
under reversible local operations (or LU-equivalence), which is proven to be the same as equivalence 
with respect to local operations and classical communication (LOCC) \cite{Bennett2000, Vidal2000}.
Moreover, in both cases we provide a direct construction for $A$ as
a function of $A_1,\ldots, A_n$.
Our second result stems from the consideration of stabilizers of states \cite{Cenci2010a} in the form of a matrix $B$ acting on one particle, and its inverse $B^{-1}$ acting on another one. Only for very specific states there are such $B$, which are non-trivial. We show that the Jordan form of $B$, disregarding the values of the eigenvalues, is an invariant for SLOCC-equivalence, and analyze it in detail, providing a coarse-grained classification of the relevant entanglement classes. If each block of the Jordan form of $B$ has a distinct eigenvalue, then there is a unique stabilized state, up to local operations. In particular, we find as entanglement class representatives a $d$-level generalization of the $n$-particle GHZ state
\begin{align}
\frac{\ket{0}^n + \ldots + \ket{d-1}^n}{\sqrt{d}},
\end{align}
and one possible generalization of the W state for $d>2$, i.e. a state with all single particle state indices adding up to $d-1$, that is
\begin{align}
\binom{n + d - 2}{d - 1}^{-1/2} \sum_{i_1+\ldots+i_n = d-1} \ket{i_1} \ket{i_2}\ldots \ket{i_n},\label{eq:excitationnormalized}
\end{align}
which we call \textit{excitation state}.

For two particles both classes coincide, as e.g.
\begin{align}
\ket{00} + \ket{11} + \ket{22} \cong \ket{02} + \ket{11} + \ket{20}
\end{align}
Table \ref{tab:n3} summarizes the entanglement classes related to Jordan blocks for the simplest non-trivial case, i.e. $n=3$ particles (a general construction is in \eqref{eq:unique_general}).
We adopt a special notation for the Jordan block structure.
Outer brackets separate eigenspaces with different eigenvalues, while the inner brackets separate different Jordan blocks of the same eigenvalue.
Each number is dimension of a single Jordan block.
Ordering of the terms does not matter, neither in inner or outer brackets.
For example: $\{ \{ 2 \}\}$ is a matrix with only one Jordan block, $\{ \{ 1, 1 \}\}$ is proportional to the identity matrix and $\{ \{ 1 \}, \{ 1 \} \}$ is a matrix with two different eigenvalues, that is ($\lambda_1 \neq \lambda_2$): 

\begin{align}\label{eq:block-notation}
\{ \{ 2 \}\} & \equiv
\left[
\begin{array}{cc}
\lambda_1 & 1\\
0 & \lambda_1
\end{array}
\right]\\
\{ \{ 1, 1 \}\} & \equiv
\left[
\begin{array}{cc}
\lambda_1 & 0\\
0 & \lambda_1
\end{array}
\right]\\
\{ \{ 1 \}, \{ 1 \} \} & \equiv
\left[
\begin{array}{cc}
\lambda_1 & 0\\
0 & \lambda_2
\end{array}
\right]
\end{align}
 
The number of different Jordan block structures for a given $d$ is given by double partitions~\cite{oeisA001970}. 

\begin{table}
\centering
\begin{tabular}{|l|l|l|}

\hline
$d$ &
Block structure &
A class representative\\

\hline
2 &
$\{ \{ 2 \}\}$ &
$\ket{001} + \ket{010} + \ket{001}$\\

\hline
 &
$\{ \{ 1, 1 \}\}$ &
(not unique) any state\\

\hline
 &
$\{ \{ 1 \}, \{ 1 \} \}$ &
$\ket{000} + \ket{111}$\\

\hline
3 &
$\{ \{ 3 \} \}$ &
$\ket{002} + \ket{020} + \ket{200}$ \\

 &
 &
$+ \ket{011} + \ket{101} + \ket{110}$\\

\hline
 &
$\{ \{ 2, 1 \} \}$ &
(not unique)\\

\hline
 &
$\{ \{ 1, 1, 1 \} \}$ &
(not unique) any state\\

\hline
 &
$\{ \{ 2 \}, \{ 1 \} \}$ &
$\ket{001} + \ket{010} + \ket{100} + \ket{222}$\\

\hline
 &
$\{ \{ 1, 1 \}, \{ 1 \} \}$ &
(not unique)\\

\hline
 &
$\{ \{ \{ 1 \}, \{ 1 \}, \{ 1 \} \} \}$ &
$\ket{000} + \ket{111} + \ket{222}$\\

\hline

\end{tabular}
\caption{
A summary of entaglement classes related to Jordan blocks, for the case of three qubits ($d=2$) and qutrits ($d=3$). The notation is explained in the main text \eqref{eq:block-notation}. The general construction for the unique states is given in \eqref{eq:unique_general}.
}
\label{tab:n3}
\end{table}

This work is organized as follows:
in Section \ref{sec:aaa-symmetry} we prove that it is sufficient to study invariance under symmetric transformations.
We supplement it with a construction of the symmetric transformation in Section \ref{sec:aaa-explicit}.
In Section \ref{sec:aaa-classes} we discuss the entanglement classes which can be obtained by
studying stabilizer operators related to one-particle transformations.

\subsection{Symmetric operations suffice}
\label{sec:aaa-symmetry}

We start with an approach similar to the one of \cite{Mathonet2010}.
Let us consider two permutation-symmetric states, $\ket{\psi}$ and
$\ket{\varphi}\in {\cal S}$, with ${\cal S}$ denoting the symmetric subspace of the full Hilbert space.
If \eqref{eq:a1a2an} holds, then any different
permutation of $A_1, \ldots, A_n$ will also work. In order to show this
property explicitly, we may use $\ket{\psi} = P_\sigma
\ket{\psi}$ and $\ket{\varphi} = P_{\sigma^{-1}} \ket{\varphi}$, where
$P_\sigma$ is a permutation matrix for the permutation of particles $\sigma$,
i.e. $P_\sigma \ket{i_1i_2 \ldots i_n}=\ket{i_{\sigma(1)} i_{\sigma(2)}\ldots
i_{\sigma(n)}}$.

Since all $A_i$ are invertible, it means in particular that
\begin{equation}
\left(A_2^{-1} \otimes A_1^{-1} \otimes \ldots \otimes A_n^{-1} \right)
\left(A_1 \otimes A_2 \otimes \ldots \otimes A_n \right) \ket{\psi} = \ket{\psi}
\end{equation}
or, equivalently, 
\begin{equation}
 \left(B \otimes B^{-1} \otimes \mathbb{I} \otimes \ldots \otimes \mathbb{I}
\right) \ket{\psi} = \ket{\psi},\label{eq:bbinv}
\end{equation}
where $B=A_2^{-1} A_1$.

From now on, we will use a subscript in parenthesis to indicate the position of
an operator in the tensor product, e.g.
\begin{equation}
 B_{(2)} \equiv \mathbb{I}  \otimes B \otimes \mathbb{I} \otimes \ldots \otimes
\mathbb{I},
\end{equation}
where the total number of factors is $n$.
Without the loss of generality, we use operations on the first and the second particle.

First, let us show that if an operation on one particle can be reversed by
applying the inverse operation on a {\em different} particle, then that
single-particle operation must preserve permutation symmetry of the state.

\begin{lemma}\label{thm:bbinv2b}
	For a symmetric $\ket{\psi} \in \mathcal{S}$ the equality \eqref{eq:bbinv}
	\begin{equation}
	B_{(1)} B_{(2)}^{-1} \ket{\psi} = \ket{\psi}
	\end{equation}
	holds if and only if 
	\begin{equation}
	B_{(1)} \ket{\psi} \in \mathcal{S}. \label{eq:b}
	\end{equation}
\end{lemma}

\begin{proof}

\begin{align}
B_{(1)} \ket{\psi} \in \mathcal{S} \Leftrightarrow B_{(1)} \ket{\psi} =
B_{(2)} \ket{\psi}\\
\Leftrightarrow B_{(1)} B_{(2)}^{-1} \ket{\psi} = \ket{\psi}
\end{align}

\end{proof}

Now we will show that the action of the aforementioned single-particle operation
$B_{(1)}$ can be expressed as an operation acting in the same way on every
particle $S^{\otimes n}$. Intuitively, we must search for an $n$-th root of
$B$. But not all such $n$-th roots will work, as the following example shows:


Consider $S=\sigma_x$, which is a square root
of $B=I$, acting on $\ket{00}\in \cal{S}$. While $B_{(1)}\ket{00}\in \cal{S}$,
$S_{(1)}\ket{00}=\ket{10}\not\in \cal{S}$, and $S \otimes
S \ket{00}=\ket{11}$, which, despite being symmetric, is not the desired
state. Thus, the relevant question is: {\em which one is the appropriate $n$-th
root?} 
Before we can proceed, we need a few lemmas.

\begin{lemma}
\label{thm:symcom}
If $\ket{\psi} \in {\cal S} $, $X_{(1)}\ket{\psi}\in {\cal S}$ and $Y_{(2)}\ket{\psi}\in{\cal S}$, then
$X_{(1)}Y_{(2)}\ket{\psi}\in {\cal S}$ $\Leftrightarrow$ the commutator
acting on the state vanishes $[X_{(1)}, Y_{(1)}]\ket{\psi}=0$.
\end{lemma}
 
\begin{proof}
If the final state is symmetric, we may permute the first two particles
without altering the result:
\begin{align}
0 &= X_{(1)}\left(Y_{(2)}\ket{\psi}\right) - Y_{(1)}\left(X_{(2)}\ket{\psi}\right)\\
&= X_{(1)}\left(Y_{(1)}\ket{\psi}\right) -
Y_{(1)}\left(X_{(1)}\ket{\psi}\right)\\
&= [X_{(1)}, Y_{(1)}]\ket{\psi}.
\end{align}
\end{proof}

To see how commutativity is important, take as an example
\begin{equation}
X = 
\left[
\begin{matrix}
1 & 1\\
0 & 1
\end{matrix}
\right], \quad
Y = 
\left[
\begin{matrix}
0 & 1\\
1 & 0
\end{matrix}
\right].
\end{equation}
acting on $\ket{\psi}=(\ket{01}+\ket{10})/\sqrt{2}$ (i.e. $\ket{\psi}$,
$X_{(1)}\ket{\psi}$ and $Y_{(2)}\ket{\psi}$ are symmetric, but $X \otimes Y
\ket{\psi} = (\ket{00}+\ket{01}+\ket{11})/\sqrt{2}$ is not).

\begin{lemma}\label{thm:comm3}
Moreover, for $n\geq3$ the commutator acting on the state always vanishes, i.e.
$[X_{(1)}, Y_{(1)}]\ket{\psi}=0$.
\end{lemma}

\begin{proof}
\begin{align}
&X_{(1)} Y_{(1)} \ket{\psi}
= X_{(1)} Y_{(3)} \ket{\psi}
= X_{(2)} Y_{(3)} \ket{\psi}\\
&= Y_{(3)} X_{(2)} \ket{\psi}
= Y_{(1)} X_{(1)} \ket{\psi}
\end{align}
\end{proof}

\begin{lemma}
If $X_{(1)}\ket{\psi}$ is symmetric, then $X^p_{(1)}\ket{\psi}$ is symmetric for
all natural $p$ (integer if $X$ is invertible).
\end{lemma}

\begin{proof}
We use mathematical induction with respect to $p$, starting at $p=0$. Since
$X$ commutes with $X^p$ (even without the restriction to a specific state), then
using Lemma \ref{thm:symcom}, $X^p_{(1)}\ket{\psi}\in{\cal S}$ implies that
$X^{p+1}_{(1)}\ket{\psi}\in {\cal S}$. If $X$ is invertible, we may use the same
argument for $X$ and $X^{-p}$, respectively.
\end{proof}

\begin{corollary}
Moreover, we get
\begin{align}
X^p_{(1)}\ket{\psi} = X^{p_1}\otimes X^{p_2}\otimes \ldots \otimes X^{p_n}
\ket{\psi},
\end{align}
for any integers $p_i$ (can be negative if $X$ is invertible) that add up to
$p$.
\end{corollary}

\begin{corollary}
\label{corol:analytic}
In particular, $f(X)_{(1)}\ket{\psi}\in {\cal S}$ for any analytic
function $f(z)$.
\end{corollary}

\begin{theorem}
\label{thm:sssb}
	For any $X$ and $\ket{\psi}\in {\cal S}$ it holds that if
	\begin{equation}
	X_{(1)}  \ket{\psi} = \ket{\phi} \in {\cal S}
	\end{equation}
	then there exists a single-particle operator $S$ such that $S^n=X$,
$S_{(1)}\ket{\psi}\in{\cal S}$ and
	\begin{equation}
	 \left(S \otimes S \otimes \ldots \otimes S \right)  \ket{\psi} =
\ket{\phi}.
	\end{equation}
\end{theorem}

\begin{proof}
The proof outline is the following: we prove that, among the $n$-th roots
of operator $X$, there is (at least) one, $S$, which {\em can be expressed as a
polynomial} of $X$;
following Corollary \ref{corol:analytic}, we get $S_{(1)}\ket{\psi}\in\cal{S}$ and the
rest of the theorem follows.

The $n$-th root function is multivalued, so we can not use it to
prove the theorem as it stands. Let us, then, prove that there exists a
polynomial function $f$, such that $[f(X)]^n=X$.

Let $\left\{\lambda_i\right\}$ be the eigenvalues of $X$, with algebraic
multiplicities $\left\{m_i\right\}$ (i.e. size of the largest Jordan block related to such eigenvalue). Matrix function theory \cite[Chapter
1]{Higham2008} states that the action of any analytical function $f$ on a matrix
$X$ is completely determined by the set of values $\left\{f(\lambda_i)\right\}$,
along with the derivatives $\left\{ f^{(k)}(\lambda_i)\right\}$, up to degree
$m_i$. Let us choose, for each $i$ separately, $f(\lambda_i)$ and
$f^{(k)}(\lambda_i)$ from the same branch of the complex $n$-th root function. It is
always possible to find a polynomial $f$ that takes exactly those values and
derivatives at the eigenvalues of $X$, e.g., via Hermite interpolation.
Thus, we can define $S\equiv f(X)$, and
we have $S^n=X$, as required.

Combining this result with the corollaries, we get that
\begin{align}
S^{\otimes n} \ket{\psi} = S^{n}_{(1)} \ket{\psi} = X_{(1)} \ket{\psi}.
\end{align}

\end{proof}

The converse of theorem \ref{thm:sssb} is false. Take,
e.g., $S=\sigma_x$ and $\ket{\psi}=\ket{00}$. It is true that $S \otimes
S \ket{00}=\ket{11}\in \cal{S}$, yet there is no $B$ such that
$B_{(1)}\ket{00}=\ket{11}$. 

\subsection{Explicit formula for symmetrization}
\label{sec:aaa-explicit}

In this section we provide the explicit form of $A$, given all $A_i$. 
Let $B_{ij} \equiv A_i^{-1} A_j$. Thus, operator $B$ in the previous section
would correspond to $B_{12}$ with the new notation. Transforming
\eqref{eq:a1a2an} we get
\begin{align}
&A_1 \otimes A_2 \otimes \ldots \otimes A_n\ket{\psi}\\
&= A_1 \otimes A_1 B_{12} \otimes \ldots \otimes A_1 B_{1n} \ket{\psi}
\\
&= A_1^{\otimes n} B_{12\ (2)} B_{13\ (3)} \ldots B_{1n\ (n)}\ket{\psi}.
\end{align}

All $B_{1j\ (j)}\ket{\psi}$ are symmetric states, similarly to
$B_{(1)}\ket{\psi}$. Consequently, the last part can be reshuffled as
\begin{align}
A_1^{\otimes n} \left(B_{11}B_{12}\ldots B_{1n} \right)_{(1)} \ket{\psi}.
\end{align}
Note that no requirements are imposed about their commutativity.
Using Lemma \ref{thm:sssb} we get $A = A_1 S$, where $S$ is an appropriate
$n$-th root of $B_{11}B_{12}\ldots B_{1n}$.

Moreover, when all $A_i$ are unitary, then $S$ is unitary, since roots of
unitary matrices can be chosen to be unitary given that $f(U X U^\dagger) = U f(X) U^\dagger$ for all unitary $U$.
This finalizes the proof of Theorem~\ref{thm:main}.

\subsection{Symmetry classes from single-particle stabilizers}
\label{sec:aaa-classes}

A well-known strategy in the search for entanglement classes theory is to study
the dimension of the stabilizers of a state \cite{Cenci2010a}, i.e.: operators
$X$ such that $X \ket{\psi} = \ket{\psi}$. In our case it is natural to consider
stabilizers in the form of $X = B_{(1)} B_{(2)}^{-1}$, and state that {\em $B$
stabilizes $\ket{\psi}\in\cal{S}$} as a convenient shorthand notation. Following
Lemma \ref{thm:bbinv2b}, $B$ stabilizes $\ket{\psi}\in\cal{S}$ if
and only if $B_{(1)}\ket{\psi}\in\cal{S}$.
Bear in mind that a set of $B$ stabilizing a particular state is
guaranteed to form a group only for $n\geq3$, as follows from Lemma \ref{thm:comm3}.

Let us consider the Jordan normal form $J$ of $B$. We have shown
that all local operations for symmetric states are equivalent to the
action of the same single-particle operation on all qudits: $A^{\otimes n}$.
Consequently, if a state is stabilized by $B$, a SLOCC transformed state is
stabilized by some $A B A^{-1}$, i.e.: {\em the Jordan form of the stabilizer is
preserved}. 

Below, we prove the following facts relating the Jordan form of $B$ to the
stabilized states. First, that the precise eigenvalues are not
important --- only their degeneracies matter (see notation from Table \ref{tab:n3}).
Second, that stabilized states
do never mix eigenspaces of different eigenvalues. In particular, it means that
the problem can be split into a direct sum over distinct eigenvalues. Third, we
will show the explicit form of a state stabilized by a single Jordan block.
Fourth, we show that when eigenvalues are non-degenerate, there is an unique state
related to it (up to SLOCC). 
Fifth, we proceed to writing down states for multiple Jordan blocks with the
same eigenvalue. This will complete the characterization of states stabilized by
any $B$.

\begin{theorem}
The set of states stabilized by $B$ does not depend on the particular values of
its eigenvalues, as long as (non-)degeneracy is preserved.
\end{theorem}

\begin{proof}
We will show that mapping eigenvalues to different ones does not break the stabilizer's condition.
Let choose a complex function $f(z)$ such that (i) for all eigenvalues $f(\lambda_i) =
\tilde{\lambda}_i$, and (ii) $f^{(k)}(\lambda_i)=\delta_{0k}$ for all $k$ up to the
algebraic multiplicity of each $\lambda_i$. Now, $f(B)_{(1)}$ is also a
stabilizer of $\ket{\psi}$, with the same Jordan blocks, but arbitrarily set
eigenvalues.
\end{proof}

In particular, for $d=2$ the only two non-trivial Jordan forms of $B$ are
related to the GHZ state (two different eigenvalues) and W state (single
eigenvalue). We proceed to showing that stabilized states never mix subspaces
with different eigenvalues. 

Given a subspace $V$, let us denote by $\mathrm{Sym}^n(V)$ the permutation-symmetric subspace of $V^{\otimes n}$.
Then:

\begin{theorem}
For a given Jordan form $J$ with generalized eigenspaces $V_1, \ldots, V_p$ for
distinct eigenvalues, stabilized states are of the form
\begin{align}
\ket{\psi} \in \bigoplus_{i=1}^p \mathrm{Sym}^{n}(V_i).
\end{align}
That is, they contain no vectors mixing components from Jordan blocks of
different eigenvalues.
\end{theorem}

\begin{proof}
Let $\ket{\mu}$ and $\ket{\nu}$ be one-particle states
($\mu,\nu\in\{0,\ldots,d-1\}$) that belong to blocks of $J$ with a different
eigenvalues.
Let us take $f(B)$ mapping all subspaces to zero, except the one to which
$\ket{\mu}$ belongs, which we map to 1.
Suppose that $\ket{\psi}$ has a component containing a product of $\ket{\mu}$
and $\ket{\nu}$ (at different sites). Then, in particular, it has
$\ket{\mu}\ket{\nu}\ket{\xi}$ and $\ket{\nu}\ket{\mu}\ket{\xi}$, for some
symmetric $\xi$ (perhaps containing $\ket{\mu}$ or $\ket{\nu}$ as well).
But
\begin{align}
&f(B)_{(1)} \left(\ket{\mu}\ket{\nu}\ket{\xi} +
\ket{\nu}\ket{\mu}\ket{\xi}\right)\\
&=  \ket{\mu}\ket{\nu}\ket{\xi}.
\end{align}
The right hand side cannot be paired with any other terms in order to make a
symmetric state. Thus $f(B)_{(1)} \ket{\psi}$ is not symmetric, which
contradicts the assumption. Thus, a stabilized state can not contain a term with
a product of elements from two Jordan subspaces with different eigenvalues.
\end{proof}

Thus, when $J$ has $d$ distinct eigenvalues, then the stabilized state is
a generalized GHZ:
\begin{equation}
\ket{\psi} = \alpha_0 \ket{0}^n + \ldots + \alpha_{d-1} \ket{d-1}^n.
\end{equation}
When we consider local unitary equivalence, then the set of $\{|\alpha_i|^2
\}_{i\in \{0,\ldots,d-1\}}$ distinguishes classes, whereas for SLOCC, the state
is equivalent to any other with the same number of non-zero $\alpha_i$.

Now, it suffices to focus on a {\em Jordan subspace} related to a single eigenvalue.
Still, for a single eigenvalue there may be more than one Jordan blocks, i.e.
invariant subspaces. We start with the analysis of a {\em single Jordan block}, then
generalize our result to more blocks with the same eigenvalue.

\begin{theorem}
Let $K$ be a $k \times k$ Jordan block with eigenvalue zero, i.e.
$\sum_{i=1}^{k-1}\ket{i-1}\bra{i}$. Its stabilized states are
\begin{equation}
\ket{\psi} = \sum_{j=0}^{k-1} \alpha_j \ket{E_j},\label{eq:ejsum}
\end{equation}
where $\ket{E_j}$ is a symmetric state with $j$ {\em excitations}, i.e.
a symmetrized sum of all basis states for which the sum of the particle
indices is $j$:
\begin{align}
\ket{E_j} = \sum_{i_1+\ldots+i_n = j} \ket{i_1} \ket{i_2}\ldots \ket{i_n}. 
\end{align}
\end{theorem}

\begin{proof}
First, let us show that all states $K_{(1)} \ket{E_j}$ are symmetric, as long as
$j < k$.

\begin{align}
K_{(1)} \ket{E_j} &= \sum_{i_1+\ldots+i_n = j} \ket{i_1 - 1} \ket{i_2}\ldots
\ket{i_n}\\
&= \sum_{i'_1+\ldots+i_n = j - 1} \ket{i'_1} \ket{i_2}\ldots \ket{i_n} =
\ket{E_{j-1}},
\end{align}
where we use $\ket{-1} \equiv 0$ as a convenient shorthand notation. Now
we can apply a substitution $i'_1 = i_1 - 1$ and change the summation limit
(thus requiring $j < k$).

Let us now show that all stabilized states $\ket{\psi}$ have the form of \eqref{eq:ejsum}.
We proceed by induction with respect to $n$, the number of particles. For
$n=1$ (inductive basis), all basis states are stabilized. Now, let us assume
that the condition works up to a given $n$. As $K_{(1)}$ reduces the total
number of excitations by one, it suffices to look at subspaces of fixed $j$.
Together with the inductive assumption (in particular, the fact that the first
$n$ particles must remain in a permutation symmetric state after application of
$K_{(1)}$) we get a general form
\begin{align}
\ket{\xi} = \sum_{l=0}^{j} \beta_l \ket{E_{j-l}} \ket{l}.
\end{align}

To find the actual constraints on $\beta_l$, we just note that the assumed
symmetry of $K_{(1)}\ket{\xi}$ implies that 
\begin{align}
 K_{(1)}\ket{\xi} &= \sum_{l=0}^{j-1} \beta_l \ket{E_{j-l-1}} \ket{l}\\
  = K_{(n+1)}\ket{\xi} &= \sum_{l=1}^{j} \beta_l \ket{E_{j-l}} \ket{j-1}.
\end{align} 
Again, with a simple shift of index, and using the orthogonality of
the components, we get $\beta_l = \beta_{l+1}$. Thus, there is only one
state (up to a factor) for a given $j$ that remains symmetric after $K_{(1)}$.

\end{proof}

When considering SLOCC-equivalence, we may take $\ket{E_{k-1}}$ as a representative
of the states stabilized by $K_{(1)}$. The reason is that all other
states (with $\alpha_{k-1}\neq0$) can be built from it via an operator
$\sum_{j=0}^{k-1} \alpha_{k-1-j} K_{(1)}^j$. This operator is invertible, since its determinant is $\alpha_{k-1}^k$.
Throughout the derivation, we work with unnormalized states for convenience.
The properly normalized excitation state is given in equation \eqref{eq:excitationnormalized}.

\begin{theorem}
There is a unique (up to SLOCC operations) state stabilized by $B$
if and only if each Jordan block of $B$ has a distinct eigenvalue. 

A $n>2$ particle state $\ket{\psi}\in\cal{S}$, stabilized by $B$, is
unique (up to SLOCC) if and only if each block of its
Jordan form has a distinct eigenvalue and no other $B'$ exists with a greater number of
eigenvalues or a lesser number of Jordan blocks.
\end{theorem}

The formulation may seem complicated, but we want to exclude {\em degenerate} states, which are also stabilized by other operators.
For the excitation state we want to ensure that amplitude of the $(d-1)$ excitations is non-zero (otherwise it is stabilized also by a matrix with two eigenvalues), or, for the GHZ states, that all amplitudes are non-zero (otherwise, two eigenvalues can be merged into one, forming a single Jordan block).
For example, a three qutrit pure state $\ket{000}+\ket{111}$ is stabilized by a matrix with its Jordan block structure
$\{ \{ 1 \}, \{ 1 \}, \{ 1 \} \}$ (as for the GHZ state). However, unlike $\ket{000}+\ket{111}+\ket{222}$ (the GHZ state), it is also stabilized by a matrix with one less Jordan block $\{ \{ 1 \}, \{ 2 \} \}$.

\begin{proof}
$'\Leftarrow'$

We have already shown that the GHZ-like state with all amplitudes
different from zero is unique, as well as the excitation state
with non-zero amplitude for the highest excitation. It follows as well for any
state without blocks of the same eigenvalue, as the problem can be split into a problem for each eigenvalue.
\begin{itemize}
	\item If any amplitude is zero in the GHZ-like case, the state is also stabilized by a $B$ with a Jordan block of dimension two.
	\item If the amplitude of for the highest excitation is zero, in the excitation state, the state is also stabilized with a $B$ with one more eigenvalue.
\end{itemize}

$'\Rightarrow'$

If there are two blocks with the same eigenvalue, then we can take two
one-particle eigenvectors $\ket{\mu}$ and $\ket{\nu}$ having the same
eigenvalue. Let us look at the projection of $\ket{\psi}$ on the subspace
spanned by $\text{Sym}^n(\text{lin}\{\ket{\mu}, \ket{\nu}\})$. Then, in
particular, a linear combination with non-zero coefficients of elements with
zero, one and two $\ket{\nu}$ states among all other $\ket{\mu}$ does not give
rise to more blocks or eigenvalues, but gives rise to some states which cannot
be interchanged with local operations.

\end{proof}

\begin{corollary}

The number of Jordan block structures with non-degenerate eigenvalues
is the same as the number of integer partitions of $d$~\cite{oeisA000041}. 

A general construction of such state is 
\begin{align}
\bigoplus_{i=1}^{\#\mathrm{blocks}} \ket{E_{k_i}},\label{eq:unique_general}
\end{align}
where $k_i$ is the dimension of the $i$-th Jordan block, in descending order.
In particular, for GHZ there are only blocks of size $k_i=1$,
whereas for the excitation state there is only one block, $k_1=d$.

\end{corollary}

It is also relevant to ask about stabilized states for $B$ whose Jordan
decomposition contains two different blocks with the same eigenvalue. 
Let us use a one-particle basis given by $\ket{i^{(b)}}$, where $i$ denotes
excitation-level (i.e. the largest $i$ such that $J^i$ acting on this vector is
non-zero) and $b$ the Jordan block to which it belongs. First, we notice
that the sum of the excitations $j$ in a given state is decreased by $1$ after
action of $K_{(1)}$. Second, we notice that the excitations can be distributed
among all Jordan subspaces which are {\em big enough} (i.e. all blocks of size
strictly lesser than $j$). Moreover, the distribution among such Jordan subspaces
needs to be permutation-invariant.

\begin{theorem}
An unnormalized state of excitation $j$ distributed among $s$
blocks (with weights $n_1,n_2,\ldots$ adding up to $n$, related to distribution of excitations among Jordan blocks) reads

\begin{equation}
\ket{E_j^{n_1,n_2,\ldots}} = \sum_{\vec{b}: \#i =
n_i}\sum_{i_1+\ldots+i_n=j}\ket{i_1^{(b_1)}}\ldots
\ket{i_n^{(b_n)}}.\label{eq:ej_multimode}
\end{equation}

We will show by induction that only states of the form
$\ket{E_j^{n_0,n_1,\ldots}}$ are stabilized by such $J$.

\end{theorem}

For example, one excitation $j=1$ among two particles, distributed among two
modes ($n_1=1$, $n_2=1$) reads
\begin{align}
\ket{E_1^{1,1}} &= \ket{0^{(1)}1^{(2)}} + \ket{1^{(1)}0^{(2)}}\\
&+ \ket{0^{(2)}1^{(1)}} + \ket{1^{(2)}0^{(1)}}.
\end{align}

\begin{proof}
The induction basis is for $n=1$
and holds trivially (as it works for all states). So let us assume that
\eqref{eq:ej_multimode} holds for $n$.

For $n+1$ particles, a generic state with fixed $j$ and $n_1,n_2,\ldots$ is
\begin{align}
\sum_{l=0}^j \sum_{b=1} \beta_{l,b}
\ket{E_{j-l}^{n_1-\delta_{b1},n_1-\delta_{b1},\ldots}} \ket{l^{(b)}}.
\end{align}
Applying $J_{(1)}$ and $J_{(n+1)}$ on the state above, we get a relation
$\beta_{l,b}=\beta_{l+1,b}=\beta_b$. Moreover, from the condition of permutation
symmetry for blocks (i.e. components with the same $(b)$) we get that all
$\beta$ need to be the same, so it is of the form \eqref{eq:ej_multimode}.
\end{proof}

This finalizes the classification of symmetric states for which \eqref{eq:bbinv} holds.

\section{Invariants as functions of creation and annihilation operators}
\label{s:polynomial-invariants}

Having shown that problem of relating bosonic states by linear optics is the same that asking whether they are equivalent with respect to local unitary operations,
we focus on specific methods for bosonic states that provide analytic invariants, basing on our work \cite{Migdal2014ffdag}.
These invariants are built upon $f^\dagger$, the homogeneous polynomial on the creation operators which transforms the vacuum into our state. We present two families of LU-invariants, i.e.: two sets of complex-valued functions on the Hilbert space which are invariant under linear optics:

\begin{itemize}
\item The spectrum of the operator $f f^\dagger$.
\item The moments: vacuum expectation values of the operators $f^k f^{\dagger k}$, for any natural $k$.
\end{itemize}

The considered invariants are both simple to calculate and, as we will show, sufficient to distinguish states in many practical situations, even some states which are generally difficult to handle.

This part of the work is organized as follows.
In Section~\ref{s:ff} we present the construction and relevance of spectral invariants related to the operator $f f^\dagger$.
We show that, despite being infinite dimensional, this operator can be easily diagonalized, as it separates into blocks of fixed numbers of particles (not related to the photon count) which are related to many-body correlators.
In Section~\ref{s:replicas} we discuss the second set of invariants: vacuum expectation values of $f^k f^{\dagger k}$.
It corresponds to the projection of the tensor power of $k$ copies of our state (in the particle basis) onto the completely symmetric Hilbert space.
In Section~\ref{s:ffdag-examples} we apply our methods in concrete examples. We show that, using our invariants, we can solve the LU-equivalence problem for two particles in two modes and for three particles in two modes. We also study which states from the four-particle singlet subspace can be reached using linear optics from another state in the same singlet subspace.
Moreover, we show that, at least in some cases, $k$-particle blocks of $f f^\dagger$ provide more invariants than $k$-particle reduced density matrices.
In Section~\ref{s:mode-product} we propose an interferometric scheme that, in principle, allows for a direct measurement of this set of invariants.
Moreover, such scheme allows direct experimental creation of states given by the polynomial $f^{k}$ for an arbitrary $k$.
Some technical discussions are left for Sec.~\ref{app:schwinger}, where we introduce Schwinger-like representation for expressing arbitrary $k$-body correlations in terms of normally ordered creation and annihilation operators.

\subsection{Reduced density matrix}
\label{sec:reduced-density-matrix}

One of straightforward methods for checking whether two states are LU-equivalent is comparing spectra of their reduced density matrices.
As we are working on permutation symmetric states, it does not matter which particle we choose, and thus we have family of density matrices parameterized by a number from $1$ to $n$.
The simplest one is the one-particle density matrix
\begin{equation}
\rho_{ij} = \bra{\psi} a_j^\dagger a_i \ket{\psi}.
\end{equation}
For two particles, two state are LU-equivalent if, and only if, they have the same spectra of the one-particle reduced density matrix, see Schmidt decomposition in Sec.~\ref{s:two_particles}.
The same condition holds for the Gaussian states, as shown in Sec.~\ref{sec:gaussian-states}.

In general it is a necessary, but not sufficient, condition for LU-equivalence.
Even for a pure state of three symmetric qubits it is no longer the case --- reduced density matrices offer $1$ invariant, whereas there are three, see Sec.~\ref{s:three-qubits}. 
For general relation of expectation values of creation and annihilation operators and reduced density matrices, see Sec.~\ref{app:schwinger}.

\subsection{Spectral method}\label{s:ff}

Let us consider the $d$-mode, $n$-particle bosonic state given in equation (\ref{def.f}), $\ket{\psi}=f^\dagger\ket{\Omega}$, where $f(a_1,\cdots,a_d)$ is a homogeneous polynomial of degree $n$ in the annihilation operators for the modes. Now, let us consider the operator $ff^\dagger$.

We will show that:
\begin{itemize}
\item its spectrum is invariant with respect to $SU(d)$ transformations (\ref{eq:u-transf}),
\item it may be decomposed into an infinite number of blocks of finite size, but
\item the first $n$ blocks suffice to reconstruct the state.
\end{itemize}

\subsubsection{Invariance of the spectrum}
\begin{theorem}
The spectrum of $ff^\dagger$ is invariant with respect to arbitrary rotations between the modes, that is,
\begin{equation}
\text{Sp}\left[f(\vec{a}) f^\dagger(\vec{a})\right] =
\text{Sp}\left[f(U\vec{a}) f^\dagger(U\vec{a})\right]
\label{ff.invariance}
\end{equation}
for every $U\in SU(d)$. 
\end{theorem}
\begin{proof}
Each unitary operator acting on the modes $U = \exp(i H)$ (with Hermitian $H$) can be promoted to act on the full Fock-space via a second quantization extension:
\begin{align}
\tilde{U} = \exp\(i \sum_{i,j=1}^{d} H_{ij} a_i^\dagger a_j \),
\end{align}
where $\tilde{U}\cong U^{\otimes n}$ on our Hilbert space ${\cal S}_n^d$. This operator $\tilde{U}$ is unitary and acts on monomials in a natural way, i.e.: $\tilde{U}^\dagger a_j \tilde{U} = \sum_i U_{ji} a_i$, which can be checked with the Hadamard lemma. Consequently,
\begin{align}
f(U\vec{a}) f^\dagger(U\vec{a}) = \tilde{U}^\dagger f(\vec{a}) f^\dagger(\vec{a}) \tilde{U},
\end{align}
i.e.: the two operators are unitarily related and, thus, they have the same spectrum.
\end{proof}

\subsubsection{Block Decomposition}
\label{s:block-decomposition}

Since operator $f$ is a {\em homogeneous} polynomial of degree $n$ on the annihilation operators, each summand in operator $ff^\dagger$ contains $n$ creation and $n$ annihilation operators. Thus, $ff^\dagger$ preserves the number of photons $k$, and decomposes into blocks $ff^\dagger|_k$. Let $\vec k$ and $\vec k'$ be multi-indices with $|\vec k|=|\vec k'|=k$. Then, matrix elements of $ff^\dagger|_k$ can be shown to correspond to {\em correlators} of our state:
\begin{align}
&\bra{\vec k'}\; f f^\dagger\; \ket{\vec k} =
\bra{\Omega}\; \tilde a_{\vec k'}\; 
f f^\dagger\; 
\tilde a_{\vec k}^\dagger \; \ket{\Omega}  \nonumber\\ =& 
\bra{\Omega}\; f\; \tilde a_{\vec k'} \tilde a^\dagger_{\vec  k} \; 
f^\dagger\; \ket{\Omega} = \bra{\psi} \; \tilde a_{\vec k'} \tilde
a^\dagger_{\vec k}\; \ket{\psi}.
\label{eq:spec2correl}
\end{align}

For example, for two modes and particle numbers $k\in\{0, 1,2\}$, the blocks are given by:
\begin{align}
f f^\dagger|_{k=0} &=
\left[%
\begin{matrix}
\bra{\psi}1\ket{\psi}
\end{matrix}
\right]
\\
f f^\dagger|_{k=1} &=
\left[%
\begin{matrix}
\bra{\psi}a_1 a_1^\dagger\ket{\psi} & \bra{\psi}a_1 a_2^\dagger\ket{\psi} \\
\bra{\psi}a_2 a_1^\dagger\ket{\psi} & \bra{\psi}a_2 a_2^\dagger\ket{\psi}
\end{matrix}
\right]
\\
f f^\dagger|_{k=2} &=
\end{align}%
\begin{equation}
\left[%
\begin{matrix}
\bra{\psi} \frac{a_1^2 a_1^{\dagger 2}}{2} \ket{\psi}
& \bra{\psi} \frac{a_1^2 a_1^\dagger a_2^\dagger}{\sqrt{2}} \ket{\psi}
& \bra{\psi} \frac{a_1^2 a_2^{\dagger 2}}{2} \ket{\psi}
\\
\bra{\psi} \frac{a_1 a_2 a_1^{\dagger 2}}{\sqrt{2}} \ket{\psi}
& \bra{\psi} a_1 a_2 a_1^\dagger a_2^\dagger \ket{\psi}
& \bra{\psi} \frac{a_1 a_2 a_2^{\dagger 2}}{\sqrt{2}} \ket{\psi}
\\
\bra{\psi} \frac{a_2^2 a_1^{\dagger 2}}{2} \ket{\psi}
& \bra{\psi} \frac{a_2^2 a_1^\dagger a_2^\dagger}{\sqrt{2}} \ket{\psi}
& \bra{\psi} \frac{a_2^2 a_2^{\dagger 2}}{2} \ket{\psi}
\end{matrix}
\right]\nonumber.
\end{equation}

The matrix elements of $ff^\dagger|_k$ are $k$-particle correlators. For $k=0$, the only matrix element is the norm of the state.
Note that the spectrum of $f f^\dagger$ is real, as each block $f f^\dagger|_k$ is a Hermitian matrix.

Unitary rotations do not change the particle count. Consequently, the block structure is preserved under rotations and, thus, the $\text{Sp}[f f^\dagger|_k]$ are invariants. If the eigenvalues for two states differ, $\text{Sp}[f_1 f_1^\dagger|_k] \neq \text{Sp}[f_2 f_2^\dagger|_k]$, then the two states {\em can not} be related by a unitary rotation of the modes.
The converse is, in general, not true --- states related by complex conjugation (of $f$), so preserving the spectrum, are not necessarily related by linear optics (see \ref{s:three-qubits} for an example).
It, however, remains an open question whether the converse (up to complex conjugation) is true.

Instead of the eigenvalues, we may compute the characteristic polynomial:
\begin{align}
w_k(\lambda) = \det\left[ f f^\dagger|_{k} - \lambda \mathbbm{I} \right].
\label{characteristic.polynomial}
\end{align}
Since its coefficients are in one-to-one correspondence with the spectrum, the method is equally powerful. Moreover, the coefficients of $w_k(\lambda)$ are polynomials in the coefficients of $f$, which is closer in spirit to formulation of Hilbert's theorem. An alternative, but equivalent, route is to investigate the moments $f f^\dagger|_k$,  $\Tr[(f f^\dagger|_k)^l]$. They are in one to one correspondence with the characteristic polynomial $w_k(\lambda)$ by the virtue of Newton identities \cite{Mead1992}.
For $k=1$, the block is related to the single-particle reduced density matrix (see Sec.~\ref{sec:reduced-density-matrix}), i.e.:
\begin{align}
\rho_1 = f f^\dagger|_{k=1} - n \mathbbm{I}.
\end{align}
For $k>1$ we do not recover the reduced $k$-particle density matrix and, as we will show,
$f f^\dagger|_k$
can provide more entanglement invariants than the spectrum of the reduced density matrices with those respective particle numbers.

Even the first block can give interesting results. We can show that no-go observation for deterministically changing one Fock state into another using with linear optics.
Let us look at $f f^\dagger|_1$. As it is a Fock state, its matrix is diagonal (i.e terms $\langle \psi | a_i a_j^\dagger |\psi \rangle$ vanish for $i\neq j$). The diagonal values, and therefore the eigenvalues, are $\bra{\psi} a_i a_i^\dagger \ket{\psi} = n_i + 1$. As they are invariants, two Fock states can be deterministically related by linear optics if and only if they have the same photon counts (up to a permutation of modes).

\subsubsection{Correlators and reconstruction}

Knowledge of $f f^\dagger |_k$ for all block particle numbers $k\leq n$ suffices to reconstruct the state $f^\dagger\ket{\Omega}$. The reconstruction strategy is to build the matrix elements of the corresponding density matrix
\begin{align}
\rho_{\vec{n} \vec{n}'} =
\bra{\psi} a_{\vec n}^\dagger a_{\vec n'} \ket{\psi},
\end{align}
which can be done by using the commutation relations in order to express the anti-normally ordered terms into terms with normal ordering.

However, we do not claim that higher blocks with $k>n$ are not important. While they are not required to reconstruct the state, there might be pairs of states whose polynomials $w_0$ up to $w_n$ coincide, yet their $w_k$ differ for some $k>n$.
That is, eigenvalues do not capture relative orientation of eigenvectors for different blocks. Eigenvalues for $k>n$ might incorporate relations between eigenvectors for $k\leq n$.

Let us provide a more straightforward way to reconstruct the state, which does not involve calculating inverting the normal ordering of the operators. Let us recall the notion of {\em frame representation} of a many qudit state \cite{Ferrie2011}. Let $\{\sigma^i\}$ be an orthogonal (in trace norm) set of generators of $SU(d)$ plus the identity (i.e. a basis for $d\times d$ Hermitian matrices). For $SU(2)$ we may just choose the Pauli matrices: $\{\mathbbm{I},\sigma^x,\sigma^y,\sigma^z \}$. Any density matrix of a $n$-qudit state can be written as:
\begin{align}
\rho = \sum_{i_1,\cdots,i_n} t_{i_1 i_2 \ldots i_n}
\sigma^{i_1} \otimes \sigma^{i_2} \otimes \ldots \otimes \sigma^{i_n}
\equiv \sum_{\vec\imath} t_{\vec\imath} \sigma^{\vec\imath},
\label{frame.rep}
\end{align}

Note that for permutation-symmetric states, $t_{i_1 i_2 \ldots i_n}$ must be permutation-symmetric. Since the $\{\sigma^i\}$ are orthogonal, the state can be reconstructed from the expectation values of {\em strings} of $\sigma^{i}$ operators:
\begin{align}
t_{i_1 i_2 \ldots i_n} = \frac{1}{2^n}\Tr\left[ \sigma^{i_1} \otimes \sigma^{i_2} \otimes \ldots \otimes \sigma^{i_n} \;\rho \right].
\label{reconstruction}
\end{align}

Expectation values of permutation-symmetric strings of $\sigma^i$ can be obtained from the correlators $ff^\dagger|_k$, as shown in Appendix (\ref{app:schwinger}). The idea behind the proof is the use of a Schwinger-like representation,
related to the one for spin systems ---see \cite[Chapter 7.2]{Auerbach1994}, and develop identities of the form
\begin{align}
\bra{\psi} \left( \sum_{perm} \sigma^{\vec{\imath}} \right) \ket{\psi} = \bra{\Omega} f A(\vec\imath) f^\dagger \ket{\Omega},
\end{align}
where $A(\vec\imath)$ is a polynomial in creation and annihilation operators. From a practical perspective it allows calculating the expectation value without immersing everything in the full Hilbert space of distinguishable particles,  which has a very high dimension.

For example, for $d=2$, we get the following relation
\begin{gather}
\bra{\psi}
\sum_{perm} (\mathbbm{I})^{\otimes n_{I}}
\otimes (\sigma^x)^{\otimes n_x}
\otimes (\sigma^y)^{\otimes n_y}
\otimes (\sigma^z)^{\otimes n_z}
\ket{\psi}\label{eq:pauli_sym}\\
=\bra{\Omega} f
: \left( a^\dagger a + b^\dagger b \right)^{n_I}
\left( a^\dagger b + b^\dagger a \right)^{n_x}\label{eq:schwinger_form}\\
\times \left( - i a^\dagger b + i b^\dagger a \right)^{n_y}
\left( a^\dagger a - b^\dagger b\right)^{n_z} :
f^\dagger \ket{\Omega},\nonumber
\end{gather}
where $n_{I} + n_x + n_y + n_z = n$ (covering all symmetric correlators), the sum is over all $n!$ permutations and :expression: stands for the normal ordering, i.e. putting the creation operators on the left and the annihilation on the right. Note that, for most of this chapter, we use anti-normal ordering, as we work with operators of the form $f^k f^{\dagger k}$.

\subsection{Symmetric component of tensor powers}
\label{s:replicas}

An alternative set of invariants can be found by studying the symmetric component of tensor copies of a given multi-photon state, taken in the particle representation.

Typically, $\ket{\psi}_P^{\otimes k}$ is not permutation-symmetric, therefore it does not describe a boson state. However, we will show that its projection on the symmetric subspace is proportional to $f^{\dagger k}\ket{\Omega}$, a $kn$-photon state in $d$ modes.

Let us give an example, with $n=2$ and $d=2$, $\ket{\psi} = \ket{1,1} = \frac{1}{g\sqrt{2}} (\ket{12}_P+\ket{21}_P)$. If we multiply it tensorially with itself, we get $\ket{\psi}_P^{\otimes 2} = \frac{1}{2}(\ket{12}_P+\ket{21}_P)\otimes (\ket{12}_P+\ket{21}_P)$. This is {\em not} a valid photon state, because it is {\em not} permutation-symmetric:
\begin{gather}
\tfrac{1}{2} \left( \ket{1212}_P + \ket{1221}_P \right.\label{eq:nonsym-prod}\\ 
\left. + \ket{2112}_P + \ket{2121}_P \right)\nonumber
\end{gather}

Nonetheless, it can be projected on the permutation-symmetric subspace, $\hbox{Sym}^{kn}(\C^d)$.
Let $\P^{(kn)}_{sym}$ stand for that projector, where the upper index represents the number of particles to be symmetrized, in this case --- $kn$.
Then,
\begin{equation}
\bra{\psi}_P^{\otimes 2} \P^{(4)}_{sym} \ket{\psi}_P^{\otimes 2} = \frac{2}{3}
\end{equation}
because \eqref{eq:nonsym-prod} contains 4 out of 6 possible permutations,
\begin{equation}
\P^{(4)}_{sym} \ket{1212}_P =
\tfrac{1}{6} \left( \ket{1122}_P + \text{permutations} \right).
\end{equation}

In order to make the LU-invariance of those values $\bra{\psi}_P^{\otimes k} \P^{(kn)}_{sym} \ket{\psi}_P^{\otimes k}$ manifest, we will show their relation to
\begin{equation}
\bra{\Omega} f^k f^{\dagger k} \ket{\Omega}
\label{fkfk}
\end{equation}
i.e.: the {\em vacuum expectation values of} $f^k f^{\dagger k}$ for all $k\in\N$. These are easy to compute and their invariance is straightforward, since the vacuum is rotation-invariant. Thus, we will prove the following:
\begin{theorem}%
For every homogeneous polynomial $f$, such that $\ket{\psi} = f^\dagger \ket{\Omega}$, the state generated by its $k$-th power is proportional to the state $\ket{\psi}_P^{\otimes k}$ projected on the fully symmetric space of all particles, that is,
\begin{equation}
f^{\dagger k} \ket{\Omega} = \tfrac{\sqrt{(kn)!}}{\sqrt{(n!)^k}} \P^{(kn)}_{sym} \ket{\psi}^{\otimes k}_P,
\label{eq:fk_and_symmetrization}
\end{equation}
so, in particular:
\begin{equation}
\bra{\Omega}f^kf^{\dagger k} \ket{\Omega} = \tfrac{(kn)!}{(n!)^k} \bra{\psi}_P^{\otimes k} \P^{(kn)}_{sym} \ket{\psi}_P^{\otimes k}.
\label{fkfk.projector}
\end{equation}
\end{theorem}
\begin{proof}
Let $\{\vi^{(1)},\cdots,\vi^{(k)}\}$ be $k$ multi-indices,
denoting photon count at each mode, i.e., for the vector with index $m$, we have 
\begin{equation}
\vi^{(m)} = \{n^{(m)}_1,\cdots,n^{(m)}_d\}.
\end{equation} 
Let us denote by $|\vi^{(m)}|=\sum_l n^{(m)}_l$ the total photon count. The monomial operator defined in \eqref{eq:multia}, $\tilde a^\dagger_{\vi^{(1)}+\cdots+\vi^{(k)}}$, can be written in terms of the individual normalized monomials as
\begin{equation}
\tilde a^\dagger_{\vi^{(1)}} \tilde a^\dagger_{\vi^{(2)}} \cdots \tilde a^\dagger_{\vi^{(k)}} = M(\vi^{(1)},\cdots,\vi^{(k)})\ \tilde a^\dagger_{\vi^{(1)}+\cdots+\vi^{(k)}}
\label{global.creator}
\end{equation}
where
\begin{equation}
M(\vi^{(1)},\cdots,\vi^{(k)})\equiv \prod_{l=1}^d
\sqrt{\frac{ (n_{l}^{(1)}+\cdots+n_{l}^{(k)})!}{(n_{l}^{(1)})! \cdots (n_{l}^{(k)})! }}
\label{combinatorial.factor}
\end{equation}
is the normalization factor. Let us express $f^{\dagger k}\ket{\Omega}$ as a sum of terms of this kind:
\begin{align}
  (f^\dagger)^k \ket{\Omega} &= \sum_{\vi^{(1)},\cdots,\vi^{(k)}} \alpha_{\vi^{(1)}} \cdots \alpha_{\vi^{(k)}}\; \nonumber \\
&\times M(\vi^{(1)},\cdots,\vi^{(k)}) \;\tilde a^\dagger_{\vi^{(1)}+\cdots+\vi^{(k)}} \ket{\Omega}
\label{eq:fk}
\end{align}
so, the coefficient for $\ket{\vec{I}}\equiv \tilde{a}^\dagger_{\vec{I}} \ket{\Omega}$ is
\begin{equation}
\sum_{\vi^{(1)}+\cdots+\vi^{(k)}=\vec I} \alpha_{\vi^{(1)}}\cdots \alpha_{\vi^{(k)}} \cdot M(\vi^{(1)},\cdots,\vi^{(k)})
\end{equation}
where $\vec I$ is a multi-index for $nk$ photons in $d$ modes.

Now, let us consider the right hand side of \eqref{eq:fk_and_symmetrization}. The tensor product $\ket{\psi}^{\otimes k}$ can be written as:
\begin{equation}
  \ket{\psi}_P^{\otimes k} =
  \sum_{\vi^{(1)},\cdots,\vi^{(k)}} 
  \alpha_{\vi^{(1)}}\cdots \alpha_{\vi^{(k)}}
  \ket{\vi^{(1)}}_P \otimes \cdots \otimes \ket{\vi^{(k)}}_P,
\label{eq:particle-power-k}
\end{equation}
Notice that the action of several partial projections on symmetric subspaces followed by a global projection on the symmetric subspace is equivalent to just the final global projection. Consequently,
\begin{align} 
&\P_{sym}^{kn} \left( \ket{\vi^{(1)}}_P\otimes\cdots\otimes\ket{\vi^{(k)}}_P
  \right) \nonumber \\
&= N(\vi^{(1)})\cdots N(\vi^{(k)}) \P^{(kn)}_{sym} \nonumber\\
&\left(
\P^{(n)}_{sym}(\ket{\vi^{(1)}}_A) \otimes \cdots \otimes
\P^{(n)}_{sym}(\ket{\vi^{(k)}}_A) \right) \\
&= N(\vi^{(1)})\cdots N(\vi^{(k)}) \P^{(kn)}_{sym} \left(
\ket{\vi^{(1)}+\cdots+\vi^{(k)}}_A \right) \nonumber \\
&= \frac{N(\vi^{(1)})\cdots N(\vi^{(k)})}{N(\vi^{(1)}+\cdots+\vi^{(k)})}
\ket{\vi^{(1)}+\cdots+\vi^{(k)}} = \nonumber \\
&= \tfrac{\sqrt{(kn)!}}{\sqrt{(n!)^k}} M(\vi^{(1)},\cdots,\vi^{(k)}) \ket{\vi^{(1)}+\cdots+\vi^{(k)}}
\end{align}

Applying the above relations to \eqref{eq:particle-power-k} we get
\begin{align}
\P_{sym}^{kn} \ket{\psi}^{\otimes k}_P &=
\sum_{\vi^{(1)},\cdots,\vi^{(k)}}
\alpha_{\vi^{(1)}}\cdots \alpha_{\vi^{(k)}}
\tfrac{\sqrt{(kn)!}}{\sqrt{(n!)^k}}\\
&\times M(\vi^{(1)},\cdots,\vi^{(k)})
\ket{\vi^{(1)}+\cdots+\vi^{(k)}}\nonumber
\end{align}
which is a state proportional to \eqref{eq:fk}, with the proportionality factor $\sqrt{(kn)!/(n!)^k}$, thus we have shown \eqref{eq:fk_and_symmetrization}. 

\end{proof}


This tensor product symmetrization trick bears resemblance to the use of Clebsch-Gordan coefficients. Indeed, already for $k=2$ the result is useful: $\ket{\psi}^{\otimes 2}$ is not permutation-symmetric unless $\ket{\psi}=\ket{\phi}^{\otimes n}$ for some single-particle state $\ket{\phi}$.

It is possible to prepare an experimental setup to measure $\langle f^k f^{\dagger k} \rangle$. We have to prepare $k$ copies of the state and project each $k$-tuple of modes into their symmetric combination. For example, if $k=2$, two modes are symmetrized using a beam-splitter. Then, $\langle f^2 f^{\dagger 2}\rangle$ is the probability amplitude for losing no photons in the procedure. In general, taking copies of bosonic states and calculating projections offers a way to measure multi-particle entanglement, since taking $k$ copies provides a way to measure R\'enyi entropy of order $k$ of the given subsystems \cite{Daley2012}.

There is another interpretation of $\langle f^k f^{\dagger k} \rangle$ in polynomial language. The quantity we are investigating is known as the {\em Bombieri norm} of homogeneous polynomials \cite{Beauzamy1990} (in this case, $f^k$), which is known to be invariant under unitary rotations of the variables.
This quantity can be expressed as an integral of $|f(\vec{a})|^{2k}$ over the (complex) unit sphere $|\vec{a}|=1$, \cite{Pinasco2012, Pinasco2005} (equivalently, see \cite[Lemma 15]{Aaronson2010}, where it is called Fock Inner Product).

\subsection{Examples}

\label{s:ffdag-examples}

The previous two sections have introduced two sets of LU-invariants for $n$-photon states in $d$-modes. The question to be addressed in this section is the following: can those invariants help us determine the LU-equivalence classes of relevant states? We will start our discussion with a benchmark problem, which can be solved in many different ways: $n=2$ photons in $d=2$ modes. Then, we will proceed to the case of $n=3$ particles, still in $d=2$ modes, which is the first non-trivial case, although it is well understood. We will show that, in that case, the right number of polynomial invariants is recovered. Our last example is a much more complicated system: $n=4$ photons in $d=8$ modes with some additional symmetries.

\subsubsection{2 particles in 2 modes}

The simplest example is $n=2$ particles in $d=2$ modes:
\begin{align}
f = \alpha_{20} \tfrac{a_1^{2}}{\sqrt{2}} + \alpha_{11} a_1 a_2 + \alpha_{02} \tfrac{a_2^{2}}{\sqrt{2}}.
\end{align}
There is just a single invariant. Let us study how we can obtain it using the methods described in this paper. In our case it suffices to look at a block of $k=1$ particles:
\begin{align}
\left[%
\begin{matrix}
3|\alpha_{20}|^2 + 2 |\alpha_{11}|^2 + |\alpha_{02}|^2 & 
\sqrt{2}(\alpha_{20}^\star \alpha_{11} + \alpha_{11}^\star \alpha_{02}) \\
\sqrt{2}(\alpha_{20} \alpha_{11}^\star + \alpha_{11} \alpha_{02}^\star) & 
|\alpha_{20}|^2 + 2 |\alpha_{11}|^2 + 3|\alpha_{02}|^2 
\end{matrix}
\right]
\end{align}
Its characteristic polynomial is
\begin{equation}
w_2(\lambda) = \lambda^2
- \Tr \left( f f^\dagger|_{k=1} \right) \lambda
+ \det \left( f f^\dagger|_{k=1} \right),
\end{equation}
where coefficients are
\begin{align}
\Tr \left( f f^\dagger|_{k=1} \right)
&= 4\left(|\alpha_{20}|^2 + |\alpha_{11}|^2 + |\alpha_{02}|^2\right),
\nonumber \\ 
\det \left( f f^\dagger|_{k=1} \right)
&= 4 \left(|\alpha_{20}|^2 + |\alpha_{11}|^2 +
|\alpha_{02}|^2\right)^2 \nonumber \\ 
- ( |\alpha_{20}|^2 &- |\alpha_{02}|^2 )^2 + 2 |\alpha_{20}^\star \alpha_{11} +
\alpha_{11}^\star \alpha_{02}|^2. 
\label{eq:twotwodet} 
\end{align}
The trace gives only the normalization, which is the same information contained in $f f^\dagger|_{k=0}$, and which we can set to $1$. The determinant, on the other hand, gives a new invariant.

Alternatively, we can factorize the (degree 2) polynomial: $f=f_1f_2$. In other terms, we can make use of the Majorana stellar representation \cite[Ch. 7]{BengtssonZyczkowski2006book}:
\begin{align}
\ket{\psi} &= \frac{1}{\sqrt{N}}f_1^\dagger f_2^\dagger
\ket{\Omega}\\ & = \frac{1}{\sqrt{2N}}\left(
\ket{\phi_1}_P \otimes \ket{\phi_2}_P
+ \ket{\phi_2}_P \otimes \ket{\phi_1}_P \right),
\end{align}
where $\sqrt{N}$ is a normalization factor and $f_i^\dagger \ket{\Omega} \equiv \ket{\phi_i}$. Since $U\in SU(2)$ acts on the representation as a simultaneous rotation of the points, for two particles the only invariant is the angle between the states, or equivalently $|\braket{\phi_1}{\phi_2}|^2$. A straightforward (albeit tedious) calculation gives
\begin{align}
|\braket{\phi_1}{\phi_2}|^2 = 
\frac{|\alpha_{20}|^2 + |\alpha_{11}|^2 + |\alpha_{02}|^2 - |\alpha_{11}^2 - 2 \alpha_{20} \alpha_{02}|}
{|\alpha_{20}|^2 + |\alpha_{11}|^2 + |\alpha_{02}|^2 + |\alpha_{11}^2 - 2 \alpha_{20} \alpha_{02}|}.
\end{align}
Along with the normalization condition it yields the invariant
\begin{align}
|\alpha_{11}^2 - 2 \alpha_{20} \alpha_{02}|^2 
&= 3  - \det \left( f f^\dagger|_{k=1} \right).
\end{align}
The above is $0$ and $1$ for orthogonal and parallel vectors $\ket{\phi_i}$, respectively.

It is also possible to find the $\langle f^k f^{\dagger k}\rangle$ invariants associated to $k$ copies. For $k=2$ we obtain:
\begin{align}
\tfrac{2^2}{4!}\bra{\Omega} f^2 f^{\dagger 2} \ket{\Omega} = 1 -
\tfrac{1}{3}|\alpha_{11}^2 - 2 \alpha_{20} \alpha_{02}|^2.
\end{align}
In particular, for each orbit under linear optics, we can give a representative, for example
\begin{align}
\frac{\cos(\theta)}{\sqrt{2}} a_1^2 + \frac{\sin(\theta)}{\sqrt{2}} a_2^2 
\end{align}
for $\theta\in[0,\frac{\pi}{4})$.

\subsubsection{Three qubits}
\label{s:three-qubits}

The case of $n=3$ photons in $d=2$ modes can be viewed as three qubits in a permutation-symmetric state, and is more involved. A full list of invariants is listed in \cite{Sudbery2000}. Disregarding mirror-reflection (i.e.: anti-unitary operators) there are 6 invariants, which reduce to 4 when we take into account normalization and permutation-symmetry. A normal form can be employed \cite{acin2000generalized, Acin2001a, Carteret2000} which, when particularized to a permutation-symmetric state, gives
\begin{align}
\ket{\psi} = &p \left(\ket{001}_P+\ket{010}_P+\ket{100}_P\right)/\sqrt{3} \nonumber\\ +
  &q \ket{111}_P + r \exp(i \varphi) \ket{000}_P,
\label{eq:acin3form}
\end{align}
where all parameters ($p$, $q$, $r$, $\varphi$) are real.
In this section, we use modes $\{0,1\}$, which are more prevalent in description of qubits, $\{1,2\}$ (in most of this paper we start enumeration from $1$).
Or, in polynomial notation:
\begin{align}
f = \alpha_{30} \tfrac{a_0^3}{\sqrt{6}}
  + \alpha_{21} \tfrac{a_0^2 a_1}{\sqrt{2}}
  + \alpha_{03} \tfrac{a_1^3}{\sqrt{6}},
\end{align}
where $\alpha_{30}$ is complex and both $\alpha_{21}$ and $\alpha_{03}$ are real parameters.

Our main result is that both the set of moments $\langle f^k f^{\dagger k}\rangle$ with $k\leq 5$ and the characteristic polynomials of the blocks $ff^\dagger|_{k\leq 2}$ {\em provide all invariants}. This can be checked by computing the matrix of partial derivatives of these invariants with respect to the parameters determining state \eqref{eq:acin3form} at, e.g., the point $(p=q=r=1, \varphi=\pi/4)$, and observing that is has maximal rank.

This result implies that blocks of $ff^\dagger$ convey more information than reduced density matrices, which are known to provide only 2 invariants, including the normalization
(note that for $1$ qubits spectra of one-particle and two-particle reduced density matrix are the same).
Beyond this dimensionality test, it is relevant to test whether those invariants can distinguish between states related by complex conjugation (or reflection, in terms of the Majorana representation), i.e.: $\ket{\psi}$ and $\ket{\psi}^*$. In general, for $n \geq 3$, such states do not need to be related by a unitary transformation
(as, in the Majorana representation, 3 indistinguishable unit vectors need not to have mirror symmetry).
Unfortunately, neither moments nor block spectra can distinguish a state from its complex conjugate (as we already noted in Sec. \ref{s:block-decomposition}).

\subsubsection{Four-particle singlet state}
\label{s:four-particle-singlet}

\begin{figure}[!htbp]
\centering
\includegraphics[width=0.35\textwidth]{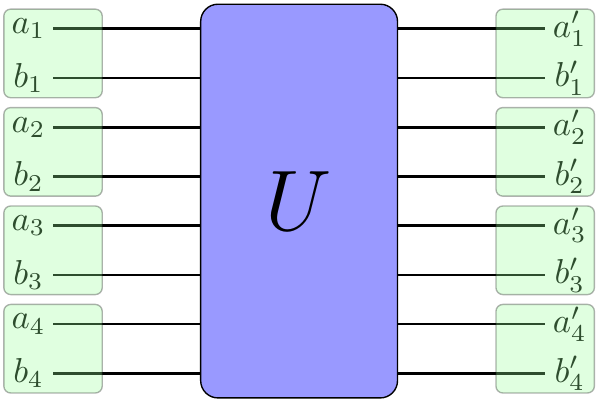}
\caption{Linear transformations for a state with 4 photons distributed among 8 modes, $\mathcal{S}^8_4$. We consider states having exactly one photon in each pair of nodes (denoted by green boxes).
This subspace is equivalent to the Hilbert space of 4 distinguishable particles, $\mathcal{H}_4$.
Furthermore, we study singlet states, i.e. states that are invariant with respect to $U=V^{\otimes 4}$, for all unitary $V$, where each $V$ acts on the respective pair of modes.}
\label{fig:4photons8modes}
\end{figure}

As a more interesting example we consider $n=4$ photons in $d=8$ modes, composing four qubits whose singlet-subspace determines a logical qubit, see Fig.~\ref{fig:4photons8modes}. There are three Hilbert spaces that are relevant for this scenario: the total Hilbert space $\mathcal{S}^8_4$, the 4-qubit subspace $\mathcal{H}_4$, and the two-dimensional singlet subspace $\mathcal{H}_s$, which determines the logical qubit, structured by the following inclusions:
\begin{equation}
\mathcal{S}^8_4 \supset \mathcal{H}_4 \supset \mathcal{H}_s \ .
\label{eq:singlet-subsubspace}
\end{equation}

We address here the following natural question: starting with a particular singlet state $\ket{\psi}\in\mathcal{H}_s$, which singlet states (also in $\mathcal{H}_s$) can be obtained from it using only linear optics? Before proceeding further, let us first describe the details of the construction of the 4-qubit and the singlet subspaces of $\mathcal{S}^8_4$.  

Let us denote the by $\{a_i,b_i\}_{i=1}^4$ the four annihilation operators required to span $\mathcal{S}^8_4$, where the $a_i$ refer to horizontal and the $b_i$ to vertical polarizations of the $i$-th beam. We define the 4-qubit subspace $\mathcal{H}_{4}\subset \mathcal{S}^{8}_4$, as a subspace spanned by states that have exactly one particle in each of the four pairs of modes: $(a_i,b_i)$. This subspace has dimension 16 and is isomorphic to the Hilbert space of four distinguishable qubits $\left(\C\right)^{\otimes 4}$. Action of the local unitary group $SU(2)^{\otimes 4}$ on $\mathcal{H}_4$ is modeled by the action of global linear optics operations that do not mix pairs $(a_i,b_i)$. The two-dimensional singlet subspace, $\mathcal{H}_s$, is defined as the subspace of $\mathcal{H}_4$, which is invariant under the action of any collective unitary rotations on all four qubits, $V^{\otimes 4}$. 

The above construction was first introduced in \cite{Zanardi1997} as the simplest example of a decoherence-free subspace for collective rotations, and it has been created experimentally \cite{Weinfurter2001}. In \cite{Migdal2011dfs} it was shown that the logical qubit is immune to one-particle loss and a protocol for quantum key distribution using such states and linear optics was provided.

Let us describe the structure of the singlet space in the mode description. For each pair of beams we can define the two-photon {\em singlet} state:
\begin{align}
s_{12} = \left( a_1 b_2 - b_1 a_2 \right) / \sqrt{2},
\end{align}
i.e. $s^\dagger_{12}\ket{\Omega} = \left( \ket{HV} - \ket{VH} \right)/\sqrt{2}$, where $\ket{H}$ and $\ket{V}$ stand for horizontal and vertical polarization, respectively. Those two-photon singlet states can be paired in three inequivalent ways in order to build a global $n=4$ state:
\begin{align}
s_{12}s_{34}, \quad s_{13}s_{42}, \quad s_{14}s_{23}.
\label{eq:pairs_of_pairs}
\end{align}
These three states are not orthogonal, since they span a two-dimensional subspace. In fact, the ordering of particles in $s_{13}s_{42}$ was selected so that the scalar product between each pair is $-1/2$. To form an orthogonal basis, we prepare two linear combinations of them, resembling circular polarization states:
\begin{align}
l &= \tfrac{\sqrt{2}}{3}(s_{12}s_{34} + \epsilon s_{13}s_{42} + \epsilon^2 s_{14}s_{23})\label{eq:singlet_lr_def}\\
r &= \tfrac{\sqrt{2}}{3}(s_{12}s_{34} + \epsilon^2 s_{13}s_{42} + \epsilon s_{14}s_{23}),
\end{align}
where $\epsilon = \exp(i 2\pi/3)$.

\begin{figure}[!htbp]
\centering
\includegraphics[width=0.40\textwidth]{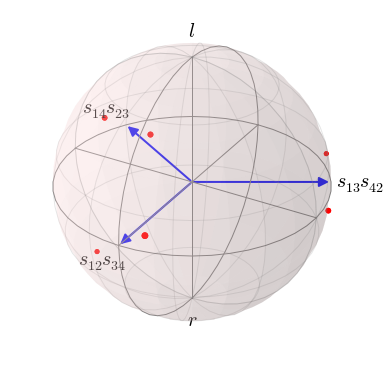}
\caption{Arrows stand for $s_{12}s_{34}$, $s_{13}s_{42}$ and $s_{14}s_{23}$. On the poles there are $l$ and $r$ states, as defined in \eqref{eq:singlet_lr_def}. Points represent a single state subjected to action related to all permutations of pairs of modes modes.}
\label{fig:2_dims_singlet_geometry}
\end{figure}

Let us introduce the following parametrization for our state
\begin{align}
f = \cos(\tfrac{\theta}{2}) l + \sin(\tfrac{\theta}{2}) e^{i\varphi} r,
\label{eq:singlet_lr_basis}
\end{align}
where $\theta\in[0,\pi)$ and $\varphi\in[0,2\pi)$, so that we can absorb the sign in $\theta$. As it is a logical qubit (i.e. a two dimensional Hilbert space), it can be represented on the Bloch sphere, see Fig.~\ref{fig:2_dims_singlet_geometry}.

Now let us compute the moments up to a few copies:
\begin{align}
\langle f^{2} f^{\dagger 2} \rangle &= \tfrac{17}{2} - \tfrac{1}{2}
\cos(2\theta), \nonumber \\
\langle f^{3} f^{\dagger 3} \rangle &= 290 - 42 \cos(2\theta) - 8
\sin^3(\theta) \cos(3 \varphi), 
\end{align}
as a side note, the normalization factors (as in \eqref{fkfk.projector}) are $1/70$ and $1/34650$, respectively. I.e.: the states are very far from being coherent. 

Consequently, we obtain two invariants:
\begin{align}
\cos(2\theta) \quad \text{and} \quad \cos(3 \varphi).
\label{eq:singlet_invariants}
\end{align}
This results restricts the allowed operations within linear optics.
If we restrict ourselves further, only to operations preserving the singlet subspace, then the only possible operations,
in the Bloch representation
(see Fig.~\ref{fig:2_dims_singlet_geometry}) are: rotation along the equator by $2\pi/3$ and $4\pi/3$, rotation around states \eqref{eq:pairs_of_pairs} by $\pi$, and mirror reflection with respect the equatorial plane. In particular, there are no continuous allowed transformations \cite{Wasilewski2007} for such singlet states.
Let show how to implement all those operations, with the exception of the mirror reflection.

What are the possible operations which hold the state within the singlet subspace? Of course, different parings can be interchanged by permuting beams. For example, $(2 \leftrightarrow 3)$ changes $s_{12}s_{34}$ into $-s_{13}s_{24}$ (and the same changing $(1 \leftrightarrow 4)$). Exchange of any two particles acting on any of the three two-singlet parings produces a state with a minus sign. Thus, permuting particles preserves the singlet subspace. 

The group of permutations of $4$ particles has $24$ elements, which can be generated by two-particle swaps:
\begin{align}
(1 \leftrightarrow 2) \text{ or } (3 \leftrightarrow 4):\quad & 
&l &\mapsto - r,\quad
&r &\mapsto - l\\
(1 \leftrightarrow 3) \text{ or } (2 \leftrightarrow 4):\quad & 
&l &\mapsto - \epsilon^2  r,\quad
&r &\mapsto - \epsilon  l\\
(1 \leftrightarrow 4) \text{ or } (2 \leftrightarrow 3): \quad & 
&l &\mapsto - \epsilon r,\quad
&r &\mapsto - \epsilon^2 l,
\end{align}
which can be checked directly by permuting particles in \eqref{eq:singlet_lr_def}. On the Bloch sphere, they are just rotations by $\pi$ around one of the states \eqref{eq:pairs_of_pairs}.
Composition of two permutations allows us to reach cyclic permutations of the three particles, e.g. ($1 \rightarrow 2 \rightarrow 3 \rightarrow 1$). It turns out that such permutations result in $\phi \mapsto \phi + 2\pi/3$ and $\phi \mapsto \phi + 4\pi/3$.

Thus we reached all operations unitary operations allowed by \eqref{eq:singlet_invariants}, with one exception. It does not cover antiunitary operations (reflections on Bloch sphere $\theta \mapsto \pi -\theta$). Thus, it is still possible that there are linear operations not preserving the singlet subspace that map some states into their complex conjugates. Nonetheless, this computation provides the most systematic study of the geometry of the simplest singlet qubit state implemented with photons, to the best of the authors' knowledge.

Alternatively, we can use the spectrum of $f f^\dagger|_k$ for different values of $k$. It suffices to check the two-particles block, i.e. $f f^\dagger |_2$, which is a $36\times 36$ matrix. The highest degree terms of its characteristic polynomial read:
\begin{align}
w_2(\lambda) &= \lambda^{36}\\
&- \lambda^{35} \tfrac{1}{4}\left( 17139 \cos(2 \theta) \right)\nonumber\\
&+ \lambda^{34} \tfrac{1}{72}\left( 9084959 + 1605 \cos(2 \theta) \right.\nonumber\\
&+ \left. 4 \cos(3 \varphi) \sin^3(\theta) \right) - \ldots,\nonumber
\end{align}
which yield the same invariants as the moments.

\subsection{Experimental recipe for tensor product in mode basis}
\label{s:mode-product}

In this section we study tensor product in the mode representation $\ket{\psi}_M^{k}$, which is different and more physically relevant than tensor product in the particle representation discussed in Sec.~\ref{s:replicas}.
Furthermore, we provide experimentally-feasible way do directly measure the invariants $\bra{\Omega} f^k f^{\dagger k} \ket{\Omega}$, defined as in \eqref{fkfk}, as related to success-rate of creation of states $f^{\dagger k}\ket{\Omega}$ from $k$ copies of state $f^\dagger \ket{\Omega}$.

To start with, let us look at example of $n=3$ particles in $d=2$ modes, raised to power $k=2$
\begin{align}
&\left( \tfrac{1}{\sqrt{2}} (\ket{0,3}_M+\ket{2,1}_M) \right)^{\otimes 2}\\
&= \tfrac{1}{2}\left( \ket{0,3,0,3}_M + \ket{0,3,2,1}_M\right.\\
&\left.+ \ket{2,1,0,3}_M+\ket{2,1,2,1}_M \right).
\end{align}
This is a valid photon state (as permutation-symmetry of particles is built-in in the mode representation), of $6$ particles in $4$ modes. 

In general, raising a bosonic state to tensor power, in the mode representation, yields in $kn$ photons in $kd$ modes (not $kd$ particles in $d$ modes, as in the tensor power for particle representation).
Tensor product in mode representation has a direct physical interpretation. If we create $k$ optical tables the same setups, each one producing state $\ket{\psi}$, then $\ket{\psi}_M^{\otimes k}$ is the quantum state produced by the laboratory.
As we see, multiplying state also multiplies number of modes, as there is one more parameter related to the number of optical table.

The question is if it this product can be related to $f^{\dagger k}$ in some way?  The answer is positive.  This time instead of symmetrizing particles (as we did for $\ket{\psi}_P^{\otimes k}$) we need to reduce number of modes from $kd$ to $d$,
by performing some symmetrization of modes.

We can write
\begin{align}
\ket{\psi}_M^{\otimes k} &= f^\dagger(a_{(1,1)},\ldots,a_{(d,1)})\label{eq:fkmanymodes}\\
&\times f^\dagger(a_{(1,2)},\ldots,a_{(d,2)}) \times \ldots\\
&\times f^\dagger(a_{(1,k)},\ldots,a_{(d,k)})\ket{\Omega}.
\end{align}
That is, if we are taking a number of copies of a bosonic state, then we in fact multiply number of modes.
The second index is related to copy.

System is symmetrized with respect to particles inside mode, by construction.
To symmetrize among modes, we need to project it on symmetric combination of respective modes
\begin{align}
b_{(i,1)} = \frac{a_{(i, 1)} + \ldots + a_{(i, k)} }{\sqrt{k}},\label{eq:sym_modes}
\end{align}
where all $b_{(i,j)}$ need to be pairwise orthogonal. 
It can be realized with linear optics, as unitary rotation of modes.
In particular, we may employ Fourier transform (i.e. $\vec{b}_i = \mathcal{F} \vec{a}_i$ for each group of modes), and we are interested in the constant term.

When inverting Fourier transform, each mode can be expressed as a linear combination of $b_{(i,j)}$,
where states with different indices are orthogonal,
and weight of $b_{(i,1)}$ is always $1/\sqrt{k}$.

Consequently,
\begin{align}
&f^\dagger(a_{(1,j)}, \ldots, a_{(d,j)})\label{eq:symmetrization_of_variables}\\
= &f^\dagger \left(\tfrac{1}{\sqrt{k}} b_{(1,1)} + \mathcal{O}, \ldots, \tfrac{1}{\sqrt{k}} b_{(d,1)} + \mathcal{O} \right)\\
= & k^{-n/2} f^\dagger(b_{(1,1)}, \ldots, b_{(d,1)}) + \mathcal{O},
\end{align}
where by $\mathcal{O}$ we denote terms containing at least one $b_{(i,j\neq 1)}$.
Thus, by using \eqref{eq:symmetrization_of_variables} for every component of \eqref{eq:fkmanymodes} we get
\begin{align}
k^{-kn/2} f^{\dagger k}(b_{(1,1)}, \ldots, b_{(d,1)}) + \mathcal{O}.
\end{align}
\begin{figure}[!htbp]
	\centering
		\includegraphics[width=0.3\textwidth]{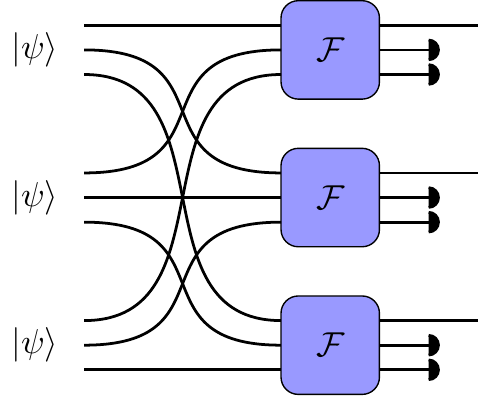}
	\caption{Experimental setup example for $d=3$ modes (and $\mathcal{F}$ operators) and $k=3$ copies (and outcome channels per operator).}
	\label{fig:fpower}
\end{figure}

Consequently, we have one more interpretation of $f^{\dagger k}\ket{\Omega}$.
It is the state you get when following the recipe, pictured in Fig.~\ref{fig:fpower}:
\begin{itemize}
\item Create $k$ copies of an $n$-photon state.
\item Perform interference on each group of respective modes.
\item Postselect results in which for each group of modes no photon was detected in non-first output mode.
\end{itemize}

Our probability to succeed is
\begin{equation}
\frac{\bra{\Omega} f^k f^{\dagger k} \ket{\Omega}}{k^{kn}}
\leq \frac{(kn)!}{(n!)^k k^{kn}}
\approx k^{-1/2} (2 \pi n)^{(1-k)/2},
\end{equation}
where the approximation is due to Stirling's formula for $(kn)!$ and $n!$.
That is, invariant $\bra{\Omega} f^k f^{\dagger k} \ket{\Omega}$ can be measured experimentally, as statistic of no clicks in detectors, in the described setting.

For the simplest case of $n=1$, $d=1$ and $k=2$, the Fourier transform becomes%
\begin{align}
\mathcal{F} =
\left[%
\begin{matrix}
\tfrac{1}{\sqrt{2}} & \tfrac{1}{\sqrt{2}}\\
- \tfrac{1}{\sqrt{2}} & \tfrac{1}{\sqrt{2}}
\end{matrix}
\right].
\end{align}
and we get Hong-Ou-Mandel interference with postselection, allowing us to produce state two photons in one mode $\ket{2,0}_M$ from two photons in two modes $\ket{1,1}_M$, with $50\%$ postselection efficiency.

Moreover, a similar experimental scheme as above can be used to produce states of the from%
\begin{align}
f_1^\dagger \cdots f_k^\dagger \ket{\Omega},
\end{align}
where all $f_i^\dagger \ket{\Omega}$ are states of a fixed number of photons (perhaps different for each $i$).
The success rate is%
\begin{align}
\frac{\bra{\Omega} f_k \cdots f_1  f_1^\dagger \cdots f_k^\dagger  \ket{\Omega}}{k^{n_1 + \cdots + n_k}}.
\end{align}
This follows directly from \eqref{eq:symmetrization_of_variables} applied to a product of functions.

\subsection{Schwinger representation of symmetric operators}\label{app:schwinger}

Below we provide technical calculations to show that \eqref{eq:pauli_sym} and \eqref{eq:schwinger_form} are the same on permutation-symmetric states, i.e.
\begin{gather*}
\bra{\psi}
\sum_{perm} (\mathbbm{I})^{\otimes n_{I}}
\otimes (\sigma^x)^{\otimes n_x}
\otimes (\sigma^y)^{\otimes n_y}
\otimes (\sigma^z)^{\otimes n_z}
\ket{\psi}\\
=\bra{\Omega} f
: \left( a^\dagger a + b^\dagger b \right)^{n_I}
\left( a^\dagger b + b^\dagger a \right)^{n_x}\\
\times \left( - i a^\dagger b + i b^\dagger a \right)^{n_y}
\left( a^\dagger a - b^\dagger b\right)^{n_z} :
f^\dagger \ket{\Omega}.
\end{gather*}
We proof a general variant of it, for qudits.

\subsubsection{Auxiliary notation}

Let us introduce the following notation:
\begin{align}
a_\mu^\dagger &=
\frac{1}{\sqrt{n+1}}
\sum_{i=0}^{n} \ket{\mu}_i\label{eq:adag_alt_def}\\
a_\mu &=
\frac{1}{\sqrt{n}}
\sum_{i=0}^{n-1} \bra{\mu}_i,
\end{align}
where $\ket{\mu}_i$ means {\it insert $\ket{\mu}$ between $i$-th and $(i+1)$-th particle}, whereas $\bra{\mu}_i$ removes $i$-th particle.
The $n$ is the total number of particles in the state it is acting on.
We show that this notation is consistent, i.e. the left hand sides of \eqref{eq:adag_alt_def} act like creation and annihilation operators, respectively.
However, the right hand side can be applied on any state, not only a permutation symmetric one.

For example:
\begin{align}
&\left( \sum_{i=0}^2 \ket{2}_i \right) \ket{01}_P\\
&= \left( \ket{2}_0 + \ket{2}_1 + \ket{2}_2 \right) \ket{01}_P\\
&= \ket{201}_P + \ket{021}_P + \ket{012}_P
\end{align}
and
\begin{align}
&\left( \sum_{i=0}^2 \bra{2}_i \right) \ket{201}_P\\
&= \left( \bra{2}_0 + \bra{2}_1 + \bra{2}_2 \right) \ket{201}_P\\
&= \braket{2}{2} \ket{01}_P + \braket{2}{0} \ket{21}_P + \braket{1}{2} \ket{20}_P\\
&= \ket{01}_P.
\end{align}

A straightforward check on $n$-particle permutation-symmetric states Dicke state show that this (abuse of) notation makes sense.
That is, let us check that:
\begin{align}
a_\mu^\dagger {\tilde a}^\dagger_{\vec{n}} \ket{\Omega}
&= \left( \frac{1}{\sqrt{n+1}} \sum_{i=0}^n \ket{\mu}_i \right) \ket{\vec{n}},\\
a_\mu {\tilde a}^\dagger_{\vec{n}} \ket{\Omega}
&= \left( \frac{1}{\sqrt{n}} \sum_{i=0}^{n-1} \bra{\mu}_i \right) \ket{\vec{n}}.
\end{align}
We proceed by writing a state in particle representation as in \eqref{eq:fock_in_particle_representation}.
For the convenience, without the loss of generality, let us pick $\mu=1$,%
\begin{align}
&\sqrt{n+1} a_1^\dagger \sqrt{\frac{n!}{n_1! \cdots n_d!}}
\ket{n_1, \cdots, n_d}\\
&= \left( \sum_{i=0}^{n} \ket{1}_i \right)
\left( \ket{1}^{n_1}_P \cdots \ket{d}^{n_d}_P
+ \text{perm.} \right)\\
&= (n_1 + 1)
\left( \ket{1}^{n_1+1}_P \cdots \ket{d}^{n_d}_P
+ \text{perm.} \right)\\
&= (n_1 + 1) \sqrt{\frac{(n+1)!}{(n_1+1)! \cdots n_d!}}
\ket{n_1+1, \cdots, n_d},
\end{align}
where \emph{perm.} means inequivalent permutations.
Factor $(n_1+1)$ in the third line comes from%
\begin{equation}
(n + 1) \frac{n!}{n_1! \cdots n_d!}
\big/ \frac{(n+1)!}{(n_1+1)! \cdots n_d!},
\end{equation}
that is, putting $n+1$ particles and comparing number of inequivalent terms in permutation, for the initial and final state.

And analogously for annihilation:%
\begin{align}
&\sqrt{n} a_1 \sqrt{\frac{n!}{n_1! \cdots n_d!}}
\ket{n_1, \cdots, n_d}\\
&= \left( \sum_{i=0}^{n-1} \bra{1}_i \right)
\left( \ket{1}^{n_1}_P \cdots \ket{d}^{n_d}_P
+ \text{perm.} \right)\\
&=  n
\left( \ket{1}^{n_1-1}_P \cdots \ket{d}^{n_d}_P
+ \text{perm.} \right)\\
&= n \sqrt{\frac{(n-1)!}{(n_1-1)! \cdots n_d!}}
\ket{n_1-1, \cdots, n_d}.
\end{align}
This time $n$ in the third line comes from%
\begin{equation}
n_1 \frac{n!}{n_1! \cdots n_d!}
\big/ \frac{(n-1)!}{(n_1-1)! \cdots n_d!}.
\end{equation}

\subsubsection{Proof}

We start the proof with the following observation.
When we remove a particle from a symmetric state,
there result does not depend which one
(state of all other particles always permutation symmetric).
That is%
\begin{align}
\bra{\mu}_i \ket{\psi}
= \bra{\mu}_j \ket{\psi}
= \frac{1}{n} \left( \sum_{i=0}^{n-1} \bra{\mu}_i \right) \ket{\psi},\label{eq:annihilation_in_one_place}
\end{align}
where the last equality is a consequence of the former (for an $n$-particle state).

Consequently, when acting on $n$-particle symmetric state we get,
we write subsequent annihilation and creation operators as a single sum:
\begin{align}
&a_{\mu_1}^\dagger \cdots a_{\mu_k}^\dagger
a_{\nu_k} \cdots a_{\nu_1} \ket{\psi}\\
= &\frac{(n-k)!}{n!} 
\sum_{i_1,\ldots,i_k} \sum_{j_1,\ldots,j_k}\\
&\Big(
\ket{\mu_1}_{i_1} \cdots \ket{\mu_k}_{i_k}
\bra{\nu_k}_{j_k} \cdots \bra{\nu_1}_{j_1}
\Big) \ket{\psi}\label{eq:aiaj2aiai}\\
= &\left( \sum_{i_1,\ldots,i_k}
\ket{\mu_1}_{i_1} \cdots \ket{\mu_k}_{i_k}
\bra{\nu_k}_{i_k} \cdots \bra{\nu_1}_{i_1}
\right) \ket{\psi}, 
\end{align}
where instead of the sum over $j_1,\ldots,j_k$ we put $j_p=i_p$ using \eqref{eq:annihilation_in_one_place}.

Note that as creation and annihilation operations add and subtract particles (respectively), indices in a product do refer to different set of particles and need to be carried out iteratively.
That is, summation over $j_p$ goes from $j_p=0$ to $n-p$.

We need to show one more thing:%
\begin{align}
&\left( \sum_{i_1,\ldots,i_k}
\ket{\mu_1}_{i_1} \cdots \ket{\mu_k}_{i_k}
\bra{\nu_k}_{i_k} \cdots \bra{\nu_1}_{i_1}
\right) \ket{\psi}
\label{eq:changing2fixed_order}\\
=
&\left( \sum_{\text{p.d. }l_1,\ldots,l_k}
\ket{\mu_1}_{l_1}\bra{\nu_1}_{l_1}
\cdots \ket{\mu_k}_{l_k}\bra{\nu_k}_{l_k}
\right) \ket{\psi},
\end{align}
where by \emph{p.d.} we mean pairwise different.
In fact the only thing we need to do is to relabel each component of the sum. 
In the first line $i_p\in{0,\ldots, n-p}$, while in the second --- $l_p\in{0,\ldots, n-1}$ but disallow repetitions.
If in the first line we relabel in such a way that we don't forget about particles that we removed with $\bra{\nu_1}_{i_p}$, then we get $l_p$.

When we combine \eqref{eq:aiaj2aiai} with \eqref{eq:changing2fixed_order} we get an important relation%
\begin{align}
&a_{\mu_1}^\dagger \cdots a_{\mu_k}^\dagger
a_{\nu_k} \cdots a_{\nu_1} \ket{\psi}
\label{eq:normal_order_and_symmetric_operators}\\
=&\left( \sum_{\text{p.d. }l_1,\ldots,l_k}
\ket{\mu_1}_{l_1}\bra{\nu_1}_{l_1}
\cdots \ket{\mu_k}_{l_k}\bra{\nu_k}_{l_k}
\right) \ket{\psi}.
\end{align}

After showing relation \eqref{eq:normal_order_and_symmetric_operators}, we proceed to the main part of the proof.
Any symmetrized product of matrices is multilinear in their matrix entries, defined by $((\mu_1, \nu_1),\ldots,(\mu_n,\nu_n))$, where each $\mu_i$ (and $\nu_i$) is in $\{0,\ldots, d-1\}$, that is%
\begin{equation}
\sum_{\vec\imath \in \sigma(\{1,\ldots,n\})}
\ket{\mu_1}_{i_1}\bra{\nu_1}_{i_1}\ldots \ket{\mu_n}_{i_n}\bra{\nu_n}_{i_n}.
\end{equation}
So we need to show that for a sum of distinct matrix elements give the corresponding normally ordered operators.
When we apply \eqref{eq:normal_order_and_symmetric_operators}, we get
\begin{equation}
: a_{\mu_1}^\dagger a_{\nu_1} \ldots a_{\mu_n}^\dagger a_{\nu_n} :,
\label{eq:all-normally}
\end{equation}
what completes the proof.

Bear in mind that in \eqref{eq:all-normally} we get $n$ creation and annihilation operators, regardless of the multi-particle operator we want to use.
When we use only a $k$-particle operator, the formula can be simplified, what we show in the examples.

\subsubsection{Examples}

Below, for the clarity, we will work with qubits and use $a$ and $b$ for the annihilation operators of $\ket{0}$ and $\ket{1}$, respectively.

First, we see that%
\begin{align}
\sum_{i=1}^n \sigma^x_i &= a^\dagger b + b^\dagger a\\
\sum_{i=1}^n \sigma^y_i &= -ia^\dagger  b + i b^\dagger a\\
\sum_{i=1}^n \sigma^z_i &= a^\dagger a - b^\dagger b,
\end{align}
which is the standard Schwinger representation of operators for symmetric states, where we directly applied \eqref{eq:aiaj2aiai}, e.g. for symmetrized $\sigma^y$%
\begin{align}
\sum_{j=1}^n \sigma^y_j &= 
\sum_{j=1}^n \left( -i \ket{0}_j\bra{1}_j + i \ket{1}_j\bra{0}_j \right)\\
&= -ia^\dagger  b + i b^\dagger a.
\end{align}

Now, let us look at symmetrized product of two operators, e.g. $\sigma^x_i$ and $\sigma^z_j$:%
\begin{align}
&\sum_{i \neq j} \sigma^x_i \otimes \sigma^z_j\label{eq:schwinger_square}\\
&= \sum_{i\neq j}
\left( \ket{0}_i \bra{1}_i + \ket{1}_i \bra{0}_i \right)
\left( \ket{0}_j \bra{0}_j - \ket{1}_j \bra{1}_j \right)\\
&= \sum_{i\neq j}
\left(
\ket{0}_i \bra{1}_i \ket{0}_j \bra{0}_j 
- \ket{0}_i \bra{1}_i \ket{1}_j \bra{1}_j \right.\\
&\phantom{=\sum_{i\neq j}(}\left.
+ \ket{1}_i \bra{0}_i \ket{0}_j \bra{0}_j 
- \ket{1}_i \bra{0}_i \ket{1}_j \bra{1}_j
\right)\\
&= \left(
a^{\dagger 2} a b 
- a^\dagger b^\dagger b^2
+ a^\dagger b^\dagger a^2
- b^{\dagger 2} a b
\right)\\
&= : 
\left( a^\dagger b + b^\dagger a \right)
\left( a^\dagger a - b^\dagger b \right) :
\end{align}
were we applied \eqref{eq:normal_order_and_symmetric_operators} to change summation to creation and annihilation operators.

\section{Singlet space for photons and information protection}
\label{s:singlet-space}

In this section we study singlet subspace implemented with bosons (as in Sec.\ref{s:four-particle-singlet}), basing on our work  \cite{Migdal2011dfs}.
We prove that for a system of qubits, subjected to collective decoherence in the form of perfectly correlated random SU($d$) unitaries, quantum superpositions stored in the decoherence free subspace are fully immune against the removal of one particle.
This provides a feasible scheme to protect quantum information encoded in the polarization state of a sequence of photons against both collective depolarization and one photon loss.
We provide a scheme for experimental demonstration with photon quadruplets using currently available technology.

We consider the DFS for an ensemble of $n$ qudits, i.e.\ elementary $d$-level systems, composed of states $\ket{\psi}$ that are invariant with respect to an arbitrary perfectly correlated SU$(d)$ transformation:
\begin{equation}
V^{\otimes n} \ket{\psi} = \ket{\psi}, \qquad V \in \text{SU}(d).
\label{Eq:invariance}
\end{equation}
Note, that in this context we consider \emph{distinguishable} particles.
That is, one we implement $n$ particles with $n$ photons in $nd$ modes.
In the context of a multi-photon states, singlet states as defined above are states invariant with respect to
\begin{equation}
U = V^{\otimes n}. 
\end{equation}

We show that this DFS features an additional degree of robustness, namely that the stored quantum information is immune to the loss of one of the qudits, regardless of the encoding. This result, specialized to the polarization state of single photons for which $d=2$, offers {\em combined} protection against two common optical decoherence mechanisms: photon loss \cite{Wasilewski2007,Lu2008} due to reflections, scattering, residual absorption, etc.\ as well as collective depolarization that occurs inevitably in optical fibers used for long-haul transmission \cite{Bartlett2007,banaszek2004experimental,Bourennane2004}. Consequently, we provide here rigorous foundations to a speculation presented in Ref.~\cite{Boileau2004} that DFS-based quantum cryptography
can be made tolerant also to photon loss. It is worth noting that another physical realization of the qubit case can be also an ensemble of spin-$\frac{1}{2}$ particles \cite{Viola2001} coupled identically to a varying magnetic field.

The section is organized as follows. First, in Sec.~\ref{Sec:SingletQubits} we briefly review the geometry of the singlet subspace for an ensemble of qubits and we explicitly show the robustness of the four qubit DFS, which spans the logical qubit space. This particular case leads us to a proposal for a proof-of-principle experiment based on currently available photonic technologies that demonstrates the robustness of DFS encoding, presented in Sec.~\ref{Sec:ExperimentalScheme}. The general proof for an arbitrary $d$ that a quantum superposition encoded in an $\text{SU}(d)$ DFS remains immune against the loss of one particle is described in Sec.~\ref{Sec:General}.

\subsection{Example with logical qubits}
\label{Sec:SingletQubits}

Because of two relevant physical realizations using photons and spin-$1/2$ particles, we will first discuss the qubit case with $d=2$. The complete Hilbert space of an ensemble of $n$ qubits, each described by a two-dimensional spin-$1/2$ space $\mathcal{H}_{1/2}$, can be subjected to Clebsch-Gordan decomposition \cite{Dicke1954}
\begin{equation}
(  \mathcal{H}_{1/2} )^{\otimes n} = \bigoplus_{j = (n \bmod 2)/2 }^{n/2} \mathbbm{C}^{K^j_n} \otimes \mathcal{H}_{j}\label{eq:clebsch},
\end{equation}
where the direct sum is taken with the step of one and
$K^j_n$ are multiplicities of spin-$j$ Hilbert spaces $\mathcal{H}_{j}$, given explicitly by
\begin{equation}
K^j_n  = \frac{2j+1}{n/2+j+1} \binom{n}{n/2+j} \label{eq:catalan}.
\end{equation}
The action of $V^{\otimes n}$, where $V$ is any $\text{SU}(2)$ transformation, affects only $\mathcal{H}_{j}$ in Eq.~(\ref{eq:clebsch}), leaving $\mathbbm{C}^{K^j_n}$ unchanged. In particular, for an even number of $n$ qubits forming the ensemble, the {\em singlet subspace} corresponding to $j=0$ is free from decoherence. Furthermore, removing one particle from that ensemble maps any initial state from the singlet subspace onto a certain state from the {\em doublet subspace} $\mathbbm{C}^{K^{1/2}_{n-1}} \otimes \mathcal{H}_{1/2}$. Because $K^{1/2}_{n-1}=K^0_n$, it is plausible that the quantum superposition will end up entirely in the decoherence-free subsystem $\mathbbm{C}^{K^{1/2}_{n-1}}$ where it will remain protected from collective depolarization.

\begin{figure}[h]
\centering
\includegraphics[width=0.7\textwidth]{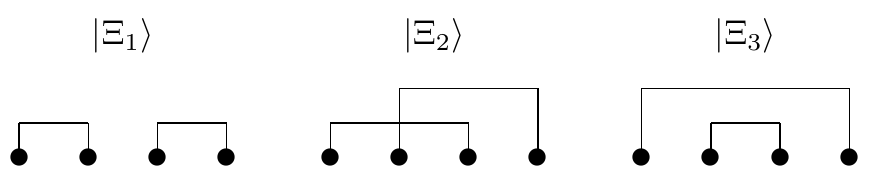} 
\caption{Diagrams depicting three non-equivalent products of two-qubit singlet states defined in Eq.~(\protect\ref{eq:dfs4}). The qubits are represented as dots with connections identifying pairs that form singlet states.}
\label{fig:spps}
\end{figure}

The simplest non-trivial case is $n=4$ physical qubits encoding one logical qubit.
Let us consider three states from the four-qubit DFS defined as products, same as in \eqref{eq:pairs_of_pairs},
\begin{align}
\ket{\Xi_1} &= \ket{\Psi^-}_{12} \ket{\Psi^-}_{34}, \nonumber\\
\ket{\Xi_2} &= \ket{\Psi^-}_{13} \ket{\Psi^-}_{42}, \label{eq:dfs4}\\
\ket{\Xi_3} &= \ket{\Psi^-}_{14} \ket{\Psi^-}_{23}, \nonumber
\end{align}
where $\ket{\Psi^-}_{ij}= (\ket{01}_{ij}-\ket{10}_{ij})/\sqrt{2}$ is the singlet state of qubits $i$ and $j$. These states, shown schematically in Fig.~\ref{fig:spps}, form an overcomplete set in the DFS. For concreteness, let us select $\ket{\Xi_1}$ and $\ket{\Xi_3}$ as a non-orthogonal basis.
Any state of the logical DFS qubit can be written as a superposition
\begin{align}
\ket{\psi} = \alpha \ket{\Xi_1}+ \beta \ket{\Xi_3},
\label{Eq:Psifourqubit}
\end{align}
where $\alpha$ and $\beta$ are complex amplitudes. Without loss of generality we can assume that the first physical qubit has been lost. The remaining three qubits are described by an equally weighted statistical mixture of two states:
\begin{align}
\ket{\psi^{(0)}}_{\bar{1}} &= \alpha \ket{1}_{2} \ket{\Psi^-}_{34} + \beta  \ket{\Psi^-}_{23} \ket{1}_{4} \nonumber \\
\ket{\psi^{(1)}}_{\bar{1}} &= \alpha \ket{0}_{2} \ket{\Psi^-}_{34} + \beta  \ket{\Psi^-}_{23} \ket{0}_{4},\label{eq:singletowyrozpisany}
\end{align}
where $\ket{\cdot}_{\bar{1}}$ denotes the state of all qubits but the first one. It is easy to see that a collective transformation
$V^{\otimes 3}$ leaves the statistical mixture $\frac{1}{2} \bigl( \ket{\psi^{(0)}}_{\bar{1}} \bra{\psi^{(0)}} +
\ket{\psi^{(1)}}_{\bar{1}} \bra{\psi^{(1)}}\bigr)$ intact.

After the loss of the first particle, the initial four-qubit state from Eq.~(\ref{Eq:Psifourqubit}) can be recovered through the following procedure.
First, one needs to measure in a non-destructive way the $z$ component of the total pseudospin operator $\sigma^{z}_{2} + \sigma^{z}_{3} + \sigma^{z}_{4}$, where $\sigma^{z} = \ket{0}\bra{0} - \ket{1}\bra{1}$, in order to discriminate $\ket{\psi^{(0)}}$ from $\ket{\psi^{(1)}}$. If the result corresponding to $\ket{\psi^{(1)}}$ is obtained, we apply a collective rotation $(\sigma^{x})^{\otimes 3}$, where $\sigma^x = \ket{0}\bra{1} + \ket{1}\bra{0}$. This yields the state $\ket{\psi^{(0)}}_{\bar{1}}$. In the second step, one replaces the lost qubit with a new one prepared in a state $\ket{+}_{1} = \frac{1}{\sqrt{2}}( \ket{0}_1 + \ket{1}_1 )$ and applies a controlled rotation which restores the original state $\ket{\psi}$:
\begin{equation}
\bigl( \ket{0}_{1} \bra{0} \otimes {\openone}^{\otimes 3} + \ket{1}_{1} \bra{1} \otimes {(\sigma^x)}^{\otimes 3}\bigr)
\bigl( \ket{+}_{1} \ket{\psi^{(0)}}_{\bar{1}}\bigr)  = \ket{\psi}
\end{equation}
Note that this rotation can be realized as a sequence of three C-NOT gates.

The robustness of DFS to particle loss can be intuitively understood in the following way. DFS states owe their invariance with respect to collective unitary transformation to a very rigid structure. In fact, if we write  a DFS state as a superposition in the computational basis for individual qubits, the state of one qubit can be determined unambiguously from the states of the remaining ones. This suggests that the loss of one particle does not destroy any information. Futher, it is always possible to repair the state as there is only one unique way to fit the lost particle such that the singlet symmetry is recovered.

\subsection{Experimental scheme}
\label{Sec:ExperimentalScheme}

We will now present a proposal a feasible experiment that demonstrates the robustness of DFS encoding using
photon quadruplets that can be generated in the process of parametric down-conversion \cite{Bourennane2004,Weinfurter2001,Gong2008}.
The basis states $\ket{0}$ and $\ket{1}$ correspond in this case to horizontal and vertical polarizations of individual photons.
Let us consider four-photon states $\ket{\Xi_k}$, $k=1,2,3$, defined in Eq.~(\ref{eq:dfs4}) as well as their orthogonal complements in the two-dimensional DFS, which we will denote as $\ket{\Xi_k^\perp}$.
The index $k$ corresponds to three non-equivalent orderings of the photons and it can be changed by suitable rerouting of the photons. As demonstrated in \cite{Bourennane2004}, the states $\ket{\Xi_1}$ and $\ket{\Xi_1^\perp}$ can be discriminated unambiguously by detecting polarizations in the horizontal-vertical basis $\ket{0}, \ket{1}$ for photons $12$ and in the diagonal basis $(\ket{0} \pm \ket{1})/\sqrt{2}$ for photons $34$. Restricted to the DFS subspace, this strategy yields the standard projective measurement.

\begin{figure}
\centering
\includegraphics[width=0.7\textwidth]{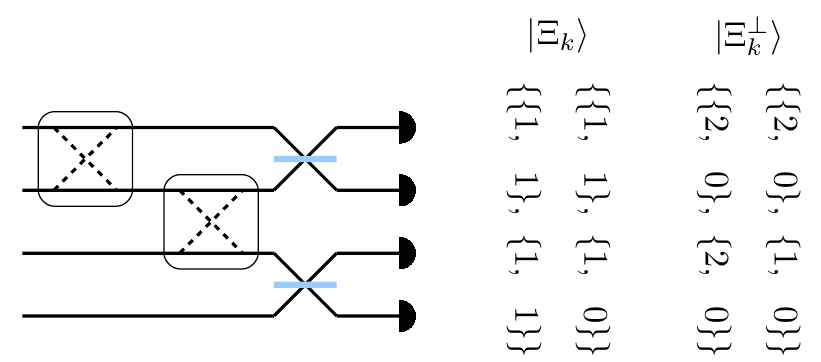} 
\caption{An experimental scheme for loss-tolerant detection of a logical qubit encoded in four photons. The projection basis $\ket{\Xi_k}$, $\ket{\Xi_k^\perp}$, where $k=1,2,3$, is selected by a suitable rerouting of input photons. Pairs of photons are interfered on two balanced beam splitters and photon numbers are counted at their outputs. Combinations of outcomes for individual detectors that correspond to unambiguous identification of $\ket{\Xi_k}$ and $\ket{\Xi_k^\perp}$ are indicated with photon numbers in curly brackets. The ordering within both inner and outer brackets does not matter.}
	\label{fig:measurement}
\end{figure}

It is easy to check that the above individual measurement no longer works if one of the photons is missing. It turns out that this problem can be solved by resorting to collective measurements. Suppose that we interfere photon pairs $12$ and $34$ on two separate balanced beam splitters, playing the role linear-optics Bell state analyzers \cite{Braunstein1995}. The state $\ket{\Xi_1}$ will yield exactly one photon in each output port of each beam splitter. In contrast, because the orthogonal state $\ket{\Xi_1^\perp}$ can be written as \cite{Kempe2001}:
\begin{equation}
\ket{\Xi_1^\perp} = \frac{1}{\sqrt{3}}\left(
\ket{00}_{12}\ket{11}_{34}
+ \ket{11}_{12}\ket{00}_{34} - \ket{\Psi^+}_{12}\ket{\Psi^+}_{34} \right),
\end{equation}
where $\ket{\Psi^+}_{ij}= (\ket{01}_{ij}+\ket{10}_{ij})/\sqrt{2}$, it will always produce two photons at the same output port for each of the two beam splitters. If one photon is lost, the states $\ket{\Xi_1}$ and $\ket{\Xi_1^\perp}$ will still give distinguishable outcomes: registering two photons at a single output unambiguously heralds $\ket{\Xi_1^\perp}$, while registering a photon pair at two different outputs of the same beam splitters detects $\ket{\Xi_1}$. The third photon will emerge separately from the second beam splitter. This detection scheme is summarized in Fig.~\ref{fig:measurement}.

An interesting question is whether the scheme described above could be exploited for quantum key distribution.
The scalar products between any two the states $\ket{\Xi_k}$ and $\ket{\Xi_l}$ with $k \neq l$ are equal to $\braket{\Xi_k}{\Xi_l} = -\frac{1}{2}$. In the Bloch representation of the two-dimensional DFS, they form a regular triangle inscribed into a great circle on the Bloch sphere,
constituting a so-called {\em trine} that warrants cryptographic security \cite{Boileau2004,Renes2004,Tabia2011}. To generate a key, the sender Alice could prepare photon quadruplets in one of randomly selected states $\ket{\Xi_1}$, $\ket{\Xi_2}$, or $\ket{\Xi_3}$. The ability to perform a projection onto any pair of orthogonal states $\ket{\Xi_k}, \ket{\Xi_k^\perp}$ would enable the receiving party Bob to tell, in the case when an outcome $\ket{\Xi_k^\perp}$ is obtained, which state has definitely not been prepared by Alice. Such correlations between Alice's preparations and Bob's outcomes can be distilled into a secure key.

We have shown that the projective measurement onto $\ket{\Xi_k}, \ket{\Xi_k^\perp}$ can be implemented in a way that tolerates the loss of one photon. In a cryptographic setting, the crucial issue is to ensure that an eavesdropper Eve does does not map the state of intercepted photons outside the DFS, which may enable eavesdropping attacks beyond those already studied  \cite{Boileau2004,Renes2004,Tabia2011}. To verify that this is not the case, Bob could perform in principle a full quantum state reconstruction on some of the transmissions, which however would be resource consuming. We conjecture that a sufficient strategy to detect such an attack would be: (i) to detect polarizations of photons emerging after the beam splitters; (ii) for a subset of transmissions to count directly received photons to ensure that no multiphoton states in individual input paths occur; (iii) for another subset of transmissions to apply before the beam splitters random and uncorrelated transformations $V \otimes V$ and $V' \otimes V'$ and check that states $\ket{\Xi_k}$ always yield the correct outcome when Bob used the matching basis for his measurement.

\subsection{General proof}
\label{Sec:General}

The reasoning presented in Sec.~\ref{Sec:SingletQubits} can be generalized to any even number of $n > 4$ qubits by considering DFS states given by products of two-qubit singlet states. Such states form an overcomplete set in the DFS \cite{Lyons2008}, which enables one to follow directly the steps described for four qubits. The robustness of DFS encoding can be shown more generally for an ensemble of $n$ qudits, i.e.\ $d$-dimensional systems.
In this case, a DFS satisfying Eq.~\eqref{Eq:invariance}  exists only when $n$ is a multiple of $d$, which follows from the structure of the Young tableux for irreducible representations of tensor products of the $SU(d)$ group \cite{Jones1998groupsrepresentations}.

As before, for concreteness we will consider removal of the first qudit. Let us consider arbitrary two states $\ket{\psi}$ and $\ket{\Phi}$ from the DFS and expand them in the form analogous to Eq.~(\ref{eq:singletowyrozpisany}):
\begin{equation}
\ket{\psi} = \frac{1}{\sqrt{d}} \sum_{i=0}^{d-1} \ket{i}_1 \ket{\psi^{(i)}}_{\bar{1}},
\quad
\ket{\Phi} = \frac{1}{\sqrt{d}} \sum_{i=0}^{d-1} \ket{i}_1 \ket{\Phi^{(i)}}_{\bar{1}}
\label{eq:sud_singletowyrozpisany}
\end{equation}
where $\ket{i}_1 , i=0,\ldots,d-1$ is an orthonormal basis in the space of the first qudit, and
$\ket{\psi^{(i)}}_{\bar{1}} = \sqrt{d}\, {}_{1}\! \braket{i}{\psi}$ and $\ket{\Phi^{(i)}}_{\bar{1}} = \sqrt{d} \, {}_{1}\! \braket{i}{\Phi}$ are states of the remaining $n-1$ qudits. We will first show that the following  general property holds:
\begin{equation}
{}_{\bar{1}} \! \braket{\Phi^{(i)}}{\psi^{(j)}}_{\bar{1}} = \delta_{ij} \braket{\Phi}{\psi}.
\label{Eq:SUd:dmuidmuj}
\end{equation}
As we will see, this property guarantees that the loss of one particle does not destroy the quantum information encoded in the DFS.

In order to show that for $i \neq j$ the states $\ket{\Phi^{(i)}}$ and $\ket{\psi^{(j)}}$  are orthogonal as implied by Eq.~\eqref{Eq:SUd:dmuidmuj}, let us consider the action of
a diagonal unitary operator $D^{\otimes n}$, where $D = \text{diag} (e^{i \phi_0}, \ldots, e^{i \phi_{d-1}} )$ with arbitrary phases $\phi_0, \ldots, \phi_{d-1}$ that sum up to zero. Invariance of $\ket{\Phi^{(i)}}_{\bar{1}}$ and $\ket{\psi^{(j)}}_{\bar{1}}$ under $D^{\otimes n}$ implies that in the basis formed by tensor products of states $\ket{0}, \cdots, \ket{d-1}$ they are composed only from terms that have exactly $n/d$ particles in each of these $d$ states. Consequently, projecting the first qudit on orthogonal states $\ket{i}_{1}$ and $\ket{j}_{1}$ leaves the remaining qudits in distinguishable states.

In order to verify the case when $i=j$ in Eq.~(\ref{Eq:SUd:dmuidmuj}) it is convenient to use the transformation of states $\ket{\psi^{(i)}}_{\bar{1}}$ under the action of $V^{\otimes (n-1)}$. In order to derive this transformation, let us rewrite the invariance condition from Eq.~\eqref{Eq:invariance} to the form
$V^\dagger \otimes \openone^{\otimes (n-1)} \ket{\psi} = \openone \otimes V^{\otimes (n-1)} \ket{\psi}$ and  project the first qudit onto $ \sqrt{d} \, {}_{1} \! \bra{i}$. This yields the identity:
\begin{equation}
V^{\otimes (n-1)} \ket{\psi^{(i)}}_{\bar{1}} =
\sqrt{d} \, \bigl( {}_{1} \! \bra{i} V^\dagger \bigr) \ket{\psi}
=
\sum_{j=0}^{d-1} \bigl( \bra{j} V \ket{i} \bigr)^\ast \ket{\psi^{(j)}}_{\bar{1}}
\label{Eq:iUdagger}
\end{equation}
Let us now specialize this result to a special unitary transformation that cyclically shifts the labelling of the basis states:
\begin{equation}
W = (-1)^{d-1} \sum_{i=0}^{d-1}  \ket{i+ 1 }\bra{i},\label{eq:unitaryW}
\end{equation}
where the addition $i+1$ is understood to be modulo $d$. Using this $W$ in Eq.~\eqref{Eq:iUdagger} implies that $\ket{\psi^{(i+1)}} =(-1)^{d-1} W^{\otimes (n-1)} \ket{\psi^{(i)}}$, i.e.\ $\ket{\psi^{(i)}}$ and $\ket{\psi^{(i+1)}}$ are related by a unitary that is independent of $\ket{\psi}$. This means that
$\braket{\Phi^{(i+1)}}{\psi^{(i+1)}} = \braket{\Phi^{(i)}}{\psi^{(i)}}$. This fact combined with expanding the scalar product $\braket{\Phi}{\psi}$ using Eq.~(\ref{eq:sud_singletowyrozpisany}) completes the proof of  Eq.~(\ref{Eq:SUd:dmuidmuj}).

With Eq.~(\ref{Eq:SUd:dmuidmuj}) in hand, further steps are straightforward.
A removal of the first qudit maps a state $\ket{\psi}$ onto a statistical mixture
\begin{equation}
\varrho_{\bar{1}}  = \Tr_1 \bigl( \ket{\psi}\bra{\psi} \bigr) =\frac{1}{d} \sum_{i=0}^{d-1} \ket{\psi^{(i)}}_{\bar{1}}\bra{\psi^{(i)}}.
\end{equation}
Eq.~(\ref{Eq:SUd:dmuidmuj}) implies that analogously to the SU(2) case the components with different $i$ occupy orthogonal subspaces. Within each subspace the state is fully preserved, which follows from applying Eq.~(\ref{Eq:SUd:dmuidmuj}) to pairs of states from an arbitrary basis in the DFS. The final step is to show that the state $\varrho_{\bar{1}}$ is invariant with respect to $V^{\otimes (n-1)}$. This is a consequence of the fact that both the initial state $\ket{\psi}$ and the procedure of tracing out a particle are invariant with respect to SU($d$) transformations. Explicitly, the invariance of $\hat\varrho_{\bar{1}}$ can be verified with a calculation based on Eq.~(\ref{Eq:iUdagger}):
\begin{align}
V^{\otimes (n-1)} \varrho_{\bar{1}} (V^\dagger)^{\otimes (n-1)} &=
\sum_{i=0}^{d-1} \bigl( {}_{1} \! \bra{i} V^\dagger \bigr) \ket{\psi}\bra{\psi} \bigl( V \ket{i}_{1} \bigr) \\
&= \Tr_1 \bigl( \ket{\psi}\bra{\psi} \bigr) = \varrho_{\bar{1}}.
\end{align}
Thus the encoded state is fully preserved.

Concluding, we have shown that DFS encoding is immune to removing one particle. Unfortunately, this property does not seem to generalize in a straightforward manner to the loss of more particles. For example, when two qubits are removed from a four-qubit DFS state, the result will be either a singlet state of the remaining two qubits, or a statistical mixture of the singlet and triplet states which does not preserve the original superposition. This observation holds also for any higher even number of qubits. Nevertheless, our result shows how to protect information in the few-photon regime from both collective depolarization and the first-order effects of linear attenuation. We have proposed an experimental demonstration of this combined protection which can provide a robust quantum cryptography protocol.

Finally, let us note that although the proof of robustness against the qudit loss was based on the assumption that Eq.~(\ref{Eq:invariance}) is satisfied for every $\text{SU}(d)$ matrix, the DFS fulfilling this condition protects quantum superpositions from any decoherence mechanism that involves a subset of $\text{SU}(d)$ transformations. Therefore our considerations apply to a range of physical systems, for example higher-spin particles in a magnetic field or multilevel atoms interacting with optical fields.

\section{Further questions}

We analyzed the problem of which states with a fixed number of photons $n$ in $d$ modes can be related using only linear optics. This problem may be mathematically formulated in terms of which homogeneous polynomials of degree $n$ in $d$ complex variables may be related by a unitary transformation between them (or linear, if we allow postselection of ancillary modes).

We relate this problem to the problem of equivalence of pure states of distinguishable particles, with respect to local operations (i.e. LU- and SLOCC-equivalence).
We show that the study of homogeneous operations, i.e.: those where the same single-particle operator acts on each particle, suffices.
Furthermore, we introduce and analyze entanglement classification by checking
which one-particle operations preserve permutation symmetry.
In that classification we obtain a sequence of states, unique up to SLOCC.
In one extreme we find the multiparticle GHZ state, whereas on the other there is a $(d-1)$ excitation state, which is a natural generalization of the W state resulting from the classification scheme.

Some questions are left open:
\begin{itemize}
\item Whether invariance under all local operations (that is, not only invertible operations) on symmetric states can be represented as the same transformation for each particle.
\item Does it work for mixed states?
\item Whether the application of $k$-particle transformations on permutation-symmetric
states which are reversible by acting on other part will give rise to different entanglement
classification.
\end{itemize}

We passive linear optics with no postselection we introduce two families of invariants.
Both are based on the global creation operator, which creates the state, $\ket{\psi}=f^\dagger(\vec a)\ket{\Omega}$, which can be written as a homogeneous polynomial on the creation operators for each mode. The first set of the invariants is just the spectrum of the operator $f f^\dagger$. The second one is the set of moments of the form $\bra{\Omega}f^k f^{\dagger k} \ket{\Omega}$. This second set of invariants can receive a physical interpretation, since they are related to the probability of not losing particles when $k$ copies of the original state are prepared, and the symmetric channel is postselected.

The main open question is whether our invariants are fine-grained enough to ensure that if two multiphoton states have the same invariants, they can be connected with linear optics and complex conjugation.
We have computed the invariants for a variety of situations, and found that they provide a complete characterization of the equivalence classes in all of them.
However, this question is not yet answered in the general case. 

Regarding future work, we would like to make the following remarks. First of all, a proof that these invariants provide a full characterization would be very desirable. Or, alternatively, a counterexample, which would lead us to find better invariants.
Second, both methods can be applied for fermions with no modifications beyond changing bosonic by fermionic operators. It deserves investigation whether this method provides new invariants in that case, or whether it simplifies the derivation of already known ones.
A third line of future research will be to extend our results to mixed states, or states without a fixed number of particles. In this last case, moments can still be used,
but the spectral method becomes impractical (as $ff^\dagger$ not longer can be decomposed into blocks). 
But perhaps the most practical open question is: if two multiphoton states $\ket{\phi_1}$ and $\ket{\phi_2}$ can not be related using only linear optics, what is the maximal efficiency for obtaining $\ket{\phi_2}$ out of $\ket{\phi_1}$ using linear optics {\em and} postselection?



%% file: quantum_sequences.tex







\chapter{Visualization of quantum sequences}
\label{ch:qubsim}

\section{Introduction}

One of the key features of quantum mechanics is that increasing the number of particles results in an exponential increase of the number of parameters we need to describe the state.
For example, a pure state of $N$ qubits needs $2^N$ complex parameters.
It is a crucial feature, related to quantum phenomena such as entanglement and some aspects of quantum computation.
However, the exponential increase of parameters makes it problematic to store, analyze, process or visualize many-particle quantum states.
Moreover, sometimes we are interested in analyzing quantum states of infinitely many particles (for example, infinite spin chain lattices).
While it is impossible even to store all parameters,
still we can work with (usually approximate) models describing the state.

The problem is not unique to quantum mechanics --- one already have it in statistical physics,
and more generally in statistics.
That is, while each state can be described with a number of parameters proportional to the number of particles,
the probability distribution requires an exponentially growing number of parameters.
However, while in classical systems we can avoid this problem by considering a single state, in quantum mechanics this is not the case \cite{Feynman1982}.

In this chapter we present methods for analysis of many-particle (and infinite particle) wavefunctions, based (or inspired) on similar methods in statistics.
While the problem is general, in this chapter we will focus of sequences --- i.e. configuration of particles, where they can be meaningfully arranged in a line.
Every pure quantum state of $N$ particles can be expressed as in the computational basis, i.e.
\begin{align}
\ket{\Psi} = \sum_{s_1,s_2,\ldots,s_N} \alpha_{s_1,s_2,\ldots s_N} \ket{s_1} \ket{s_2} \cdots \ket{s_N},
\label{eq:state_in_computational_basis}
\end{align} 
where $\alpha_{s_1,s_2,\ldots s_N}$ are complex parameters and the sum is over respective number of states for each particle, i.e. $s_i \in \{0, \ldots, d_i - 1 \}$. 

We focus on systems of distinguishable particles of the same number of states, i.e. $d \equiv d_1 = \ldots = d_N$.
We will put special emphasis on translationally-invariant states.
That is, let us define the shift operator by
\begin{align}
T \ket{s_1} \ket{s_2} \cdots \ket{s_{N-1}} \ket{s_N} = \ket{s_2} \ket{s_3} \cdots \ket{s_N} \ket{s_1},
\end{align}
then translationally invariant states are the states fulfilling
\begin{align}
T \ket{\Psi} = \ket{\Psi}.\label{eq:transl-inv}
\end{align}
The dimension of translationally invariant states still grows exponentially \cite{oeisA000031}.
To see that, let us take computational basis of $N$ qudits, which is of the dimension $d^N$,
and construct abstractions classes of basis states related by $T^k$, for some $k$. 
As an orbit of $T^k$ has at most $N$ elements, the dimension of translationally invariant subspace is at least $d^N/N$.
Nonetheless sometimes this symmetry simplifies substantially properties of the state.

In this chapter we present a pictorial representation of quantum many-body wavefunctions, for which we have coined the name \emph{qubism}\footnote{The name \emph{qubism} (inspired by Cubism, the art movement) should not be confused with \emph{QBsim} (quantum Bayesianism) \cite{Caves2002,Fuchs2010}.} \cite{Rodriguez-Laguna2011}.
In this visualization, a wavefunction characterizing a pure state of a chain of $N$ qudits is mapped to an image with $d^{N/2} \times d^{N/2}$ pixels.
It is presented in a few flavors and applied to analyze properties of ground states of commonly used Hamiltonians in condensed matter and cold atom physics, such as the Heisenberg or the Ising model in a transverse field (ITF).
The main property of the plotting scheme is recursivity: increasing the number of qubits reflects in an increase in the image resolution.
Thus, the plots are typically fractal-like, at least for translationally-invariant states.
The two-dimensional structure is especially capable of capturing correlations between neighboring particles.
Many features of the wavefunction, such as magnetization, correlations and criticality, are represented by visual properties of the images.
In particular, factorizability can be easily spotted: entanglement entropy turns out to be the deviation from exact self-similarity.
Furthermore, we use similar a scheme to visualize density matrices and operators.

We show that some properties of \emph{qubistic} plots do not depend on particular graphical representation, but are related to information theoretic properties of the state.
Once the measurement basis is chosen, we analyze outcomes as classical probabilistic sequences.
We use tools such as (classical) conditional entropy and mutual information, as well as R\'enyi fractal dimension, to describe the state.

\subsection{Classical sequence analysis}

Analysis of probabilistic sequence is one of important problems in classical information theory.
Initial considerations on how much information can be sent as a probabilistic sequence of letters 
gave raise to Shannon entropy \cite{Shannon1948original} and related tools such as conditional and mutual information.
These concepts have proven to be crucial in communication --- as they provide rigorous bounds both on how to avoid redundancy by efficiently compressing information and how to add minimal redundancy, so that the message can be decoded, even if it is subjected to noise \cite{CoverThomas2006book,MacKay2003}.
Moreover, they remain one of the main general-purpose approaches to data analysis, as these tools deal with abstract information and require little assumptions.

Information theory is widely used for analysis of stationary processes, that is, probabilistic sequences of letters over an alphabet, with probabilities being invariant under translation.
They are a direct analogue of quantum translationally-invariant states \eqref{eq:transl-inv}.
Stationary processes are applied to as diverse topics as analysis of the structure of languages \cite{Norvig2009}, DNA sequences \cite{Almeida2001}, heart arrhythmias \cite{Jones2008} and correlations for grounds states of a Hamiltonian \cite{Bialek2001}.
One of key techniques for simplification and modeling of stationary processes are hidden Markov models \cite{Rabiner1989}.
That is, certain processes can be simulated as a memoryless stochastic process on the internal states (a random walk on a fixed graph) and a \emph{observation matrix} mapping the internal states to probabilities of observing particular outcomes.
Nonetheless, for some stationary processes memory properties are crucial \cite{Jones2008}.

\subsection{Data visualization}

It is not uncommon for a communication in technical sciences to involve presenting data, whether derived from an experiment, a numerical simulation or an exact formula.
It can be conveyed in the form of a table with numbers, a histogram, a line plot, a scatter plot or a density map --- to name only a few ways of visualizing data.
However, using plots to present data should not be taken for granted.
Even typical plots such as bar plots or scatter plots appeared for the first time in late 18th century \cite{Tufte2007}.

When we interact with data (especially data coming from an experiment or simulation), it is useful to have at the same time access to raw data and a representation enabling us to get further insight.
For example, when we are studying the correlation between two variables, a scatter plot is often a better way to show the data than just only the linear correlation coefficient.
First, from raw data presented as such a plot is easy to see correlations.
Second, it also allows to see why such correlation happen (maybe it is only due to a few outliers, or there is no correlation, but the data is still highly dependent in a non-linear way).

While most of such plots are multi-purpose tools can be applied to various kinds of data, some are more specific, with the visualization being deeply related to properties of the visualized object.
Perhaps the most beautiful example, Mendeleev's periodic table of elements, arranges elements in a way related to their nuclear (number of protons) and chemical (electric structure of orbitals) properties \cite{Marchese2012}. 

It is important to remember that every data visualization puts emphasis on some aspects of data at the expense of others.
For example, scatter and bar plot are good at showing relative differences, and put emphasis on values standing out of the crowd. 
Yet they may mask small but crucial changes, for example:
\begin{itemize}
\item Prices $\$5.00$ and $\$4.99$ convey a different message to the consumer \cite{Schindler2006}.
\item In some voting models \cite{Migdal2010mafia,Migdal2011twochoice} the parity of the number of participants may matter even in the limit of infinitely many participants --- i.e. adding two participants changes the value less than adding one participant.
\item Numerical value $1.57$ is close to $\pi/2$, but does not have the unique properties of the later.
\end{itemize}
Consequently, depending both on our data and the features we want to put the emphasis on,
we need to choose, tweak or create visualization schemes according to our needs.
It is a choice we cannot avoid as, all in all, even presenting numbers using Arabic numerals (e.g. $0.231 + 0.150i$) is a form of data visualization (and often abstraction, if we round numbers with fixed point precision).

\subsection{Visualizing sequences}

Analysis of the statistical distribution of sequences is important in a few fields of science.
In natural language processing texts are cut into so called $N$-grams --- sequences of $N$ consecutive characters or words).
Their distribution is being applied for language recognition and for various statistical interferences about language \cite{Norvig2009}. 

Another application is in molecular genetics --- analysis of deoxyribonucleic acid (DNA) sequences.
From the information theory perspective, each DNA sequence is a word over the alphabet of 4 letters, $\{A,C,G,T\}$,
denoting nucleobases --- adenine, cytosine, guanine and thymine, respectively.
A triple of nucleobases encodes an amino acid, the building block of a protein.
Thus, presence and absence of sequences of nucleobases is related to the structure of the proteins that are being encoded.
To visualize that, in 1990 Jeffrey \cite{Jeffrey1990,Jeffrey1992} used the so-called \emph{chaos game representation} to plot different sequences on the same graph.
We describe the scheme, as it is directly related to \emph{qubism}.

The chaos game representation applied to DNA sequences works as follows.
Fist, we plot a square and put ($\{A,C,G,T\}$) on the edges, for example:
\begin{align}
\vec{A} = (0,0),\quad
\vec{C} = (1,0),\quad
\vec{G} = (0,1),\quad
\vec{T} = (1,1).
\label{eq:acgt_positions}
\end{align}

\begin{figure}[!htbp]
	\centering
		\includegraphics[width=0.40\textwidth]{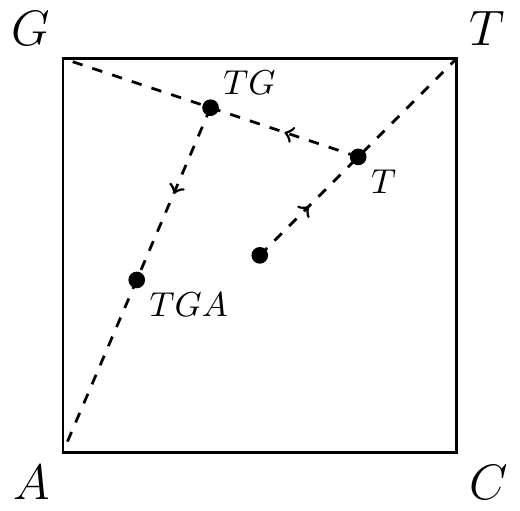}
	\caption{Construction of the chaos game representation for a DNA sequence.
	Points for a null sequence, $T$, $TG$ and $TGA$.}
	\label{fig:chaos_games}
\end{figure}

Then for each sequence ($s_1 s_2 s_3,\ldots,s_N$) we find its position with the following iterative procedure:
\begin{align}
\vec{r}_0 &= (1/2, 1/2)\\
\vec{r}_{i} &= (\vec{r}_{i-1} + \vec{s_i})/2. 
\end{align}
That is, we start in the middle of the square and for each consecutive symbol we move the the position in the middle-way between its current position and the symbol's corner. See Fig.~\ref{fig:chaos_games}.

Written in other way, it is just
\begin{align}
\vec{r}_{N} = \sum_{i=1}^N 2^{i-N-1} \vec{s_i},
\end{align}
the two coordinates are related to the binary expansion of the reversed sequence, i.e.
\begin{align}
\vec{r}_{N} = (&0.(s_{N})_x (s_{N-1})_x \ldots (s_{2})_x (s_{1})_x 1,\label{eq:chaos_position}\\
         &0.(s_{N})_y (s_{N-1})_y \ldots (s_{2})_y (s_{1})_y 1),\nonumber
\end{align}
where $x$ and $y$ mean the first and the second coordinate of the symbols,
as in \eqref{eq:acgt_positions}. Because of the reversed order, typically it does not converge for infinite sequences.

If the sequence distribution is uniform, it gives raise to a uniform distribution of points (up to the discretization) on square, see \eqref{eq:chaos_position}. 
If it is not, it typically looks fractal, showing the presence (or absence) of some particular subsequences.
For example, we can cut a DNA into non-overlapping sequences of 6 nucleobases (each encoding 2 amino acids).
Then the presence of particular strings says which pairs of amino acids are being encoded.

Chaos game representation for DNA sequences was used as a starting point to compare genes and calculate their information content \cite{Almeida2001} and multifractal properties \cite{Yu2001}.
Moreover, it was applied to compare proteins basing on their structure \cite{Liu2007}.
The idea was rediscovered by \cite{Hao2000}, with mapping slightly different from \eqref{eq:chaos_position}. 
The order of symbols in this formula is reversed, thus the first symbols carry more weight in $\vec{r}$ that the last ones.
In particular, it allows every sequence to be convergent at the price of restricting ourselves to the analysis of sequences of the same length.

Unbeknownst of the previous works, in 2005 Latorre \cite{Latorre2005} used this mapping to encode an image as a quantum state.
These quantum states were written down as states of a spin chain and then expressed it as matrix product states (MPS).
This proof-of-principle encoding was called \emph{qpeg} compression.


\section{Qubism}\label{s:qubism}

\subsection{Basic mapping}

To plot a pure quantum state of many qubits, let us start by writing it in the computational basis \eqref{eq:state_in_computational_basis}.
Similarly to the DNA sequence, we want to map each sequence to a particular position (or region) on a unit square.
Then we will color the region depending on its amplitude.

For simplicity, in this section we concentrate on qubits, with an even number of particles.
Generalization for qudits, particles of different dimension and an odd number of particles is straightforward.

\begin{figure}[!htbp]
	\centering
		\includegraphics[width=0.60\textwidth]{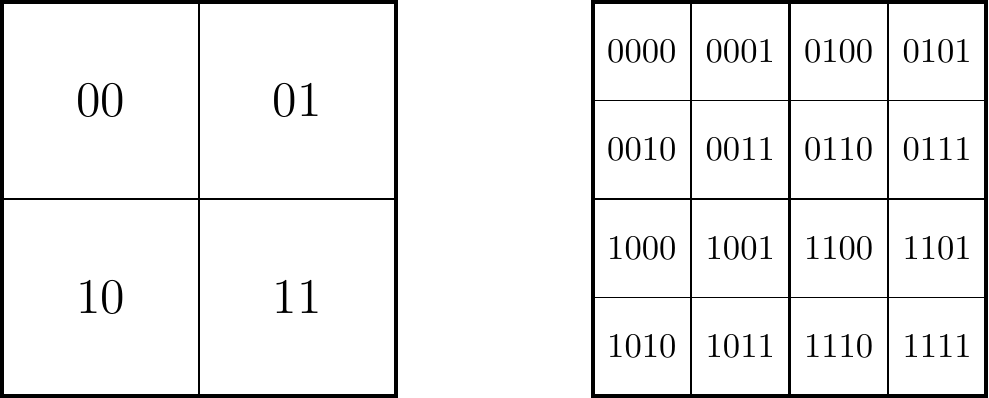}
	\caption{Qubism mapping for $N=2$ and $N=4$ particles. Adding more particles results in recursive splitting.}
	\label{fig:qubism_simplest}
\end{figure}

We proceed as in Fig.~\ref{fig:qubism_simplest}, constructing the mapping recursively.
We start with a unit square. We divide it into four quadrants. Depending on the first two bits, we pick a quadrant, according to:
\begin{align}
\begin{matrix}
00 \to \hbox{upper left} & 01 \to  \hbox{upper right} \\
10 \to \hbox{lower left} & 11 \to  \hbox{lower right}.
\end{matrix}
\end{align}
Then for each quadrant, we proceed recursively with the remaining part of the sequence.

After mapping sequences to squares, we create a complex function on the unit square,
[$0,1]\times[0,1] \rightarrow \mathbb{C}$,
that has values taken from the wavefunction amplitudes.

To be specific, for each sequence $y_1 x_1 y_2 x_2 \ldots y_{N/2} x_{N/2}$ we create a square with edge size $2^{-N/2}$ and with position (i.e. its top left corner)
\begin{align}
x = \sum_{i=1}^{N/2} 2^{-i} x_i, \qquad
y = \sum_{i=1}^{N/2} 2^{-i} y_i,\label{eq:qubism_position},
\end{align}
where we plot the $x$ coordinate from left to right and the $y$ coordinate from up to down.

\begin{figure}[!htbp]
	\centering
		\begin{tabular}{cc}
			\includegraphics[width=0.40\textwidth]{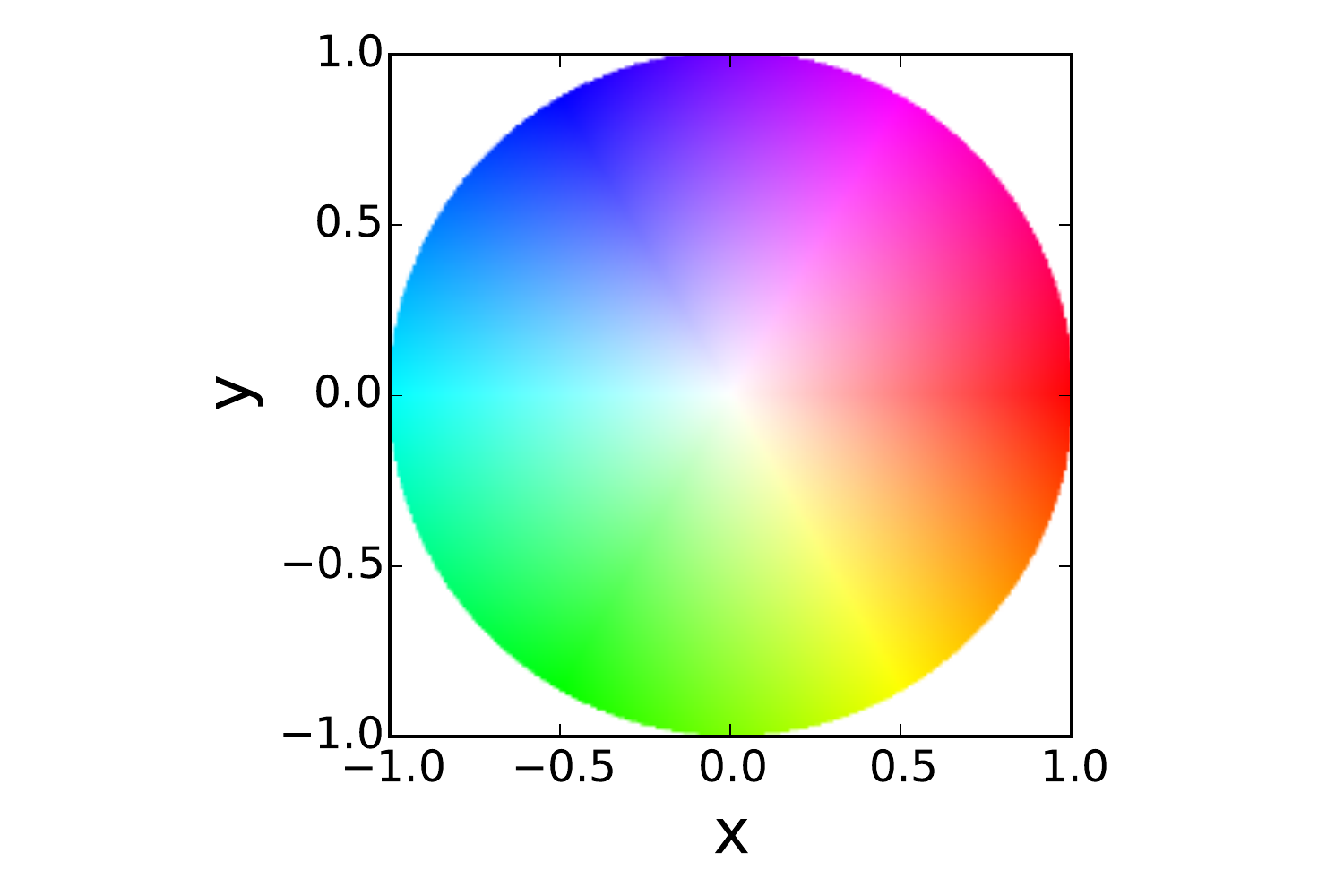} &
			\includegraphics[width=0.40\textwidth]{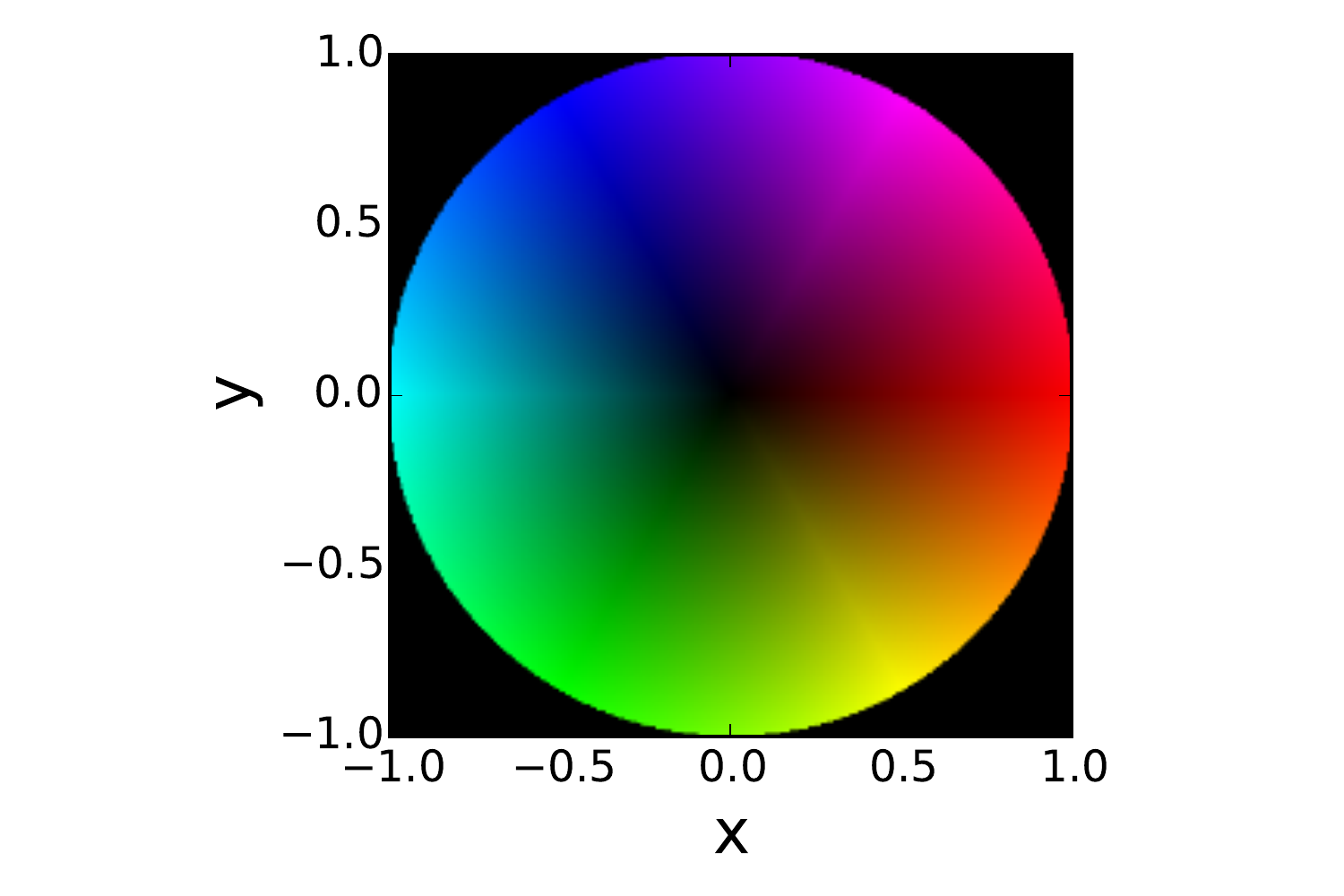} 
		\end{tabular}
	\caption{Examples of color mappings for a complex number $z=x + i y$. For clarity, only the unit disk is shown.
	Values can be scaled, so that full saturation takes place $|z|_{max}$.}
	\label{fig:complex_colors}
\end{figure}

We map complex numbers to colors \cite{Wegert2011,Petrisor2014plotting_complex}, using the absolute value $|z|$ for lightness or saturation and the phase ($\arg(z)$) for hue.
To be more specific, we use two mappings, defined in hue-saturation-value (HSV) coordinates as follows:
\begin{equation}
	\text{light: }
	\begin{bmatrix}
		H\\S\\V
	\end{bmatrix}
	=
	\begin{bmatrix}
		\arg(z)/(2\pi)\\
		\max(|z|, 1)\\
		1
	\end{bmatrix}
	\qquad
	\text{dark: }
	\begin{bmatrix}
		H\\S\\V
	\end{bmatrix}
	=
	\begin{bmatrix}
		\arg(z)/(2\pi)\\
		\max(|z|, 1)\\
		1
	\end{bmatrix}
\label{eq:complex_to_hsv}
\end{equation}
as in Fig~\ref{fig:complex_colors}.
The mapping of the phase to hue is standard. However, there are various convention for mapping of $|z|$ to lightness or saturation;
we adopt mapping as above, without going into details.

\begin{figure}[!htbp]
	\centering
		\includegraphics[width=0.30\textwidth]{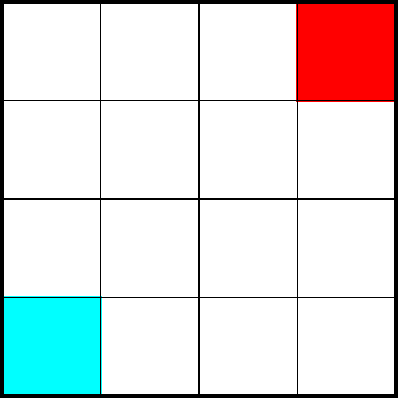}
	\caption{Plot for the state $(\ket{0101}-\ket{1010})/\sqrt{2}$. The red square denotes amplitude of $\ket{0101}$, while the teal --- of $\ket{1010}$.}
	\label{fig:qubism_example_0101}
\end{figure}

For example, for state $\ket{0101}-\ket{1010}$, the qubistic plot is as in Fig.~\ref{fig:qubism_example_0101}.

\subsection{Properties}

\begin{figure}
	\centering
		\includegraphics[width=0.8\textwidth]{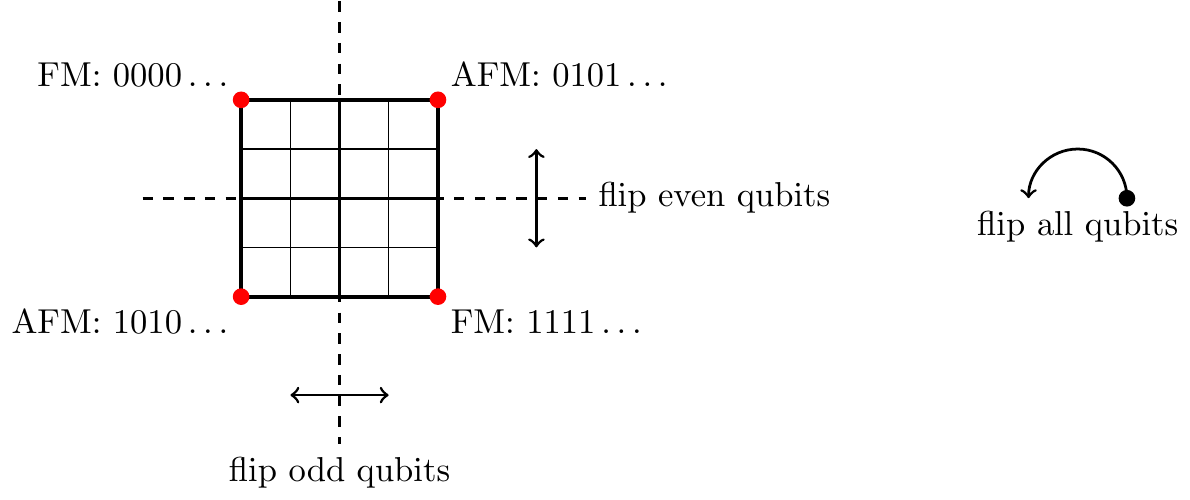}
	\caption{Basic geometrical symmetries of the plotting scheme as in Fig~\ref{fig:qubism_simplest}.}
	\label{fig:qubism_symmetries}
\end{figure}

The visualization scheme described above has some interesting geometrical properties,
which can be translated into symmetries of the state, as shown in Fig.~\ref{fig:qubism_symmetries}.

\begin{itemize}
	\item Corners correspond to:
	\begin{itemize}
		\item ferromagnetic states (i.e. $\ket{0000\ldots}$ for upper left and $\ket{1111\ldots}$ for lower right), and
		\item antiferromagnetic states (i.e. $\ket{0101\ldots}$ for upper right and $\ket{1010\ldots}$ for lower left).
	\end{itemize}
	\item Rotation of the plot by $180^\circ$ corresponds to $0\leftrightarrow1$ (changing zeros into ones and vice versa), or equivalently: application of bit swap on all particles $(\sigma^x)^{\otimes N}$).
	\item Horizontal reflection flips every even qubit ($(\mathbb{I} \otimes \sigma^x)^{\otimes N/2}$).
	\item Vertical reflections flips every odd qubit ($(\sigma^x \otimes \mathbb{I})^{\otimes N/2}$).
\end{itemize}

The recursive structure is related to the state of consecutive pairs of qubits.
Each quadrant defines a subplot, related to the projection of a wavefunction on a certain state of the first two qubits.
For example, if we measure the first two qubits in the basis $\{\ket{0}, \ket{1}\}$, and obtain $10$, then our new wavefunction is $\bra{10}_{12} \ket{\Psi}$ up to its normalization. 
But it is already in the plot --- it is just the lower left quadrant.
If we measure the two last particles, then $\bra{10}_{N-1,N} \ket{\Psi}$ is the same as taking every second pixel in both $x$ and $y$ direction, i.e. taking all pixels corresponding to sequences ending with $\ldots10$.
In particular, if a state is translationally invariant then the two above coincide, i.e. $\bra{10}_{12} \ket{\Psi} = \bra{10}_{N-1,N} \ket{\Psi}$.


The plotting scheme is valid for an arbitrary number of qubits.
However, once the number of particles gets bigger, it does make little sense to plot anything but translationally-invariant states \eqref{eq:transl-inv}.

\subsection{Examples}

\subsubsection{Product state}

\begin{figure}[!htbp]

\centering
\includegraphics[width=8cm]{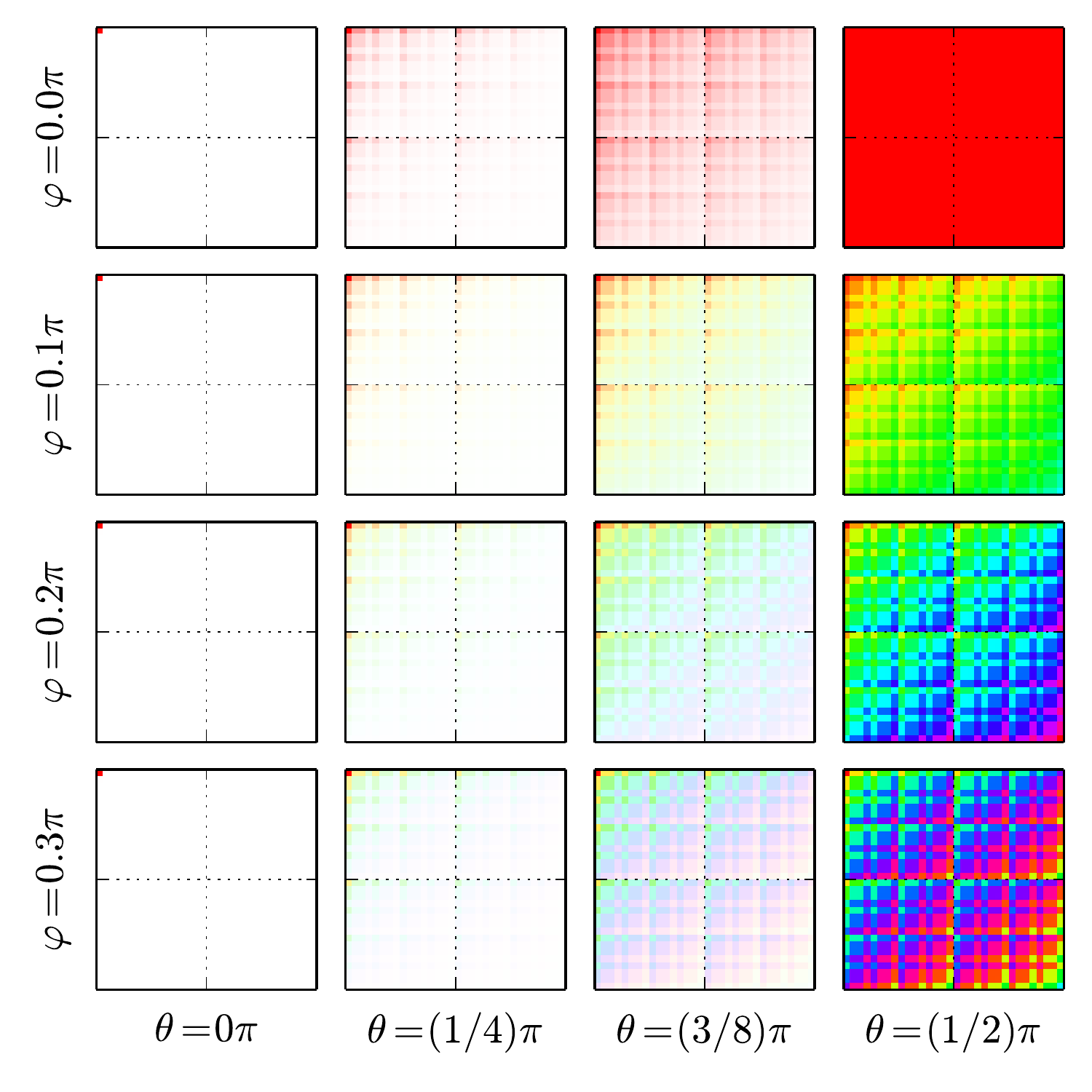}
\caption{\label{fig:product_state} An array of product states for $N=10$ particles,
for $\alpha=\cos(\theta/2)$ and $\beta=\sin(\theta/2) \exp(i \varphi)$.}
\end{figure}

Let us start with the simplest possible state --- a product state of the form 
\begin{align}
\ket{\Psi} = (\alpha \ket{0} + \beta \ket{1})^{\otimes N},
\end{align}
which is depicted in Fig.~\ref{fig:product_state}.
The pattern is self-similar, but in some sort of trivial way --- each subplot is proportional to other subplots.
For example $\bra{00}_{12} \ket{\Psi} \propto \bra{10}_{12} \ket{\Psi}$.
It is directly related to the fact that by measuring the state of one particle we do not disturb the results for others.
Or, in other words, that particles are not correlated in any way.

\subsubsection{Dicke states}

\begin{figure}[!htbp]
\centering
	\includegraphics[width=8cm]{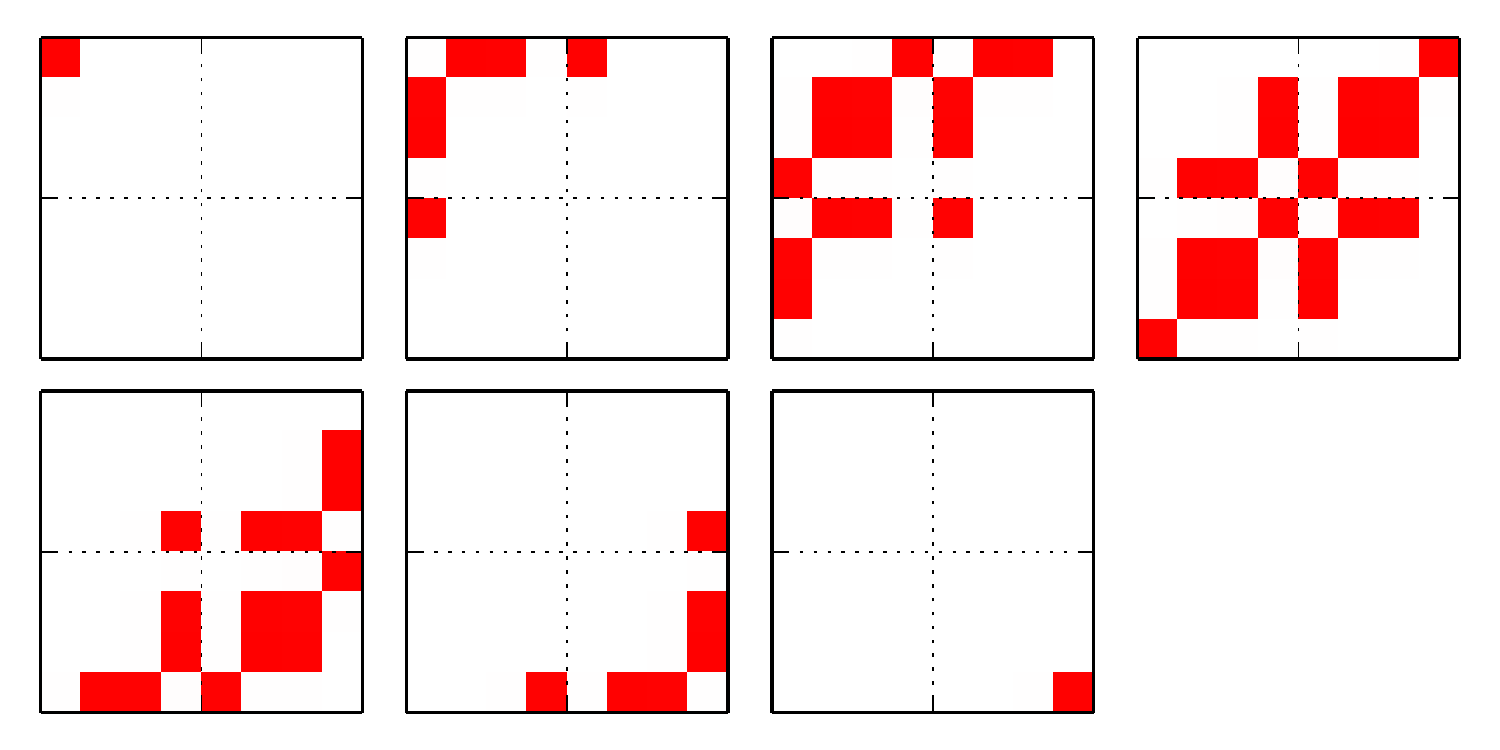}
	\caption{\label{fig:dicke-states-6}Dicke states for $N=6$ particles, with $k=0,1,\ldots,N$. Notice how the plot changes from all zeros $\ket{000000}$, going through a state with the same number of 0s and 1s ($k=3$) to all ones $\ket{111111}$.} 
\end{figure}

The next state we would like to plot is the \emph{Dicke state} \cite{Dicke1954}, that is
\begin{align}
D^N_k = \binom{N}{k}^{-1/2} \sum_{\text{inequiv. perm.}} \ket{0}^{\otimes (N-k)} \ket{1}^{\otimes k},
\end{align}
or, in other words, a state defined by all linear combinations of basis states with fixed number of $1$s.
In particular, for $k=1$ we get the \emph{W states}, for $N=3$ particles is
\begin{equation}
\frac{\ket{001} + \ket{010} + \ket{100}}{\sqrt{3}}.
\end{equation}
For six particles we plot all Dicke states, in Fig.~\ref{fig:dicke-states-6}.

\begin{figure}[!htbp]
\centering
	\includegraphics[width=3cm]{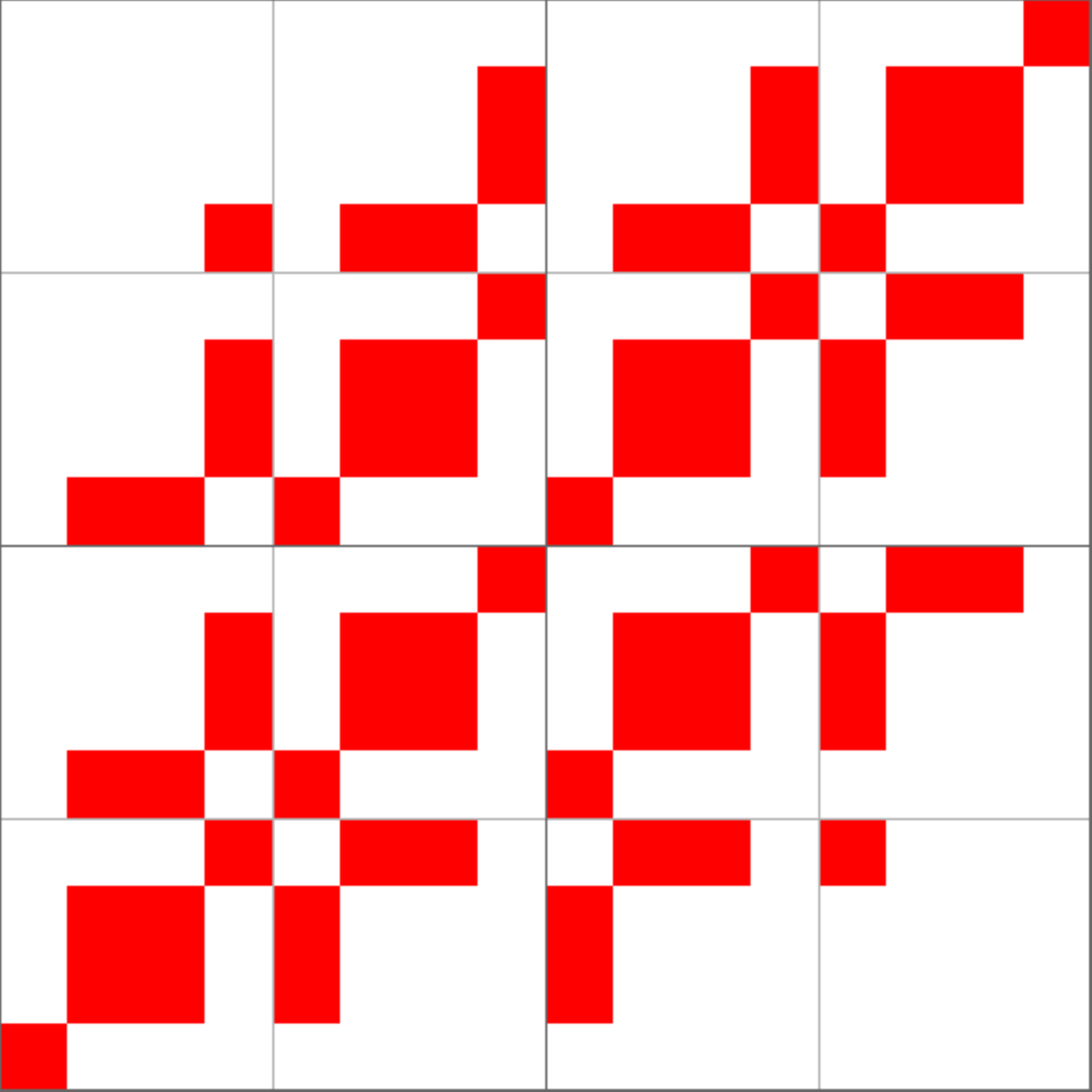}
	\includegraphics[width=3cm]{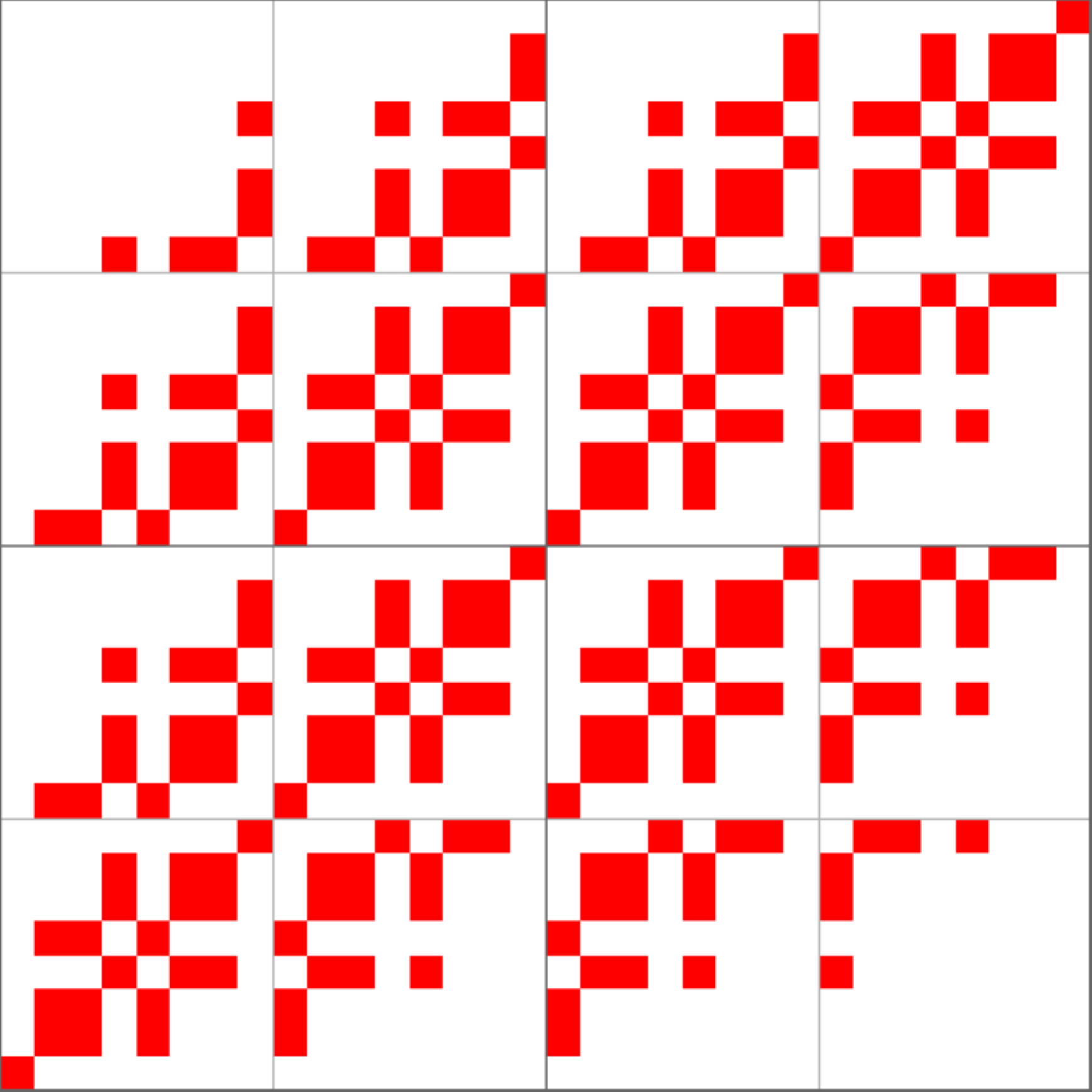}
	\includegraphics[width=3cm]{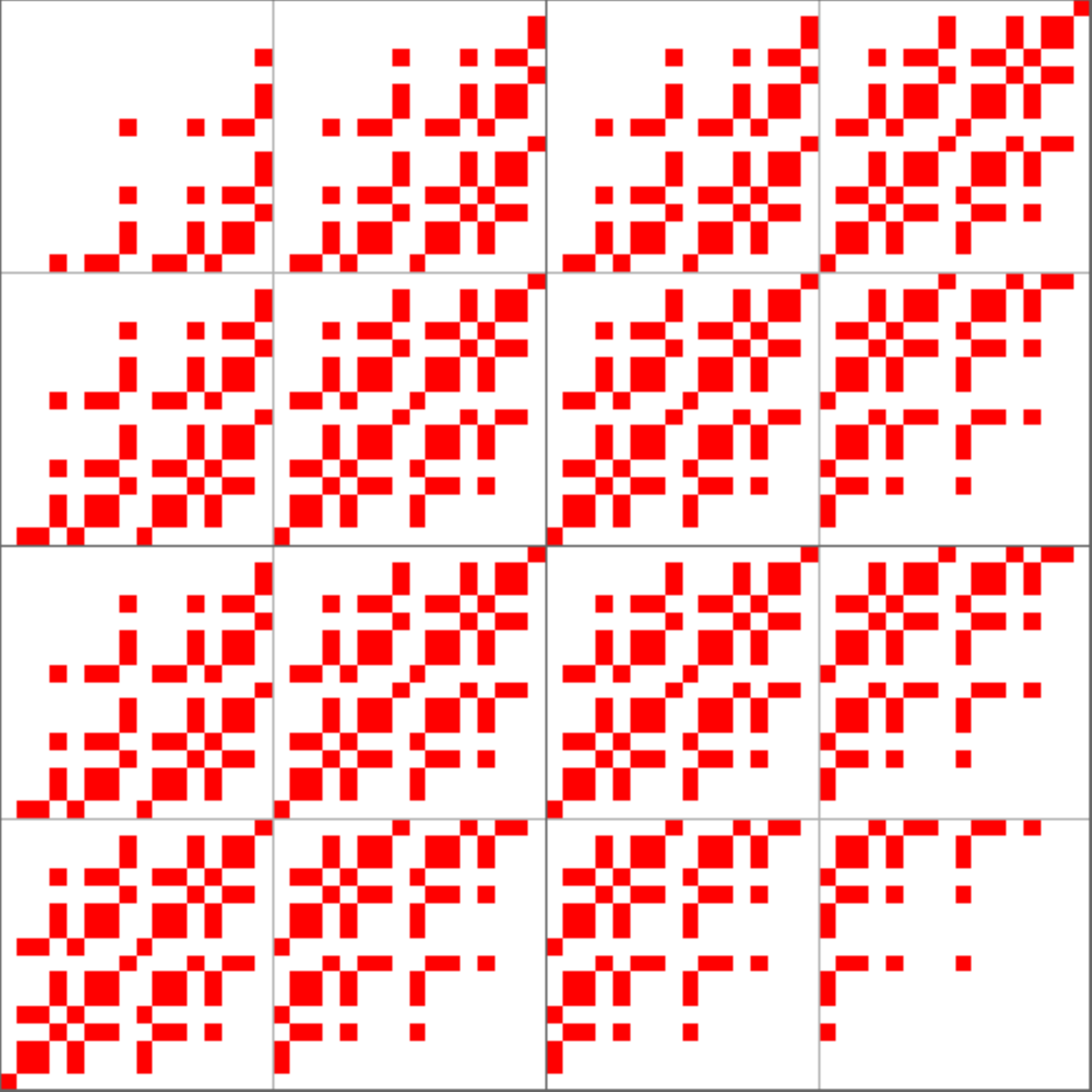}
	\includegraphics[width=3cm]{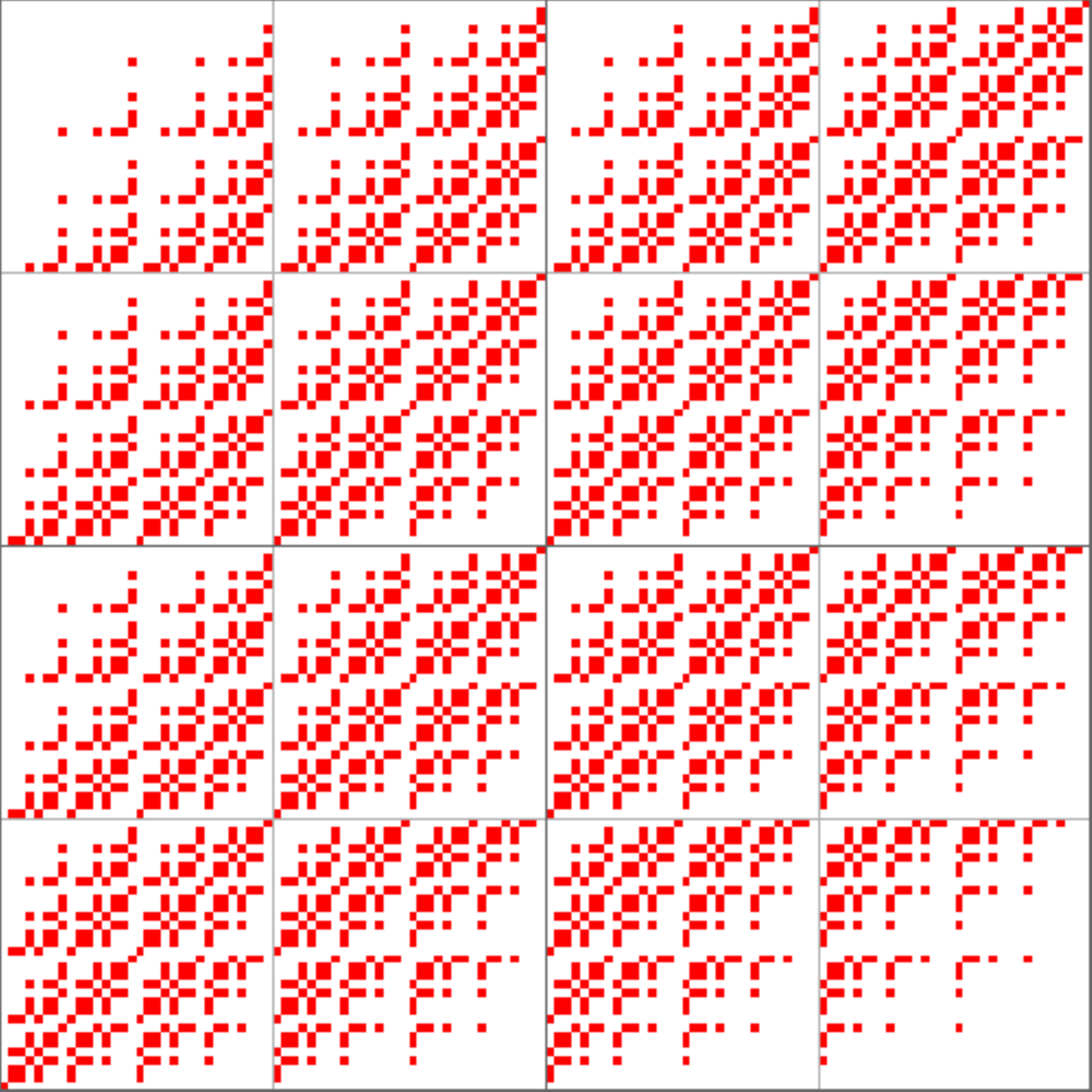}
	\caption{\label{fig:dicke-states-hf}Dicke states with half-filling for $N=8$, $10$, $12$
	and $14$ qubits. Notice how the fractal structure develops.} 
\end{figure}

For even $N$ we can consider Dicke states with the same number of $0$s as $1$s, that is, with $k=N/2$.
They can be related to the ground state of a fermionic system at half-filling,
where every fermion interacts with every other with the same coefficient.
This state is plotted in Fig.~\ref{fig:dicke-states-hf}, for various particle numbers.

Every Dicke state is permutation-symmetric, i.e. permutation of the particle order leaves it unchanged.
In fact, they form a basis for the permutation-symmetric subspace of qubits;
or equivalently ---- for bosonic states in two modes, written in the particle basis.
This means that whenever we find a qubistic plot being a superposition of shapes as in Fig.~\ref{fig:dicke-states-6},
the state is permutation symmetric.

\subsubsection{Ising model in a transverse field}
\label{s:ising_model}

Let us consider spin-$1/2$ antiferromagnetic Ising model in a transverse field, in a 1D spin chain
\begin{equation}
H = \sum_{i=1}^N \sigma^z_i \sigma^z_{i+1} - \Gamma \sum_{i=1}^N \sigma^x_i,
\label{eq:itf_model}
\end{equation}
where $\Gamma$ is a parameter describing the strength of the transverse field.
Let us use periodic boundary conditions, i.e. $\sigma^z_{N+1} \equiv \sigma^z_{1}$.
That is, spins of neighboring particles are coupled through the $z$ component of their spins,
while at the same time a perpendicular field tries to align spins along its axis.

\begin{figure}[!htbp]
\centering
	\includegraphics[width=\textwidth]{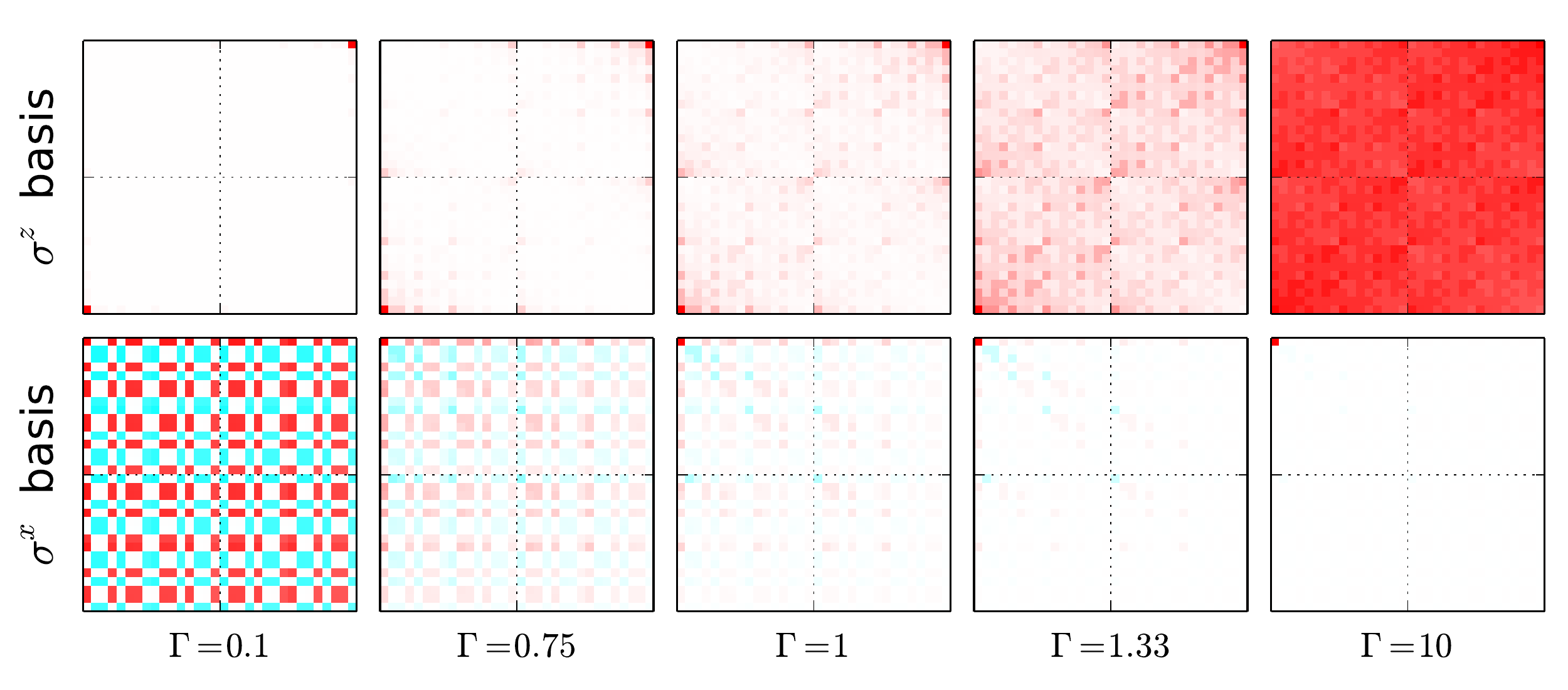}
	\caption{\label{fig:itf_transition}Ground state of the Ising model with transverse field Hamiltonian with $N=10$ qubits and periodic boundary conditions.
	Values of the transverse field are $\Gamma=0.1$, $0.75$, $1.0$, $1.33$, and $10$.
 	The critical point, $\Gamma_c=1$, corresponds central panel.
 	We show results both in $\sigma^z$ and $\sigma^x$ basis.} 
\end{figure}

Depending on the strength of the transverse field, one of two alignments dominate. 
For $\Gamma=0$ the ground state consists only of two N\'eel states
(i.e. $\ket{0101\ldots}$ and $\ket{1010\ldots}$).
For $\Gamma \to \infty$ the ground state is a product of states pointing in the $x$ direction, i.e. 
$\ket{+}=(\ket{0} + \ket{1})/\sqrt{2}$.
But the most interesting is what happens in between. At $\Gamma=1$ there is a quantum phase transition.
The transition is plotted in Fig.~\ref{fig:itf_transition}.

\subsubsection{Heisenberg Hamiltonian and Majumdar-Ghosh model}

The next system we want to study is the 1D Majumdar-Ghosh model \cite{Majumdar1969}
\begin{equation}
H = \sum_{i=1} \vec{\sigma}_i \cdot \vec{\sigma}_{i+1}
+ J \sum_{i=1} \vec{\sigma}_i \cdot \vec{\sigma}_{i+2},\label{eq:majumdar_ghosh_hamiltonian}
\end{equation}
that is, an antiferromagnetic model with spin-spin interactions between nearest neighbors and second nearest neighbors.
The later are parametrized by $J$.
We can consider different boundary conditions:
\begin{itemize}
\item periodic (spin chain forms a circle, summations in \eqref{eq:majumdar_ghosh_hamiltonian} are up to $N$; we identify $\vec{\sigma}_{N+1} \equiv  \vec{\sigma}_{1}$ and $\vec{\sigma}_{N+2} \equiv  \vec{\sigma}_{2}$),
\item open (spin chain does not form a circle, summations in \eqref{eq:majumdar_ghosh_hamiltonian} are up to $N-1$ and $N-2$).
\end{itemize}

Isotropic spin-spin interaction
\begin{equation}
\tfrac{1}{2}\vec{\sigma}_i \cdot \tfrac{1}{2}\vec{\sigma}_j
= \tfrac{1}{4}\left(\sigma_i^x \sigma_j^x + \sigma_i^y \sigma_j^y + \sigma_i^z \sigma_j^z\right)
\end{equation}
is invariant with respect to collective rotation, i.e. $U^{\otimes N}$ for any $U\in \text{SU}(2)$.
Consequently, all eigenstates of the Hamiltonian built from these operators can be labeled by their total spin number.
In this case, for an even number of particles $N$, we expect the ground state to be a singlet, i.e. to have total spin $0$.

\begin{figure}[!htbp]
\centering
\includegraphics[width=8cm]{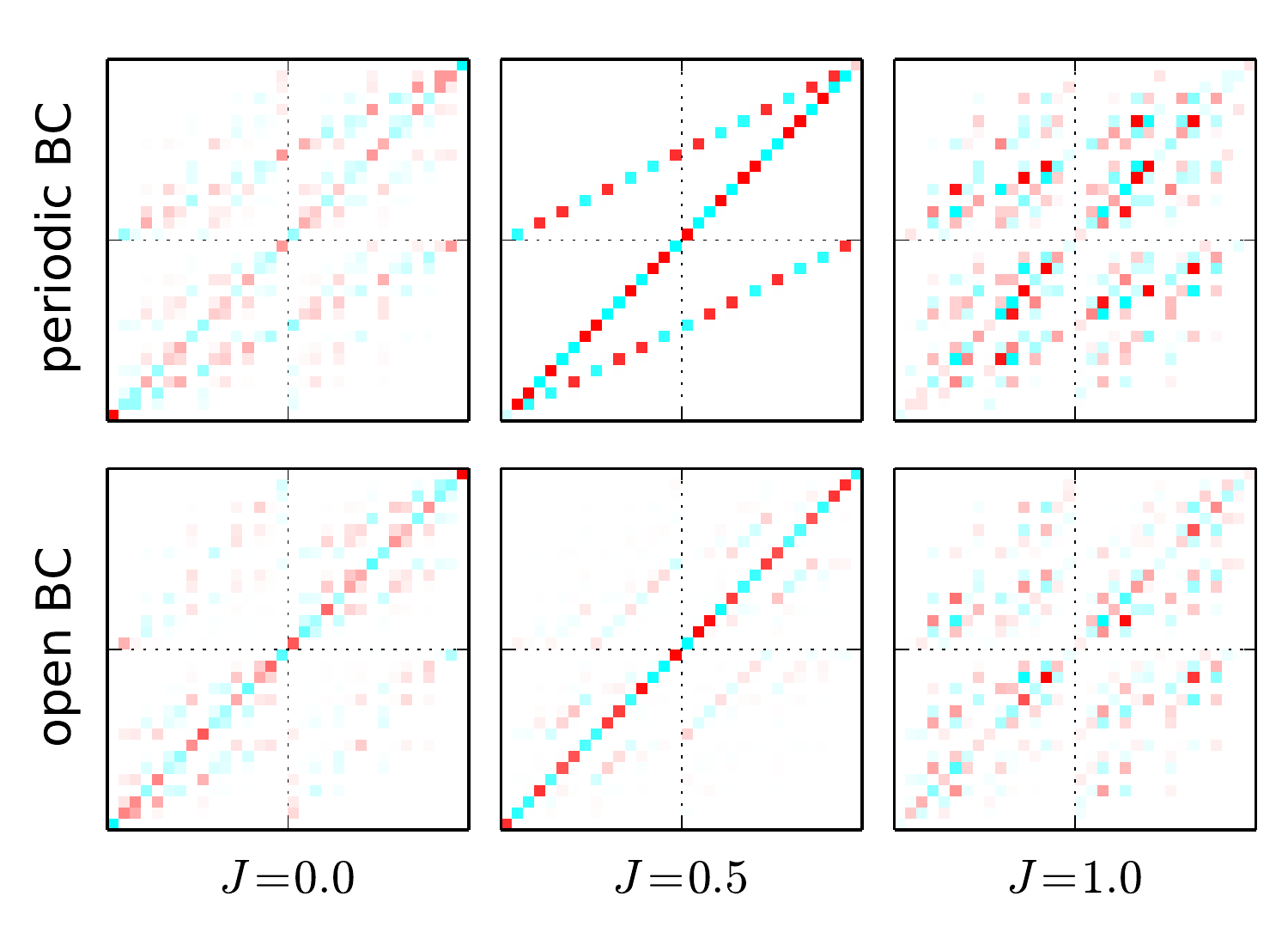}
\caption{\label{fig:heisenberg_majumdar_ghosh}
Majumdar-Ghosh model for periodic and open boundary conditions for $N=10$ qubits.
For $J=0$ it corresponds to the Heisenberg model, while for $J=1/2$ its ground state is the Majumdar-Ghosh state.
The qubistic plot is drawn only in one basis, as in other bases it is the same, due to the ground state being a singlet.
Notice the characteristic Z-like shape and how it is affected by changing boundary conditions. See also Fig.~\ref{fig:heisenberg_majumdar_ghosh_skewed}.
}
\end{figure}

For $J=0$ we have only nearest neighbor interactions --- the Heisenberg model.
For $J=0.5$ the ground state can be exactly found and is called the Majumdar-Ghosh state \cite{Majumdar1970}.
In Fig.~\ref{fig:heisenberg_majumdar_ghosh} we plot the ground state for various $J$,
both for periodic and open boundary conditions.

One of the striking features of this qubistic plot is the Z-like shape.
For now, let us focus only on the anti-diagonal line.
From their position in the plot, these states are of the form:
\begin{equation}
\{\ket{01}, \ket{10} \}^{N/2}.
\end{equation}
As we see, absolute values of their amplitudes are the same, but their sign varies.
Colors in the upper-right quadrant ($\ket{01}_{12}$) are complementary to colors in the lower-left quadrant ($\ket{10}_{12}$).
Consequently, we can write the state as
\begin{equation}
\ket{\psi_{ad}} = \frac{\ket{01}_{12} - \ket{10}_{12}}{\sqrt{2}} \ket{\psi_{ad}}_{34\ldots N}.
\end{equation}
Noticing that the plot is recursive (either graphically or from the fact that we deal with translation-invariant state, at least for periodic boundary conditions),
we see that the the anti-diagonal is a product of two-particle singlets $(1,2)(3,4)\ldots(N-1,N)$ --- with the bracket $(i,j)$ meaning the two-particle singlet state of $i$-th and $j$-th particle --- or
\begin{equation}
\ket{\psi_{ad}} = \left(\frac{\ket{01} - \ket{10}}{\sqrt{2}} \right)^{N/2}\label{eq:singlet_product_even}.
\end{equation}

But how can we interpret the two remaining lines in the Z-like shape? 
For periodic boundary conditions, the ground state needs to be translationally invariant.
After shifting \eqref{eq:singlet_product_even} by $1$ particle, we get a product of singlet pairs for $(2,3)(4,5)\ldots(N,1)$. 
It should be not surprising that this state has very low amplitude for open boundary conditions.
In fact, in \cite{Majumdar1970} it was shown that the ground state of \eqref{eq:majumdar_ghosh_hamiltonian} for $J=1/2$ and periodic boundary conditions
is exactly a superposition of \eqref{eq:singlet_product_even} and its shift, i.e. $(\ket{\psi_{ad}} + T\ket{\psi_{ad}})\sqrt{2}$.

What may remain puzzling is why, in the qubistic plot, there are two lines for $T\ket{\psi_{ad}}$.
It is related to the fact, that for plotting we use as our ``alphabet'' consecutive pairs of spins. 
Position of a single amplitude is, in the binary system,
\begin{align}
x &= 0.s_2 s_4 \ldots s_N,\\
y &= 0.s_1 s_3 \ldots s_{N-1}.
\end{align}
So, for the singlet pairs $(1,2)(3,4)\ldots(N-1,N)$ we have
\begin{equation}
s_{2k} = 1-s_{2k-1} \quad \text{so} \quad x \quad \approx 1 - y,\label{eq:mg_antidiag}
\end{equation}
where the approximation is up to plot resolution.
For the singlet pairs $(2,3)(4,5)\ldots(N,1)$ we have
\begin{equation}
s_{2k+1} = 1-s_{2k} \quad \text{so} \quad x \approx (1 - 2y) \mod 1,
\end{equation}
as multiplying by $2$ shifts $y$ into the already solved instance \eqref{eq:mg_antidiag}.

\subsubsection{Spin-1 and AKLT states}

The qubistic plotting scheme is by no means restricted to qubits. 
While we are presenting more general theory in Sec.~\ref{s:qubism_general},
it is straightforward to make a generalization for quantum states built out of qudits ($d$-level systems).

\begin{figure}[!htbp]
	\centering
		\includegraphics[width=0.20\textwidth]{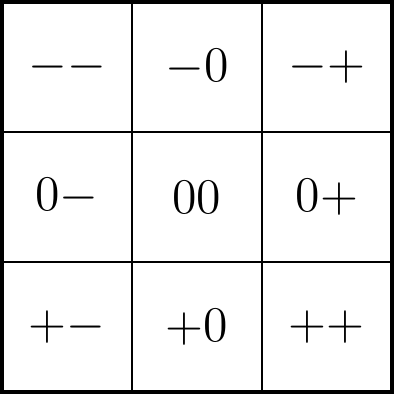}
	\caption{Qubism plotting scheme for qutrits, analogous to Fig.~\ref{fig:qubism_simplest}.}
	\label{fig:qubism_qutrits}
\end{figure}

For example, let us focus on $d=3$ in terms of a spin-1 system.
As a basis, we can use eigenstates of the spin operator in the $z$-th direction.
Then, the local basis is $\{-1,0,1\}$ or, for the sake of simplicity,
$\{-,0,+\}$.
The only difference from $d=2$ (or qubits), is that instead of dividing the square into $2 \times 2$ quadrants, we divide it into $3 \times 3$ quadrants, see Fig.~\ref{fig:qubism_qutrits}.

As an example, let us choose the Affleck-Kenedy-Lieb-Tasaki (AKLT) state \cite{Affleck1987},
i.e. the ground state of the following Hamiltonian: 
\begin{equation}
H = \sum_{i=1}^N \vec{S}_i \cdot \vec{S}_{i+1}
+ \frac{1}{3} (\vec{S}_i \cdot \vec{S}_{i+1})^2,
\label{eq:haldane_hamiltonian}
\end{equation}
where $\vec{S}$ is the spin-1 operator,
i.e. $\vec{S}=(S_x, S_x, S_z)$ and
\begin{align}
S_x =
\frac{1}{\sqrt{2}}
\left[
\begin{matrix}
0 & 1 & 0\\
1 & 0 & 1\\
0 & 1 & 0
\end{matrix}
\right]
\quad
S_y =
\frac{1}{\sqrt{2} i}
\left[
\begin{matrix}
0 & 1 & 0\\
-1 & 0 & 1\\
0 & -1 & 0
\end{matrix}
\right]
\quad
S_z =
\left[
\begin{matrix}
1 & 0 & 0\\
0 & 0 & 0\\
0 & 0 & -1
\end{matrix}
\right].
\end{align}

This state is an example of a valence bond solid, and has attracted considerable attention because of its relation to the Haldane conjecture \cite{Haldane1983},
its non-local order parameter \cite{DenNijs1989} and as a source of inspiration for tensor-network states \cite{Perez-Garcia2006}.

\begin{figure}
\centering
\includegraphics[width=6cm]{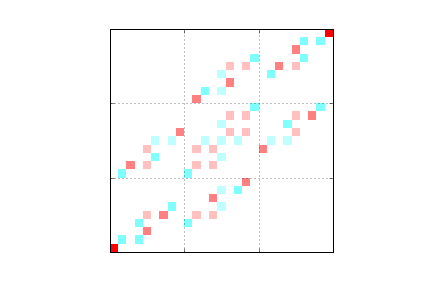}
\caption{\label{fig:aklt} Ground state of the AKLT spin-1 Hamiltonian,
for $N=6$ spins.}
\end{figure}
\begin{figure}
\centering
\includegraphics[width=4cm]{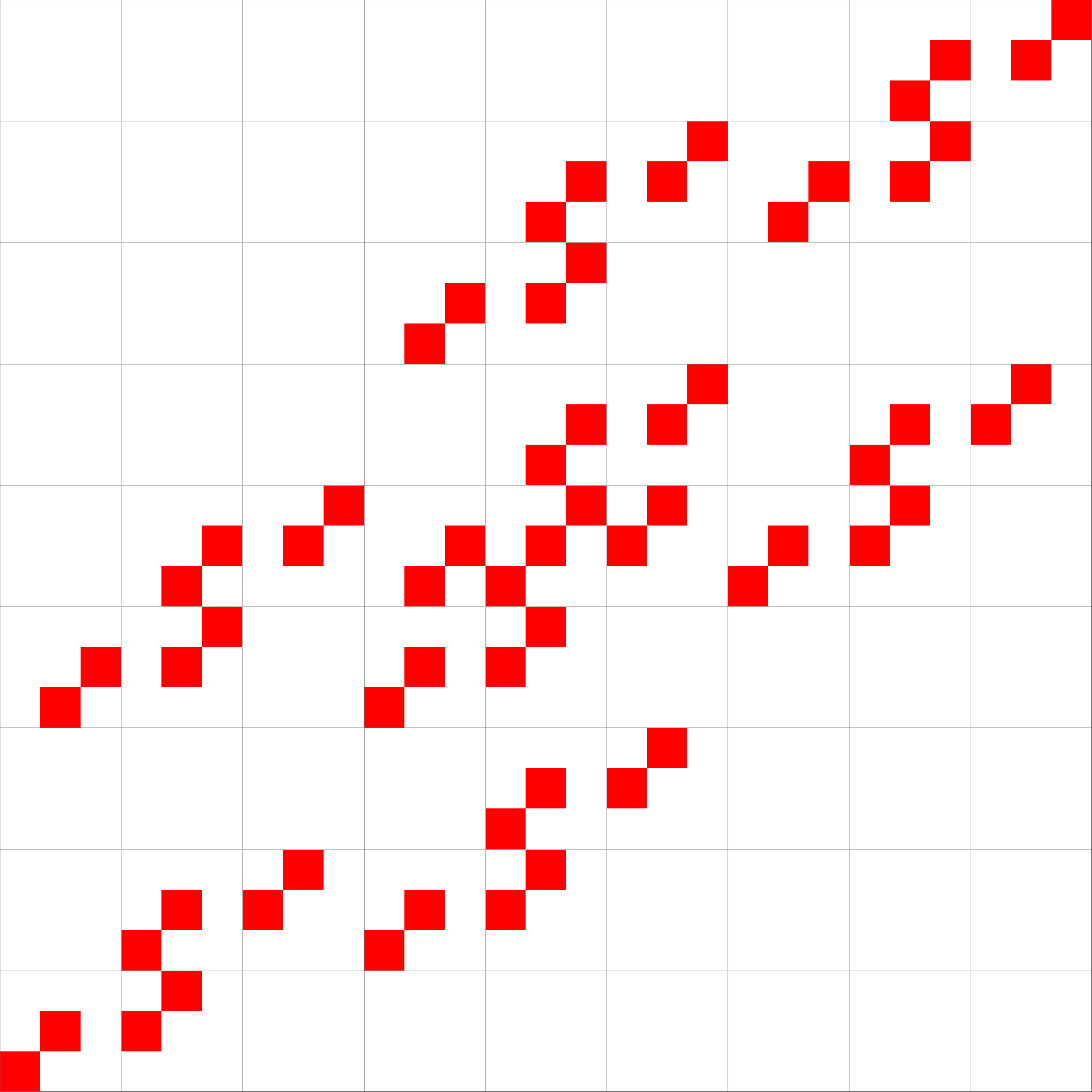}
\includegraphics[width=4cm]{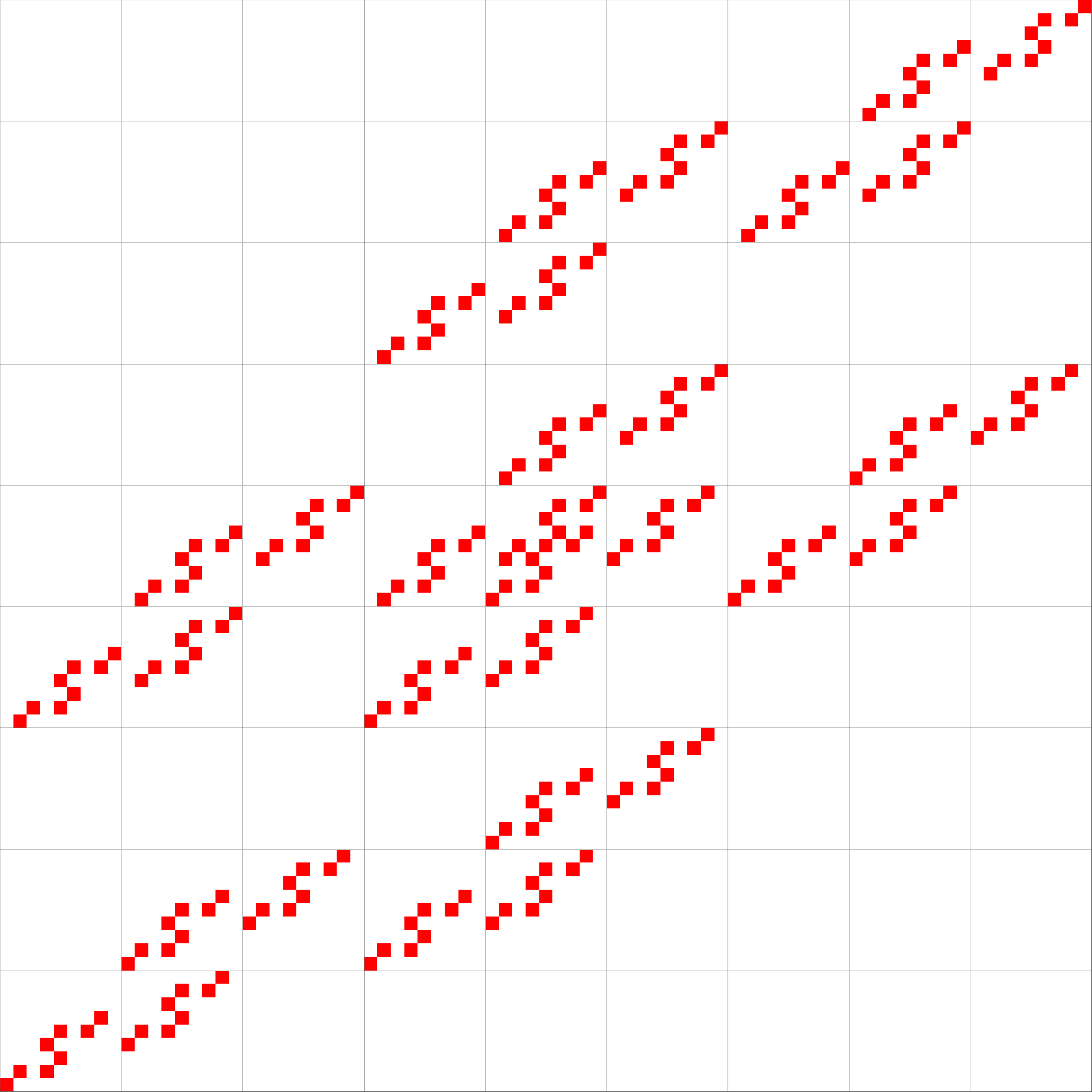}
\includegraphics[width=4cm]{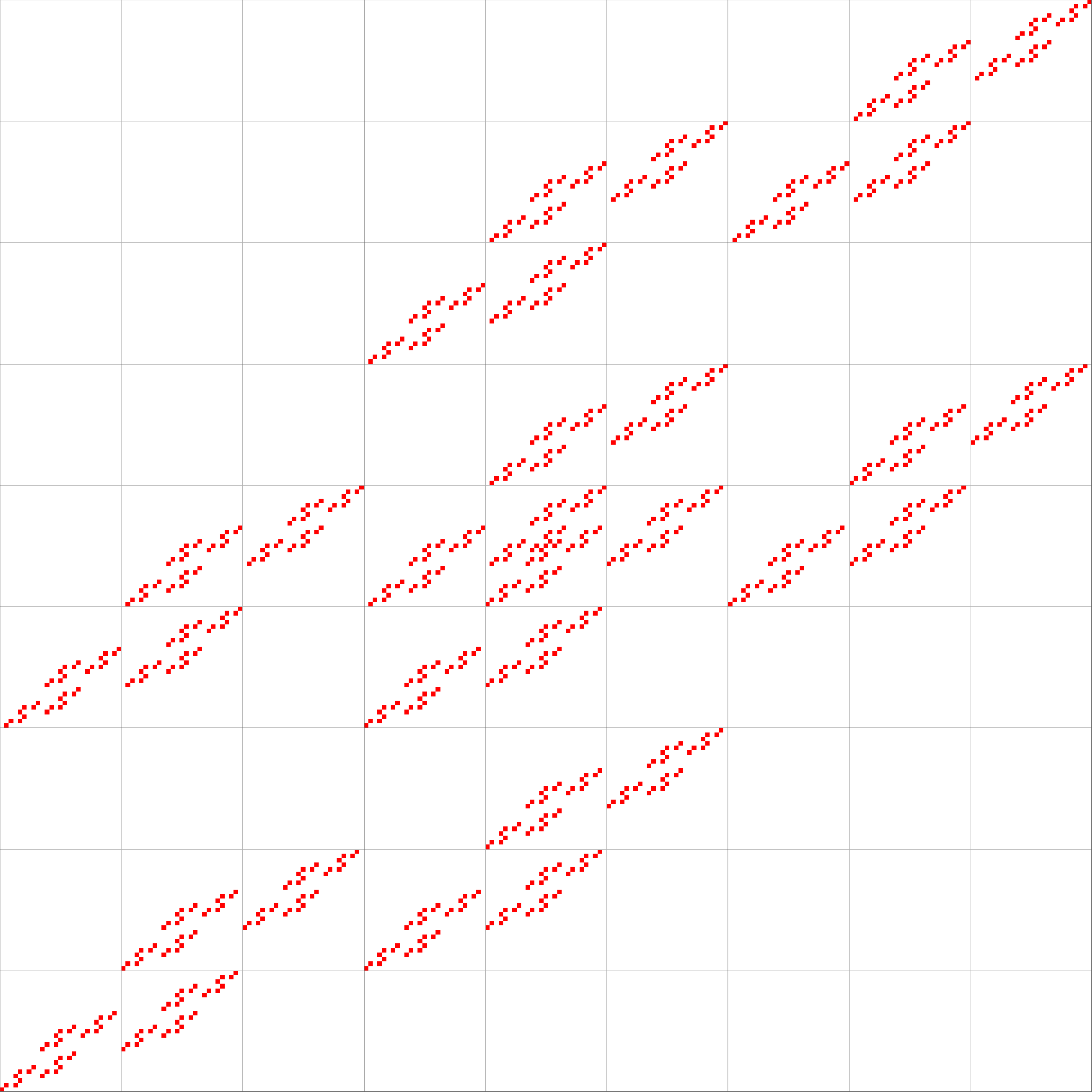}
\caption{\label{fig:aklt_nonzero} Ground state of the AKLT spin-1 Hamiltonian,
for $N=6$, $8$ and $10$ spins. The same color is used to every non-zero amplitude.
  Notice how the characteristic fractal structure of a snowflake develops.}
\end{figure}

We plot the ground state in Fig.~\ref{fig:aklt},
and shows its fractal structure in Fig.~\ref{fig:aklt_nonzero},
where we abstract the wavefunction to zero and non-zero values.
The qubistic plot shows that there are no consecutive $+$ or $-$ in any sequence.
Furthermore, we see that when the last entry was $+$ then the next one cannot start with $+$ or be $0+$ (and analogously for $-$).
Thus, we arrive at a rather accurate description of AKLT state, which contains all sequences with alternating $+$ and $-$ separated by an arbitrary number of $0$ states.

It is worth noting that the AKLT state is a prototypical matrix product state.
That is, it can be written \cite[Sec. 4.1.5.]{Schollwock2011} as
\begin{equation}
\alpha_{s_1 s_2 \ldots s_N} = \Tr \left[ A^{s_1} A^{s_2} \cdots A^{s_N} \right],
\end{equation}
where matrices are
\begin{equation}
A^{-} =  -\sqrt{\tfrac{2}{3}}
\left[
\begin{matrix}
0 & 0\\
1 & 0
\end{matrix}
\right]
\quad
A^{0} =  \sqrt{\tfrac{1}{3}}
\left[
\begin{matrix}
-1 & 0\\
0 & 1
\end{matrix}
\right]
\quad
A^{+} =  \sqrt{\tfrac{2}{3}}
\left[
\begin{matrix}
0 & 1\\
0 & 0
\end{matrix}
\right].
\end{equation}
or with the notation from \cite{Crosswhite2008}, i.e. using
\begin{equation}
A \equiv \ket{-} A^{-} + \ket{0} A^{0} + \ket{+} A^{+}
\end{equation}
we get
\begin{equation}
A =
\begin{bmatrix}
\tfrac{-1}{\sqrt{3}} \ket{0} & \tfrac{\sqrt{2}}{\sqrt{3}} \ket{+}\\
\tfrac{\sqrt{2}}{\sqrt{3}} \ket{-} & \tfrac{1}{\sqrt{3}} \ket{0}
\end{bmatrix}.
\end{equation}
That is, we use pure states as matrix entries, and use tensor product when multiplying matrices.
For instance
\begin{align}
\Tr [ A A ]
&= \Tr
\begin{bmatrix}
\tfrac{1}{3} \ket{00} + \tfrac{2}{3} \ket{+-} &
- \tfrac{\sqrt{2}}{3} \ket{0+} + \tfrac{\sqrt{2}}{3} \ket{+0} \\
- \tfrac{\sqrt{2}}{3} \ket{-0} + \tfrac{\sqrt{2}}{3} \ket{0-}&
\tfrac{2}{3} \ket{-+} + \tfrac{1}{3} \ket{00}
\end{bmatrix}\\
&= \tfrac{2}{3} \left( \ket{00} + \ket{+-} + \ket{-+}\right).
\end{align}

\subsection{General framework}\label{s:qubism_general}

From a very abstract point of view,
a set of all possible visualization of a $N$-qudit wavefunctions onto a unit square is:
\begin{align}
\left( \{0,1,\ldots,d-1\}^N \rightarrow \mathbb{C} \right)
\rightarrow \left( [0,1]^2 \rightarrow \mathbb{R}_{\geq 0}^3 \right),
\end{align}
where $\mathbb{R}_{\geq 0}^3$ stands for intensities of red, green and blue components.
This formula is very general --- it also includes writing amplitudes with fixed precision numbers.

However, we want to focus on specific visualizations, where
\begin{itemize}
\item all amplitudes are shown, 
\item each amplitude is represented by a color,
\item the position of the region in which a certain amplitude is drawn does not depend on any of amplitude values.
\end{itemize}
That is, we restrict ourselves to visualizations which can be formulated as
\begin{align}
[0,1] \times[0,1] &\to \{0,1,\ldots,d-1\}^N\label{eq:geometric_mapping}\\
\mathbb{C} &\to \mathbb{R}_{\geq 0}^3,\nonumber
\end{align}
that is, visualization schemes for which each position is related to some sequence $\vec{s}$, and the color there is the color for amplitude $\alpha_{\vec{s}}$.

When it comes to the spatial mapping, we would like to add assumptions related to their recursive structure (making it a \emph{qubistic} visualization, not --- any ordering of amplitudes on a square).
For a function $f$ as in \eqref{eq:geometric_mapping}, we take the inverse image $f^{-1}(\{\cdot\})$, which for every spin sequences gives the region domain it is mapped to:
\begin{align}
f^{-1}(\{s_1 s_2 \ldots s_N\})
= A_{s_1} f^{-1}(\{s_2 s_3 \ldots s_N\}),
\end{align}
where $A_{s}$ is an affine transform and they are both complete and non-intersecting:
\begin{align}
\bigcup_{s_1} A_{s_1} f^{-1}(\{s_2 s_3 \ldots s_N\})
= f^{-1}(\{s_2 s_3 \ldots s_N\})\\
s_i \neq s_j \Rightarrow
\left( A_{s_i} f^{-1}(\{s_2 s_3 \ldots s_N\}) \right)
\cap
\left( A_{s_j} f^{-1}(\{s_2 s_3 \ldots s_N\}) \right) 
= \emptyset.
\end{align}
The second condition can be relaxed --- it does need to be empty, measure zero is enough.

Moreover, instead of using states of one particle, as an alphabet, we can use states of a small number of consecutive particles
e.g. $\{\ket{00}, \ket{01}, \ket{10}, \ket{11}\}$,
though we will not do it for all visualizations we study here.
This recipe can be easily generalized for qudits, and for higher dimensional representations.
For example, 3D representations  can show relations between 3 consecutive particles easily.

\subsubsection{Technical remarks}

Since the global phase has no physical meaning, we can fix it by setting the phase according to one of these recipes:
\begin{itemize}
\item Ensure that a certain selected amplitude is positive (arbitrary and not always possible),
\item Ensure that the sum of the wavefunction entries is positive (not always possible; for singlet states it is always impossible).
\item For a sequence of wavefunctions, ensure that $\braket{\psi_i}{\psi_{i+1}}$ is positive (works only for sequences, with consecutive entries being non-orthogonal; the starting global phase remains arbitrary). 
\end{itemize}
For real wavefunctions it is somehow easier, as only the sign can change. Yet, even in this case, when for example,
tracking how the ground state changes when Hamiltonian parameters are being modified, it is better to have coherent colors.
It is especially important for processes where changes of the phase are important, for example --- the Berry phase \cite{Berry1984} acquired for a state evolving in an adiabatically changed setting.

As another remark, recursive structure allows us to find the position of ferromagnetic and antiferromagnetic states as a limit of
\begin{align}
\lim_{N \to \infty} A_{s_1 s_2}^N
\left[
\begin{matrix}
1/2\\
1/2\\
1
\end{matrix}
\right].
\end{align}

\subsection{Mappings}

\subsubsection{Typical mapping for qubits}

\begin{figure}[!htbp]
	\centering
		\includegraphics[width=0.30\textwidth]{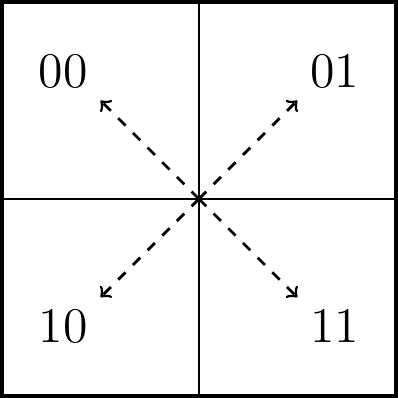}
	\caption{Qubism mapping with affine transforms.}
	\label{fig:qubism_mapping_simplest}
\end{figure}

The typical mapping for qubits, defined as in Fig.~\ref{fig:qubism_mapping_simplest} and which we use in the previous examples, can be defined with the following affine transformations:
\begin{align}
A_{s_1s_2} &=
\left[
\begin{matrix}
B & \vec{r}_{s_1s_2}\\
0 & 1
\end{matrix}
\right]
\end{align}
where $B$ is a matrix scaling down by factor $2$, and $\vec{r}_{s_1s_2}$ is a translation dependent on two consecutive spins, i.e.:
\begin{equation}
B =
\left[
\begin{matrix}
1/2 & 0\\
0 & 1/2
\end{matrix}
\right]
\end{equation}
\begin{equation}
\vec{r}_{00} =
\left[
\begin{matrix}
-1/4\\
1/4
\end{matrix}
\right]
\quad
\vec{r}_{01} =
\left[
\begin{matrix}
1/4\\
1/4
\end{matrix}
\right]
\quad
\vec{r}_{10} =
\left[
\begin{matrix}
-1/4\\
-1/4
\end{matrix}
\right]
\quad
\vec{r}_{11} =
\left[
\begin{matrix}
1/4\\
-1/4
\end{matrix}
\right]
\end{equation}

or as depicted in Fig.~\ref{fig:qubism_mapping_simplest} (compare it to Fig.~\ref{fig:qubism_simplest}).

\subsubsection{Alternate mapping for qubits}

We can define an alternative mapping that puts emphasis on the difference between ferromagnetic and antiferromagnetic states.
It is similar to the original one, but positions for $01$ and $10$ are swapped.
That is:
\begin{equation}
\vec{r}_{00} =
\left[
\begin{matrix}
-1/4\\
1/4
\end{matrix}
\right]
\quad
\vec{r}_{01} =
\left[
\begin{matrix}
-1/4\\
-1/4
\end{matrix}
\right]
\quad
\vec{r}_{10} =
\left[
\begin{matrix}
1/4\\
1/4
\end{matrix}
\right]
\quad
\vec{r}_{11} =
\left[
\begin{matrix}
1/4\\
-1/4
\end{matrix}
\right]
\end{equation}

In this mapping ferromagnetic states are on the left, while antiferromagnetic are on right.

In general, for qubits there are only three inequivalent qubistic square plotting schemes.
That is, there are $4!=24$ permutations of $\{ \ket{00}, \ket{01}, \ket{10}, \ket{11}\}$,
but the symmetry group of square has $8$ elements (identity, 3 rotations, 4 reflections).
Or, in other words,
a square scheme for qubits can be defined by what is the square is on the opposite site of $\ket{00}$.
However, if we consider visualization of states up to translations, then we end up with only two schemes.

Note that this alternate mapping is the same as the typical mapping of a state subjected to a product of controlled swaps, i.e.
\begin{equation}
\ket{\psi} \mapsto
\Big( \ket{0}\bra{0} \otimes \mathbb{I} + \ket{1}\bra{1} \otimes \sigma^x \Big)^{N/2} \ket{\psi}.
\end{equation}

\begin{figure}[!htbp]
\centering
\includegraphics[width=8cm]{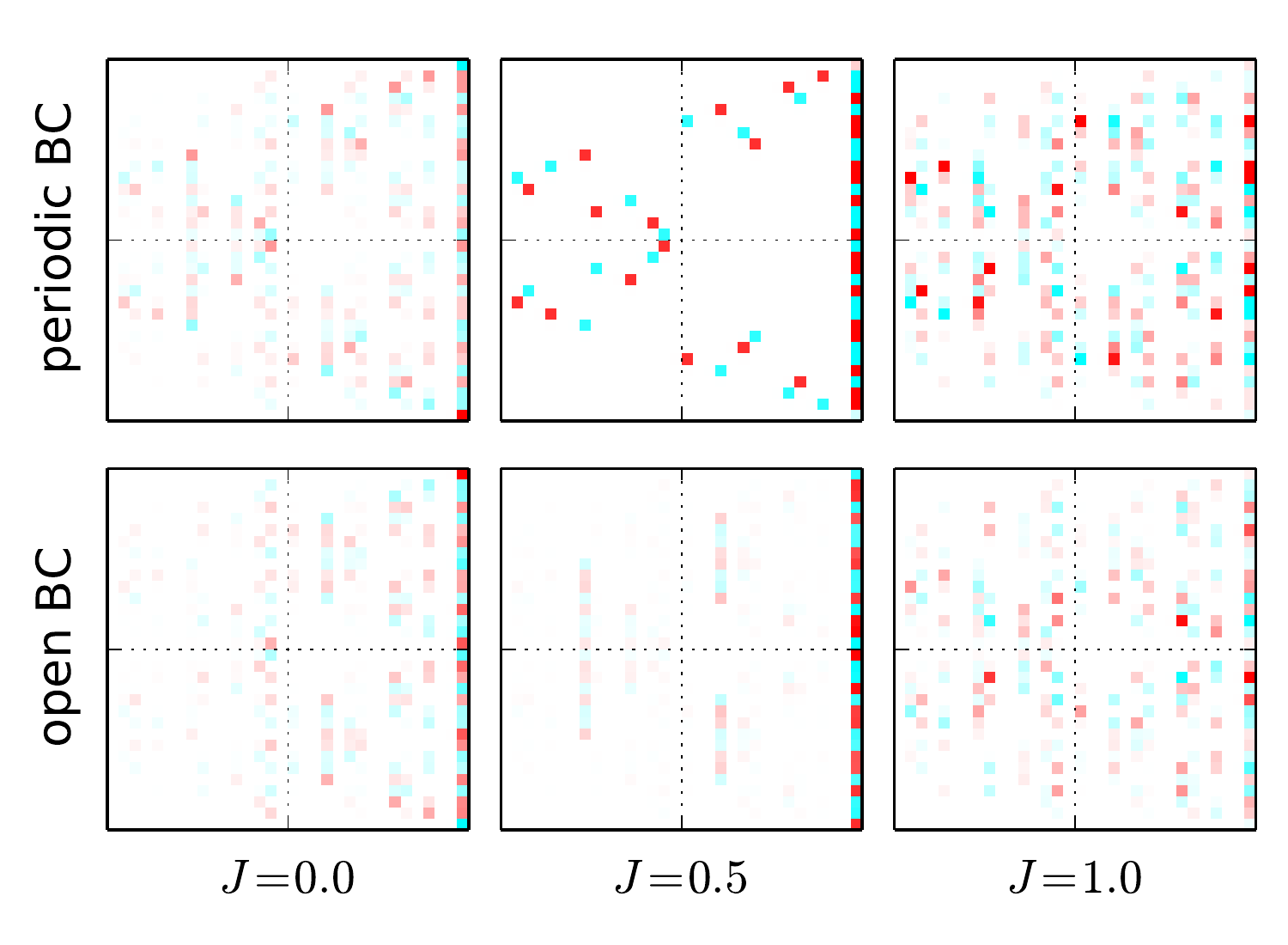}
\caption{\label{fig:heisenberg_majumdar_ghosh_skewed}
Majumdar-Ghosh model for periodic and open boundary conditions for $N=10$ qubits,
plotted using alternate mapping for qudits. Cf. Fig.~\ref{fig:heisenberg_majumdar_ghosh}.
Note that for products of singlets of the form $(1,2)(3,4)\ldots(N-1,N)$, as in \eqref{eq:singlet_product_even}, we obtain a line on the right (instead of the diagonal line).
}
\end{figure}

For example, in Fig.~\ref{fig:heisenberg_majumdar_ghosh_skewed} we show the grounds states of the Majumdar-Ghosh model using the alternate mapping. The physical content is the same as in Fig.~\ref{fig:heisenberg_majumdar_ghosh}.

\subsubsection{Square mapping for qudits}

\begin{figure}[!htbp]
	\centering
		\includegraphics[width=0.80\textwidth]{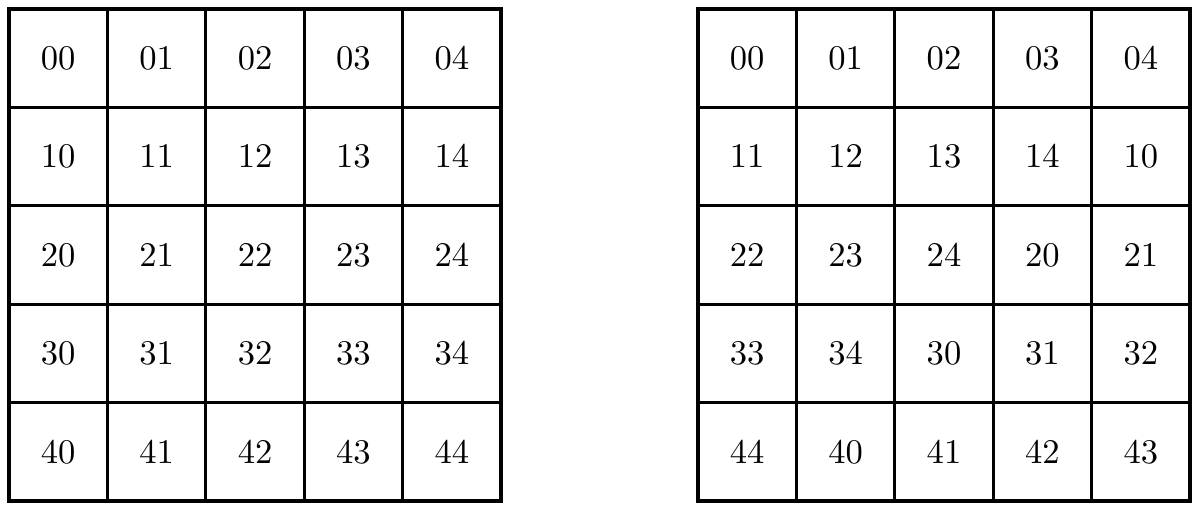}
	\caption{Qubism mapping for qudits in two variants: typical (right) and alternate; example for $d=5$.}
	\label{fig:qudits_mapping}
\end{figure}

The square visualizations can be generalized for $d$-level systems, as exemplified in Fig.~\ref{fig:qudits_mapping}.
In this case we have:
\begin{equation}
B =
\left[
\begin{matrix}
1/d & 0\\
0 & 1/d
\end{matrix}
\right]
\end{equation}
and
\begin{equation}
\vec{r}_{s_1 s_2} =
\left[
\begin{matrix}
\frac{2 s_1 + 1}{2 d} - \frac{1}{2}\\
\frac{2 f(s_1, s_2) + 1}{2 d} - \frac{1}{2}
\end{matrix}
\right]
\end{equation}
for $s_1,s_2 \in \{0, 1, \ldots, d-1 \}$,
where for the typical scheme we have
\begin{equation}
f(s_1, s_2) = s_2
\end{equation}
and for the alternate one:
\begin{equation}
f(s_1, s_2) = (s_2 - s_1) \mod d.
\end{equation}
In general, $f$ can be any permutation of $s_2$ as a function of $s_1$.
Such mapping can be also understood in terms of coordinates as \eqref{eq:qubism_position},
where instead of base $2$ we use base $d$.
In fact, we have already used this mapping (for $d=3$) in Figures \ref{fig:aklt} and \ref{fig:aklt_nonzero}.

\subsubsection{Triangular scheme}

Square plots are not the only possibility.
One of their shortcomings is that they look differently for an even and odd number of particles.
That is for an odd number of particles they pixel is a rectangle, instead of a square.

\begin{figure}[!htbp]
\centering
\includegraphics[width=10cm]{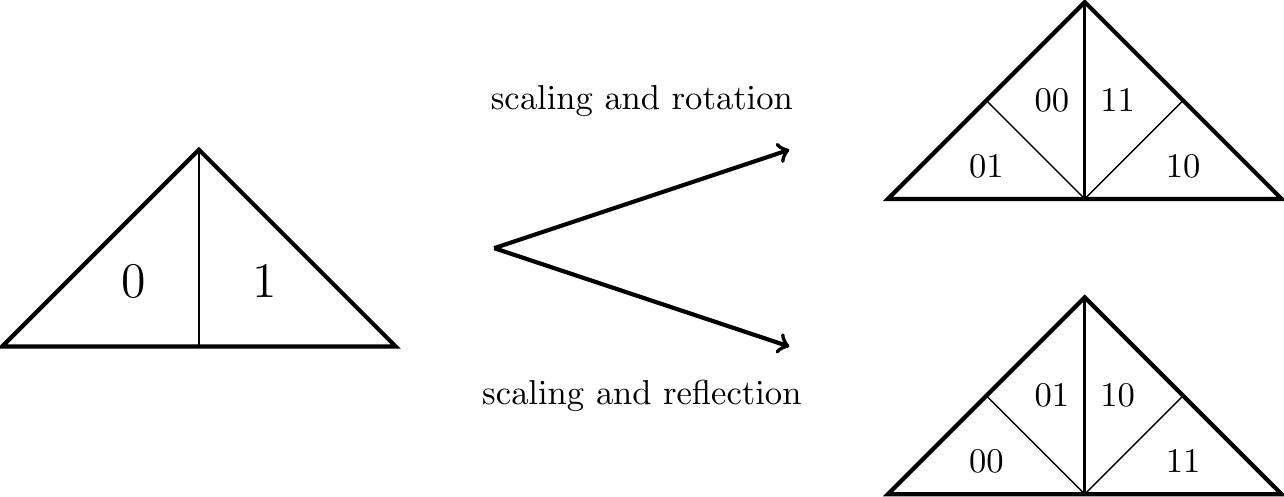}
\caption{\label{fig:qubism_scheme_triangle} Two variants for the triangular qubistic plotting schemes.}
\end{figure}

\begin{figure}
	\centering

	\includegraphics[width=6cm]{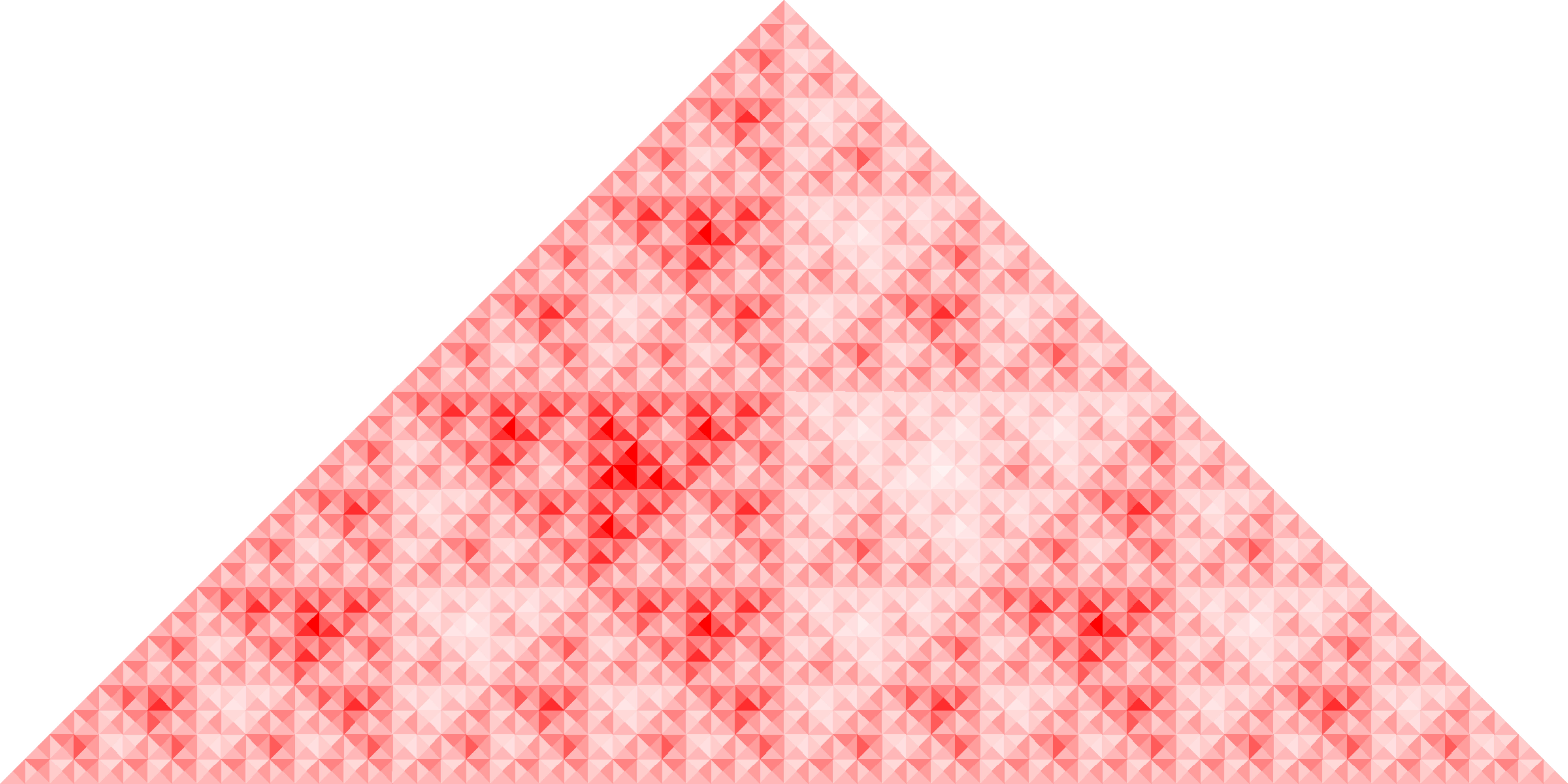}\\
	product state
	\vspace{5mm}

	\includegraphics[width=6cm]{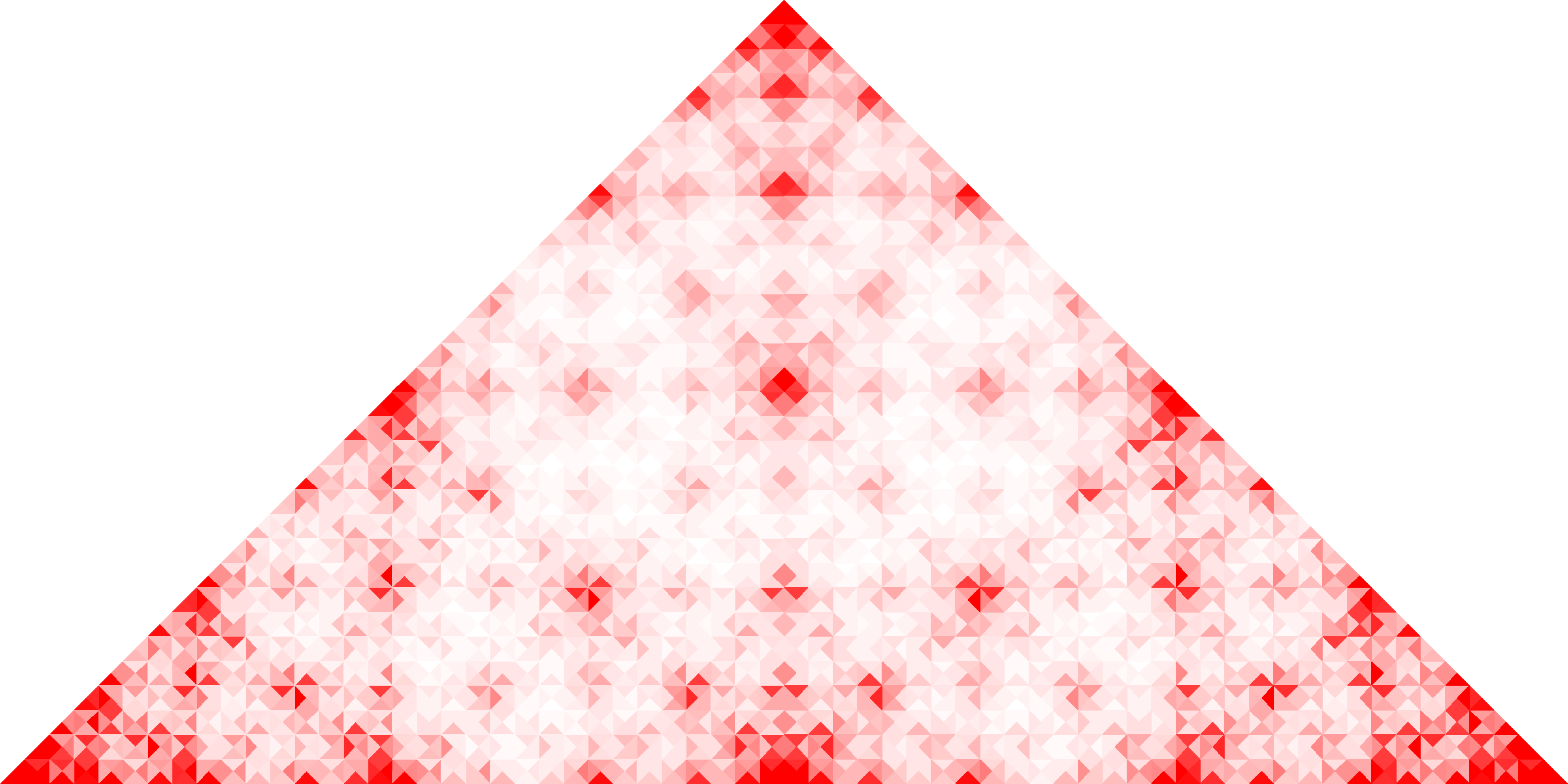}\\
	ITF ground state for $\Gamma=1$
	\vspace{5mm}

	\includegraphics[width=6cm]{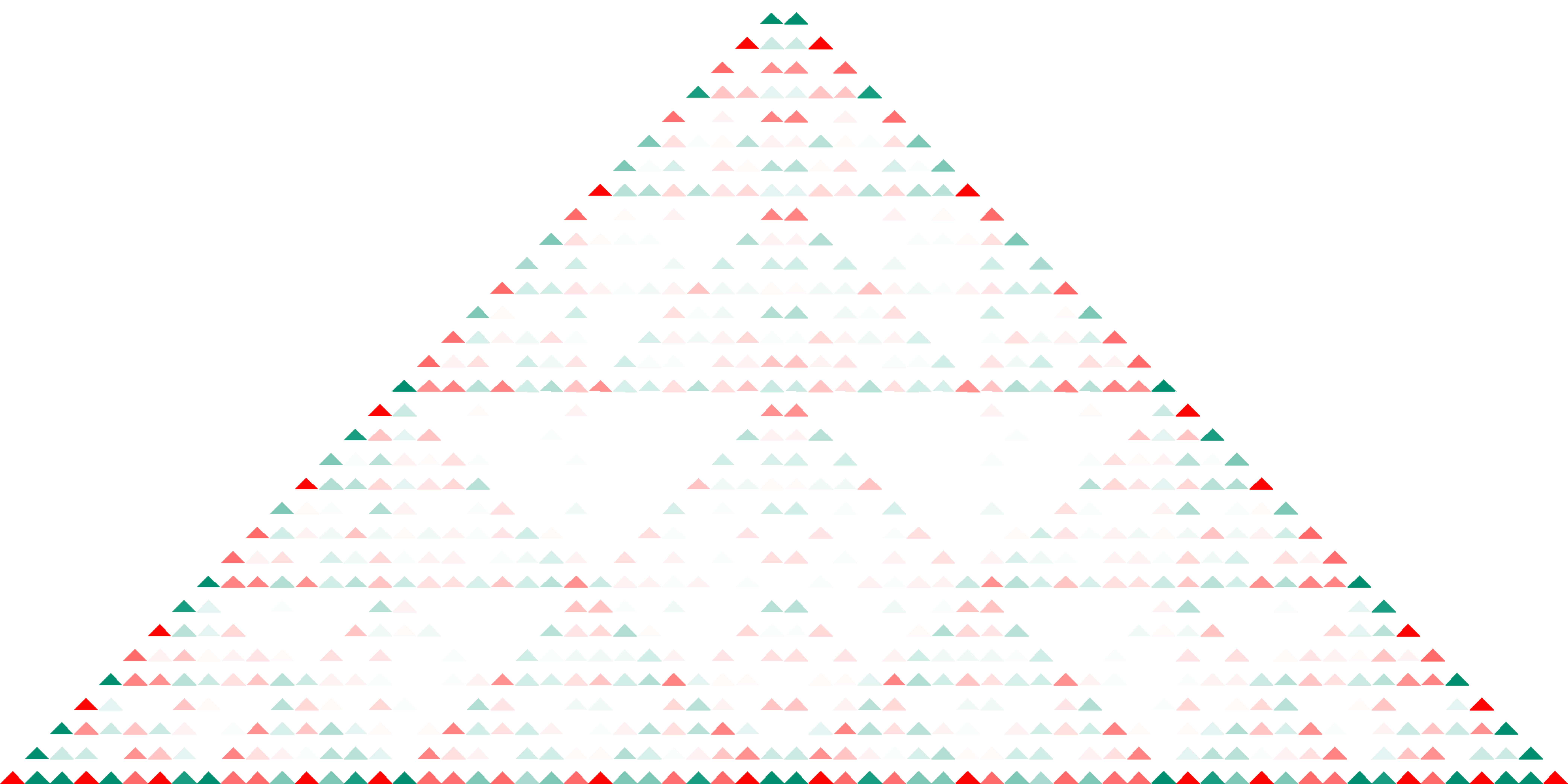}\\
	Heisenberg ground state

	\caption{\label{fig:triangular_examples}Triangular representations of many-body wavefunctions, all for $N=12$ qubists.}
\end{figure}

Let us create a plot starting from a right triangle, with vertices at $(-1,0)$, $(0,1)$ and $(1,0)$.
It can be split into two similar triangles, scaled by a factor $1/\sqrt{2}$.
The shifts, starting from the middle of the basis of the triangle, are
\begin{equation}
\vec{r}_0 =
\begin{bmatrix}
-\tfrac{1}{2}\\
+\tfrac{1}{2}
\end{bmatrix}
\quad
\vec{r}_1 =
\begin{bmatrix}
\tfrac{1}{2}\\
\tfrac{1}{2}
\end{bmatrix}
\end{equation}
and the linear transformations are
\begin{equation}
B_0 =
\begin{bmatrix}
-\tfrac{a}{2} & -\tfrac{1}{2}\\
 \tfrac{a}{2} & -\tfrac{1}{2}
\end{bmatrix}
\quad
B_1 =
\begin{bmatrix}
-\tfrac{a}{2} &  \tfrac{1}{2}\\
-\tfrac{a}{2} & -\tfrac{1}{2}
\end{bmatrix},
\end{equation}
where the parameter $a=1$ is for rotation and $a=-1$ for reflection.
Both variants are depicted in Fig.~\ref{fig:qubism_scheme_triangle}.
We provide example plots in Fig.~\ref{fig:triangular_examples}.

\subsubsection{Other qubistic schemes}

One can design other visualizations.
For example, a qubistic scheme based on splitting a equilateral triangle into $4$ similar triangles would produce a pattern similar to that of Sierpiński triangle.

Furthermore, while presented visualizations emphasize two-body correlations, it is possible to devise a qubistic scheme capturing a few-body relations.
For instance, doing a 3-dimensional visualization analogous to Fig.~\ref{fig:qubism_simplest} would reveal 3-particle correlations. 
Alternatively, particles can gather in tuples, so instead of considering a system of $n$ particles of $d$ levels, we can consider $n/k$ particles of $d^k$ levels.

\subsubsection{Schmidt plot}
\label{sec:schmidt_plot}

Besides qubistic plotting schemes, with their recursive structure,
we would like to discuss one more type of plots --- \emph{Schmidt plots}.
Justification for the name will be given later, related to the Schmidt decomposition.
A pure state of $N$ particles can be decomposed into correlated systems of $k$ and $N-k$ particles.
That is, parameterizing the wavefunction by indices related the respective sets of particles, we effectively get a matrix $\ket{\psi}_{\mu,\nu}$.
We plot this matrix as a density plot, with the same color scheme as discussed throughout this chapter.

We call this kind of plot the \emph{Schmidt plot}, as it is related to the Schmidt decomposition (i.e. the Singular Value Decomposition of the matrix $\ket{\psi}_{\mu,\nu}$).
A product state with respect to a given partition is a state
\begin{equation}
\ket{\psi}_{\mu,\nu} = \ket{\phi_1}_\mu \ket{\varphi_1}_\nu.
\end{equation}
Consequently, we can see entanglement by observing structure of the plot.

An alternative description of the Schmidt plot is plotting amplitudes of a wavefunction in a similar manner to that of \eqref{eq:qubism_position}, but using the $y$ coordinate for first $k$ particles and $x$ for the last $N-k$ particles.
Also note that the typical qubistic scheme is equivalent to the Schmidt plot for odd vs even particles.
In particular, if the qubistic scheme is a product of horizontal and vertical lines, it means there is no entanglement between subsets
\begin{equation}
	(1,3,5,\ldots,N-1) \qquad \text{and} \qquad (2,4,6,\ldots,N).
\end{equation}
Furthermore, the Schmidt plot is is related to a variant of qubistic scheme, where instead of starting with the first particles, we start with the middle particles, i.e. with the consecutive pairs being $(N/2, N/2 + 1)$, $(N/2 - 1, N/2 + 2)$, $(N/2 - 2, N/2 + 3)$, $\ldots$.
This plot is the same as the Schmidt plot for ordering the following ordering of particles:
\begin{equation}
	(N/2, N/2 - 1, \ldots, 1, N/2 +1, N//2 + 2, \ldots N). 
\end{equation}
Thanks to these similarities, some Schmidt plots look the same as their qubistic variants: for example plots of permutation-symmetric states, for which the ordering of particles is irrelevant.

\subsection{Entanglement visualization}
\label{sec:entanglement-visualization}

One of the hallmark properties of quantum mechanics is the existence of entanglement \cite{Horodecki2009,Amico2008,Eisert2010} --- many-particle correlations that cannot be described by classical models.
In this section we present a general way to visualize quantum entanglement.
While, given a pure state, it is easy to compute whether a system split into two parties is entangled, is it possible to plot a state in a way that entanglement is visible?

For a pure state $\ket{\psi}$ we can perform the Schmidt decomposition:
\begin{equation}
\ket{\psi} = \sum_i \lambda_i \ket{\phi_i} \ket{\varphi_i},
\label{eq:schmidt_decomposition}
\end{equation}
and entanglement is stored in the Schmidt coefficients $\lambda_i$.
The Schmidt decomposition is Singular Value Decomposition of the matrix $\ket{\psi}_{\mu,\nu}$,
where indices $\mu$ and $\nu$ are related to the first and the second subsystem, respectively.
A straightforward way to visualize such system would be to show the matrix using a Schmidt plot, as in Sec.~\ref{sec:schmidt_plot}.
However, it allows us to visualize entanglement only for one particular splitting.
We will show that with qubism it is possible to show entanglement for various splittings within one plot.

Alternatively to \eqref{eq:schmidt_decomposition}, we can perform the partial trace of one of the subsystems, and get a reduced density matrix
\begin{equation}
\rho_1 = \sum_i \lambda_i^2 \ket{\phi_i}\bra{\phi_i},
\label{eq:schmidt_partial_trace}
\end{equation}
where the states $\ket{\phi_i}$ and numbers $\lambda_i$ are exactly as in \eqref{eq:schmidt_decomposition}

A general way to assess bipartite entanglement is to use the R\'enyi entropy of the Schmidt coefficients
\begin{equation}
H_q(\{\lambda_i \}) = \tfrac{1}{1-q}
\ln \sum_i \lambda_i^q
= \tfrac{1}{1-q} \ln \Tr \left(\Tr_1 \rho\right)^q.
\label{eq:schmidt_entropy}
\end{equation}
In particular, the most important entanglement measures can be expressed in terms of \eqref{eq:schmidt_entropy} as follows:
\begin{itemize}
\item The von Neumann entropy is the Shannon entropy of the Schmidt coefficients squared --- i.e. $H_{q \to 1}$. It plays an important role in quantum information.
\item The Schmidt rank (number of non-zero Schmidt components) is $\exp(H_{q \to 0})$. It is important for $GL$ transformations of states, i.e. which states can be obtained (with any non-zero probability) form a given state, when one can use any local operations.
\item State purity is $1-\exp(H_2)=1-\sum_i \lambda_i^2$. It is often used, as it is easy to calculate it and relate to other quantities.
\end{itemize}

When there is only one non-zero Schmidt coefficient, the state is not entangled --- it is a product state:
\begin{equation}
\ket{\psi} = \ket{\phi_1} \ket{\varphi_1}
\end{equation}
and has all entanglement entropies equal to $0$.
The maximally entangled state is
\begin{equation}
\ket{\psi} = \tfrac{1}{\sqrt{m}} \sum_{i=1}^m \ket{\phi_i} \ket{\varphi_i},
\end{equation}
where $m$ is the smaller of the two dimensions. 
For such state all entropies are $H_{q}=\ln m$.

It is easy to compute \eqref{eq:schmidt_entropy}, but how to visualize it?
Let us go back to writing the state as partitioned between two parties (but not Schmidt-decomposed, as in \eqref{eq:schmidt_decomposition}; this time we used a fixed local basis)
\begin{align}
\ket{\phi} &= \sum_{ik} \alpha_{ik} \ket{i} \ket{k}\\
&\equiv \sum_k \ket{\Xi_k} \ket{k}.\nonumber
\label{eq:xi-parition}
\end{align}
So, the reduced density matrix for the first subsystem reads
\begin{align}
\rho_1 &= \sum_{ijk}  \alpha_{ik}
\alpha_{jk}^* \ket{i} \bra{j}\\
&= \sum_k \ket{\Xi_k} \bra{\Xi_k}.
\end{align}
Bear in mind that vectors $\ket{\Xi_k}$ are neither normalized not orthogonal to each other. 
The Schmidt number is, equivalently, the rank of $\rho_1$, i.e.
the dimension of the subspace spanned by
\begin{equation}
X = \left\{ \bra{\Xi_k} \right\}_{k}
\end{equation}
and the number of linearly independent components in $X$.
All other entanglement measures can be described by the set of vectors $X$.
In particular, the purity is
\begin{align}
\Tr( \rho_1^2 ) &= \sum_{kk'}  \left| \braket{\Xi_k}{\Xi_{k'}} \right|^2\\
&= \sum_{k}  \left| \braket{\Xi_k}{\Xi_{k}} \right|^2
+ \sum_{k \neq k'}  \left| \braket{\Xi_k}{\Xi_{k'}} \right|^2.\nonumber
\label{eq:tiles_purity}
\end{align}

As a side note, such vectors are related to classical probabilities, once off-diagonal terms in $\rho_1$ are removed.
Then instead of the number of linearly independent terms we get the number of non-zero terms, and in the case of purity --- we get only the $k=k'$ term in \eqref{eq:tiles_purity}. 
Intuitively speaking, we lose the interference between different vectors in $X$.

\begin{figure}[!htbp]
	\centering
		\includegraphics[width=0.40\textwidth]{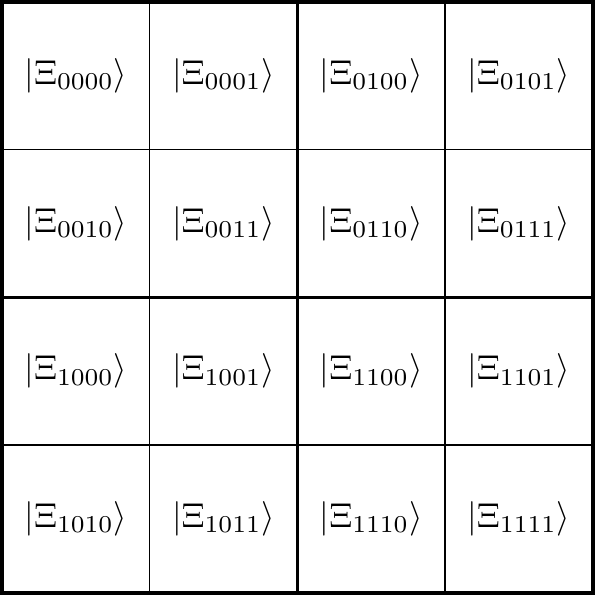}
	\caption{Schematic for presenting a qubistic plot for $N$ qubits as $2^4$ tiles, each being a plot of $N-4$ qubits.}
	\label{fig:tiles_entanglement}
\end{figure}

Describing the system as two subsystems --- one with the first $k$ particles, and the other --- with the last $N-k$ particles \eqref{eq:xi-parition} can be readily presented graphically, see Fig.~\ref{fig:tiles_entanglement}.
The same presentation makes it easy to perform a projective measurement.
If we measure the state of, say, the first $4$ qubits, and the outcome is $0110$
(what happens with probability $\braket{\Xi_{0110}}{\Xi_{0110}}$),
the final state is
$\ket{\Xi_{0110}}/\sqrt{\braket{\Xi_{0110}}{\Xi_{0110}}}$.

\begin{figure}
	\begin{center}
	\begin{tabular}{cccccc}

		\raisebox{0.65cm}{$\{\ket{0},\ket{1}\}^{4}$} &
		\includegraphics[width=1.3cm]{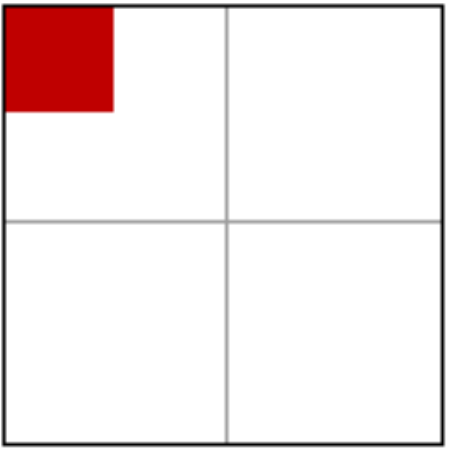} &
		\includegraphics[width=1.3cm]{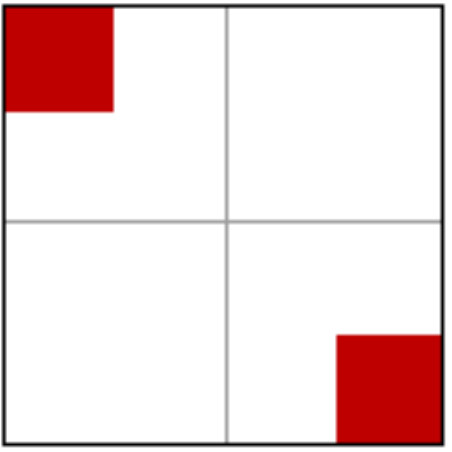} &
		\includegraphics[width=1.3cm]{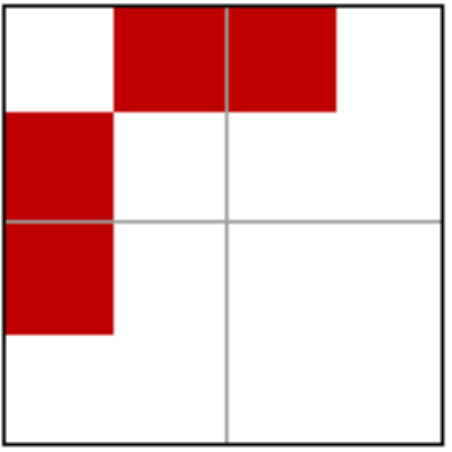} &
		\includegraphics[width=1.3cm]{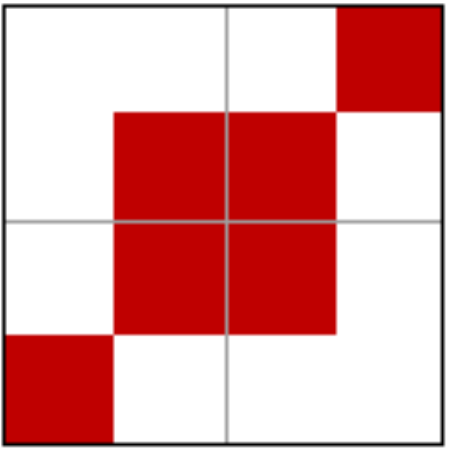} &
		\includegraphics[width=1.3cm]{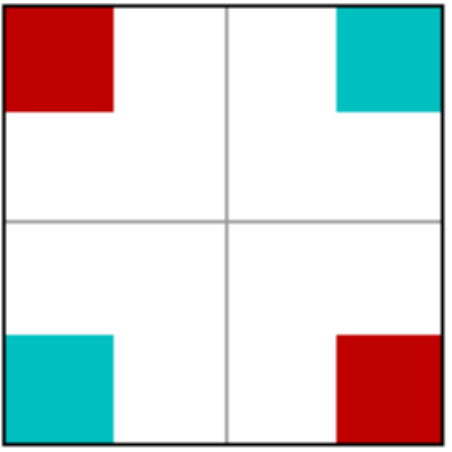}\\
		\raisebox{0.65cm}{$\{\ket{P},\ket{M}\}^{4}$} &
		\includegraphics[width=1.3cm]{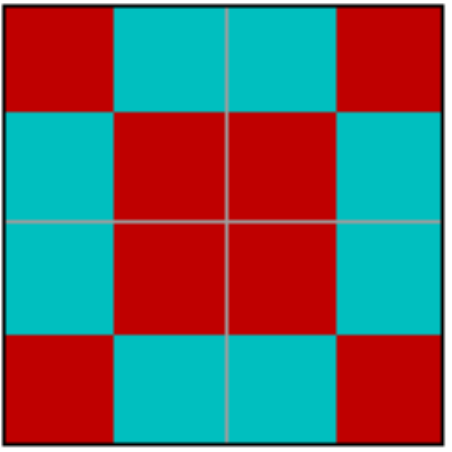} &
		\includegraphics[width=1.3cm]{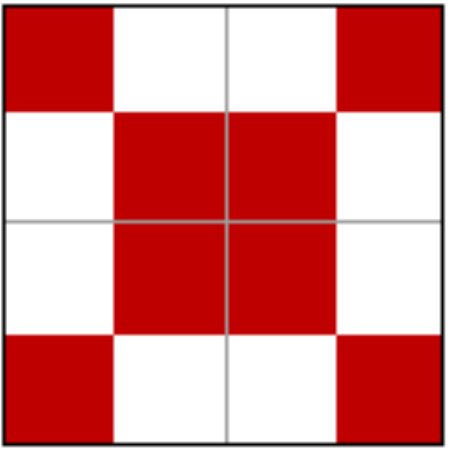} &
		\includegraphics[width=1.3cm]{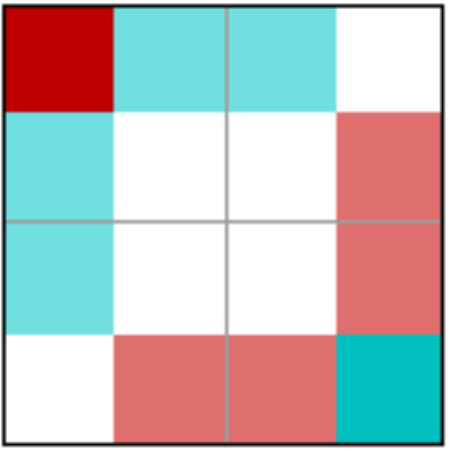} &
		\includegraphics[width=1.3cm]{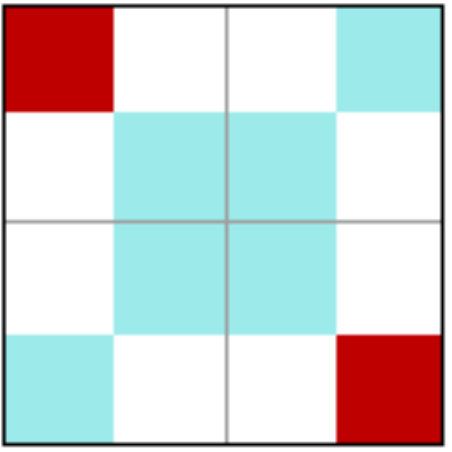} &
		\includegraphics[width=1.3cm]{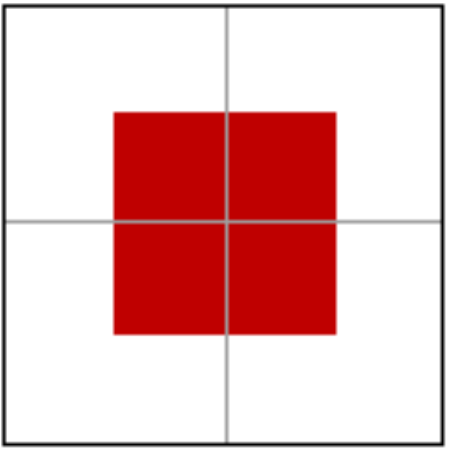}\\
		 &
		$\ket{0000}$ &
		$\ket{GHZ}$ &
		$\ket{W}$ &
		$\ket{D^4_2}$ &
		$\ket{\chi}$ \\
		Schmidt rank: &
		1 &
		2 &
		2 &
		3 &
		4 \\
		ent. entropy: &
		$0$ &
		$\log 2 = 1$  &
		$\log 2 = 1$  &
		$\log 3-\frac{1}{3}$  &
		$\log 4 = 2$ \\
		&
		&
		&
		&
		$\approx 1.25$ &

	\end{tabular}
	\end{center}
	\caption{\label{fig:tiles_entanglement_2x2} Entanglement estimation for 2--2 partition of four-qubit states. As examples we use a separable state $\ket{0000}$, the Greenberger–Horne–Zeilinger state $\ket{GHZ}$, the W state (i.e. the Dicke state with one excitation) $\ket{W}$, the Dicke state with two excitations $\ket{D^4_2}$ and $\ket{\chi}=(\ket{0000}-\ket{0101}-\ket{1010}+\ket{1111})/2$. They are presented in two different bases, where $\ket{P}=(\ket{0}+\ket{1})/\sqrt{2}$ and $\ket{M}=(-\ket{0}+\ket{1})/\sqrt{2}$. Dividing in blocks is related to separating the first two particles from the last two. The Schmidt rank ($m$) equals the number of linearly independent blocks, which can be counted by a naked eye. While the result is basis-invariant, in some bases the task may be simpler than in others. For example, for $\ket{D^4_2}$ in the computational basis one easily sees that there are $3$ different blocks while in the other one needs to spot that top left and bottom right blocks are linearly independent. Entanglement entropy is bounded from above by $\log m$.}
\end{figure}

As hinted, our goal is to visualize quantum entanglement. In Fig.~\ref{fig:tiles_entanglement_2x2} we provide plots for some $4$-qubit state and describe how entanglement between the first and last two particles can be spotted, with no calculations.

\subsection{Fractal dimension of the state}

\begin{figure}
\centering
\includegraphics[width=6cm]{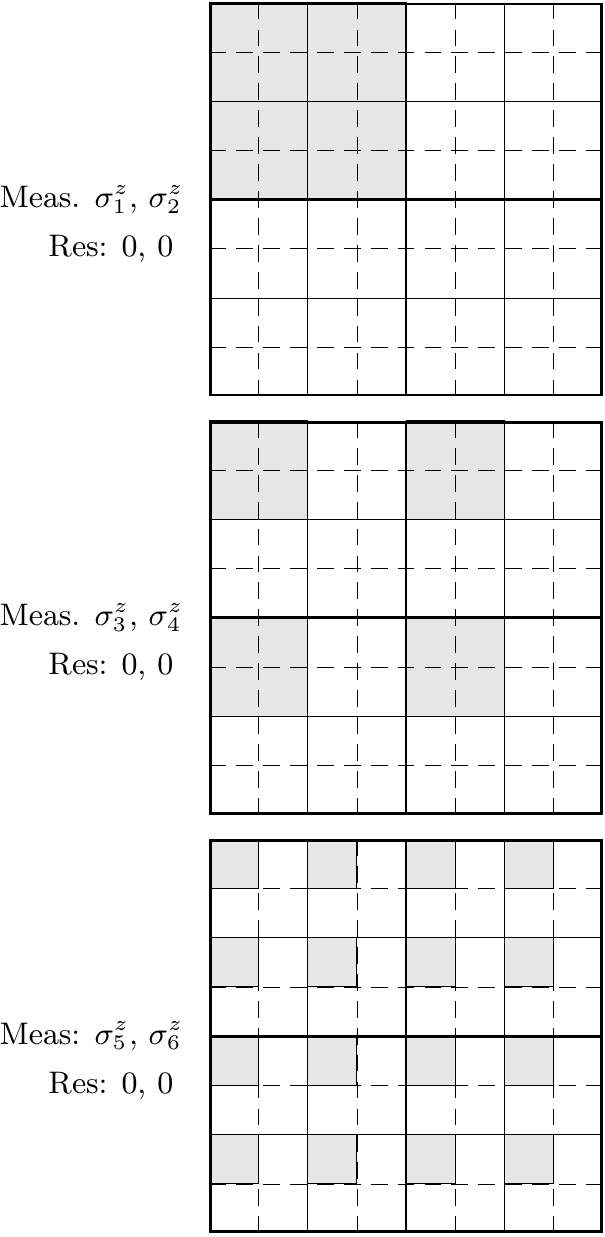}
\caption{\label{fig:translation_sym} The figures show what happens if we measure two particles in the computational basis $\sigma^z$ and get results $(+1,+1)$ (or state $\ket{00}$).
For particles $(1,2)$ we are in upper left quadrant, alike in the construction of a qubistic scheme.
For particles $(3,4)$ or $(5,6)$ we get tiles as in the picture.
For translationally invariant states all 3 subplots (once put together) are the same.}
\end{figure}

Qubistic plots often look fractal-like. It arises directly from the recursive nature of the plotting scheme, see Fig.~\ref{fig:translation_sym}.
In this section we will show that indeed some plots are fractals \cite{Schroeder1991book} and calculate their fractal dimensions.

For a probability distribution $X=(p_1, \ldots, p_n)$,
the R\'enyi entropy of order $q$ \cite{Renyi1961,Beck1993} is defined as 
\begin{align}
H_q(X) = \frac{1}{1-q} \log \left( \sum_{k=1}^n p_i^q \right).
\end{align}
Throughout this work we use $\log \equiv \ln$, so we measure information in \emph{nits} instead of \emph{bits}.
Some entropies have particular names:
\begin{align}
H_0(X) &= \log \left( \#\{p_i > 0\}_i \right)
\quad& \text{Hartley entropy}\\
H_1(X) &= - \sum_{k=1}^n p_i \log(p_i)
\quad& \text{Shannon entropy}\\
H_2(X) &= - \log\left(\sum_{k=1}^n p_i^2\right)
\quad& \text{collision entropy}\\
H_\infty(X) &= \log \left( \#\{p_i = p_{max} \} \right)
\quad& \text{max entropy}
\end{align}
Furthermore, as already discussed in Sec.~\ref{sec:entanglement-visualization}, 
R\'enyi entropies can be used for quantum states, in which case we use probabilities of pure components of a density matrix, see also \cite{Wehrl1978,Muller-Lennert2013}.

Fractal dimension can be defined for a set, being a subset of a hypercube $[0,1]^n$.
There are a few different definitions of the fractal dimension.
One numerical way for defining the fractal dimension is called \emph{box counting}. 
We divide the hypercube into boxes (smaller hypercubes) of the linear size $\varepsilon = 2^{-n}$.
Then we count the number of non-empty boxes, $b(\varepsilon)$.
The box counting dimension is defined as
\begin{align}
d = \lim_{\epsilon \to 0} \frac{\log(b(\varepsilon))}{\log(1/\varepsilon)},
\end{align}
provided the limit exists.
For example, as a function of $n$, for a point it stays constant, for a line it grows linearly and for a square --- quadratically.
But what if instead of a set we have a probability distribution?
Then we can define the dimension, parametrized by a real number $q$, using the R\'enyi entropy:  
\begin{align}
d_q = \lim_{\epsilon \to 0} \frac{H_q(X(\varepsilon))}{\log(1/\varepsilon)},
\end{align}
where $X(\varepsilon)$ is the probability distribution coarse-grained by boxes of size $\varepsilon$, i.e.
\begin{align}
p_i = \int_{i\text{th box}} p(x) d^n x.
\end{align}

It is a generalization of the box counting dimension, since by mapping a set to a probability distribution being non-zero on the set, and zero everywhere else we get $d = d_0$.
In general $d_0$ is dimension of the support. 
To calculate the fractal dimension of a qubistic plot,
in the first step we change amplitudes into probabilities, so that we can use the above methods:
\begin{equation}
p_{s_1\ldots s_N} = |\alpha_{s_1\ldots s_N}|^2.
\end{equation}

The next one is to see how does the \Renyi{} entropy scale with coarse-graining.
In the case of qubism, spatial coarse-graining is the same as coarse graining with respect to particles --- i.e. tracing out probabilities.   
Let us have
\begin{equation}
P_k \equiv \{ p_{s_1\ldots s_k} \} 
\end{equation}
where
\begin{equation}
p_{s_1\ldots s_k} \equiv \sum_{s_{k+1},\ldots,s_N}p_{s_1\ldots s_k \ldots s_N}.
\end{equation}
That is $P_k$ is the set of probabilities (in a selected basis) if we forget about the state of the $N-k$ last particles.

The fractal dimension \cite{Halsey1986,Theiler1990} is
\begin{equation}
d_q = \lim_{k \to \infty} \frac{H_q(P_k)}{\log(1/\varepsilon)},
\label{eq:dq_linear}
\end{equation}
where $\varepsilon$ is the linear box size.
As we operate with two-dimensional visualizations, $\varepsilon = d^{-k/2}$.
Or, alternatively, we can use l'Hôpital rule to get an alternative formula
\begin{equation}
d_q = \lim_{k \to \infty} 2\left(H_q(P_k) - H_q(P_{k-1})\right).
\label{eq:dq_diff}
\end{equation}
In other words --- we can either look at the slope for linear fit \eqref{eq:dq_linear} or the derivative \eqref{eq:dq_diff}.
For ideal fractals $H_q(P_k)$ grows linearly with $k$, so both formulas give the same result. 
In practical cases we work with systems of fixed size (for example, $N=12$, which is feasible for exact diagonalization).
In this case formula \eqref{eq:dq_linear} is sensitive to short-range correlations, whereas \eqref{eq:dq_diff} is to long range correlations.
What seems to be the best trade-off is to take the derivative in the middle $k=N/2$.

\begin{figure}
\centering
\includegraphics[width=8cm]{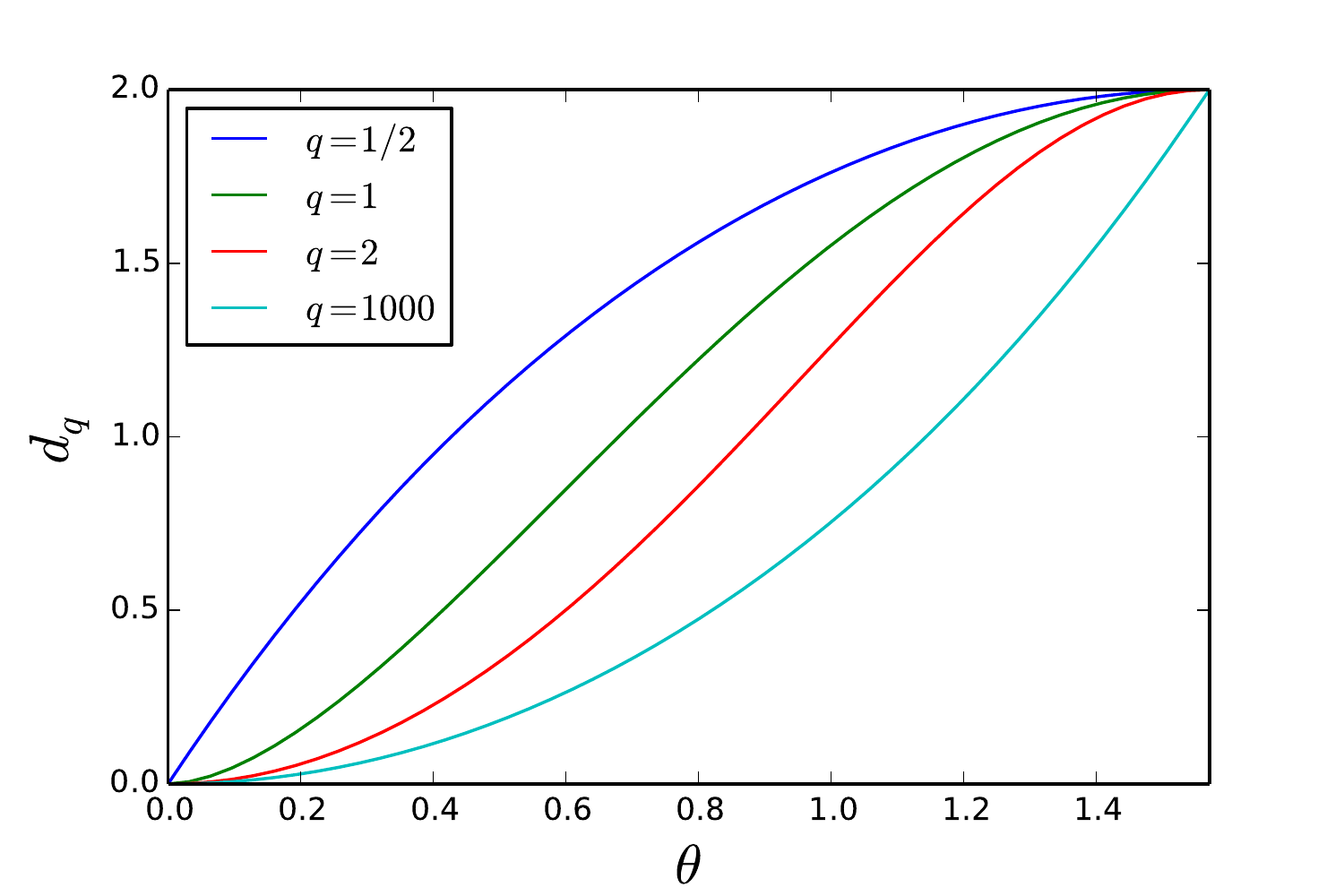}
\caption{\label{fig:fractal_dim_product_state} Fractal dimension \eqref{eq:product-theta-dq} of the product state \eqref{eq:product-theta} changes with the basis.
Look at Fig.~\ref{fig:product_state} for the respective qubistic plots.}
\end{figure}

This definition of $d_q$ is basis-dependent. For example, for the product state
\begin{equation}
\left(
\cos(\tfrac{\theta}{2})\ket{0} + \cos(\tfrac{\theta}{2}) \exp(i \varphi) \ket{1},
\right)^N
\label{eq:product-theta}
\end{equation}
a rotation of the local basis (equivalently, changing $\theta$) results in a fractal dimension changing from $0$ to $2$.
For this state, $H_q(P_k) = k H_q(P_1)$, so the fractal dimension is
\begin{equation}
	d_q = \tfrac{2}{1-q} \log\left( \cos^{2q}(\tfrac{\theta}{2}) + \sin^{2q}(\tfrac{\theta}{2}) \right),
	\label{eq:product-theta-dq}
\end{equation} 
see Fig.~\ref{fig:fractal_dim_product_state}.
So, in general, fractal dimension alone does not suffice to tell much about entanglement or any other properties which are basis-independent. 
Moreover, for states that are not translationally invariant, qubistic plots typically are not fractal-like and we do not have a well-defined fractal dimension as the relevant limit does not exist.

\begin{figure}
\centering
\includegraphics[width=5cm, angle=270]{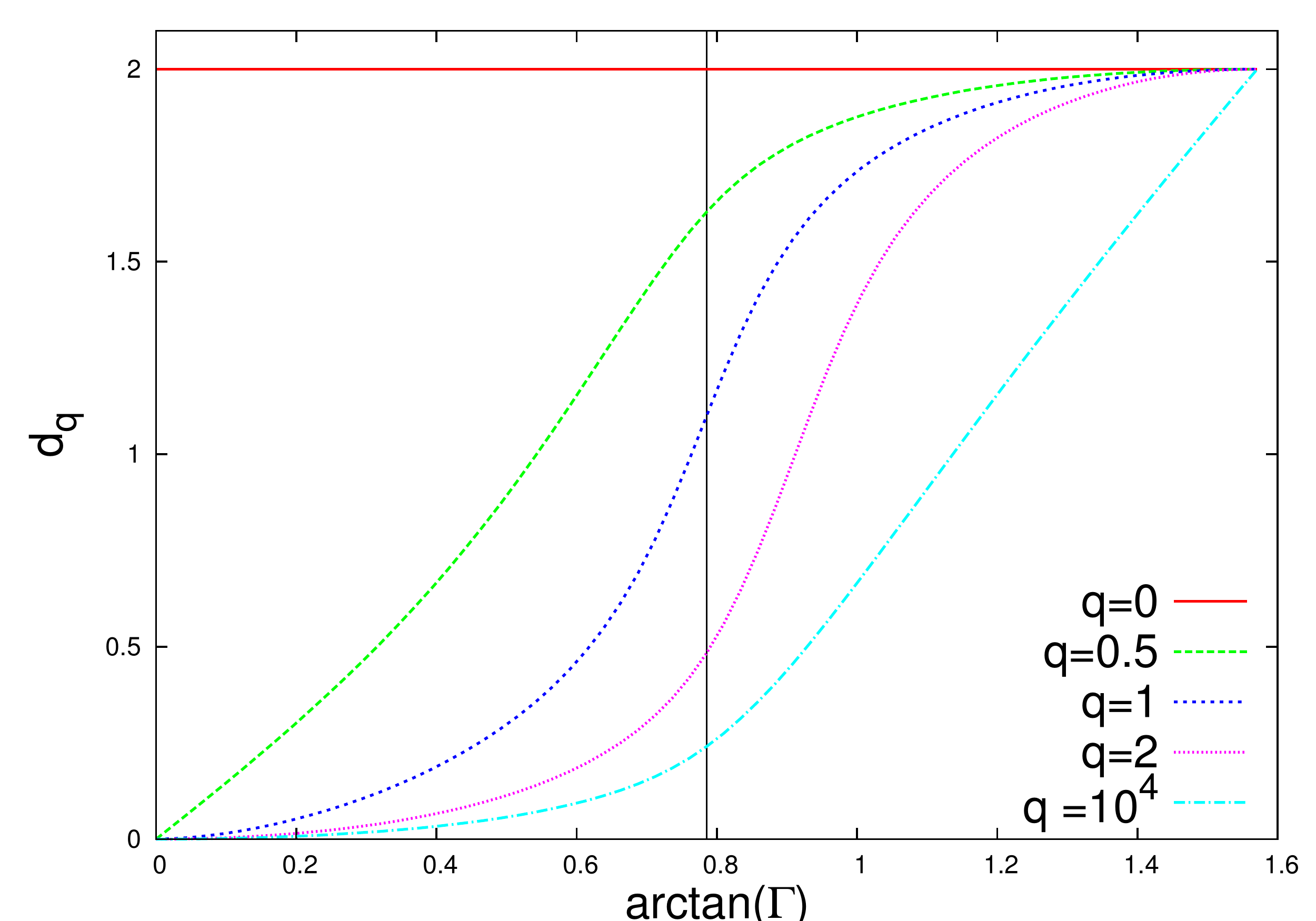}
\caption{\label{fig:fractal_dim_itf} Fractal dimension of the Ising model in the transverse field.
Note that for the phase transition, $\Gamma=1$, the fractal dimension related to the Shannon entropy ($q=1$) seems to be close to $1$.}
\end{figure}

A more interesting example is the ITF model, already discussed in Sec.~\ref{s:ising_model}.
As we already saw in Fig.~\ref{fig:itf_transition}, the plot changes with parameter $\Gamma$, from two points (Ne\'el state) to uniform color (as all particle point in the $x$ direction).
We quantify these changes in Fig.~\ref{fig:fractal_dim_itf}.

\subsection{Qubism for mixed states and operators}

Above we described the qubistic visualization for pure states.
Below, we show a representation of mixed states and operators.
As both of them are Hermitian matrices, we can propose a single visualization scheme suitable for both of them.
Note, that the density matrix has twice as many coordinates as the wavefunction --- so, unless we are operating in four dimensions, we cannot straightforwardly use qubism for mixed states.
Also, a typical two-dimensional plot of a density matrix, i.e. $\rho_{ij}$ plotted as a density plot, does not show multiparticle relations. 

\begin{figure}[!htbp]
	\centering
		\includegraphics[width=0.40\textwidth]{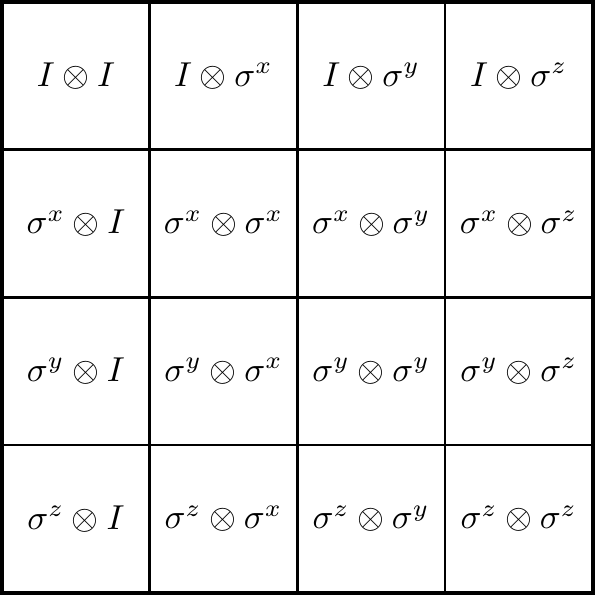}
	\caption{Scheme for presenting mixed qubit states with qubism.}
	\label{fig:scheme_mixed}
\end{figure}

\begin{figure}
	\begin{center}
	\begin{tabular}{cc}
		\includegraphics[width=5cm]{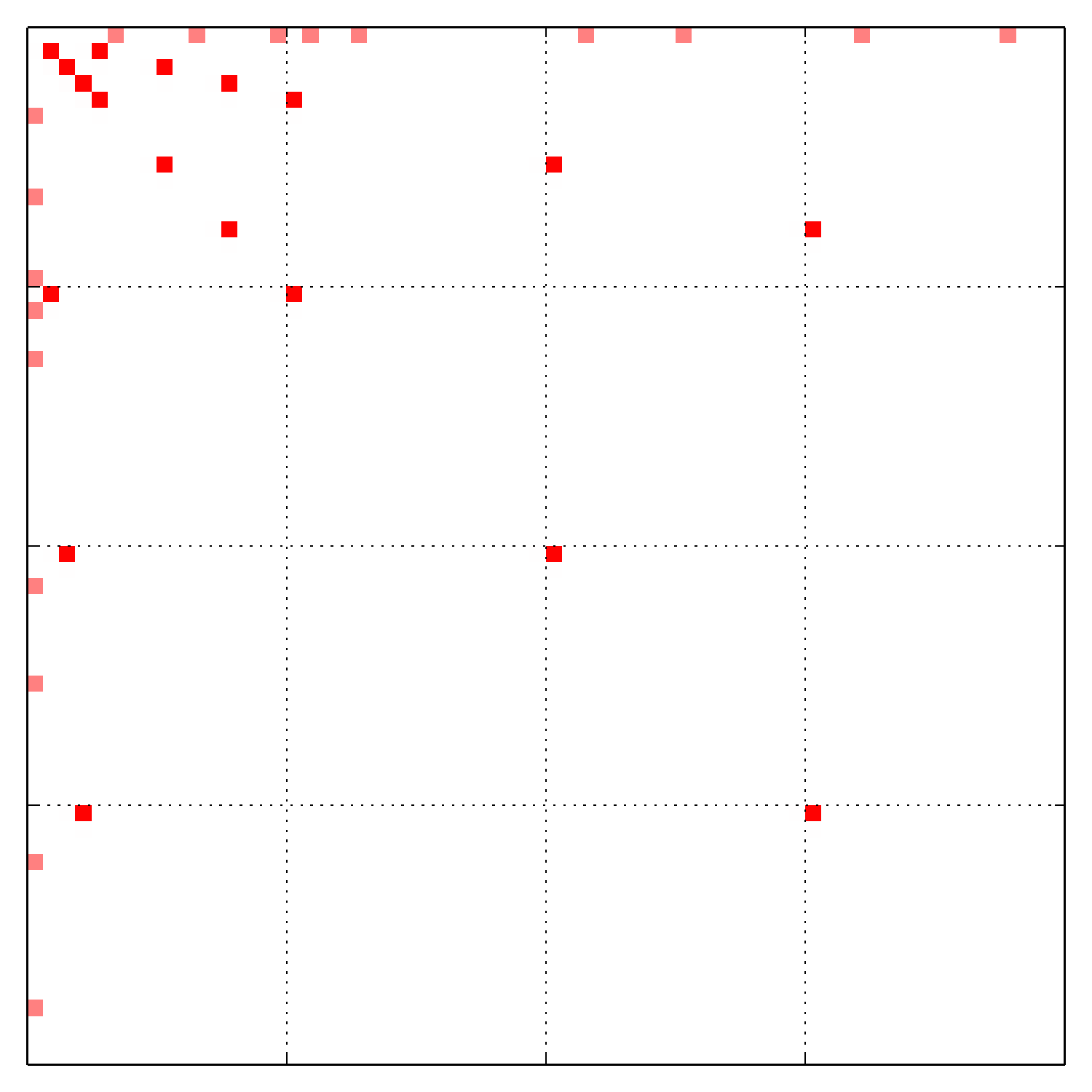}
		&
		\includegraphics[width=5cm]{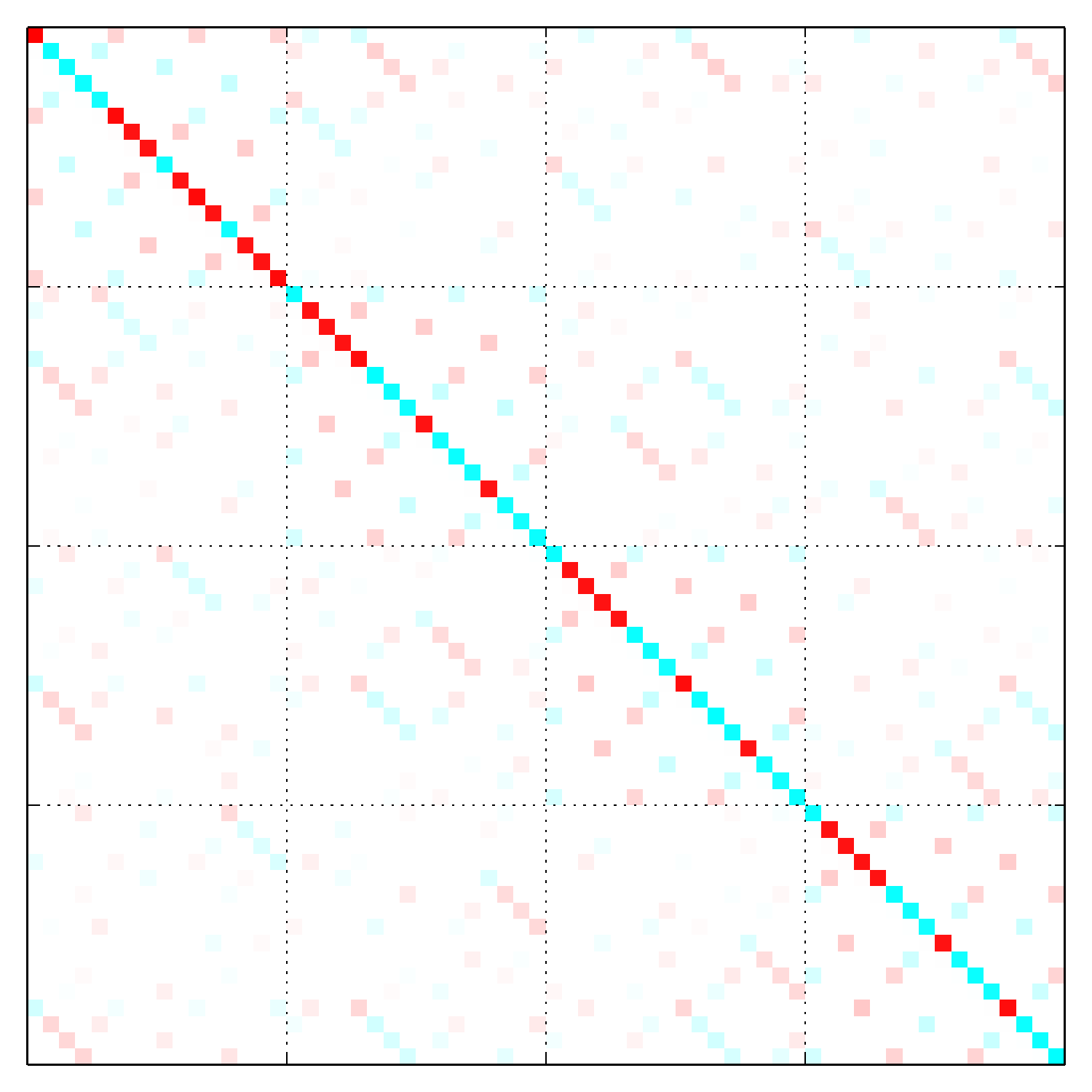} 
	\end{tabular}
	\end{center}
	\caption{\label{fig:qubism-opertors-mixed} A plot of the Majumdar Ghosh Hamiltonian \eqref{eq:majumdar_ghosh_hamiltonian} for $J=1/2$ (left) and its ground state, expressed as a density matrix (right).}
\end{figure}

Let us introduce so-called frame representations, that is, the expression of a density matrix as the following sum
\begin{equation}
	\rho = \sum_{\vec{i}} t_{\vec{i}}
	\sigma^{i_1} \otimes \sigma^{i_2} \otimes \cdots \otimes \sigma^{i_N},
\end{equation} 
where $t_{\vec{i}}$ are real numbers and $\sigma^{s}$ are generators of $d$-dimensional density matrix. For example for qubits ($d=2$), they are the identity and the three Pauli matrices. The scheme, for qubits, is presented in Fig.~\ref{fig:scheme_mixed} (cf. Fig.~\ref{fig:qubism_simplest}), with examples for the Majumdar Ghosh model \eqref{eq:majumdar_ghosh_hamiltonian} provided in Fig.~\ref{fig:qubism-opertors-mixed}.
So, it can be seen as a qubistic visualization of sequence-like objects, but where instead of $d$ symbols we have $d^2$ symbols.
Unfortunately, $t_{\vec{i}}$ can be interpreted as neither amplitudes nor probabilities. 
They are related to purity, though,
\begin{equation}
\sum_{\vec{i}} t_{\vec{i}}^2 = d^N \Tr [\rho^2].
\end{equation}

\subsection{Discussion}

Qubistic plots allow to visualize any pure state of $N$ qudits and, with modifications, of any operator and mixed state of particles in a finite number of levels.
Moreover, it makes it possible to show:
\begin{itemize}
\item two-particle correlations between nearest neighbors,
\item entanglement between the first $k$ particles and the rest,
\item some other patterns: for example permutation invariance or the structure of a singlet state. 
\end{itemize}

However, it has its own limitations. For example:
\begin{itemize}
\item states which lack any symmetry may produce very cluttered plots,
\item three and more particle correlations are not always visible,
\item it is basis-dependent (as any representation of a state in a given basis),
\item for the square plot there is a difference between having even and odd number of particles,
\item adding particles changes the plot, although only in the resolution for translation-invariant states.
\end{itemize}

The last two remarks make it harder to compare, say, the ground states of a given system as a function of number of particles.
Moreover, plots of state look the same when we add $\ket{+}=(\ket{0}+\ket{1})/\sqrt{2}$ state to its end, in other words, $\ket{\Psi}$ and $\ket{\Psi}\otimes \ket{+}^{k}$ yield the same plot.
Furthermore, dependence on the local basis implies also dependence on phase.
It is a feature of the wavefunction the description of quantum states, so it should not be surprising.
If we want to plot numerical or experimental data, it may be useful to know how to disregard phase.
However, due to phase-to-hue mapping it should be not that hard even visually.

We proposed a few variants of qubistic plots. 
It seems that for most applications the typical, square qubistic plotting scheme should be the standard choice. 
However, other plotting schemes may be useful for putting emphasis on certain features of the wavefunction, e.g. ferromagnetism.










%% file: networks.tex








\chapter{Quantum walks on complex networks}
\label{ch:networks}

\section{Introduction}

In this chapter we develop a complex networks \cite{Newman03thestructure,Newman2010book,Albert2002} approach to the unitary evolution of a single particle, which we interpret as a quantum walk.
We study the analogies between this process and related classical walks.
The focus is on studying the long time probability distribution and the coherence between nodes,
which brings tools for analyzing properties of quantum walk on a complex network.
In particular, we consider a splitting of a complex networks into independent pieces that are not related by quantum superposition.


\subsection{Networks and quantum walks}

Study of quantum walks goes back to the Feynman checkerboard \cite{feynman1965quantum}, a toy model in which a particle travels as the speed of light on a one-dimensional lattice, while being subjected to reversal with some amplitude.
The effective behavior of this particles turns out to be the same as for a massive particle, evolving as described by the Sch{\"o}dinger equation. 
Another seminal model is a quantum walk with a coin \cite{Aharonov1993}, where the path taken by a particles is specified by tossing a quantum coin.
For an overview of quantum walks see \cite{Kempe2003,Venegas-Andraca2012}.

In general, the quantum dynamics of any discrete system can be re-expressed and interpreted as a single particle quantum walk~\cite{PhysRevLett.103.240503,PhysRevA.81.032327}, which is capable of performing universal quantum computation~\cite{childs2009universal}.
One dimensional walks can be simulated with photons,  both in the single particle variant \cite{Schreiber2011} and the walk with a coin variant \cite{Schreiber2010}.

Quantum walks are used to study transport properties in physical systems~\cite{Mulken2011,FG98,caruso09,MRLA08,CF09}, such as transport of energy through  biological complexes or artificial solar cells.
Additionally, there have been theoretical proposals for speed-up of algorithms for large social and links networks~\cite{Faloutsos:1999:PRI:316194.316229,albert1999internet},
for example for
PageRank \cite{Page1999}, the famous ranking algorithm based on the simulation of a random walk through Internet.
It has been studied used quantum annealing~\cite{Burillo2012,paparo2012google,garnerone2012pagerank,Garnerone2012google},
that is, a simulation of the procedure by which a quantum system is driven to its ground state, chosen so that the same ground state represents the Google ranking vector.

While analytical results have been obtained for some specific topologies, such as star-like \cite{muelken2007inefficient,mulken2006coherent,cai1997rouse}, regular or semi-regular \cite{salimi2010continuous} networks, progress in analyzing quantum walks on complex networks has largely been based on numerical analysis.
More general analytic results, applicable to real-world complex systems, can be brought by
studying the probability distribution of finding the walker at each node in the long time limit of a certain continuous-time unitary quantum walk.
For unitary quantum walks, even for arbitrary long times there are oscillations rather than a steady state; this is not necessarily the case for open quantum walks~\cite{spohn1977algebraic,Whitfield2010}.
Consequently, we work with the long time averages, which are equivalent to removing oscillations \cite{mulken2005asymmetries,muelken2007inefficient}.
We show that the result can be approximated by the steady state of a classical random walk. 
Moreover, we measure the quality of this approximation, by studying a certain parameter, which is called \emph{quantumness}~\cite{Faccin2013} --- a number in the unit interval quantifying the strength of quantum effects.
In classical random walks on undirected, connected graphs, there is a unique steady state, so the long time limit does not depend on the choice of the initial state.
However, in the quantum walk the final state depends on the initial conditions. 
In particular, we show how these quantum effects are related to the energy of a given state and the degree distribution of the underlying network.


As a case study, we investigate quantum walks on a range of model complex network structures, including the \ac{ba}, \ac{er}, \ac{ws} and \ac{rg} networks.
We repeat this analysis for several real-world networks, specifically a \ac{kc} social network~\cite{zachary1977information}, the \ac{em} network of the URV university~\cite{guimera2003self}, the \emph{Caenorhabditis elegans} 
network~\cite{duch2005community}, and a \ac{ca} network of scientists~\cite{newman2006finding}.
Let us make a brief introduction to the models we use as benchmarks.
They are parametrized by the number of nodes $N$ and some other parameter that can be mapped to the number of edges $M$.

The \acf{er} model is one of the first random graph models \cite{Erdos1959,ER60}.
We create a random graph with $M$ edges, that is, out of all possible graphs with $N$ nodes and $M$ edges we select one.

The \acf{ws} \cite{Watts1998} is a model showing how addition of a few links changes graph behavior from short to long range.
We start with $N$ nodes connected as a circle, that is, with each node connected to its two neighbors.
Then we add further $N-M$ edges, similarly as for the \acf{er} model, so as to have $M$ at the end.

In the \acf{rg} model on a square \cite{penrose2003} we start creating $N$ points, each of them from the uniform probability distribution on a unit square.
Then we connect all pairs of nodes, which are closer than a certain distance cutoff $r$,
which can be adjusted so that we obtain $M$ edges.

The \acf{ba} model \cite{BA99} is based on preferential attachment and serves as a key example for scale-free behavior of real-world networks.
We start with a few nodes, connected with each other.
New nodes are added and linked to the old ones with a probability proportional to their degree.
This way nodes having many edges get new edges easier than others.

\subsection{Community detection}

Real-world complex networks are typically not homogeneous --- some of their regions are much more connected internally than with the rest of networks.
These regions are called \emph{communities}.
The identification of the community structure within a network addresses the 
problem of characterizing the mesoscopic boundary between the microscopic scale of basic network components 
(herein called nodes) and the macroscopic scale of the whole 
network~\cite{girvan2002,porter2009communities,Fortunato2010}. 
The detection of community structures dates back to 1927~\cite{rice1927identification}, when index of cohesion within a community was introduced to study behavior of political parties in the United States. 
The analysis of the community structure has revealed countless important hierarchies of community groupings within real-world complex networks.
Salient examples can be found in social networks such as
human~\cite{zachary1977information} or animal relationships~\cite{lusseau2004}, 
biological~\cite{jonsson2006cluster,pimm1979structure,krause2003compartments,guimera2005functional}, biochemical~\cite{holme2003subnetwork} and
technological~\cite{flake2002self,gauvain2013communities} networks, as well as numerous others
 \cite{girvan2002}.
In quantum networks, as researchers explore networks of an increasingly 
complex geometry and large 
size~\cite{allegra2012,plenio2008dephasing,renger2006}, 
the tractability of their analysis and understanding may rely on identifying 
relevant community structures. 

An interesting application is the quantum simulation of electric excitation transport in biological dissipative networks
\cite{ringsmuth2012,fleming10,IF12,CF09,caruso09,MRLA08,Scholak2011}.
The major light harvesting complex of plants, photosystem II (LHCII) \cite{Croce2014}, is of particular interest.
In past works, researchers have divided this complex \emph{by hand} in order to gain more insight into the
system dynamics~\cite{pan2013architercture,novoderezhkin2005lhcii,fleming2009lhcii}.
We have devised methods that optimize the task of identifying communities within
a quantum network \emph{ab initio} and, as we will show, the resulting
communities consistently point towards a structure that is different to those
previously identified for the LHCII \cite{Faccin2013community}.
We also consider larger networks,
for which an automatic method would appear to be the only
feasible option.

\begin{figure}[b!]
  \centering
  \begin{center}
      \includegraphics[width=\textwidth]{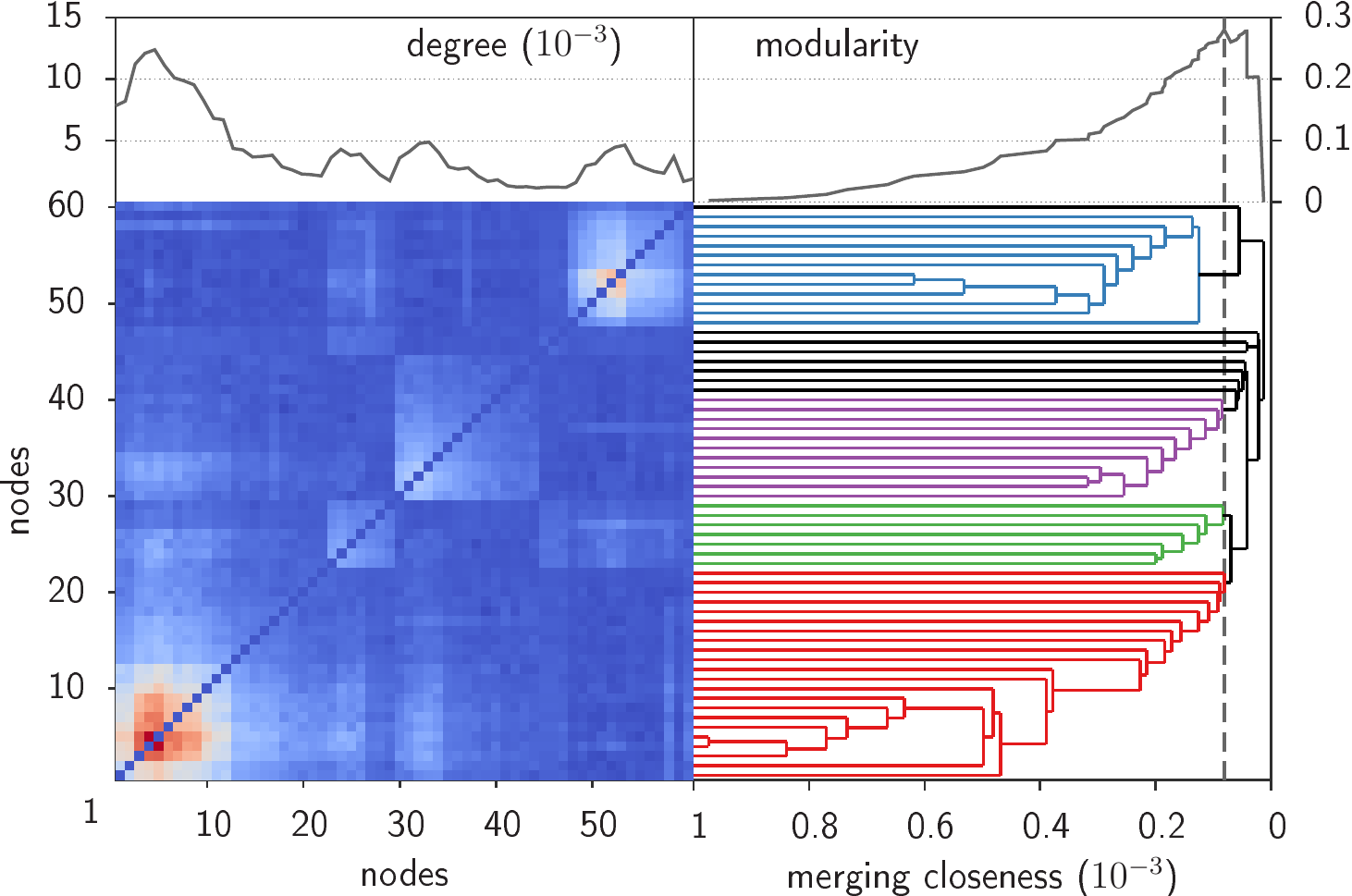}
  \end{center}
\caption{Hierarchical community structure arising from a quantum evolution. 
Left: the closeness matrix $c(i,j)$ between $n=60$ nodes.
Right: the dendrogram showing the resulting hierarchical community 
structure.
The dashed line shows the optimum level within this hierarchy,
according to the maximal modularity criterion.
The particular example shown here is the one corresponding 
to~\fir{fig:art-fidelity}.
}
\label{fig:dendro}
\end{figure}

We introduce a set of novel methods based on community detection for quantum walks \cite{Faccin2013community}.
As in typical classical methods, the backbone our approach is a hierarchical aggregation of communities \cite{carlsson2010characterization}.
That is, we start with $N$ communities, each of them consisting of a single node.
We define a closeness function between each pair of nodes.
In each iteration, we merge the two closest communities into a new one
and proceed until all communities are merged into a single one.
The output of the algorithm can be either the splitting a network into a given number of communities, or the splitting that maximizes some target function.
The procedure is depicted in Fig.~\ref{fig:dendro}.

Unlike the classical case, where classical \emph{modularity} \cite{Newman2004} is used both for measuring closeness and establishing the target function, we introduce a few modularity-like functions based on coherence and transport properties of a quantum walk \cite{zimboras2013quantum}.
All our methods are based on the full unitary dynamics of the system,
as described by the Hamiltonian, and account for quantum effects such as coherent evolution and interference.
In fact, phases are often fundamental to characterize the system evolution.
For example, in \cite{harel2012quantum} it was shown that in light harvesting
complexes interference between pathways is important even at room temperature.
We use our community detection methods to automatically find communities, which turn out to be in good agreement with communities picked by hand by experts studying this system.
As with the case of classical community structure, there are many possible
definitions of a quantum community.
We restrict ourselves to two broad
classes based on transport properties and fidelity under unitary evolution.

The use of community detection in quantum systems addresses an open
challenge in the drive to unite quantum physics and complex network
science.
We expect such partitioning, based on our definitions or
extensions such as above, to be used extensively in making the large
quantum systems currently being targeted by quantum physicists tractable to
numerical analysis.

\subsection{Structure}

This chapter is structured as follows.
First, in Sec.~\ref{sec:walks}, we look at similarities between a classical random walk and quantum walk on a graph \cite{Faccin2013}.
Sec.~\ref{sec:walks-framework} an introduction to the dynamics for a continuous-time random walk and a continuous time unitary quantum walk.
We describe the long time averaged probability distribution of the quantum walk.
In Sec.~\ref{sec:degree-quantum} we introduce quantity called \emph{quantumness} to assess the difference between the classical and quantum behavior on a given graph.
Sec.~\ref{sec:walks-numericalresults} is dedicated to numerical studies of this quantity 
on a range of artificial and real-world complex network topologies.

Second, in Sec.~\ref{sec:community}. we design community detection algorithms for quantum system \cite{Faccin2013community}.
They are based on quantum walk and depend on properties such as interference different paths, thus cannot be replicated by any classical random walk.
Moreover, some of our definitions of communities are directly to quantum informational properties of the equilibrium state.
In Sec.\ref{sec:comdet} we begin by recalling several common notions from classical community
detection that we rely on in this work. 
This sets the stage for the development of a quantum treatment of community detection in Section~\ref{sec:quantumcom}. 
We then turn to several examples in Section~\ref{sec:performance} including the LHCII complex mentioned previously.
Some technical details of community detection are left for Sec.~\ref{sec:comdet-appendix}.

\section{Classical and quantum walk}\label{sec:walks}

\subsection{Walks framework}\label{sec:walks-framework}
We consider a walker moving on a connected network of $N$ nodes, with each
weighted undirected edge between nodes $i$ and $j$ described by the element 
$A_{ij}$ of the off-diagonal adjacency matrix $A$.
The matrix is symmetric ($A_{ij}=A_{ji}$) and has real, non-negative entries, with zero entries being equivalent to absence of an edge.
We use Dirac notation and represent $A = \sum_{ij} A_{ij} \ket{i} \bra{j}$ in terms of $N$ orthonormal vectors $\ket{i}$. 

The network gives rise to both a quantum walk and a corresponding classical walk.
There is no unique mapping from a network to evolution. 
However, there is a number of conditions required to be kept.

For a classical random walk, the infinitesimal generator needs to ensure that for any state:
\begin{itemize}
  \item probabilities sum up to one,
  \item probabilities are non-negative.
\end{itemize}
Operators fulfilling these criteria are called infinitesimal stochastic operators, and are defined by
\begin{itemize}
  \item all columns sum up to zero,
  \item all off-diagonal entries are positive.
\end{itemize}
Additionally, one more property is added --- requirement that rate of leaving a node is the same for all nodes. Thus, it has an interpretation of a random walk, rather than any probability flow.
It translates to property, that all diagonal values of the generator are the same.

On contrary, for quantum evolution the only property we need for infinitesimal stochastic operators is Hermitian symmetry, so that generated evolution is unitary.
It is not possible to ensure that rate of leaving each nodes is the same, due to interference.
However, at least we can set diagonal terms to be of the same value and normalize amplitudes on edges.

The classical stochastic walk $S(t) = \ee^{ -H_C t}$ we consider is generated by the infinitesimal stochastic (see e.g.~Refs.~\cite{BB12, johnson2010,BF13}) operator
\begin{equation}
H_C = \LL D^{-1},
\end{equation}
where $D = \sum_i d_i \ket{i} \bra{i}$ is a diagonal matrix of the node degrees, $d_i = \sum_j A_{ij}$ and $\LL$ is the graph Laplacian, defined as $\LL = D - A$.
For this classical walk, the total rate of leaving each node is identical, what is ensured by the normalization by multiplying by $D^{-1}$.

The corresponding unitary quantum walk $U(t) = \ee^{-\ii H_Q t}$ is generated by the Hermitian operator
\begin{equation}
H_Q =  D^{-\nicefrac 12} \LL D^{-\nicefrac 12}.
\end{equation}
For this quantum walk, the energies $\brackets{i}{H_Q}{i}$ at each node are identical.

The generators $H_C$ and $H_Q$ are similar matrices, related by
\begin{equation}
H_Q = D^{-\nicefrac 12} H_C D^{\nicefrac 12}. 
\end{equation}
This mathematical framework, represented in \fir{fig:scheme}, underpins our analysis.
As we will describe in \secr{sec:classical}, the long time behavior of 
the classical walk generated by $H_C$ has been well 
explained in terms of its underlying network properties, specifically 
the degrees $d_i$.
Our goal in \secr{sec:quantum} is to determine the role this concept plays in the quantum walk generated by $H_Q$.

\begin{figure}[!htb]
\begin{center}
  \includegraphics[width=0.5\textwidth]{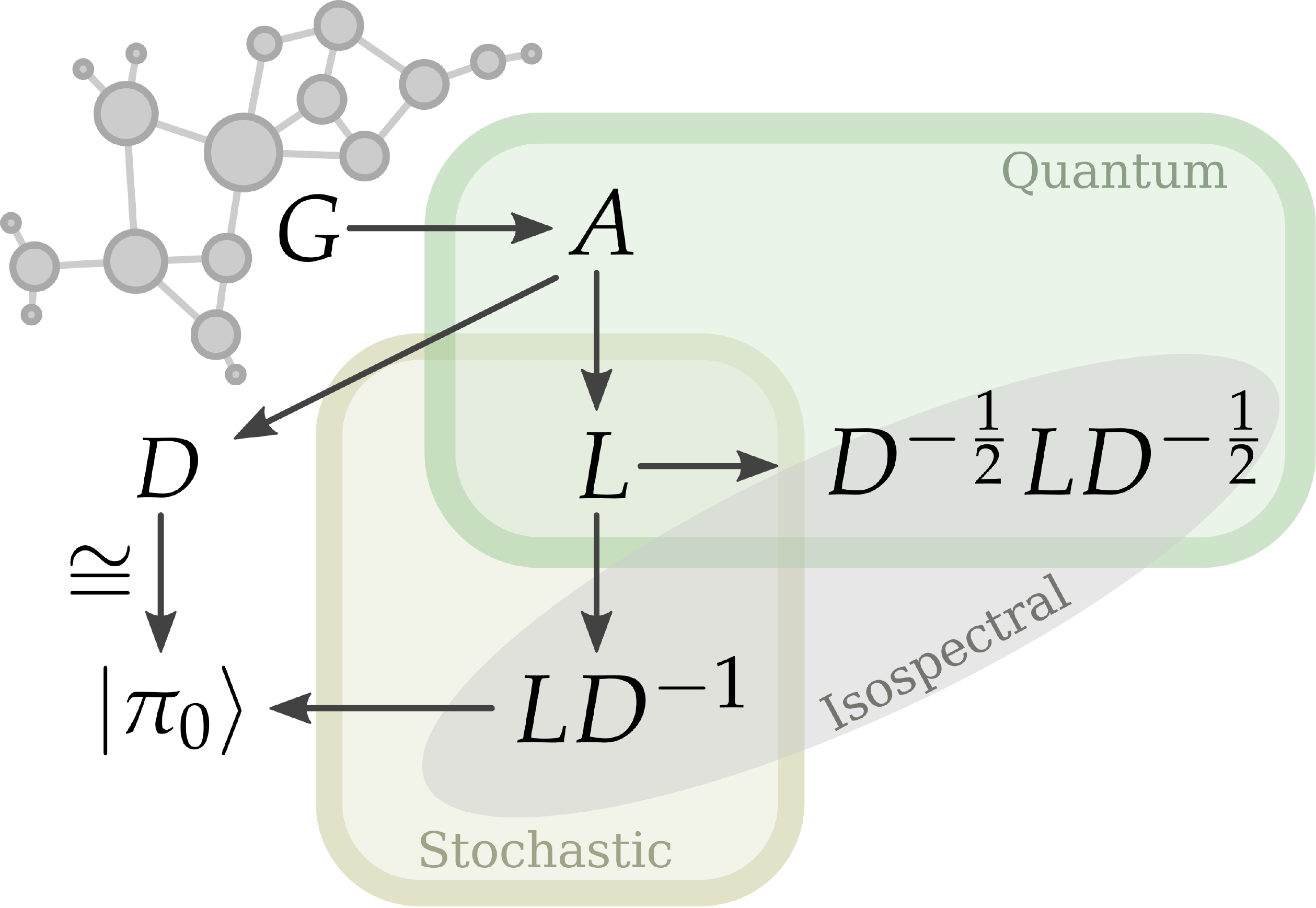}
\end{center}
\caption{
Relating stochastic and quantum walks.
An undirected weighted network (graph) $G$ is represented by a symmetric, off-diagonal and non-negative adjacency matrix $A$. 
There is a mapping from $A$ (by summing columns) to the diagonal matrix $D$ with entries given by the weighted degree of the corresponding node.
The node degrees are proportional to the steady state probability distribution of the continuous-time stochastic walk (with uniform escape rate from each node) generated by $H_C = \LL D^{-1}$, where $L = D - A$ is the Laplacian. 
The steady state probabilities, represented by the vector $\ket{\pi_0}$, are proportional to the node degrees.
We generate a corresponding continuous-time unitary quantum walk by the Hermitian operator $H_Q = D^{-\nicefrac12} \LL D^{-\nicefrac 12}$, which is similar to $H_C$. 
The probability of being in a node in the stochastic stationary state $\ket{\pi_0}$ and the probability arising from the ground state of the quantum walk are the same. 
}
\label{fig:scheme}
\end{figure}

\subsubsection{Classical walks} \label{sec:classical}
In the classical walk the probability $P_i (t)$ of being at node $i$ at time $t$ evolves as $\ket{P(t)} = S(t) \ket{P(0)}$, where $\ket{P(t)} = \sum_i P_i (t)  \ket{i}$.
The stationary states of the walk are described by eigenvectors $\ket{\pi_i^k}$ of $H_C$ with eigenvalues $\lambda_i$ equal to zero.
We assume throughout this work that the walk is connected, i.e., it is possible 
to transition from any node to any other node through some series of allowed 
transitions. In this case there is a unique eigenvector $ \ket{\pi_0} = 
\ket{P_C}$ with $\lambda_0 = 0$, and $\lambda_i > 0$ for all $i \neq 
0$~\cite{keizer1972steady,lancaster1985theory,norris1998markov,BB12}. 
This (normalized accordingly) eigenvector $\ket{P_C} = \sum_i (P_C)_i \ket{i}$ describes the steady state probability distribution 
\begin{equation}\label{eq:classical}
(P_C)_i = \frac{d_i }{\sum_j d_j}.
\end{equation} 
In other words, the process is ergodic and after long times the probability of 
finding the walker at any node $i$ is given purely by the importance of the degree $d_i$ of that 
node in the network underlying the process. 

\subsubsection{Quantum walks} \label{sec:quantum}

When considering quantum walks on networks, it is natural to ask what is 
the long time behavior of a quantum walker \cite{Burillo2012,Mulken2011,aharonov2001quantum,mulken2005asymmetries}.
This problem is similar to some thermalization problems \cite{Polkovnikov2011}, in which
the unitary evolution does not drive the system towards a steady state.
Therefore, to obtain a static picture we consider the long time average 
probability $(P_Q)_i$ of being on node $i$, which reads
\begin{equation}\label{eqn:pq}
  ( P_Q )_i = 
  \lim_{T\to\infty} \frac 1T \int_0^T {\textrm d}t\  \brackets{i}{U(t) \rho(0) U^\dagger (t)}{i} .
\end{equation}
For ease of comparison with $\ket{P_C}$ we will also write the distribution in \eqr{eq:pn} as a ket $\ket{P_Q} = \sum_i  ( P_Q )_i \ket{i}$. Unlike the classical case, \eqr{eqn:pq} depends on the initial state $\rho(0)$.
The long time average can be carried out 
\begin{align}
  ( P_Q )_i &= 
  \lim_{T\to\infty} \frac 1T \sum_{kl} \int_0^T {\textrm d}t\ 
  \braket{i}{\phi_k} e^{- i E_k t} \bra{\phi_k} \rho(0) \ket{\phi_l} e^{i E_j t} \braket{\phi_l}{i}\\
  &=
  \lim_{T\to\infty} \frac 1T \sum_{kl} \int_0^T {\textrm d}t\ 
  \braket{i}{\phi_k}\bra{\phi_k} \rho(0) \ket{\phi_l} \braket{\phi_l}{i} e^{i (E_j - E_k) t}\\
  &= \sum_{kl:\ E_k=E_l} \braket{i}{\phi_k}\bra{\phi_k} \rho(0) \ket{\phi_l} \braket{\phi_l}{i}.
\end{align}
That is, interference between subspaces of different energy vanish in the long time average, so we obtain an expression for the probability $ ( P_Q )_i$ in terms of the energy eigenspace projectors $\Pi_j$ of the Hamiltonian $H_Q$, 
\begin{align}
  ( P_Q )_i = 
     \sum_j \brackets{i}{ \Pi_j \rho (0) \Pi_j}{i} .
     \label{eq:pn}
\end{align}
Here $\Pi_j = \sum_k\ket{\phi_j^k}\bra{\phi_j^k}$ projects onto the subspace spanned by the eigenvalues $\ket{\phi_j^k}$ of $H_Q$ corresponding to the same eigenvalue $\lambda_j$. 
In other words, the long time average distribution is a mixture of the 
distributions obtained by projecting the initial 
  state onto each eigenspace.

Due to the similarity transformation $H_Q = D^{-\nicefrac 12} H_C D^{\nicefrac 12}$ the classical $H_C$ and quantum $H_Q$ generators share the same eigenvalues $\lambda_i \geq 0$, and have eigenvectors related by $\ket{\phi_i^k} = D^{-\nicefrac 12} \ket{\pi_i^k}$ up to their normalizations. 
In particular, the unique eigenvectors corresponding to $\lambda_0 = 0$ are $\ket{\pi_0} = D \ket{\identity}$ and $\ket{\phi_0} 
= D^{\nicefrac 12} \ket{\identity}$ up to their normalizations, with $\ket{\identity} = \sum_i \ket{i}$. Therefore the probability vector describing the outcomes of a measurement of 
the quantum ground state eigenvector $\ket{\phi_0} $ in the node basis is the classical steady state distribution $\ket{\pi_0} = \ket{P_C}$. 

The state vector $\ket{P_C}$ appears in \eqr{eq:pn} for the quantum long time average distribution $\ket{P_Q}$ with weight $ \brackets{ \phi_0 }{ \rho (0) }{\phi_0}$.
Accordingly we split the sum in \eqr{eq:pn} into two parts, the first we call the ``classical term'' $\ket{P_C}$ and the rest we call the ``quantum correction'' $\ket{\tilde{P}_Q}$, as 
\begin{align}
  \ket{P_Q} = (1-\varepsilon) \ket{P_C} +\varepsilon \ket{ \tilde{P}_Q} .
    \label{eq:twoterms}
\end{align}
The normalized quantum correction $\ket{\tilde{P}_Q} = \sum_i ( \tilde{P}_Q )_i \ket{i}$ is given by 
\begin{align} \label{eq:quantumcorrection}
( \tilde{P}_Q )_i &= \frac{1}{ \varepsilon} \sum_{j \neq 0} \brackets{i}{ \Pi_j \rho (0) \Pi_j}{i} ,
\end{align}
and the weight
\begin{equation}
\varepsilon = 1 - \brackets{ \phi_0 }{ \rho (0) }{\phi_0},\label{quantumnessformula}
\end{equation}
we call \emph{quantumness} is a function both of the degrees, through $\ket{\phi_0}$, and the initial state.

We can think of the parameter $\varepsilon$, which controls the classical-quantum mixture,  
as the quantumness of $\ket{P_Q}$ for the following three reasons. 
First, the proportion of the elements in $(P_Q)_i$ \eqref{eq:twoterms} that corresponds to the genuinely quantum 
correction is $\varepsilon$. 
Second, the trace distance between the normalized distribution $(P_C)_i$ and the 
unnormalized distribution $(1-\varepsilon) (P_C)_i$ forming the classical part 
of the quantum result is also $\varepsilon$. 
Last, using a triangle inequality, the trace distance between the normalized 
distributions $(P_C)_i$ and $(P_Q)_i$ is upper bounded by $2 \varepsilon$.

This expression for the quantumness in \eqr{quantumnessformula} enables us to make some physical statements about a general initial state.
By realizing that $\ket{\phi_0}$ is the ground state of zero energy $\lambda_0 = 0$ and the gap $\Delta = \min_{i \neq 0} \lambda_i$ in the energy spectrum is non-zero for a connected network~\cite{keizer1972steady,lancaster1985theory,norris1998markov,BB12},
the above implies a bound $E / \Delta \ge \varepsilon $ for the quantumness $\varepsilon$ of the walk in terms of the energy $E = \tr \{ H_Q \rho \}$ of the initial state. The bound is obtained through the following steps
\begin{align}
E &= \tr \{ H_Q \rho \} = \sum_{j \neq 0} \lambda_j \tr \{ \Pi_j \rho (0) \} \nonumber \\
&\geq \Delta \sum_{j \neq 0} \tr \{ \Pi_j \rho (0) \} = \Delta \left( 1 - \tr \{  \Pi_0  \rho (0) \} \right)  = \Delta \varepsilon  \label{eq:EnergyBound}. 
\end{align}
 
The above demonstrates that the classical stationary probability distribution will be recovered for low energies.
A utility of this result is that it connects the long time average distribution 
to a simple physical property of the walk, the energy, which provides a total ordering of all possible initial states.

\subsection{Degree distribution and quantumness} \label{sec:degree-quantum}

Quantumness is both a function of the degrees of the network nodes and the initial state.
To compare the quantumness of different complex networks, we fix the initial state $\rho(0)$.
For our example we choose the even superposition state $\rho(0) = \ket{\Psi(0)} \bra{\Psi(0)}$ with $\ket{\Psi(0)} = \ket{\identity} / \sqrt{N}$. 
This state has several appealing properties, for example, it is invariant under node permutations and thus independent of the arrangement of the network.
In this case the quantumness is given by the expression
\begin{align}\label{eq:quantumnessdegree}
\varepsilon= 1 -  \frac{ \langle \sqrt{d} \rangle^2 }{  \langle d \rangle} ,
\end{align}
where $\langle d \rangle = \sum_i d_i / N$ is the average degree and $\langle 
\sqrt{d} \rangle = \sum_i \sqrt{d}_i / N$ is the average root degree of the 
nodes. As such, the quantumness depends only on the degree distribution of the network and increases with network heterogeneity. 

This statement is quantified by writing the quantumness
\begin{align}
\varepsilon = 1 - \frac{1 }{ N}\exp \left[ H_{\nicefrac 12} \left( \left \{ \frac{ d_i }{ \sum_j d_j } \right \}  \right) \right] ,
\end{align}
in terms of the R\'{e}nyi entropy
\begin{align} \label{eq:renyientropy}
H_q ( \{ p_i \}) = \frac{1}{1-q} \ln\left(\sum_i p_i^q \right),
\end{align}
where $ d_i / \sum_j d_j = (P_C)_i$ are the normalized degrees.

To obtain an expression in terms of the more familiar Shannon entropy $H_1$ (obtained by taking the $q\rightarrow1$ limit of \eqr{eq:renyientropy}), we recall that the R\'{e}nyi entropy is non-increasing with $q$ \cite{Beck1993thermodynamics}.
This leads to the upper bound
\begin{align}\label{eq:entropybound}
\varepsilon \leq 1 - \frac{ 1}{ N} \exp \left[ H_{1} \left( \left \{ \frac{ d_i }{ \sum_j d_j } \right \}  \right) \right]  .
\end{align}
The quantumness approaches this upper bound in the limit that $M$ nodes have uniform degree $d_i = M \langle d \rangle/ N$ and all others have $d_i = 0$. This limit is never achieved unless $M = N$ and $\varepsilon = 0$, e.g., a regular network.
Physically, $\varepsilon = 0$ for a regular network because the symmetry of the Hamiltonian $H_Q$ implies its eigenvectors are evenly distributed.
The only eigenvector of this type that is positive is the initial state $\ket{\Psi (0)}$, which due to the Perron-Frobenius theorem must also be the ground state $\ket{\Psi (0)} = \ket{\phi_0}$. Therefore $E=0$ and so, from \eqr{eq:EnergyBound}, $\varepsilon = 0$.

In another limit, the quantumness takes 
its maximum value $\varepsilon = (N-2)/N \approx 1$ when the degrees of two 
nodes are equal and much larger than those of the others (note that the 
symmetry of $A$ prevents the degree of a single node from dominating). In the case 
that $A_{ij} \in \{ 0 , 1 \}$, i.e., the network 
underlying the walks is not weighted, the quantumness of a connected network is more restricted. It is maximized by a walk based on a star network---where a 
single node is connected to all others. For a walk of this type $\varepsilon = 
1/2 - \sqrt{N-1}/N \approx 1/2$.

\begin{figure*}[!htb]
\begin{center}
    \includegraphics[width=0.75\textwidth]{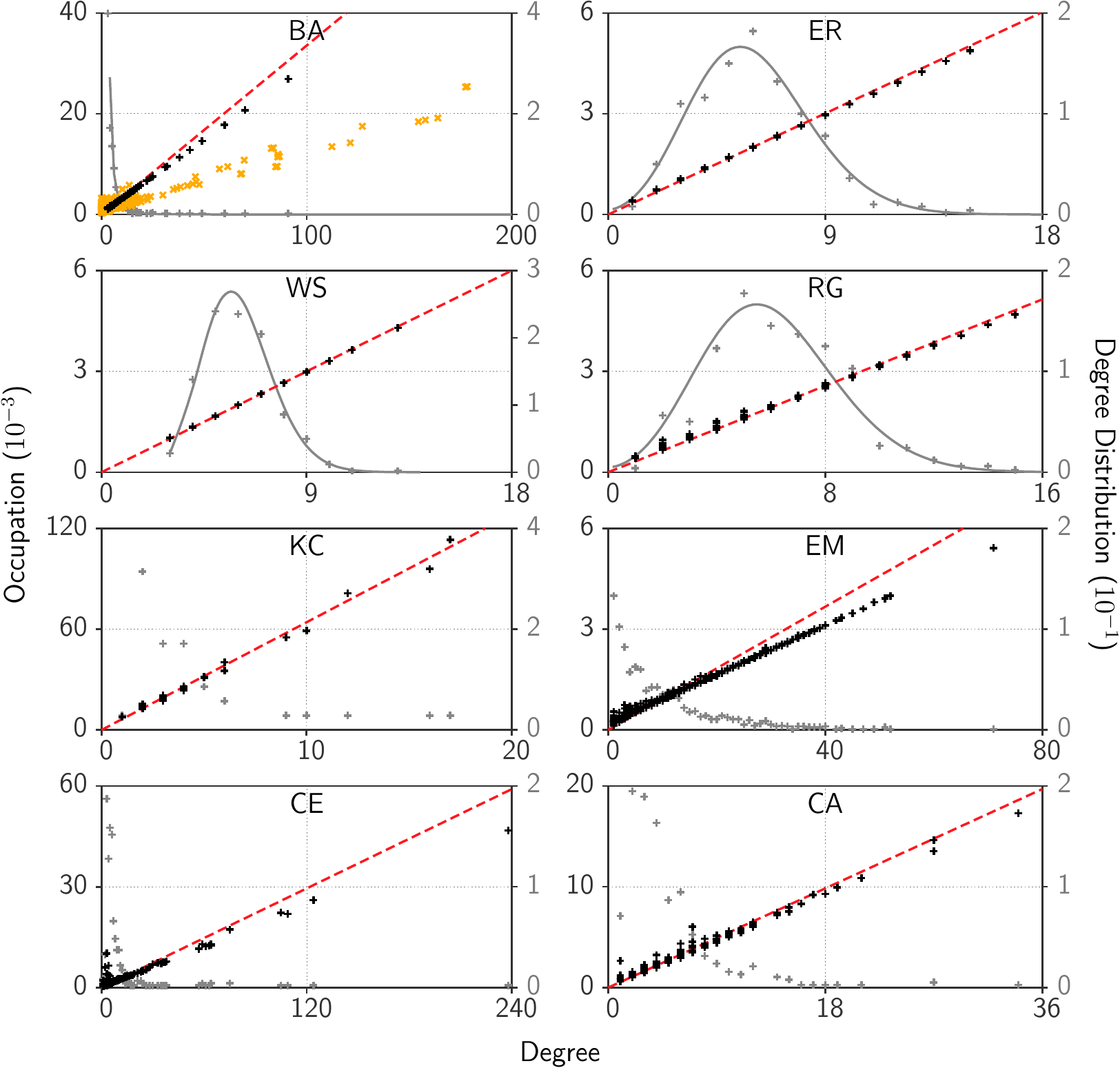}
\end{center}
\caption{
  Long time average probability and degree for nodes in a complex network.
  Eight networks are considered: \acf{ba},
  \acf{er}, \acf{ws}, \acf{rg}, \acf{kc}, \acf{em}, \acf{ce} and \acf{ca}.
  We plot the classical $( P_C )_i$ (red dashed line) and quantum $( P_Q )_i$ 
  (black $+$) probabilities against the degree $d_i$ for every node $i$.
  We overlay this with a plot of the average degree distribution $P(d)$ against 
  $d$ for each network type (grey full line), when known, along with the distribution for the specific realization used (grey $+$).
Alongside the \ac{ba} network we also plot $( P_Q )_i$ for the optimized BA (BA-opt) network, in which the internode weights of the BA network are randomly varied in a Monte Carlo algorithm to reach
  $\varepsilon = 0.6$ (orange $\times$). We do not include a plot of the degree distribution for this network.
}
\label{fig:average}
\end{figure*}

Next, in \secr{sec:walks-numericalresults} we study the form of the quantum 
correction $\ket{\tilde{P}_Q}$ given by \eqr{eq:quantumcorrection} for a range 
of complex network topologies.

\subsection{Numerical results} \label{sec:walks-numericalresults}

To obtain numerical results, we consider non-weighted binary networks $A_{ij} \in \{ 0, 1\}$
with various complex network topologies.
Specifically we consider the 
\ac{ba} scale free network~\cite{BA99}, the \ac{er}~\cite{ER60}
and the \ac{ws}~\cite{Watts1998} small world networks, and the 
\ac{rg} (on a square)~\cite{penrose2003}, a network without the scale 
free or small world characteristics.
We set number of nodes to be $N = 500$ and the average degree $\langle d \rangle \approx 6$.
If a disconnected network is obtained, only the giant component is considered.

The long time average probability of being on each node $i$ is plotted against 
its degree $d_i$ for a quantum ($P_Q$) and stochastic ($P_C$) walk in 
\fir{fig:average}.
The two cases are nearly identical for these binary networks and the evenly distributed initial state, illustrating that the quantumness $\varepsilon$ is small, below $0.13$.
See Tab.~\ref{tab:en-bound} for the comparison of values.
Within these, the \ac{ba} network shows the highest quantum correction.
This is expected since the \ac{ba} network has the higher degree heterogeneity.
The \ac{ws} network, which is well known to have quite uniform 
degrees~\cite{barrat2000}, is accordingly the network with the lowest quantum correction.

For many of the network types the typical quantumness can be obtained from 
the expected (thermodynamic limit) degree distribution. In the \ac{ba} network, 
the degree distribution approximately obeys the continuous probability density $P(d)= \langle 
  d\rangle^2 / 2 d^3 $~\cite{BA99}. Integrating 
this to find the moments, results in $\varepsilon = 1/9$, which is independent 
of the average degree $\langle d \rangle$ and is compatible with our numerics.
The degree distributions of the \ac{er} and \ac{rg} networks both approximately 
follow the Poissonian distribution 
$P(d) \approx  \langle d\rangle^d \ee^{-\langle d\rangle} / d! $ for large 
networks, which explains the similarity of their quantumness $\varepsilon$ 
values. 
For $\langle d\rangle = 6$ we 
recover $\varepsilon \approx 0.046$, which is 
compatible with the values for the particular networks we generated.
From the general form, calculating the quantumness numerically and performing a best fit we find 
that $\varepsilon \approx \kappa_1 \langle d\rangle^{-\kappa_2}$, with 
fitting parameters $\kappa_1 = 0.429$ and $\kappa_2 = 1.210$.

The size of the quantum effects can be enhanced by introducing heterogeneous 
weights $A_{ij}$ within a network. We have done this for a \ac{ba} network using several iterations of the following procedure.
A pair of connected nodes is randomly selected then the associated weight is doubled of halved at random.
As anticipated, the effect is to increase the discrepancy between 
the classical and quantum dependence of the long time average probability on 
degree, illustrated in \fir{fig:average}.
As the number of iterations is increased, the quantumness follows 
the bound given in \eqr{eq:entropybound}.
In fact, most  networks are found close to saturating this bound, especially for 
low quantumness. 

\begin{table}
\centering
\begin{tabular}{p{4cm}cr}\hline
  type     & $\varepsilon$ & $E /\Delta$\\\hline
  \ac{ba}     &  0.1299 &    0.5583\\
  \ac{er}     &  0.0431 &    0.1734\\
  \ac{rg}     &  0.0396 &   11.2875\\
  \ac{ws}     &  0.0164 &    0.0846\\
  \ac{ba}-opt &  0.6092 &  844.9181\\
  \ac{kc}     &  0.1204 &    1.3471\\
  \ac{ce}     &  0.2247 &    4.7622\\
  \ac{em}     &  0.1987 &    1.5449\\
  \ac{ca}     &  0.1138 &   39.8535\\\hline
\end{tabular}
\caption{Quantumness, energy and gap. The quantumness $\varepsilon$ and its upper bound $E/\Delta$, the ratio of energy and gap, for each of the nine networks considered in \fir{fig:average}.
Note that the energy gap can be arbitrarily small, and zero for disjoint networks;
this gives raise to very high $E/\Delta$ values.}
\label{tab:en-bound}
\end{table}

Further, the energy $E = \brackets{\Psi_0}{H_Q}{\Psi_0}$ of the given initial state 
  has a simple expression $E = 1 - (1/N) \sum_{ij} A_{ij} / \sqrt{d_i d_j}$, 
  which allows us to determine the extent to which the bound $E / \Delta \geq 
  \varepsilon$ is saturated by comparing the values of $E/ \Delta$ and 
  $\varepsilon$. We find that for some networks, e.g., the BA, ER and WS 
  networks, the bound is quite restrictive and reasonably saturated. However for 
  the other networks we find that quantumness takes a low value without this being ensured by the bound only, see Table~\ref{tab:en-bound}.

Finally, our numerical calculations reveal the behavior of the quantum part 
$\tilde{P}_Q$ of the long time average node occupation. 
We find that the quantum part enhances the long time average probability of being at nodes with small degree relative to the classical part. More precisely $ ( 
\tilde{P}_Q )_i / ( P_C )_i$ exhibits roughly $( d_i )^{-\kappa_3}$ scaling, 
with $\kappa_3 \approx 1$, as shown in 
\fir{fig:diff}. Interestingly, there is a correlation between the amount of 
enhancement, given by $\kappa_3$, and the type of complex network. The network 
types with smaller diameters (order of increasing diameter: \ac{ba}, then 
\ac{er} and \ac{ws}, then \ac{rg}) have the smallest $\kappa_3$, and the quantum 
parts enhance the low degree nodes least. Moreover, the enhancement $\kappa_3$ seems to be quite independent of the internode weights. Thus our numerics show a qualitatively common quantum effect for a range of complex network types. Quantitative details vary between the network types, but appear robust within each type. 

\begin{figure*}[!htb]
\begin{center}
    \includegraphics[width=0.8\textwidth]{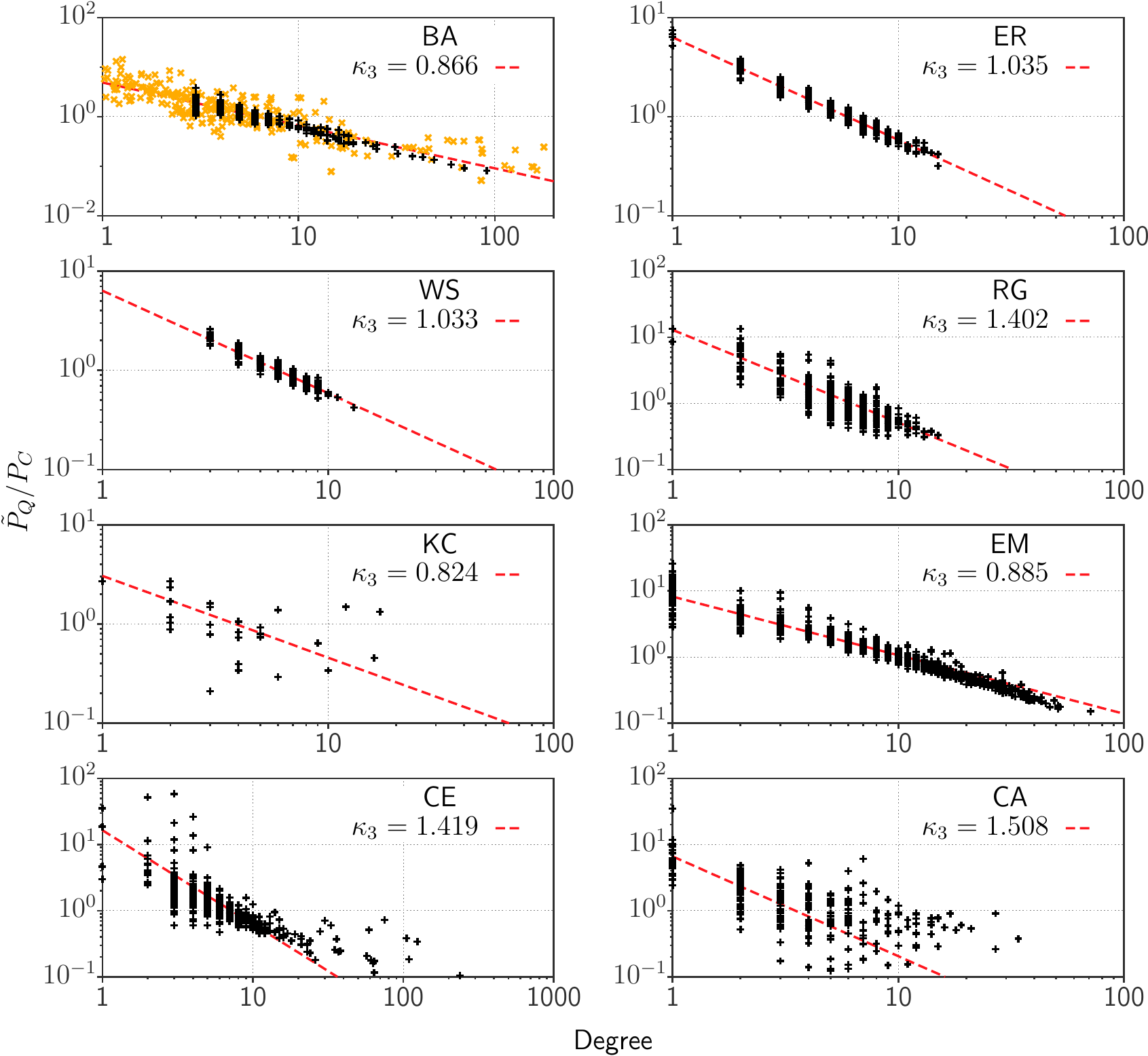}
\end{center}
\caption{Quantum effects.
  The ratio of the quantum $(\tilde{P}_Q)_i$ and classical 
  $(P_C )_i$ probabilities plotted against degree $d_i$ (black 
  $+$) for every $i$, for the the networks considered in \fir{fig:average}.
  We also plot the best fitting curve (red dashed line) to this data of the form $(\tilde{P}_Q)_i / (P_C )_i \propto ( d_i )^{\kappa_3}$ whose exponent $\kappa_3$ is given in the plot.}
\label{fig:diff}
\end{figure*}

The models of networks examined in the previous subsection have very 
specific topologies and therefore degree distributions,
and do not capture the topological properties of all real-world networks
(for details see chapter 9 of Ref.~\cite{estrada2011structure}).
We therefore now study the behavior of the quantumness and 
gap for topologies present in a variety of real-world networks: 
a \acf{kc} social network~\cite{zachary1977information}, the \acf{em} network 
of the URV university~\cite{guimera2003self}, the 
\acf{ce} network~\cite{duch2005community}, and a \acf{ca} network of 
scientists~\cite{newman2006finding}.

Despite the variety of topologies, we again find that the quantumness is consistently small, see  Tab.~\ref{tab:en-bound}. Therefore the classical and quantum distributions are very close, as shown in \fir{fig:average}. Additionally, the quantum correction exhibits the same generic behavior as observed for the  artificial networks.
Interestingly, the quantumness of real-world networks is appreciably smaller than enforced by the bound of \eqr{eq:EnergyBound}, with $E/\varepsilon \Delta$ taking large values.

\section{Community detection for quantum walks}\label{sec:community}

\subsection{Community detection}\label{sec:comdet}

Community detection is the partitioning of a set of nodes~$\NNN$ into non-overlapping
and non-empty subsets 
$\AAA,~\BBB,~\CCC,\ldots~\subseteq~\NNN$,
called communities, that together sum up to $\NNN$.

There is usually no agreed upon optimal partitioning of nodes into communities. 
Instead there is an array of approaches that differ in both the definition of 
optimality and the method used to achieve, exactly or approximately, this 
optimality (see~\cite{Fortunato2010} for a recent review). In classical 
networks optimality is, for example, defined 
statistically~\cite{lancichinetti2011oslom}, e.g.\ in terms of 
connectivity~\cite{girvan2002} or 
communicability~\cite{estrada2011community,estrada2009community}, or increasingly, and sometimes 
relatedly~\cite{meila2001random}, in terms of stochastic random 
walks~\cite{Delvenne2010,rosvall2011infomap,pons2005computing}. Our particular focus is on the 
latter, since the concept of transport (e.g.~a quantum walk) is central to
nearly all studies conducted in quantum physics.  As for achieving optimality,
methods include direct maximization 
via simulated annealing~\cite{guimera2004modularity,guimera2005functional} or, 
usually faster, iterative division or agglomeration of 
communities~\cite{hastie2001elements}. 
We focus on the latter since it provides a simple and effective 
way of revealing a full hierarchical structure of the network, requiring only 
the definition of the closeness of a pair of communities.

Formally, hierarchical community structure detection methods are
based on a (symmetric) closeness function
$c(\AAA,\BBB) = c(\BBB,\AAA)$
of two communities $\AAA \neq \BBB$. 
In the agglomerative approach,
at the lowest level of the hierarchy, the nodes are
each assigned their own communities. An iterative procedure then
follows, in each step of which the closest pair of communities
(maximum closeness $c$) are merged. This procedure ends at the highest level, where
all nodes are in the same community.
To avoid instabilities in this agglomerative procedure, the closeness
function is required to be non-increasing under the merging of two communities,
$c(\AAA \cup \BBB, \CCC) \le \max(c(\AAA,\CCC), c(\BBB,\CCC))$,
which allows the representation of the community
structure as a linear hierarchy indexed by the merging closeness.
The resulting structure is often represented as a dendrogram (as shown in 
\fir{fig:dendro}).%
\footnote{%
In general it may happen that more than one pair of communities are at the 
maximum closeness. In this case the decision on which pair merges first can 
influence the structure of the dendrogram, 
see~\cite{jain1988algorithms,carlsson2010characterization}.
In~\cite{carlsson2010characterization} a permutation invariant formulation of 
the agglomerative algorithm is given, where more than two clusters can be merged 
at once. In our work we use this formulation unless stated otherwise.
}

This leaves open the question of which level of the hierarchy yields
the optimal community partitioning. If a
partitioning is desired for simulation, for example, then there may be
a desired maximum size or minimum number of communities. However, without such
constraints, one can still ask what is the best choice of communities
within those given by the hierarchical structure.

A type of measure that is often used to quantify the quality of a community partitioning
choice for this purpose is
modularity~\cite{newman2004pre,Newman2004,Clauset2004}, denoted $Q$.
It was originally
introduced in the classical network setting, in which a network is
specified by a (symmetric) adjacency matrix of (non-negative) elements
$A_{ij} = A_{ji} \ge 0$ ($A_{ii} = 0$), each off-diagonal element giving the weight of
connections between nodes $i$ and $j\neq i$ \footnote{As will become apparent, we need 
  only consider undirected networks without self-loops.}.
The modularity attempts to
measure the fraction of weights connecting elements in the same
community, relative to what might be expected. Specifically, one takes the fraction of intra-community
weights and subtracts the average fraction obtained when the
start and end points of the connections are reassigned randomly,
subject to the constraint that the total connectivity $k_i = \sum_j
 A_{ij}$ of each node is fixed. The modularity is then given
by
\begin{align}
\label{eq:modularity}
  Q = \frac{1}{2m} \tr \left \{ C^{\mathrm{T}} B C \right \} ,
\end{align}
where $m = \half \sum_i k_i$ is the total weight of connections, $B$
is the modularity matrix with elements $B_{ij} = A_{ij}-k_i k_j / 2 m$, and
$C$ is the community matrix, with elements $C_{i \AAA}$ equal to unity if $i \in
\AAA$, otherwise zero. 
The modularity then takes values strictly less than one, possibly negative, and exactly zero in the case that the nodes form a single community.

As we will see, there is no natural adjacency matrix associated with
the quantum network and so for the purposes of modularity we use
$A_{ij} = c (i , j)$ for $i \neq j$. The modularity $Q$ thus measures
the fraction of the closeness that is intra-community, relative to
what would occur if the inter-node closeness $c (i , j)$ were randomly
mixed while fixing the total closeness $k_i = \sum_{j\neq i} c (i ,
j)$ of each node to all others. Thus both the community structure and
optimum partitioning depend solely on the choice of the closeness
function.

Finally, once a community partitioning is obtained it is often desired
to compare it against another. Here we use the common 
normalized mutual information
(NMI)~\cite{ana2003normalized,strehl2003cluster,danon2005}
as a measure of the mutual dependence of two community partitionings.
Each partitioning
$\CC = \{\AAA, \BBB,\dots \}$
is represented by a probability distribution
${P_\CC = \{|\AAA|/|\NNN|\}_{\AAA \in \CC}}$, where
$|\AAA| = \sum_i C_{i \AAA}$ is the
number of nodes in community~$\AAA$. The similarity of two community
partitionings $\CC$ and $\CC'$ depends on the joint
distribution
$P_{\CC \CC'} = \{ |\AAA \cap \AAA'| / |\NNN| \}_{\AAA \in \CC, \AAA' \in \CC'}$,
where
$|\AAA \cap \AAA'| = \sum_i C_{i \AAA} C_{i \AAA'}$
is the number of nodes that belong to both communities $\AAA$ and $\AAA'$. Specifically, NMI is defined as
\begin{equation}\label{eq:nmi}
\operatorname{NMI}(\CC,\CC') = 
\frac{2\,I(\CC,\CC')}{H(\CC)+H(\CC')}.
\end{equation}
Here $H(\CC)$ is the Shannon entropy of $P_\CC$, and the
mutual information $I(\CC,\CC')=
H(\CC)+H(\CC')-H(\CC,\CC')$ depends on the
entropy $H(\CC,\CC')$ of the joint distribution
$P_{\CC \CC'}$.
The mutual information is the average of the amount of information about
the community of a node in $\CC$ obtained by learning its
community in $\CC'$.
The normalization
ensures that the NMI has a minimum value of zero and takes its maximum value of unity for two identical
community partitionings. 
The symmetry of the definition of NMI follows from that of mutual
information and Eq.~\eqref{eq:nmi}.
  
\subsection{Quantum community detection}\label{sec:quantumcom}

The task of community detection has a particular interpretation in a quantum
setting. The state of a quantum system is described in terms of a
Hilbert space $\HHH$, spanned by a complete orthonormal set of basis
states $\{ \ket{i} \}_{i \in \NNN}$. Each basis state~$\ket{i}$ can be associated
with a node~$i$ in a network and often, as in the case of single
exciton transport, there is a clear choice of basis states that makes
this abstraction to a spatially distributed network natural.

The partitioning of nodes into communities then corresponds to the
partitioning of the Hilbert space $\HHH = \bigoplus_{\AAA \in \CC} \VV_\AAA $
into mutually orthogonal subspaces $\VV_\AAA = \Span_{i \in \AAA} \{\ket{i} \}$.
As with classical networks, one can then imagine an assortment of
optimality objectives for community detection, for example, to identify a
partitioning into subspaces in which inter-subspace transport is
small, or in which the state of the system remains relatively unchanged within each subspace.
In the next two subsections we introduce two classes of community closeness
measures that correspond to these objectives.
A more detailed derivation can be found in Sec.~\ref{sec:comdet-appendix}.

Our closeness measures
take into account the full unitary evolution of an isolated
system
governed by its Hamiltonian~$H$. Rather than being applicable to 
isolated systems only however, this type of community partitioning could be 
used, among other things, to guide the simulation or analysis of a more complete
model in the presence of an environment, where this more complete
model may be much more difficult to describe.

\subsubsection{Inter-community transport}
\label{sec:mixing}
Several approaches to detecting communities in classical networks are
based on the flow of probability through the network during a
classical random walk~\cite{meila2001random,eriksen2003modularity,pons2005computing,weinan2008optimal,Delvenne2010,rosvall2011infomap}.
In particular, many of these methods seek communities for which the
inter-community probability flow or transport is small. 
A natural approach to quantum community detection is thus to
consider the flow of probability during a continuous-time quantum
walk, and to investigate the \emph{change} in the probability of observing
the walker within each community:
\begin{align}
\T_{\CC}(t) &= \sum_{\AAA \in \CC} \T_\AAA (t)
= \sum_{\AAA \in \CC} \frac{1}{2}\left| p_\AAA \left \{ \rho (t) \right \} - p_\AAA \left \{ \rho (0) \right \} \right|,
\end{align}
where
$
\rho (t) = \mathrm{e}^{-\ii H t} \rho (0) \mathrm{e}^{\ii H t}
$
is the state of the walker, at time~$t$, during the walk generated by $H$, and
\begin{align}
p_\AAA \left \{ \rho \right \} = \tr \left \{ \SPr_\AAA \rho \right \}
\end{align}
where $\SPr_\AAA=\sum_{i\in\AAA} \ket{i}\bra{i}$ is the projector on the
$\AAA$ subspace,
is the probability of a walker in state~$\rho$ being found in
community~$\AAA$ upon a von Neumann-type measurement.\footnote{%
Equivalently, $p_\AAA \left \{ \rho \right \}$ is
the norm of the projection (performed by projector $\SPr_\AAA$) of the
state $\rho$ onto the community subspace $\VV_\AAA$.}

The initial state~$\rho (0)$ can be chosen freely.
The change in inter-community transport is clearest when the process begins either entirely inside or entirely outside each community. Because of this, we choose the walker to be initially localized
at a single node $\rho (0) = \proj{i}$ and then, for symmetry, sum
$\T_\CC (t)$ over all $i \in \NNN$. This results in the particularly
simple expression
\begin{align}
\T_\AAA (t) = \sum_{i \in \AAA, j \notin \AAA} \frac{\TM_{ij}(t)+\TM_{ji}(t)}{2}
= \sum_{i \in \AAA, j \notin \AAA} \sym{\TM}_{ij}(t),
\end{align}
where $\TM(t)$ is the doubly stochastic transfer matrix whose elements
$\TM_{ij}(t) = |\brackets{i}{\mathrm{e}^{-\ii H t}}{j}|^2$ give
the probability of transport from node~$j$ to node~$i$,
and $\sym{\TM}(t)$ its symmetrization.
This is reminiscent of classical community detection methods,
e.g.~\cite{pons2005computing}, using closeness measures based on the
transfer matrix of a classical random walk.

We can thus build a community structure that seeks to reduce $\T_\CC (t)$
at each hierarchical level by using the closeness function
\begin{align}
\label{eq:closeness_transport}
\notag
c^\T_t(\AAA , \BBB) &= \frac{\T_\AAA (t) + \T_\BBB (t) -
  \T_{\AAA\cup\BBB}(t)}{|\AAA||\BBB|}\\
&= 
\frac{2}{|\AAA||\BBB|} \sum_{i \in \AAA, j \in \BBB} \sym{\TM}_{ij}(t)
\end{align}
where the numerator is the decrement in $\T_\CC (t)$ caused by merging
communities $\AAA$ and $\BBB$.
The normalizing factor in Eq.~\eqref{eq:closeness_transport} avoids the
effects due to the uninteresting scaling of the numerator with the community
size.

Since a quantum walk does not converge to a stationary state, a
time-average of the closeness defined in Eq.~\eqref{eq:closeness_transport}
is needed to obtain a quantity that eventually converges with
increasing time. 
Given the linearity of the formulation, this corresponds to replacing
the transport probability~$\TM_{ij}(t)$
in \eqr{eq:closeness_transport}
with its time-average
\begin{align}
\label{eq:avtransfer}
\tave{\TM}_{ij}(t) = \frac{1}{t} \int_0^t \TM_{ij}(t') \:\dd t'.
\end{align}

It follows that, as with similar classical community detection methods~\cite{Delvenne2010},
our method is in fact a class of approaches, 
each corresponding to a different time $t$. 
The appropriate value of $t$ will depend on the specific
application, for example, a natural
time-scale might be the decoherence time. 
Not wishing to lose generality and focus on a particular system, 
we focus here on the short and long time limits. 

In the short time limit $t \to 0$, relevant if $t H_{ij} \ll 1$ for
$i \neq j$, the averaged transfer matrix $\tave{T}_{ij} (t)$ is simply
proportional to $|H_{ij}|^2$.
Note that in the short
time limit there is no interference between
different paths from $\ket{i}$ to $\ket{j}$, and therefore for
short times $c^\T_t (i , j)$ does not depend on the on-site energies $H_{ii}$ or the phases of the
hopping elements $H_{i j}$.
This is because, to leading order in time, interference
does not play a role in the transport out of a single node. 
For this reason we can refer to this approach as ``semi-classical''.

In the long time limit $t \to \infty$, relevant if $t$ is much larger
than the inverse of the smallest gap between distinct
eigenvalues of $H$, the probabilities are elements of the
mixing matrix~\cite{godsil2013},
\begin{align}
\lim_{t\to \infty} \tave{\TM}_{ij}(t) = \sum_k | \bracket{i}{\HPr_k}{j}|^2 ,
\label{eq:mixing-matrix}
\end{align}
where $\HPr_k$ is the projector onto the $k$-th eigenspace of $H$. This
thus provides a simple spectral method for building the community
structure.

Note that, unlike in a classical infinitesimal stochastic walk where
each $\tave{\TM}_{ij} (t)$
eventually becomes proportional to the connectivity $k_j$ of the final
node $j$, the long time limit in the quantum setting is non-trivial and,
as we will see, $\tave{\TM}_{ij}(t)$ retains a strong impression of the
community structure for large~$t$.%
\footnote{Note that, apart from small or large
times $t$, there is no guarantee of symmetry $\TM_{ij}(t) = \TM_{ji}(t)$ in the
transfer matrix for a given
Hamiltonian. See~\cite{zimboras2013quantum}. Hamiltonians featuring
this symmetry, e.g., those with real $H_{ij}$, are called
time-symmetric.}

\subsubsection{Intra-community fidelity}
\label{sec:coher}
Classical walks, and the community detection methods based on them, are fully 
described by the evolution of the probabilities of the walker occupying each 
node. The previous quantum community detection approach is based on the 
evolution of the same probabilities but for a quantum walker.
However, quantum walks are richer than this, they are not fully
described by the evolution of the node-occupation probabilities. We
therefore introduce another community detection method that captures
the full quantum dynamics within each community subspace.

Instead of reducing merely the change in probability within the
community subspaces, we reduce the change in the projection of the
quantum state in the community subspaces. This change is measured
using (squared) fidelity, a common measure of distance between two
quantum states.
For a walk beginning in state $\rho (0)$ we therefore focus on the quantity
\begin{align}
  \F_\CC (t) &= \sum_{\AAA \in \CC} \F_\AAA (t)
= \sum_{\AAA \in \CC}  F^2 \left \{ \SPr_\AAA \rho(t) \SPr_\AAA, \SPr_\AAA \rho(0) \SPr_\AAA \right \},
\end{align}
where $\SPr_\AAA \rho \SPr_\AAA$ is the projection of the state $\rho$ onto the subspace $\VV_\AAA$ and
\begin{align}
  F \left\{\rho ,\sigma \right\} = \tr \left\{ \sqrt{\sqrt{\rho} \sigma \sqrt{\rho}} \right\} \in  [0, \sqrt{\tr \{\rho\} \tr \{\sigma\}}]
\end{align}
is the fidelity, which is symmetric between $\rho$ and $\sigma$.

We build a community structure that seeks to maximize 
the increase in $\F_\CC (t)$ at 
each hierarchical level by using the closeness measure
\begin{align}
\label{eq:fidelitydist}
c^\F_t (\AAA , \BBB) = \frac{\F_{\AAA \cup \BBB}(t) -\F_\AAA(t) -\F_\BBB(t)
}{|\AAA| |\BBB|} \in [-1 ,1],
\end{align}
i.e., the change in $\F_\CC (t)$ caused by merging communities $\AAA$ and~$\BBB$.
Our choice for the denominator prevents uninteresting size scaling, 
as in Eq.~\eqref{eq:closeness_transport}.

The initial state~$\rho(0)$ can be chosen freely. Here we choose the
pure uniform superposition state $\rho(0)=\ket{\psi_0}\bra{\psi_0}$ satisfying
$\brakets{i}{\psi_0} = 1/\sqrt{n}$ for all~$i$.
This state was used to 
investigate the effects of the connectivity on the dynamics of a quantum walker
in~Ref.~\cite{Faccin2013}. 

As for our other community detection approach, we consider the time-average of 
Eq.~\eqref{eq:fidelitydist} which yields
\begin{align}
c_t^\F (\AAA,\BBB) =
\frac{2}{|\AAA||\BBB|} \sum_{i\in\AAA, j\in\BBB} 
\real(\tave{\rho}_{ij}(t)\rho_{ji}(0)),
\end{align}
where $\tave{\rho}_{ij}(t) = \frac 1t \int_0^t \dd t' \rho_{ij}(t')$.
In the long time limit, the time-average of the density matrix takes a particularly simple 
expression:
\begin{align}
  \lim_{t \to \infty} \tave{\rho}_{ij}(t) = \sum_k \HPr_k \rho_{ij}(0) \HPr_k,
\end{align}
where $\HPr_k$ is as in the previous Sec.~\ref{sec:mixing}.

The definition of community closeness given in
Eq.~\eqref{eq:fidelitydist} can exhibit negative values. 
In this case the usual definition of modularity fails~\cite{traag2009}
and one must extended it.
In this work we use the definition of modularity proposed
in~\cite{traag2009}, which coincides with Eq.~(\ref{eq:modularity}) in
the case of non-negative closeness. 
The extended definition treats negative and positive 
links separately, and tries to minimize intra-community negative
links while maximizing intra-community positive links.

\subsection{Performance analysis}\label{sec:performance}

To analyze the performance of our quantum community detection methods
we apply them to three different networks.
The first one (Sec.~\ref{sec:quantumnet}) is a simple quantum network, 
which we use to highlight how some intuitive notions in 
classical community detection do not necessarily transfer over 
to quantum systems.
The second example (Sec.~\ref{sec:artificial}) is an artificial quantum
network designed to exhibit a clear classical community structure, 
which we show is different from the quantum community structure obtained and 
fails to capture significant changes in this structure induced by quantum
mechanical phases on the hopping elements of the Hamiltonian.
The final network (Sec.~\ref{sec:lhcii}) is a real world quantum
biological network, the LHCII light harvesting complex, for which we find a
consistent quantum community structure differing from the
community structure cited in the literature.
These findings confirm that a quantum mechanical treatment of community
detection is necessary as classical and semi-classical methods
cannot be reproduce the structures that appropriately capture quantum effects.

Below we will compare quantum community structures
against more classical community structures, such the one given
by the semi-classical method based on the short time transport and,
in the case of the example of Sec.~\ref{sec:artificial},
the classical network from which the quantum network is constructed.
Additionally we use a traditional classical
community detection algorithm, OSLOM~\cite{lancichinetti2011oslom}, an algorithm based on the maximization
of the statistical significance of the proposed partitioning,
whose input adjacency matrix~$A$ must be real.
For this purpose we use the absolute
values of the Hamiltonian elements in the site basis: $A_{ij} = |H_{ij}|$. 

\subsubsection{Simple quantum network}
\label{sec:quantumnet} 

\begin{figure}
    \centering
    \includegraphics[width=0.3\columnwidth]{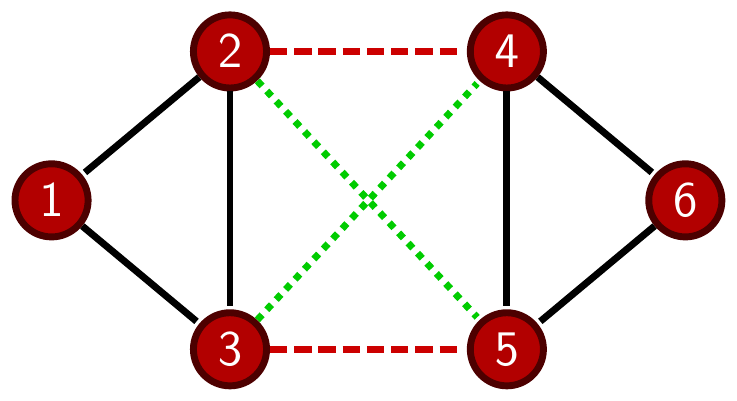}\\ \medskip
    Disconnected components:\\
    \vspace{-10pt}
    \subfloat[Transport]{\label{fig:phase-null-transport}
      \includegraphics[height=0.15\columnwidth]{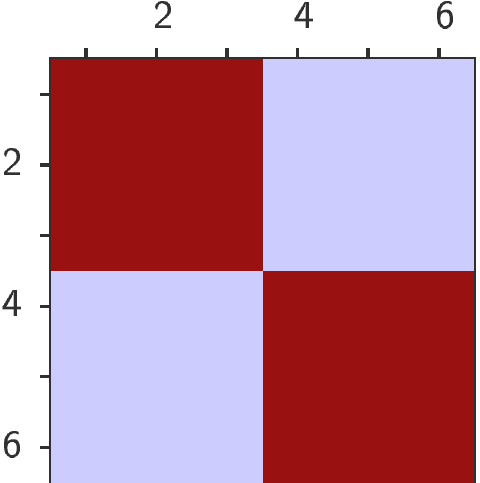}}
    \hfill
    \subfloat[Fidelity]{\label{fig:phase-null-fidelity}
      \includegraphics[height=0.15\columnwidth]{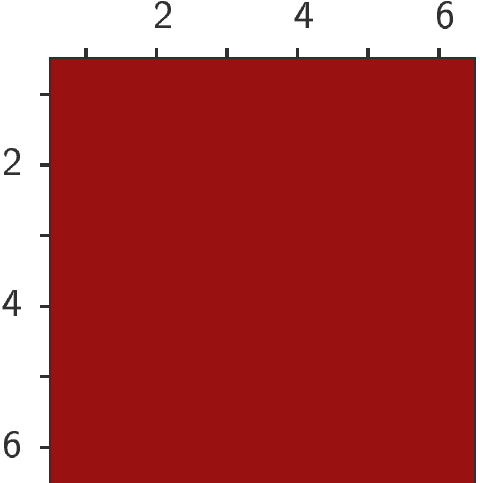}}
    \hfill
    \subfloat[Fidelity (Perturbed)]{\label{fig:phase-null-fidelity-pert}
      \includegraphics[height=0.15\columnwidth]{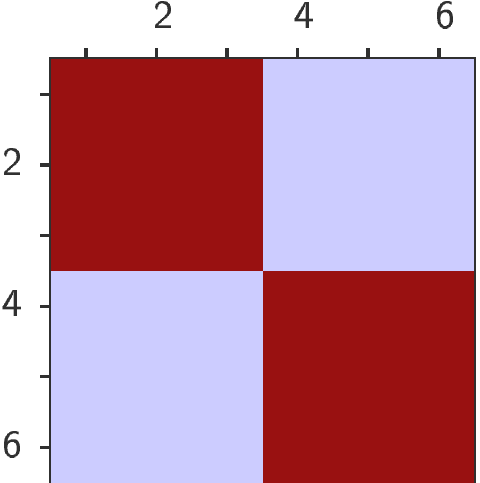}
      \includegraphics[height=0.15\columnwidth]{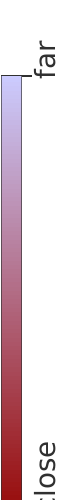}}\\  \medskip
    Phases' effect on transport:\\
    \vspace{-10pt}
    \subfloat[Coherent phases]{\label{fig:phase-coher-transport}
      \includegraphics[height=0.15\columnwidth]{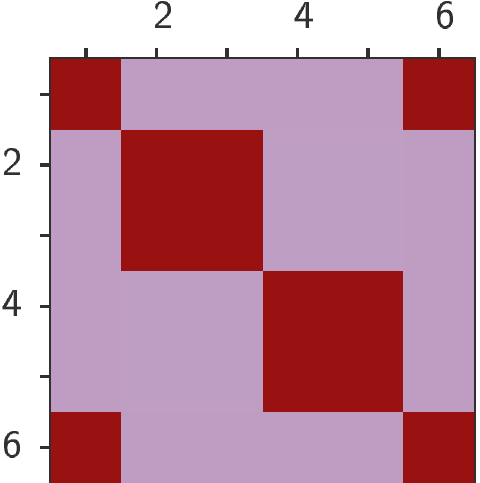}}
    \hfill
    \subfloat[Random phases]{\label{fig:phase-rand-transport}
      \includegraphics[height=0.15\columnwidth]{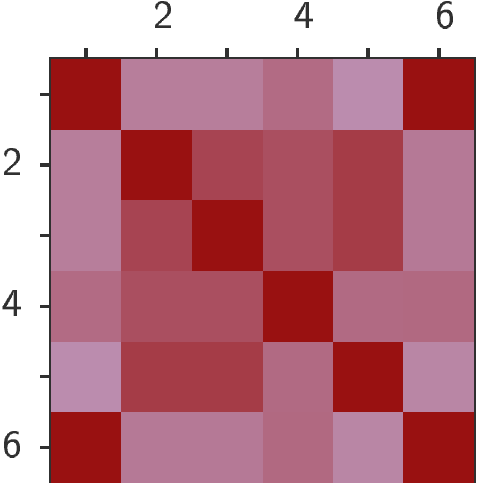}}
    \hfill
    \subfloat[Cancelling phases]{\label{fig:phase-canc-transport}
      \includegraphics[height=0.15\columnwidth]{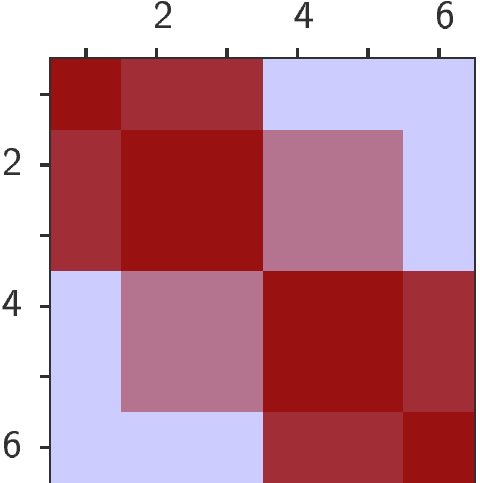}
      \includegraphics[height=0.15\columnwidth]{figs/article/communities/colorbox}}\\  \medskip
    Phases' effect on fidelity:\\
    \vspace{-10pt}
    \subfloat[Coherent phases]{\label{fig:phase-coher-fidelity}
      \includegraphics[height=0.15\columnwidth]{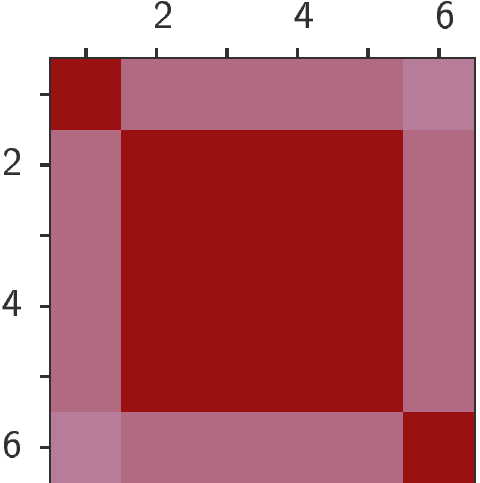}}
    \hfill
    \subfloat[Random phases]{\label{fig:phase-rand-fidelity}
      \includegraphics[height=0.15\columnwidth]{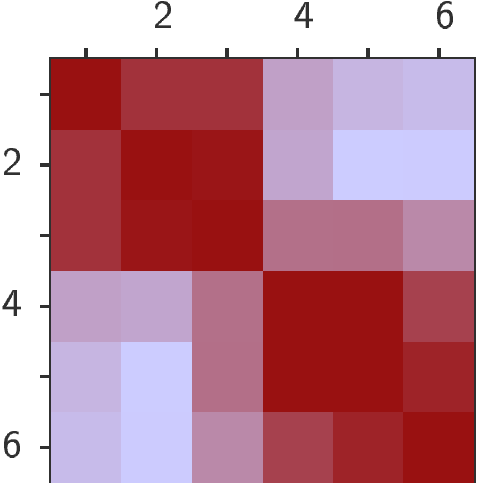}}
    \hfill
    \subfloat[Cancelling phases]{\label{fig:phase-canc-fidelity}
      \includegraphics[height=0.15\columnwidth]{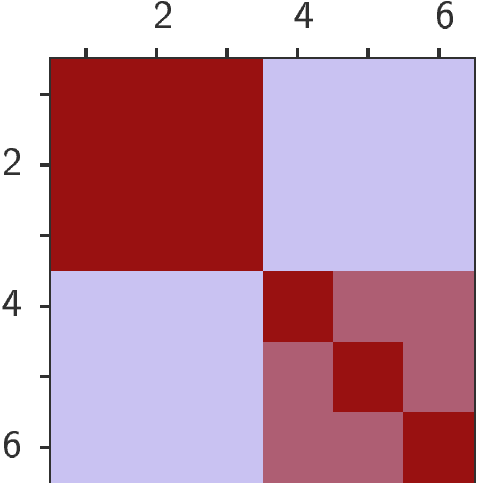}
      \includegraphics[height=0.15\columnwidth]{figs/article/communities/colorbox}}\\
  \caption{
  Simple quantum network --- a graph with six nodes.
  Each solid line represents transition amplitude $H_{ij}=1$.
  For dashed and dotted lines the transition amplitude can be either zero (a, b 
  and c) or the absolute value is the same $|H_{ij}|=1$ but phase is
  (d and g) coherent (all ones),
  (e and h) random $\exp(i \varphi_k)$ for each link,
  (f and i) canceling (ones for dashed red and minus one for dotted green).
  Plots show the node closeness for both methods based on transport and fidelity 
  (only the long-time-averages are considered, in plots (g), (h) and (i) we
  used a perturbed Hamiltonian to solve the eigenvalues degeneracy, this
  explains the non-symmetric closeness in (i)).
  }
  \label{fig:phase-sensitive-graph}
\end{figure}

Here we use a simple six-site network model to study ways in which quantum 
effects lead to non-intuitive results, and how methods based on different quantum 
properties can, accordingly, lead to very different choices of communities.

We begin with two disconnected cliques of three nodes each, 
where all Hamiltonian matrix elements within the groups are identical and real.
\fir{fig:phase-sensitive-graph} illustrates this highly 
symmetric topology.
The community detection method based on quantum transport identifies the two 
fully-connected groups as two separate communities 
(\fir{fig:phase-null-transport}), as is expected. 
Contrastingly, the methods based on fidelity predict
counter-intuitively only a single community; two disconnected nodes can retain 
coherence and, by this measure, be considered part of the same community 
(\fir{fig:phase-null-fidelity}). 

This symmetry captured by the fidelity-based community structure
breaks down if we introduce random perturbations into the Hamiltonian.
Specifically, the fidelity-based closeness~$c_t^F$
is sensitive to perturbations of the order~$t^{-1}$, above
which the community structure is divided into the two groups of three
(\fir{fig:phase-null-fidelity-pert})
expected from transport considerations. Thus we may tune the resolution of
this community structure method to asymmetric perturbations by varying~$t$.

Due to quantum interference we expect that the
Hamiltonian phases should significantly affect the quantum community partitioning.
The same toy model can be used to demonstrate this effect.
For example, consider adding four elements to the Hamiltonian corresponding to
hopping from nodes 2 and 3 to 4 and 5 (see diagram in
\fir{fig:phase-sensitive-graph}). 
If these hopping elements are all identical to the others, it
is the two nodes, 1 and 6, that are not directly connected for which the
inter-node transport is largest (and thus their inter-node closeness is the
largest). However, when the phases of the four additional elements are
randomized, this transport is decreased.
Moreover, when the phases are canceling, the
transport between nodes 1 and 6 is reduced to zero, and the closeness between
them is minimized
(see Figs.~\ref{fig:phase-coher-transport}--\ref{fig:phase-canc-transport}).  

The fidelity method has an equally strong dependence on the phases (see 
Figs.~\ref{fig:phase-coher-fidelity}--\ref{fig:phase-canc-fidelity}), with
variations in the phases breaking up the network from a large central community
(with nodes 1 and 6 alone)
into the two previously identified communities.

\subsubsection{Artificial quantum network}
\label{sec:artificial}

\begin{figure*}[t!]
  \centering
  \begin{minipage}{0.9\textwidth}
    \subfloat[Original data]{\label{fig:art-theo}
      \includegraphics[angle=0,width=0.2\textwidth]{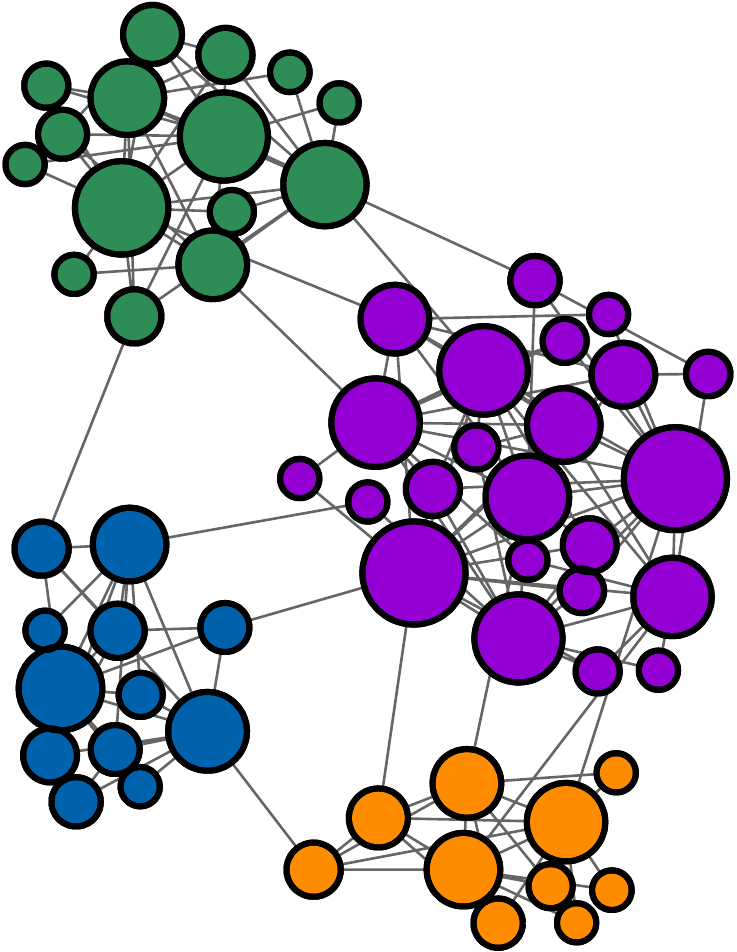}}
    \hfill
    \subfloat[Transport; $t\to 0$]{\label{fig:art-transport_short}
      \includegraphics[angle=0,width=0.2\textwidth]{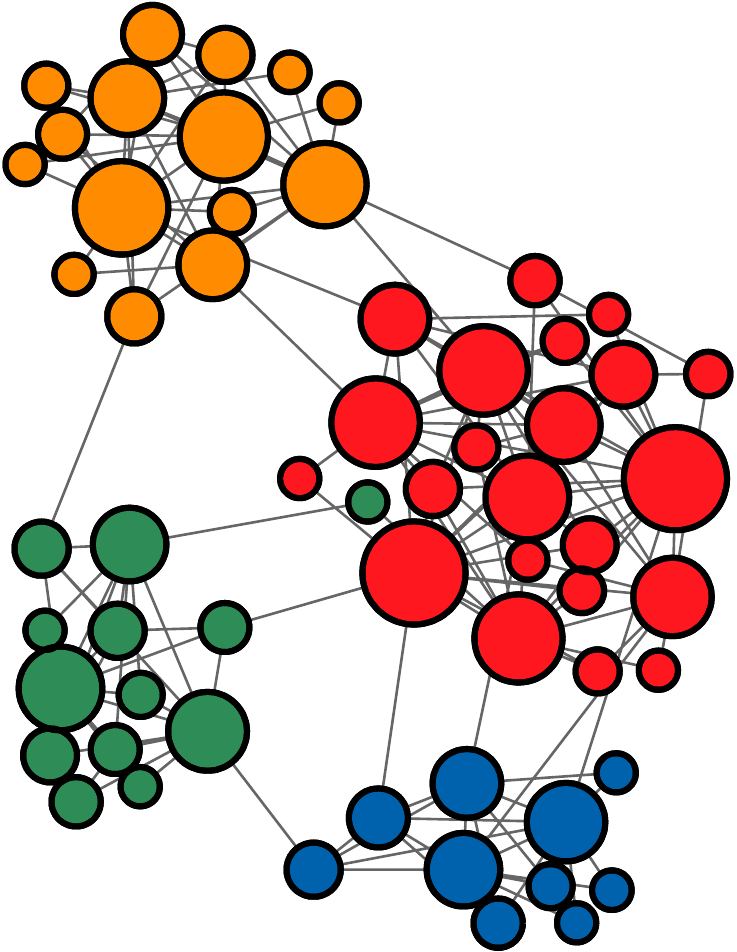}}
    \hfill
    \subfloat[Transport; $t\to\infty$]{\label{fig:art-transport_long}
      \includegraphics[angle=0,width=0.2\textwidth]{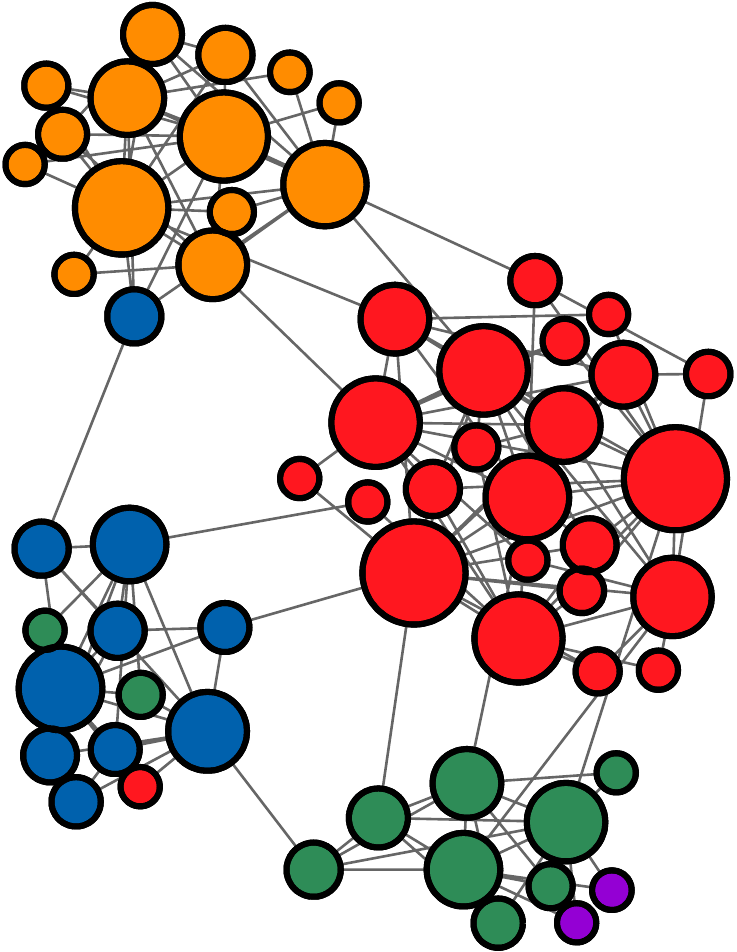}}
    \hfill
    \subfloat[Fidelity; $t\to\infty$]{\label{fig:art-fidelity}
      \includegraphics[angle=0,width=0.2\textwidth]{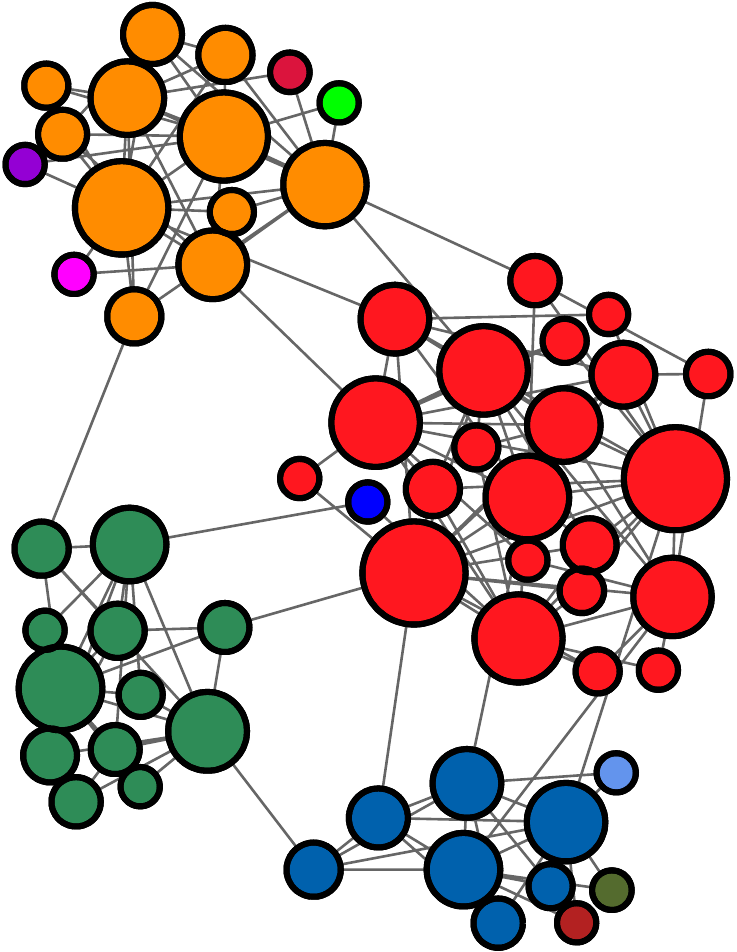}} 
    \vspace{10pt}
    \\
    \subfloat[OSLOM]{\label{fig:art-oslom}
      \includegraphics[angle=0,width=0.2\textwidth]{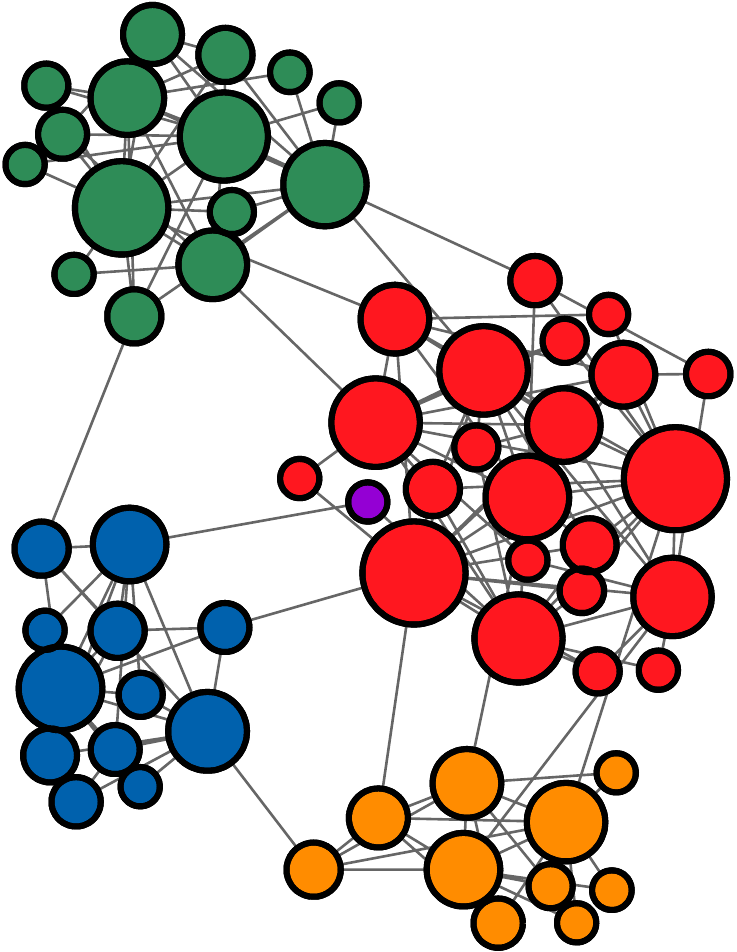}}
    \hfill
    \subfloat[Phases dependence (original partitioning)]{
      \label{fig:art-phases-nophases} 
      \includegraphics[width=0.3\textwidth]{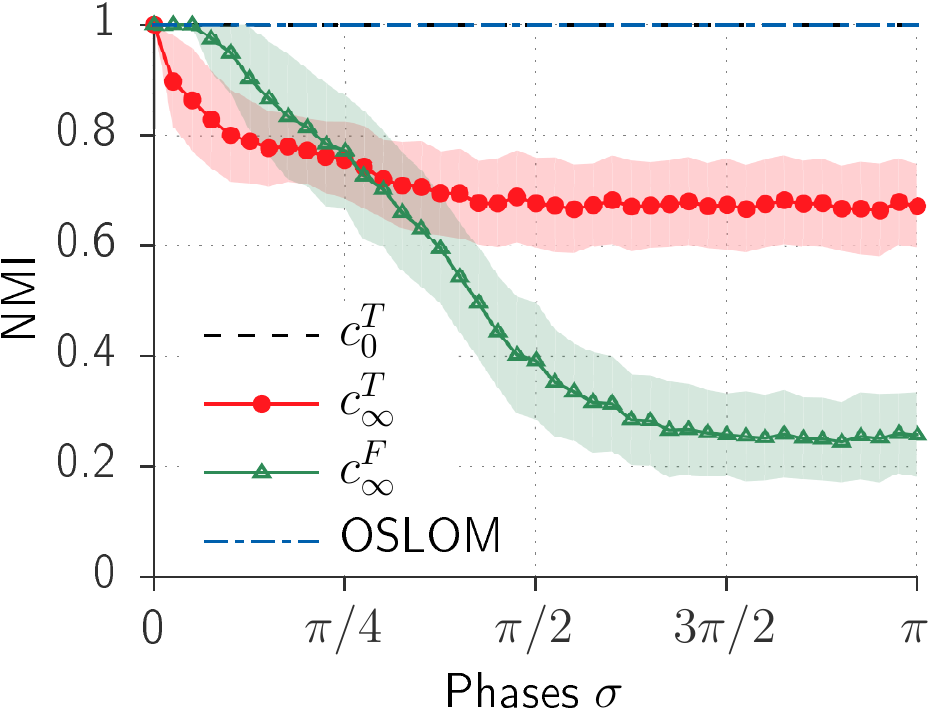}
    }
    \hfill
    \subfloat[Phases dependence (classical model)]{
      \label{fig:art-phases-classical} 
      \includegraphics[width=0.3\textwidth]{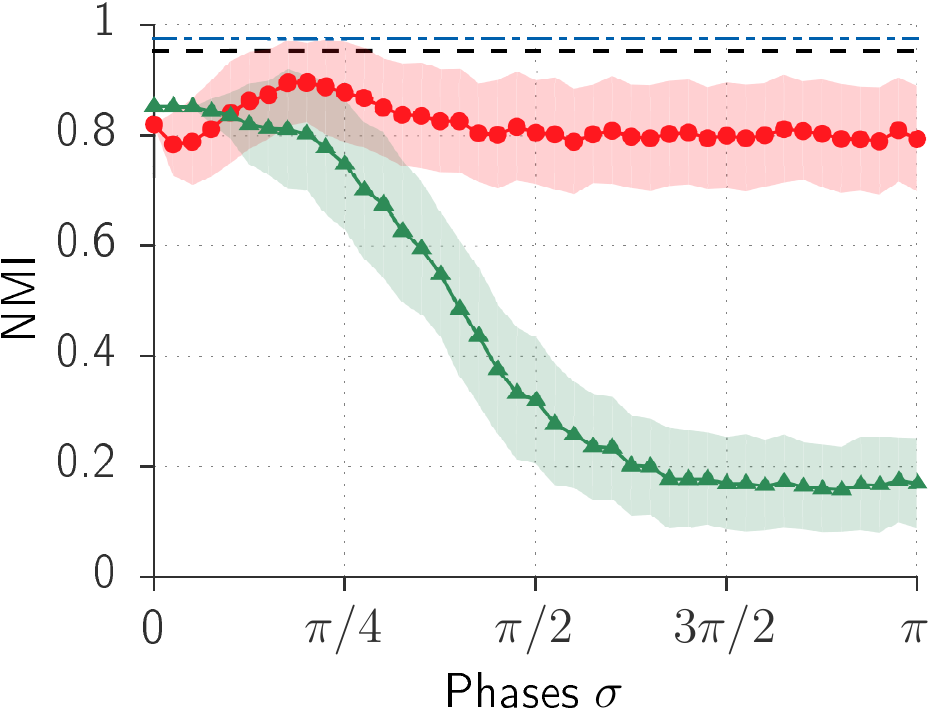}
    }
  \end{minipage}
\caption{Artificial community structure.
  (a) Classical community structure used in creating the network.
  (b--e) Community partitionings found using the
  three quantum methods and OSLOM.
  (f,g) 
  Behavior of the approaches as the phases of the Hamiltonian elements are randomly
  sampled from a Gaussian distribution of width~$\sigma$. The mean
  NMI, compared with zero phase partitioning (f) and the classical
  model data (g),
  over 200 samplings of the phase distribution is plotted. The
  standard deviation is indicated by the shading.
  Both OSLOM and $c^{\T}_0$ are insensitive to phases and thus do not
  respond to the changes in the Hamiltonian.
}
\label{fig:artificial}
\end{figure*}

The Hamiltonian of our second quantum network is constructed from the
adjacency matrix $A$ of a classical unweighted, undirected network
exhibiting a clear classical partitioning,
using the relation $H_{ij} = A_{ij}$.
We construct~$A$ using the algorithm proposed by Lancichinetti {\em et
al.}~in~\cite{lancichinetti2008}, which provides a method
to construct a network with heterogeneous distribution both for the node
degree and for the communities dimension and a controllable inter-community
connection.  We start with a rather small network of 60 nodes with average
intra-community connectivity $\langle k \rangle=6$, and only 5\% of the
edges are rewired to join communities.
The network is depicted in \fir{fig:art-theo}.
To confirm the expected, the known classical community structure is indeed
obtained by the semi-classical short-time-transport algorithm%
\footnote{In the case of short-time transport, a small
  perturbation was also added to the closeness
  function in order to break the symmetries of the system.}
and the OSLOM
algorithm (see Figs.~\ref{fig:art-transport_short}--\ref{fig:art-oslom}),
achieving $\text{NMI}=0.953$ and $\text{NMI}=0.975$ with the known
structure, respectively.

The quantum methods based on the long-time average of both transport and fidelity
reproduce the main features of the original community structure
while unveiling new characteristics.  The transport-based long-time average
method ($\text{NMI}=0.82$ relative to the classical partitioning)
exhibits disconnected communities, i.e.\ the
corresponding subgraph is disconnected. This behavior can be explained
by interference-enhanced quantum walker dynamics, as exhibited by the toy
model in the previous subsection.
The
long-time average fidelity method ($\text{NMI}=0.85$) returns the four main
classical communities plus a number of single-node communities.
Both methods demonstrate that the quantum and classical community
structures are unsurprisingly different, with the quantum community
structure clearly dependent on the quantum property being
optimized, more so than the different classical partitionings.

\subsubsection*{Adjusted phases}

As shown in Sec.~\ref{sec:quantumnet},
due to interference the dynamics of the quantum system can change
drastically if the phases of the Hamiltonian elements are non-zero. This is
known as a chiral quantum walk~\cite{zimboras2013quantum}. Such walks exhibit, for example, time-reversal
symmetry breaking of transport between sites~\cite{zimboras2013quantum} and
it has been proposed that nature might actually make use of phase
controlled interference in transport processes~\cite{harel2012quantum}.
OSLOM, our semi-classical short-time transport algorithm and other
classical community partitioning methods are insensitive to changes in the
hopping phases. Thus, by establishing that the quantum community structure
is sensitive to such changes in phase, as expected from above, we show that
classical methods are inadequate for finding quantum community structure.

To analyze this effect we take the previous network
and adjust the phases of the Hamiltonian terms while preserving their
absolute values. Specifically, the phases are sampled randomly from a
normal distribution with mean zero and standard deviation $\sigma$.
We find that, typically, as the standard deviation $\sigma$ increases, 
when comparing quantum communities and the corresponding communities
without phases the NMI between them decreases, as shown in
\fir{fig:art-phases-nophases}.
A similar deviation reflects on the comparison with the classical
communities used to construct the system, shown in
\fir{fig:art-phases-classical}.
This sensitivity of the quantum community structures to phases, as revealed
by the NMI, confirms the expected inadequacy of classical methods.
The partitioning based on long-time average fidelity seems to be the most sensitive
to phases.

\subsubsection{Light-harvesting complex}
\label{sec:lhcii} 

\begin{figure*}[t!]
  \centering
  \includegraphics[height=0.24\textwidth]{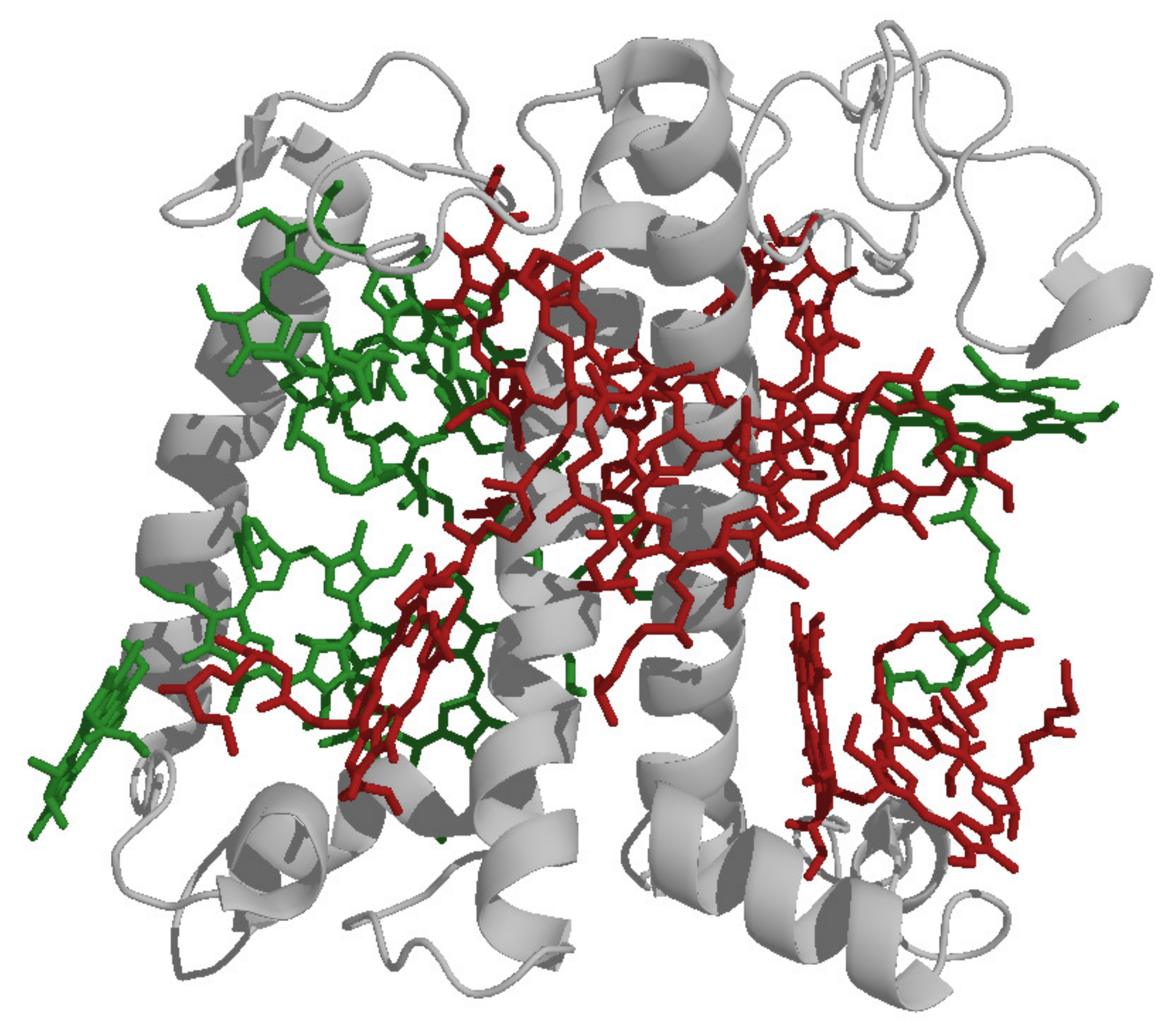}\hfill
  \includegraphics[height=0.24\textwidth]{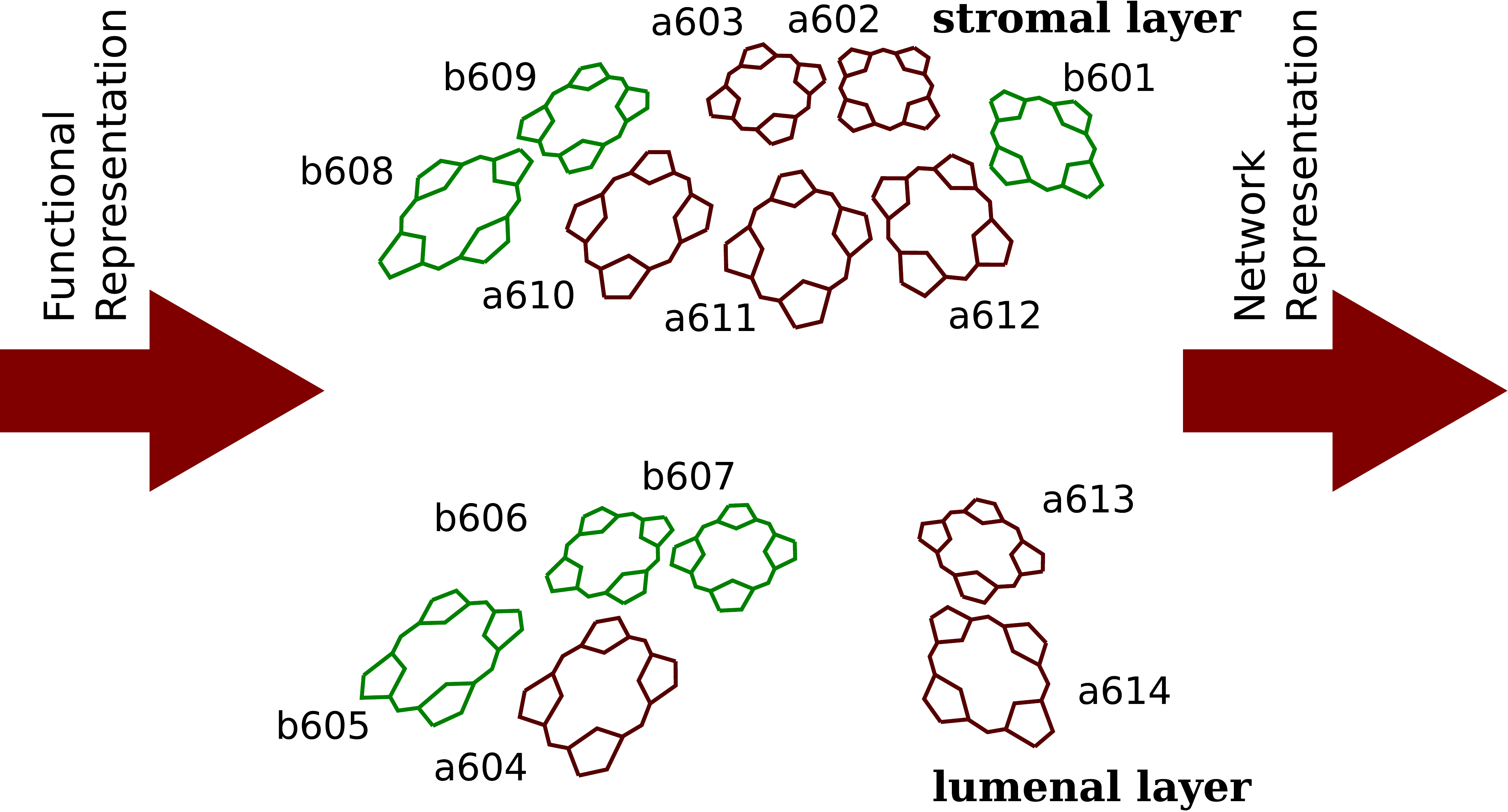}\hfill
  \includegraphics[height=0.24\textwidth]{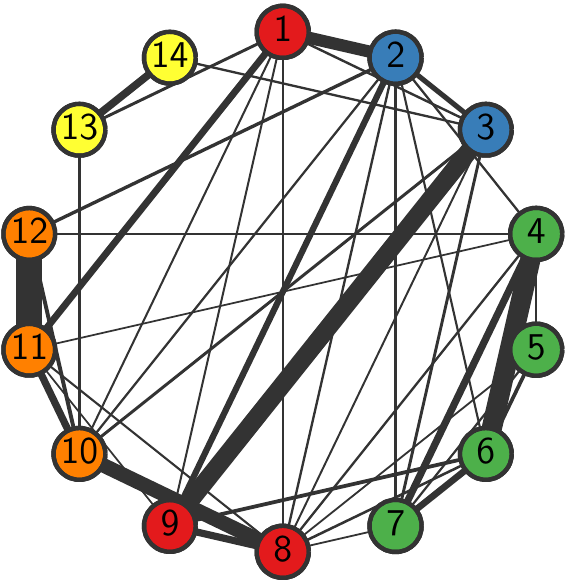}\\
  \bigskip
  \subfloat[Transport; $t\to 0$]{\label{fig:lhcii-ampli}
    \includegraphics[height=0.24\textwidth]{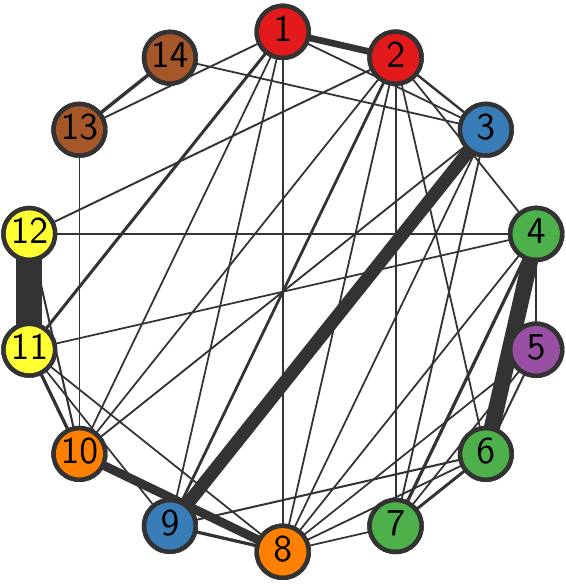}}
  \hfill
  \subfloat[Transport; $t\to\infty$]{\label{fig:lhcii-mixing}
    \includegraphics[height=0.24\textwidth]{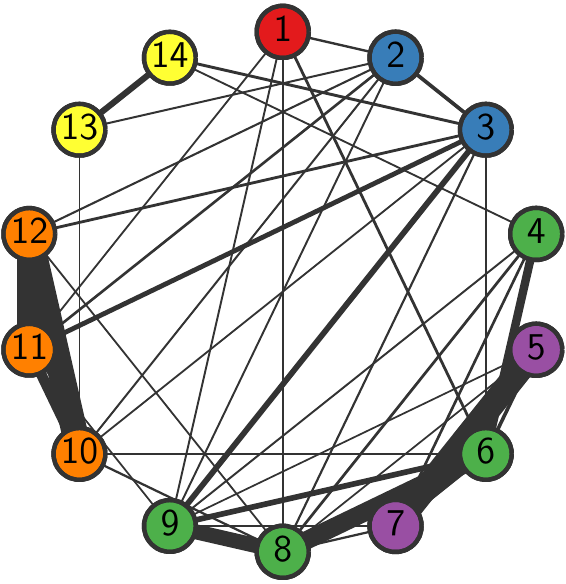}}
  \hfill
  \subfloat[Fidelity; $t\to\infty$]{\label{fig:lhcii-purity}
    \includegraphics[height=0.24\textwidth]{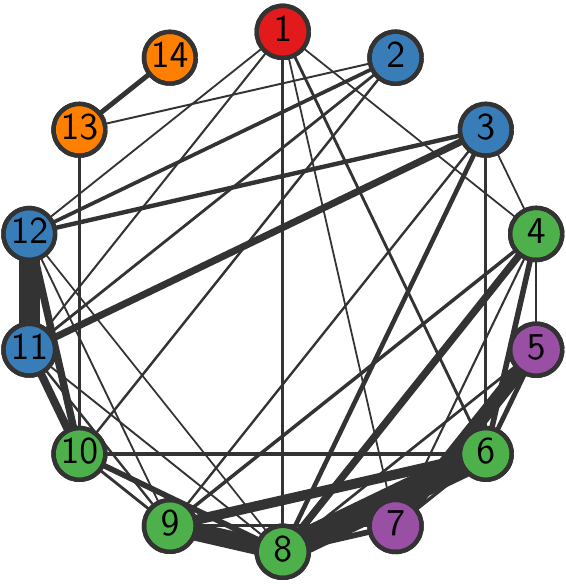}}
\caption{Light harvesting complex II (LHCII).
  (top left) Monomeric subunit of the LHCII complex with pigments Chl-a (red) and 
  Chl-b (green) packed in the protein matrix (gray).
  (top center) Schematic representation of Chl-a and Chl-b in the monomeric 
  subunit, here the labeling follows the usual nomenclature (b601, a602\dots).
  (top right) Network representation of the pigments in circular layout, colors represent 
  the typical partitioning of the pigments into communities. The widths of the links 
  represent the strength of the couplings~$|H_{ij}|$ between nodes.
  Here the labels maintain only the ordering (b601$\to$1, a602$\to$2,\dots).
  (a,b,c) Quantum communities as found by the different quantum community detection methods.
  Link width denotes the pairwise closeness of the nodes.
}
\label{fig:lhcii}
\end{figure*}

An increasing number of biological networks of non-trivial  topology are
being described using quantum mechanics.  For example, light harvesting
complexes have drawn significant attention in the quantum information
community.

One of these is the LHCII, a two-layer 14-chromophore complex embedded into
a protein matrix (see \fir{fig:lhcii} for a sketch) that
collects light energy and directs it toward the reaction center where it is
transformed into chemical energy.  The system can be described as a network
of 14 sites connected with a non-trivial topology.
The single-exciton subspace is spanned by 14
basis states, each corresponding to a node in the network, and the
Hamiltonian in this basis was found in Ref.~\cite{fleming2009lhcii}.

In a widely adopted chromophore community
structure~\cite{novoderezhkin2005lhcii}, the sites are partitioned
\emph{by hand} into communities according to their physical closeness
(e.g. there are no communities spanning the two layers of the complex), and
the strength of Hamiltonian couplings
(see the top right of \fir{fig:lhcii}). Here, we
apply our \emph{ab initio} automated quantum community detection algorithms
to the same Hamiltonian.

All of our approaches predict a modified partitioning to that commonly used
in the literature. The method based on short-time transport returns
communities that do not connect the two layers. 
This semi-classical approach relies only on
the coupling strength of the system, without considering interference
effects, and provides the closest partitioning to the one provided by the literature
(also relying only on the coupling strengths).
Meanwhile, the methods
based on the long-time transport and fidelity return very similar community
partitionings, in which node 6 on one layer and node 9 on the other are in
the same community.
These two long-time community partitionings are
identical, except one of the communities predicted by the fidelity based
method is split when using the transport based method. It is therefore a
difference in modularity only. 

The classical OSLOM algorithm fails spectacularly: it gives only one significant community
involving nodes 11 and 12 which exhibit the highest coupling strength. If
assigning a community to each node is forced, a unique community with all
nodes is provided.

\section{Appendix}\label{sec:comdet-appendix}

\subsection{Definitions}
\subsubsection{Modularity}
Assume we have a directed, weighted graph (with possibly negative
weights) and self-links, described by a real adjacency matrix~$A$.
The element~$A_{ij}$ is the weight of the link from node~$i$ to
node~$j$.

The in- and outdegrees of node~$i$ are defined as
\begin{equation}
  k^{\text{in}}_i = \sum_j A_{ji}, \qquad
  k^{\text{out}}_i = \sum_j A_{ij}.
\end{equation}
For a symmetric graph~$A$ is symmetric and the indegree is equal to the
outdegree.
The total connection weight is
$m = \sum_i k^{\text{in}}_i = \sum_i k^{\text{out}}_i = \sum_{ij} A_{ij}$.

The community matrix~$C$ defines the membership of the nodes in
different communities. The element $C_{i \AAA}$ is equal to unity if $i \in
\AAA$, otherwise zero.%
\footnote{For a fuzzy definition of membership we could
require $C_{i \AAA} \geq 0$ and $\sum_\AAA C_{i \AAA} = 1$ instead.}
The size of a community is given by $|\AAA| = \sum_i C_{i \AAA}$.
For strict (non-fuzzy) communities we can define $C$ using
an assignment vector~$\sigma$ (the entries being the communities of
each node): $C_{i \AAA} = \delta_{\AAA, \sigma_i}$. This yields
$(C C^T)_{ij} = \delta_{\sigma_i, \sigma_j}$.

There are many different ways of partitioning a graph into communities.
A simple approach is to minimize the \emph{frustration} of the partition,
defined as the sum of the absolute weight of positive links between communities and negative links within
them:
\begin{equation}
F = -\sum_{ij} A_{ij} \delta_{\sigma_i, \sigma_j} = -\tr\left(C^T A C\right).
\end{equation}
Frustration is inadequate as a goodness measure for partitioning nonnegative graphs
(in which a single community containing all the nodes minimizes it).
For nonnegative graphs we can instead maximize another measure called \emph{modularity}:
\begin{equation}
Q = \frac{1}{m}\sum_{\AAA, ij} (A_{ij}-p_{ij}) C_{i \AAA} C_{j \AAA}
= \frac{1}{m}\tr\left(C^T (A-p) C\right),
\end{equation}
where $p_{ij}$ is the ``expected'' link weight from $i$ to~$j$, with~$\sum_{ij} p_{ij} = m$,
and is what separates modularity from plain frustration.
Different choices of the ``null model''~$p$ give different modularities.
Using degrees, we can define $p_{ij} = k^{\text{out}}_i k^{\text{in}}_j / m$.

For graphs with both positive and negative weights the usual definitions of degrees do not make much sense,
since usually negative and positive links should not simply cancel each other out.
Also, plain modularity will fail e.g.\ when~$m=0$.
This can be solved by treating positive and negative links separately~\cite{traag2009}.

\subsubsection{Hierarchical clustering}

All our community detection approaches share a common theme.
For each (proposed) community~$\AAA$
we have a goodness measure~$M_\AAA(t)$ that depends on the system Hamiltonian, the initial state, and~$t$.
This induces a corresponding measure for a partition~$\CC$:
\begin{align}
M_{\CC}(t) = \sum_{\AAA \in \CC} M_\AAA(t).
\end{align}
Using this, we define a function for comparing two partitions, $\CC$
and $\CC'$, which only differ in a single merge that combines
$\AAA$ and~$\BBB$:
\begin{align}
M_{\AAA, \BBB}(t)
= M_{\CC'}(t)-M_{\CC}(t)
= M_{\AAA \cup \BBB}(t) -M_\AAA(t)-M_\BBB(t).
\end{align}
We can make $M_{\AAA,\BBB}(t)$ into a symmetric closeness measure $c(\AAA,\BBB)$
by fixing the time~$t$ and normalizing it with~$|\AAA||\BBB|$.
Using this closeness measure together with the agglomerative
hierarchical clustering algorithm (as explained
in Sec.~\ref{sec:comdet}) we then obtain a community hierarchy.
The goodness of a specific partition in the hierarchy is given by its
modularity, obtained using the adjacency matrix given by
$A_{ij} = c(i,j)$.

The standard hierarchical clustering algorithm requires closeness to fulfill the \emph{monotonicity property}
\begin{align}
\label{eq:max}
\min(c(\AAA,\CCC), c(\BBB,\CCC)) \le c(\AAA \cup \BBB, \CCC) \le \max(c(\AAA,\CCC), c(\BBB,\CCC)).
\end{align}
for any communities~$\AAA, \BBB, \CCC$.
If this does not hold, we may encounter a situation where the
merging closeness sometimes increases, which in turn means that the
results cannot be presented as a dendrogram indexed by decreasing closeness.
The real downside of not having the monotonicity property, however, is stability-related.
The clustering algorithm should be stable, i.e. a small change in the system
should not dramatically change the resulting hierarchy.
Assume we encounter a situation where all the pairwise closenesses between a
subset of clusters $S = \{\AAA_i\}_i$
are within a given tolerance. A small perturbation can now change the
pair $\{\AAA,\BBB\}$ chosen for the merge.
If Eq.~\eqref{eq:max} is fulfilled, then 
the rest of $S$ is merged into the same new cluster during subsequent
rounds, and hence their relative merging order does not matter.

\subsubsection{Notation}

Let the Hamiltonian of the system have the spectral decomposition
$H = \sum_k \E_k \HPr_k$.
The unitary propagator of the system decomposes as
$U(t) = \mathrm{e}^{-\ii H t} = \sum_k e^{-i \E_k t} \HPr_k$.
We denote the state of the system at time~$t$ by
\begin{align}
\rho (t) =  U(t) \rho(0) U(t)^\dagger.
\end{align}
Sometimes we make use of the state obtained
by measuring in which community subspace $\VV_\AAA$ the quantum
state is located, and then discarding the result. The resulting state is
\begin{align}
\label{eq:measure}
\rho_\CC(t) &= \sum_{\AAA \in \CC} \SPr_\AAA \rho(t) \SPr_\AAA.
\end{align}
This state is normally not pure even if~$\rho(t)$ is.

The probability of transport from node~$b$ to node~$a$, the transfer matrix,
is given by the elements
\begin{align}
\TM_{ab}(t) = |\brackets{a}{U(t)}{b}|^2.
\end{align}
$\TM(t)$ is doubly stochastic, i.e.\ its rows and columns all sum up to unity.
We use $\sym{\TM} = (\TM+\TM^T)/2$ to denote its symmetrization.

The time average of a function $f(t)$ is denoted using~$\tave{f}(t)$:
\begin{align}
\tave{f}(t) = \frac{1}{t} \int_0^t f(t') \: \dd t'.
\end{align}
Now we have
\begin{align}
\tave{\TM}_{ab}(t)
&= \sum_{jk} \frac{1}{t} \int_0^t e^{-i(\E_j-\E_k)t'} \: \dd t' \brackets{a}{\HPr_j}{b} \brackets{b}{\HPr_k}{a}.
\end{align}
The $tH \ll 1$ and $t \to \infty$ limits of this average are
\begin{align}
\label{eq:T_ave_lims}
\notag
\tave{\TM}_{ab}(t \to 0) &= \delta_{ab}\left(1-\frac{t^2}{3}(H^2)_{aa}\right) +\frac{t^2}{3}|H_{ab}|^2 +O(t^3),\\
\tave{\TM}_{ab}(t \to \infty)
&= \sum_{jk} \delta_{jk} \brackets{a}{\HPr_j}{b} \brackets{b}{\HPr_k}{a}
= \sum_{k} |\brackets{a}{\HPr_k}{b}|^2.
\end{align}

The time average of the state of the system is given by
\begin{align}
\tave{\rho}(t)
= \sum_{jk} \frac{1}{t} \int_0^t e^{-i(\E_j-\E_k)t'} \: \dd t' \HPr_j \rho(0) \HPr_k.
\end{align}
It can be interpreted as the density matrix of a system that has
evolved for a random time, sampled from the uniform distribution on
the interval~$[0,t]$.
Again, in the short- and infinite-time limits this yields
\begin{align}
\label{eq:rho_ave_lims}
\notag
\tave{\rho}(t \to 0)
=& \rho(0) -\frac{it}{2} \left[H, \rho(0)\right]
+\frac{t^2}{3}\left(H \rho(0) H -\frac{1}{2}\left\{H^2, \rho(0)\right\} \right)
+O(t^3),\\
\tave{\rho}(t \to \infty)
=& \sum_{k} \HPr_k \rho(0) \HPr_k.
\end{align}

\subsection{Closeness measures}
\subsubsection{Inter-community transport}
\label{sec:S:mixing}

Considering the flow of probability during a continuous-time quantum
walk, let us investigate the \emph{change} in the probability of observing
the walker within a community:
\begin{align}
\T_\AAA (t)
= \frac{1}{2}\left| p_\AAA \left \{ \rho (t) \right \} - p_\AAA \left \{ \rho (0) \right \} \right|,
\end{align}
where
$p_\AAA \left \{ \rho \right \} = \tr \left( \SPr_\AAA \rho \right)$
is the probability of a walker in state~$\rho$ being found in
community~$\AAA$ upon a von Neumann-type measurement.\footnote{
Equivalently, $p_\AAA \left \{ \rho \right \}$ is
the norm of the projection (performed by projector $\SPr_\AAA$) of the
state $\rho$ onto the community subspace $\VV_\AAA$.}
A good partition should intuitively minimize this change, keeping the walkers as localized to the communities as possible.
$\T_\CC= \sum_{\AAA \in \CC}\T_{\AAA}$ is of course minimized by the trivial choice of a single
community, $\CC = \{\AAA\}$, and any merging of communities can only decrease~$\T_{\CC}$.
Therefore we have
$\T_{\AAA \cup \BBB}(t) \le \T_{\AAA}(t) +\T_{\BBB}(t)$.

The initial state~$\rho (0)$ can be chosen freely.
For a pure initial state $\rho(0) = \ket{\psi}\bra{\psi}$ we obtain
\begin{align}
\T_\AAA (t) = \frac{1}{2} \left| \bracket{\psi}{U^\dagger(t) \SPr_\AAA U(t)}{\psi} -\bracket{\psi}{\SPr_\AAA}{\psi}  \right|.
\end{align}
The change in inter-community transport is clearest when the process begins either entirely inside or entirely outside each community. Because of this, we choose the walker to be initially localized
at a single node $\rho (0) = \proj{b}$ and then, for symmetry, sum (or average)
$\T_\AAA (t)$ over all $b \in \NNN$:
\begin{align}
\label{eq:T}
\notag
\T_\AAA (t) &= \frac{1}{2} \sum_b  \left| \bracket{b}{U(t)^\dagger \SPr_\AAA U(t)}{b} -\bracket{b}{\SPr_\AAA}{b}  \right|\\
\notag
&= \frac{1}{2} \sum_b \left|\sum_{a \in \AAA} (\TM_{ab}(t) -\delta_{ab}) \right|\\
\notag
&= \frac{1}{2} \left(\sum_{b \in \AAA} \left|1 -\sum_{a \in \AAA} \TM_{ab}(t) \right| 
  +\sum_{b \notin \AAA} \left|\sum_{a \in \AAA} \TM_{ab}(t) \right| \right)\\
\notag
&= \frac{1}{2} \left(\sum_{a \notin \AAA,b \in \AAA} \TM_{ab}(t) 
  +\sum_{a \in \AAA, b \notin \AAA} \TM_{ab}(t)\right)\\
&= \sum_{a \in \AAA, b \notin \AAA} \frac{\TM_{ab}(t)+\TM_{ba}(t)}{2}
= \sum_{a \in \AAA, b \notin \AAA} \sym{\TM}_{ab}(t),
\end{align}
since $\TM(t)$ is doubly stochastic.
Now we have
\begin{align}
\T_{\AAA,\BBB}(t) = \T_{\AAA}(t) +\T_{\BBB}(t) -\T_{\AAA \cup \BBB}(t)
= 2 \sum_{a \in \AAA, b \in \BBB} \sym{\TM}_{ab}(t)
\end{align}
with $0 \le \T_{\AAA,\BBB}(t) \le 2 \min(|\AAA|, |\BBB|)$.
The short- and long-time limits of the time-averaged $\T_{\AAA,\BBB}(t)$
can be found using Eqs.~\eqref{eq:T_ave_lims}:
\begin{align}
\T_{\AAA, \BBB}^{t \to 0}
&= 2 \sum_{a \in \AAA, b \in \BBB}
\left( \delta_{ab} +\frac{t^2}{3}\left(|H_{ab}|^2 -\delta_{ab}(H^2)_{aa}\right) +O(t^3)\right),\\
\T_{\AAA, \BBB}^{t \to \infty}
&= 2 \sum_{a \in \AAA, b \in \BBB} \sum_k |(\HPr_k)_{ab}|^2.
\end{align}

\subsubsection{Intra-community fidelity}
\label{sec:S:coher}
Our next measure aims to maximize the ``similarity'' between the
evolved and initial states when projected to a community subspace.
We do this using the squared fidelity
\begin{align}
\F_\AAA (t) = F^2 \left \{ \SPr_\AAA \rho(t) \SPr_\AAA, \SPr_\AAA \rho(0) \SPr_\AAA \right \},
\end{align}
where $\SPr_\AAA \rho \SPr_\AAA$ is the projection of the state $\rho$ onto the subspace $\VV_\AAA$ and
\begin{align}
F \left \{ \rho , \sigma \right \} = \tr \left \{ \sqrt{ \sqrt{\rho} \sigma \sqrt{\rho} } \right \} \in  [0, \sqrt{\tr \{ \rho \} \tr \{ \sigma \}} ],
\end{align}
is the fidelity, which is symmetric between $\rho$ and $\sigma$.
If either $\rho$ or $\sigma$ is rank-1, their fidelity reduces to
$
F \left \{ \rho, \sigma \right \} = \sqrt{\tr \{\rho \sigma\}}
$.
Thus, if the initial state~$\rho(0)$ is pure, we have
\begin{align}
\F_\AAA (t) = \tr \left( \SPr_\AAA \rho(t) \SPr_\AAA \rho(0) \right).
\end{align}
This assumption makes
$\F_\CC(t)$ equivalent to the squared fidelity between $\rho_\CC(t)$ and a pure~$\rho(0)$:
\begin{align}
\label{eq:FX}
\notag
\F_\CC(t)
&= \sum_{\AAA \in \CC} \tr\left(\SPr_\AAA \rho(t) \SPr_\AAA \rho(0)\right)
= \tr\left(\rho_\CC(t) \rho(0)\right)\\
&= F^2\{\rho_\CC(t), \rho(0)\}
= F^2\{\rho(t), \rho_\CC(0)\},
\end{align}
and yields
\begin{align}
\label{eq:f2}
\notag
\F_{\AAA,\BBB}(t)
&= \F_{\AAA \cup \BBB}(t)-\F_\AAA(t)-\F_\BBB(t)\\
\notag
&= 2 \real \tr \left(\SPr_\AAA \rho(t) \SPr_\BBB \rho(0) \right)\\
&= 2 \sum_{a \in \AAA, b \in \BBB} \real \left(\rho_{ab}(t) \rho_{ba}(0) \right).
\end{align}
We use as the initial state the uniform superposition of all the basis states with arbitrary phases:
$\ket{\psi} = \frac{1}{\sqrt{n}}\sum_k e^{i \theta_k} \ket{k}$, which gives
\begin{align}
\F_{\AAA,\BBB}(t)
&= 
\frac{2}{n^2} \sum_{a \in \AAA, b \in \BBB} \sum_{xy}
\real\left(e^{i(\theta_x -\theta_y +\theta_b -\theta_a)} U_{ax} \overline{U_{by}}\right).
\end{align}
In this case the short-term limit does not yield anything interesting.
The long-time limit of the time-average of $\F_{\AAA,\BBB}(t)$ is
\begin{align}
\notag
\F_{\AAA, \BBB}^{t \to \infty}
&= \frac{2}{n^2} \sum_{a \in \AAA, b \in \BBB} \sum_{xy,k} \real \left(e^{i(\theta_x -\theta_y +\theta_b -\theta_a)}(\HPr_k)_{ax} (\HPr_k)_{yb} \right).
\end{align}
We may now (somewhat arbitrarily) choose all the phases~$\theta_k$ to be the same,
or average the closeness measure over all possible phases~$\theta_k \in [0, 2\pi]$.

\subsubsection{Purity}
The coherence between any communities $\CC = \{ \AAA,\BBB,\dots \}$ is completely destroyed
by measuring
in which community subspace $\VV_\AAA$ the quantum
state is located, see Eq.~\eqref{eq:measure}. If the measurement outcome is not revealed, the
purity of the measured state~$\rho_\CC(t)$ is,
due to the orthogonality of the projectors,
\begin{align}
\notag
\P_X(t) &= \tr\left(\rho_\CC^2(t)\right)
= \sum_{\AAA \in \CC} \tr\left((\SPr_\AAA \rho(t))^2\right)
= \sum_{\AAA \in \CC} \P_\AAA(t),
\end{align}
where
\begin{align}
\P_\AAA(t) &= \tr\left((\SPr_\AAA \rho(t) \SPr_\AAA)^2\right) = \tr\left((\SPr_\AAA \rho(t))^2\right).
\end{align}
If $\rho(t)$ is pure, we have (cf. Eq.~\eqref{eq:FX})
\begin{align}
\P_\CC(t)
= \sum_{\AAA \in \CC} \tr(\SPr_\AAA \rho(t) \SPr_\AAA \rho(t))
= F^2\{\rho_\CC(t), \rho(t)\}.
\end{align}

The change in purity of the state after a projective measurement
locating the walker into one of the communities is
\begin{align}
\notag
\P_{\AAA, \BBB}(t) &= \P_{\AAA \cup \BBB}(t) -\P_\AAA(t)-\P_\BBB(t)\\
\notag
&= 2\tr\left(\SPr_\AAA \rho(t) \SPr_\BBB \rho(t)\right)\\
&= 2 \sum_{a \in \AAA, b \in \BBB} | \rho_{ab} (t) |^2 \ge 0.
\end{align}

Again, we will use the initial state 
$\ket{\psi}~=~\frac{1}{\sqrt{n}}\sum_k e^{i \theta_k} \ket{k}$:
\begin{align}
\P_{\AAA,\BBB}(t)
&= 
\frac{2}{n^2} \sum_{a \in \AAA, b \in \BBB}
\left|\sum_{xy} e^{i(\theta_x-\theta_y)} U_{ax}(t) \overline{U_{by}(t)}\right|^2.
\end{align}
As with the fidelity-based measure, the short-time limit is uninteresting.
The long-time limit of the time-average of $\P_{\AAA,\BBB}(t)$ is
\begin{align}
\notag
\P_{\AAA,\BBB}^{t \to \infty}
&= 2 \sum_{a \in \AAA, b \in \BBB}
\left(|\bracket{a}{\tave{\rho}(\infty)}{b}|^2 +\sum_{k \neq m}
|\bracket{a}{\HPr_k \rho_0\HPr_m}{b}|^2\right)\\
&=
2 \sum_{a \in \AAA, b \in \BBB} \left( 
|\sum_{kxy} e^{i(\theta_x -\theta_y)} (\HPr_k)_{ax} (\HPr_k)_{yb}|^2\right.\\
&\left.+\sum_{k \neq m}|\sum_{xy} e^{i(\theta_x -\theta_y)} (\HPr_k)_{ax} (\HPr_m)_{yb}|^2
\right).
\end{align}



%% file: conclusion.tex
\chapter{Conclusion}
\label{ch:conclusion}

This PhD thesis is devoted to three threads:
\begin{itemize}
\item \textbf{\nameref{ch:invariants}},\\
about relations between permutation symmetry of a state and its other properties related to quantum information. The main focus was on local unitary equivalence of states, transformations achievable with linear optics and polynomial invariants.

\item \textbf{\nameref{ch:qubsim}},\\
about a plotting scheme for many-body states, \emph{qubism}.
This tool allows to show entanglement and phase transitions, as well as discover other symmetries of a pure state.

\item \textbf{\nameref{ch:networks}},\\
about using complex network approach to study quantum systems, with the special emphasis on community detection.
We use it to asses the range of quantum effects in a biochemical system.
\end{itemize}

Each of these topics give raise to further questions and lines of investigation.
However, this combination of topics is a source of creativity and open paths to further developments, particularly:
\begin{itemize}
\item Geometric representations for mixed states.\\
Majorana representation for symmetric qubit states serves both as a visualization and a mathematical isomorphism giving rigorous insight into properties of the state.
A variant for mixed states would be beneficial. 

\item General visualization schemes putting emphasis on symmetries of a given state.\\
We plotted amplitudes, which represent all knowledge about the state, but also are susceptible to ``unimportant'' changes (e.g. local basis).
Directly showing symmetries of a state, whether rigorous or approximate, may be fruitful.

\item Special visualizations for symmetric and antisymmetric states.\\
Qubism representation, while can be used for any state, focuses on translationally invariant states.
It is likely that there are plotting schemes that are more suitable for states with different symmetries. 

\item Relation of quantum community detection to other hierarchical schemes.\\
Splitting a system into subsystem that are weakly correlated is the key principle standing behind matrix product states (MPS) and projected entangled pair states (PEPS).
There are analogies between these techniques and quantum community detection, which are worth pursuing.

\item Community detection methods for many-body systems.
We performed splitting of a one-particle systems into subsystem not sharing coherent quantum superposition.
A natural extension would be to work on multiparticle systems and, instead of coherence, work on entanglement.

\end{itemize}